\documentclass[11pt]{article}
\usepackage{jheppubmod}
\pdfoutput=1
\usepackage{mathtools}
\mathtoolsset{showonlyrefs}
\usepackage{psfrag}
\usepackage{array}
\usepackage{amssymb}
\usepackage{amsmath}
\usepackage{amsthm}
\usepackage{graphicx}
\usepackage[font={small,it}]{caption}
\usepackage[labelsep=quad]{subcaption}
\usepackage[punctsep]{collref}
\collectsep[]{;}

\newcommand{\psig}{\Psi} 
\newcommand{\psisup}{\Psi_s} 
\newcommand{\psient}{\Psi_{\text{en}}} 
\newcommand{\psigen}{\Psi_{\text{gen}}} 
\newcommand{\psisupent}{\psisup} 
\newcommand{\psiqub}{\Psi_{\text{qub}}} 
\newcommand{\psimd}{\Psi_{\text{md}}} 
\newcommand{\psiexc}{\Psi_{\text{ex}}} 
\newcommand{\psidual}[1]{\Psi_{#1}}  
\newcommand{\psidualexc}[1]{\Psi_{#1}^{\text{ex}}} 
\newcommand{\psine}{\Psi^{\text{ne}}} 
\newcommand{\psican}{\Psi_{\text{can}}}
\newcommand{\psimicro}{\Psi_{\text{mic}}}
\def\psinesub[#1]{\psine_{#1}}
\def\psisupne{\psinesub[s]}
\newcommand{\tfd}{\Psi_{\text{tfd}}} 
\newcommand{\tfdk}{|\Psi_{\text{tfd}}\rangle} 
\newcommand{\tfdbra}{\langle \Psi_{\text{tfd}}|}
\newcommand{\tfdT}{\Psi_{\text{T}}} 
\newcommand{\tfdvar}[1]{\Psi_{#1}}
\newcommand{\tfdmT}{|\Psi_{\text{-T}} \rangle} 

\newcommand{\cop}[1]{\boldsymbol{#1}}
\newcommand{\al}{\cop{A}} 
\newcommand{\alset}{{\cal A}} 
\newcommand{\alsetgff}{{\cal A}_{\text{gff}}}
\newcommand{\alseth}{{\cal A}_{\hcft}}
\newcommand{\alsetent}{{\cal A}_{\text{product}}} 
\newcommand{\op}{\cop{\cal O}} 
\newcommand{\hcft}{\cop{H}} 
\newcommand{\hleft}{\cop{H}_L}
\def\ulgen{\cop{U}_{L,g}}
\def\comm{C}
\def\tal{\widetilde{\al}} 
\def\Oright[#1]{\op_{#1}} 
\def\Oleft[#1]{\op_{L#1}} 
\def\anc{\cop{a}}
\def\tashm{\widetilde{a}}
\def\an[#1]{\anc_{#1}} 
\def\aleft[#1]{\cop{a}_{L#1}} 
\def\arelr[#1]{\cop{a}^{\text{rel}}_{R#1}} 
\def\phirelr{\phi} 

\def\ta{\widetilde{\cop{a}}}
\def\paul{\sigma}
\def\pault{\widetilde{\sigma}}

\def\tN{\widetilde{N}}
\def\alen[#1]{\al_{\text{L}, #1}} 
\def\timerev{{\cal T}}
\def\hmodr{\cop{H}^R_{\text{mod}}}
\def\enrange[#1,#2]{{\cal R}_{#1}}
\def\dimrange[#1,#2]{{\cal D}_{#1}}
\def\hilb[#1]{{\cal H}_{#1}}
\newcommand{\projrangenew}[2]{\cop{P}_{#1}}
\newcommand{\dimrangenew}[2]{{\cal D}_{#1}}
\def\projrange[#1, #2]{\cop{P}_{#1}}
\def\projh[#1]{\cop{P}_{\hilb[#1]}}
\def\projf{\cop{P}_F}
\def\meas{\mu}
\def\microen{{\cal R}_{E}}
\def\dimicro{{\cal D}_{E}}

\def\numinus{{\cal R}_{-}}

\def\diminus{{\cal D}_{-}}

\def\projminus{\cop{P}_{-}}
\def\hinf{{\cop{H}_{0}}}
\def\nb{\cop{N}_{\omega_{n}}}
\def\capt{T}
\def\ang{m}	
\def\lt{t_L}
\def\rt{t_R}
\def\rhor{r_0}
\def\normdelta{\kappa_{N}}
\def\eigens{O}
\def\dima{{\cal D}_{\alset}}
\def\diff{\xi}
\def\distphase{\upsilon}
\def\tarelr[#1]{\widetilde{a}^{\text{rel}}_{R#1}}
\def\coeff[#1]{\alpha_{#1}}

\def\lvac{\langle 0 |}
\def\rvac{|0 \rangle}

\def\pb[#1,#2]{\{#1, #2\}}
\def\deb[#1,#2]{[#1,#2]_{\text{D.B.}}}
\def\tO{\widetilde{\op}}
\def\ohar{\widetilde{\cop{X}}^H}

\def\cn{{\mathcal N}}
\def\tr{{\rm Tr}}

\def\Or[#1]{{\text{O}}\left({#1}\right)}
\def\dotl[#1,#2]{\left\langle #1,\, #2 \right\rangle}
\def\dotlb[#1,#2]{\left\langle #1,\, #2 \right\rangle}
\def\dotlm[#1,#2]{\left[ #1,\, #2 \right]}
\def\dotp[#1,#2]{(\vect{#1} \cdot\vect{#2})}
\def\aff[#1,#2]{\hat{#1}(#2)}
\def\nc{{\cal N}}
\def\n4sym{{\cal N}=4 SYM}
\def\>{\rangle}
\def\<{\langle}
\def\weight[#1,#2,#3]{\{(#1),#2,#3\}}
\def\ads[#1]{$\text{AdS}_{#1}$}

\def\rtor{r_*}
\def\rtors[#1]{r_{*#1}}
\def\gads{g_{\mu \nu}^{\text{ads}}}
\def\tband{T_{\text{b}}}
\def\capt{T}
\def\cutoffT{T_{\text{cut}}}
\def\intT{T_i}

\newcommand{\be}{\begin{equation}}
\newcommand{\ee}{\end{equation}}
\newcommand{\ba}{\begin{align}}
\newcommand{\ea}{\end{align}}
\newcommand{\bs}{\begin{split}}

\def\sess\end{split}
\newcommand{\vect}[1]{{\vec{#1}}}
\newcommand{\vectb}[1]{{\vect{#1}^b}}

\def\DDelta{c}
\keywords{AdS-CFT, Information Paradox, Black Holes}
\title{Comments on the Necessity and Implications of State-Dependence in the Black Hole Interior}
\author[a,b]{Kyriakos Papadodimas}
\author[c,d]{and Suvrat Raju}
\affiliation[a]{Theory Group, Physics Department, CERN, CH-1211 Geneva 23,
Switzerland.}
\affiliation[b]{Centre for Theoretical Physics, University of Groningen, Nijenborgh 4, 9747 AG, The Netherlands.}
\affiliation[c]{International Centre for Theoretical Sciences, Tata Institute of Fundamental Research, IISc Campus, Bengaluru 560012, India.}
\affiliation[d]{Center for Mathematical Sciences and Applications, Harvard University, 1 Oxford Street, Cambridge, MA 02138, USA.}
\emailAdd{kyriakos.papadodimas@cern.ch}
\emailAdd{suvrat@icts.res.in}
\abstract{
We revisit the ``state-dependence'' of the map that we proposed recently between bulk operators in the
interior of a large AdS black hole and operators in the boundary CFT.  By refining recent versions of the information paradox, we show that 
this feature is necessary for the CFT to successfully describe local physics  behind the horizon  --- not only for single-sided black holes but even in the eternal black hole.  We show that state-dependence is invisible to an infalling observer who cannot differentiate these operators from
those of ordinary quantum effective field theory.  Therefore the infalling observer does not observe any violations of quantum mechanics. We successfully resolve a large class of potential ambiguities in our construction. We analyze states  where the CFT is entangled with another system and show that the ER=EPR conjecture emerges from our construction in a natural and precise form.  We comment on the possible semi-classical origins of state-dependence.}
\begin{document}
\maketitle

\section{Introduction}
Recent work by Mathur \cite{Mathur:2009hf}, Almheiri et al. \cite{Almheiri:2012rt,Almheiri:2013hfa} and then by Marolf and Polchinski \cite{Marolf:2013dba} has 
sharpened the information paradox \cite{Hawking:1974sw,Hawking:1976ra} and highlighted some of the difficulties in analyzing questions about local bulk physics in the AdS/CFT correspondence. Put briefly, these authors argued that the CFT does not contain operators with the right properties to play the role of local field operators behind the black hole horizon. Their arguments were phrased in terms of various paradoxes, and they
interpreted these apparent contradictions to mean 
that generic high energy states in the CFT do not have
a smooth interior; and even if they do, the CFT cannot describe it meaningfully.  

If correct, this conclusion would be a striking violation of effective field theory. A semi-classical analysis performed by quantizing fluctuations about the classical black hole solution would suggest that for a large black hole, quantum effects detectable within effective field theory are confined to the neighbourhood of the singularity. However, the papers above suggested that the range of quantum effects, visible to a low energy observer, may spread out all the way to the horizon.

In previous work \cite{Papadodimas:2012aq,Papadodimas:2013wnh,Papadodimas:2013jku,Papadodimas:2013kwa}, we analyzed these arguments in detail. We found that they made two tacit assumptions. The first, which was important
for the strong subadditivity paradox of Mathur \cite{Mathur:2009hf} and the first paper of Almheiri et al.\cite{Almheiri:2012rt}, was that locality holds exactly in quantum gravity. We showed how a precise version of black hole
complementarity, where the commutator of operators outside and inside the black 
hole vanishes within low point correlators but is not exactly zero as an operator, allowed one to resolve this paradox. We will review this resolution briefly below.

However, in \cite{Almheiri:2013hfa} Almheiri et al. argued that even large black holes in AdS should contain firewalls.  To make this argument they had to make a second tacit assumption, which was that
local bulk observables like the metric are represented by fixed linear operators in the CFT. More precisely, this is the idea that  even in two different states one may use the same CFT operator to represent the metric at a ``given point''.

By identifying and discarding this assumption in \cite{Papadodimas:2013wnh,Papadodimas:2013jku}, we were able to resolve all the paradoxes alluded to above. Furthermore, we were able to explicitly identify
 CFT observables that were dual to local correlation functions in the 
 black hole interior. This construction allowed us to probe the geometry of the horizon
and show that the horizon was smooth --- as predicted by effective field
theory, and in contradiction with the firewall and fuzzball proposals.

The operators in our construction are state-dependent. This means that they act  correctly about a given state,
and in excitations produced on that state by performing low energy experiments. If one moves far in the Hilbert space --- even just by changing the microscopic and not the macroscopic degrees of freedom --- then one has to use a different operator to represent the ``same'' local degrees of freedom. 

Our resolution to the firewall paradox has encountered two kinds of objections. A technical point is that our construction relies on a notion of equilibrium. It was
first noticed by van Raamsdonk \cite{VanRaamsdonk:2013sza} that our equilibrium conditions were necessary but not sufficient; Harlow \cite{Harlow:2014yoa} later elaborated on this point. This leads to a potential ``ambiguity'' in our construction where, at times, we cannot definitively identify the right operators in the black hole interior. 

The second is more fundamental. Is it acceptable at all, within quantum mechanics, to use state-dependent bulk to boundary maps so that the metric at a ``given point'' in space may be represented by different operators in different microstates and backgrounds?
\vskip10pt
This is the context for our paper. In this work we  make the following advances.
\begin{enumerate}
\item
In section \ref{secparadoxes}, we revisit and sharpen the arguments of Almheiri et al. \cite{Almheiri:2013hfa}. We believe that this strongly suggests that
there is no alternative to firewalls except for a state-dependent construction of the black hole interior.
In fact, we show in section \ref{seceternal} that the paradoxes of \cite{Almheiri:2013hfa} also arise for the eternal black hole.  We show that it is necessary to use state-dependent operators, which we construct explicitly, to rule out a scenario where even the eternal black hole does not have a smooth interior.
\item
In section \ref{secremoveamb}, we resolve a large class of ambiguities in our construction by refining our notion of an equilibrium state, including all of those pointed out by van Raamsdonk \cite{VanRaamsdonk:2013sza}.  We point out  difficulties with Harlow's analysis that invalidate the attempt made in \cite{Harlow:2014yoa} to accentuate these ambiguities.
\item
We show how our analysis extends naturally to superpositions of states in section \ref{secdefnmirror}. We reiterate and expand on the point, already made in \cite{Papadodimas:2013jku,Papadodimas:2013wnh} that the infalling observer does not observe any violations of quantum mechanics or the ``Born rule''.
\item
In section \ref{secentangled}, we show how our construction extends naturally to entangled systems. This leads to a new and interesting outcome: a precise version of the ER=EPR conjecture \cite{Maldacena:2013xja} emerges automatically from our analysis. In particular our construction shows --- without any additional assumptions --- why one should expect a geometric wormhole in the thermofield double state, and a somewhat ``elongated'' wormhole in states with less entanglement. Our analysis also shows why there is no geometric wormhole in a generic entangled state of two CFTs, or when the CFT is entangled with a system of a few qubits.\footnote{We limit our assertions to wormholes that can be probed geometrically using effective field theory.  Therefore we do not have any comment on the strong form of the ER=EPR conjecture, which posits that any entanglement should be accompanied by a wormhole.}
\end{enumerate}

We also initiate an investigation into the semi-classical origins of state-dependence in Appendix \ref{appcoherentgravity}. We show that local observables
like the metric are well defined classical functions on the phase space of canonical gravity. Ordinarily such functions would lift to state-independent operators in the quantum theory. However, our analysis of state-dependence in the eternal black hole suggests an interesting obstacle to this map: the inner product between states in the CFT representing different geometries does not die off as fast as a naive analysis of coherent states in canonical gravity would suggest. Instead it saturates at a nonperturbatively small but finite value. We present some evidence that it is
this overcompleteness that prevents the existence of state-independent operators behind the horizon.\footnote{A similar idea was earlier suggested by Motl \cite{luboscoherentblog}.}

Apart from the new results mentioned above, we also present some material
that we hope will help to clarify some conceptual issues and be of pedagogical utility. For example, in section \ref{secstatedepvsindep} we present a discussion of relational observables in AdS quantum gravity. This concept is important throughout this paper to understand the geometric properties of operators behind the horizon, but we believe that it
may be of broader significance. This idea has often been used in discussions of the subject (and was first described to us by Donald Marolf) but we attempt to present a pedagogical and precise definition here. 

We also present a derivation of the properties of operators behind the horizon from a pedagogically new perspective in section \ref{secsmoothint}. We consider the two point function of a massless scalar field propagating in the geometry. By using the properties of this two point function, when the two points are almost null to each other, we are able to derive the correct formula for the entanglement of modes behind and in front of the horizon. One concern about our previous analysis \cite{Papadodimas:2012aq} was that even though a black hole in a single CFT does not have a second asymptotic region, we had to appeal to the analogy with the thermofield double, to derive the properties of our operators behind the horizon. We now perform this derivation from a purely local calculation.

We believe that the results of this paper present compelling evidence in favour
of the claim that there are no firewalls in generic states, and also that
the map between bulk and boundary operators is state-dependent behind the horizon.

The recent literature on the information paradox is extensive \cite{Bousso:2012as,  Susskind:2012uw,  Mathur:2012jk, Chowdhury:2012vd, Susskind:2012rm, Bena:2012zi, Giveon:2012kp, Banks:2012nn, Ori:2012jx, Hossenfelder:2012mr, Hwang:2012nn, Avery:2012tf, Larjo:2012jt,   Rama:2012fm, Page:2012zc,  Saravani:2012is, Jacobson:2012gh, Susskind:2013tg,  Kim:2013fv, Park:2013rm, Hsu:2013cw, Giddings:2013kcj,  Lee:2013vga, Avery:2013exa,   Kang:2013wda, Chowdhury:2013tza,Page:2013mqa,  Axenides:2013iwa,  Gary:2013oja, Chowdhury:2013mka, delaFuente:2013nba,  Barbon:2013nta,  Lloyd:2013bza, Hsu:2013fra, Page:2013dx,Giddings:2013noa,Mathur:2013qda,  Mathur:2013gua, Balasubramanian:2014gla,Freivogel:2014dca,Akhoury:2013bia,Lashkari:2014pna,Barbon:2014rma}. In particular, Erik Verlinde and Herman Verlinde also reached the conclusion that state-dependence is required to construct the black hole interior from a different perspective \cite{Verlinde:2013vja,Verlinde:2012cy,Verlinde:2013uja}. We direct the reader to \cite{Verlinde:2013qya} for a discussion of the relation between our approach and theirs. The effects of the back-reaction of Hawking radiation were discussed in \cite{Nomura:2012sw,Brustein:2012jn,Brustein:2013ena,Brustein:2013xga,Brustein:2013qma,Silverstein:2014yza}, and Nomura et al. also presented another perspective in \cite{Nomura:2012cx,Nomura:2012ex,Nomura:2013gna}. For a precursor of the firewall paradox, see \cite{braunstein2009v1} and for approaches using complexity see \cite{Harlow:2013tf,Susskind:2013lpa}.


\section{Summary}
In this section, we briefly summarize the contents of various sections
and suggest different paths that could be taken through the paper.

\paragraph{Reconstructing the bulk and state-dependence \\ } 
Section \ref{secstatedepvsindep} is partly devoted to clarifying some 
conceptual issues related to bulk to boundary maps. We quickly review what 
it means for such a map to be state-independent or state-dependent. We also 
point out that all existing methods of extracting bulk physics from the 
boundary are state-dependent. Experts in the subject may wish to look only 
at \ref{subsecrelational} where we define the relational observables that 
we use in the rest of the paper and at \ref{secgeomfromentang} where 
we describe the state-dependence of prescriptions to relate geometric
quantities to entanglement.

\paragraph{Need for operators behind the horizon\\} Section \ref{secsmoothint} is largely devoted to a detailed derivation of the
fact that we require new modes
that can play the role of ``right moving'' excitations behind the horizon to describe the interior of a black hole.  We derive
the two point function of these modes with modes outside the horizon from a local 
calculation, thereby removing the need to make an analogy to the thermofield 
double state and also sidestepping the trans-Planckian issues in Hawking's original computation.  In this section, we also review
the standard construction of local operators outside the horizon.
 Experts may be interested in  \ref{secstateindoutside} where we describe a state-independent construction of local
operators outside the horizon in the mini-superspace approximation. 

\paragraph{Either state-dependence or firewalls\\} The objective of section \ref{secparadoxes} is to try and show that we must 
accept one of two possibilities: either the black hole interior is mapped to the CFT by a state-dependent map, or generic microstates have firewalls.  Our arguments here are extensions and refinements of the arguments presented in \cite{Almheiri:2013hfa,Marolf:2013dba}. In particular, we strengthen the argument of \cite{Marolf:2013dba} by bounding potential errors in that calculation.  We also rephrase the ``counting argument'' of \cite{Almheiri:2013hfa} entirely within the context of two-point correlation functions to remove potential loopholes. This section
can be skipped, at a first reading, by a reader who already accepts the validity of the arguments of \cite{Almheiri:2013hfa,Marolf:2013dba}. 

\paragraph{State-dependence for the eternal black hole\\} In section \ref{seceternal} we show that these versions of the information 
paradox also appear in the eternal black hole.  Therefore it is inconsistent to adopt the position that the eternal black hole in AdS has a smooth interior whereas the large single sided black hole does not.  We would urge
the reader to consult \cite{Papadodimas:2015xma} --- where a concise version
of these arguments has already appeared --- in conjunction with this section, which contains some additional details. Since there is substantial evidence that the interior of the eternal black hole is smooth, this provides strong support
for state-dependence behind the black hole horizon.

\paragraph{Definition of mirror operators; consistency with superposition principle\\} In section \ref{secdefnmirror}, we review the state-dependent construction
of the black hole interior that was first presented in \cite{Papadodimas:2013jku,Papadodimas:2013wnh}. Experts may be interested in \ref{secsmallsuperpositions} where we check the linearity of this map for superpositions of a small number of states. In \ref{secexpliciteternal} we construct the interior of the eternal black hole. This construction is of interest since it provides some insight into state-dependence
as arising from the ``fat tail'' of the inner product between different microstates of a black hole.

\paragraph{Detecting unitaries behind the horizon\\} In section \ref{secremoveamb}, we show how to remove some of the ambiguities
in our definition of equilibrium. This section will be of interest to experts. We point out that by using the CFT Hamiltonian, we can detect excitations
behind the horizon in states that we might otherwise have classified as being in equilibrium. We also point out, in some detail, that the effort made in \cite{Harlow:2014yoa} to sharpen this ambiguity by considering a new class of excitations is based on an erroneous analysis of local operators in the eternal 
black hole. While, for this reason, the analysis of \cite{Harlow:2014yoa} does not have direct physical significance, it does point to an interesting new class of excited states that we discuss in some detail. 

\paragraph{Entangled systems and relation to ER=EPR\\}
In section \ref{secentangled}, we extend our construction to account for
 cases where the CFT is entangled with another system. The equations that describe modes in the interior do not change at all. The only new element that need to introduce is that the ``little Hilbert space'' of excitations about a base state may get enlarged since we can also act with operators in the other system. 
Surprisingly we show that a precise version of the ER=EPR conjecture emerges
automatically from our analysis. We are able to show that when two entangled
CFTs are in the thermofield state the modes observed by the 
right infalling observer inside the black hole are the same as those
observed by the left observer outside. However,  when the CFTs are entangled
in a generic manner this is no longer true.

We also consider cases where the CFT is entangled with a small system --- say a collection of qubits. Our analysis of this setup, together with our verification
of linearity in section \ref{secdefnmirror} establishes that the infalling observer cannot detect any  departures from ordinary linear quantum mechanics.


\section{Generalities: State-dependent vs state-independent operators }\label{secstatedepvsindep}
Since this paper focuses on state-dependent bulk-boundary maps, it is useful to first clarify the meaning of state-dependence and, conversely, what we would require of a putative ``state-independent'' operator. Since this issue has been the cause of significant confusion --- some of which has arisen because of the use of imprecise terminology ---  we have tried to make this section as precise and detailed as possible.  

A brief summary of this section is as follows. We define state-dependence. We point out that state-dependent bulk-boundary maps are already common in the AdS/CFT literature. Finally we explain the origin of the naive expectation that the bulk and boundary are related in a state-independent manner, and also indicate why this intuition fails.

Apart from the pedagogical definitions, we also pay some attention to the techniques of extracting bulk physics using entanglement entropy. These are all state-dependent since entanglement entropy does not correspond to a linear operator on the boundary.  This includes, for example, the well known Ryu-Takayanagi relation \cite{Ryu:2006bv,Ryu:2006ef}  between the entanglement entropy of a region on the boundary and the corresponding area of an extremal surface in the bulk. As we will emphasize repeatedly in this paper, as a result of very robust statistical  properties of the Hilbert space of the CFT at large $\nc$,\footnote{In this paper we adopt notation that is consistent with \cite{Papadodimas:2013wnh,Papadodimas:2013jku}. So $\nc$ is proportional to the central charge of the CFT. In the commonly considered case of the maximally supersymmetric SU(N) Yang Mills theory, we would have $\nc \propto N^2$.} it is perfectly natural for such a state-dependent formula to emerge within effective field theory, and its use does not lead to any violation of quantum mechanics.

While the use of state-dependent operators may be common in AdS/CFT, 
from a broader viewpoint it is true that this is a rather special situation in physics. So it would be incorrect to go to the other extreme and dismiss
state-dependence as mundane or unremarkable.

In this section, we point out that
based on intuition from canonical gravity, one may have naively expected that the is some overarching linear operator in the CFT that includes, in various limits, all these state-dependent prescriptions. If one were to obtain gravity
through phase space quantization, then one may naively expect that many reasonable functions on the phase space of gravity --- such as the metric at a point ---  would lift to operators. We show why this naive intuition runs into difficulty in the context of AdS/CFT. We complete this analysis in greater detail in Appendix \ref{appcoherentgravity}. The semi-classical origins of state-dependence that we outline in this section and in the Appendix are, we believe, an important and interesting subject of study.

In this section, and later in the paper we will often speak of CFT operators that also have a dual geometric interpretation. To avoid confusion, we adopt the following notational convention.
\paragraph{Notation:} A CFT operator is denoted with a bold-symbol; for example an operator in the CFT corresponding to the bulk metric would be denoted by  $\cop{g}_{\mu \nu}$, as opposed to the value of the semi-classical metric for a geometry $g_{\mu \nu}$, which is written in ordinary font. 

\subsection{State-independent operators}
We consider an AdS/CFT duality, where we expect a number of ``effective fields'' to propagate in the bulk. One of these is the metric $\cop{g}_{\mu \nu}$ but in general there will be other fields, which can include scalars but also fields of higher spin. We will collectively denote these fields by $\cop{\phi}$. We then have the following definition. 

\paragraph{Definition of a state-independent bulk-boundary map \\}
We will say that there is a state-independent map between the bulk and the boundary if there exist CFT operators $\cop{g}_{\mu \nu}(\vect{x})$ and $\cop{\phi}(\vect{x})$ parameterized by $d+1$ real numbers, which we denote by $\vect{x}$, so that in {\em all CFT} states  that are expected to be dual to a semi-classical geometry, which we denote by $|\psig \rangle$, the CFT correlators involving both the metric and other light fields 
\be
\label{stateindcorrs}
C(\vect{x}_1, \ldots \vect{x}_{m+p}) = \langle \psig| \cop{g}_{\mu_1 \nu_1}(\vect{x}_{1}) \ldots \cop{g}_{\mu_m \nu_m}(\vect{x}_m) \cop{\phi}(\vect{x}_{m+1}) \ldots \cop{\phi}(\vect{x}_{m+p}) |\psig\rangle,
\ee
have the right properties to be interpreted as ``effective field theory correlators''. 

This definition has many parts that we unpack below, where we explain what it means for a state to be dual to a semi-classical geometry, and what one expects from effective field theory. 

An immediate issue --- but one that does not have significant physical ramifications --- is that the bulk theory has diffeomorphism invariance. The $d+1$ real numbers above play the role of coordinates in the bulk. Given any valid diffeomorphism, $\vect{x} \rightarrow \diff(\vect{x})$, the {\em distinct} CFT operators  $\cop{\phi}(\diff^{-1}(\vect{x}))$ give an equally valid bulk to boundary map. So we must always discuss equivalence classes of bulk-boundary maps. Maps that are related by diffeomorphisms belong to the same equivalence class. Later in this section, we also describe various physical choices of gauge that help to remove this redundancy, and pick a preferred element of the equivalence class. We now turn to other aspects of the definition above.

\paragraph{Semi-classical States \\}

We now explain what we mean by semi-classical states in the definition above. In the AdS/CFT duality, we often identify certain states with dual bulk geometries.  These maps have been developed as a result of various calculations. Schematically, we may represent this process of identifying a metric dual to a state by
\be
\label{statetometric}
|\Psi_g \rangle \leftrightarrow g_{\mu \nu}(\vect{x}).
\ee

Two examples may help in elucidating this concept. Consider the vacuum of the CFT, $\rvac$. In this case, the expectation is that 
\be
\rvac \leftrightarrow \gads,
\ee
where the metric on the right hand side is the metric of {\em empty global AdS.}

In this paper, we will be particularly interested in a second example of such maps: a generic state at high energies in the CFT is believed to be dual to a large black hole in the bulk. 

Consider a set of energy eigenstates centered around a high energy $E_0 \gg \nc$, and with a width $\Delta \ll \nc$. The set of all energy eigenstates in this range is called
\be
\enrange[E_0,\Delta] \equiv\{|E_i \rangle : \quad E_0 - \Delta \leq E_i \leq E_0 + \Delta\}.
\ee
We denote the dimension of this space by $\dimrange[E_0, \Delta]$. By taking all linear combinations of these states, we get a subspace of the Hilbert space of the CFT
\be
\label{genericstate}
|\psig \rangle = \sum \coeff[i] |E_i \rangle, \quad |E_i \rangle \in \enrange[E_0, \Delta].
\ee
We assume above (and whenever we use $\coeff[i]$ to take superpositions of states) that they are chosen so that the state is correctly normalized. We can place an additional restriction on $|\psig \rangle$ above that it has vanishing  $SO(d)$ and $R$-charges. 

Next, we consider the set of unitary matrices that act entirely within this subspace. This is a very  large unitary group $U(\dimrange[E_0, \Delta])$. For $\Delta = \Or[1]$, we expect that $\dimrange[E_0, \Delta] = \Or[e^{\nc}]$.  The Haar measure on this unitary group now defines a measure for the coefficients $\coeff[i]$ in \eqref{genericstate}, and we can pick a ``typical''
state in the microcanonical ensemble by using this measure.
Then the expectation is that almost all states chosen in this manner, except for an exponentially small fraction of states,  correspond to a dual Schwarzschild black hole geometry in the bulk:
\be
|\psig\rangle \leftrightarrow g^{\text{bh}}_{\mu \nu}.
\ee
We can get other kinds of black holes by varying the other charges. This is the central class of ``semi-classical states'' that we will be interested in, in this paper.  

The example above also points to an additional important fact, which the reader should keep in mind. While we write $|\psidual{g} \rangle$ to prevent the notation from becoming unwieldy  the state dual to a geometry is far from unique. There are several microstates that represent the same geometry.

Two additional classes of states will be of some interest to us, and are entirely derivative from the class above.
\begin{enumerate}
\item{\bf Superpositions of semi-classical states \\}
First, given states corresponding to different metrics $|\psidual{{g_1}} \rangle \leftrightarrow g_{1,\mu \nu}, \ldots,  |\psidual{{g_m}} \rangle \leftrightarrow g_{m,\mu \nu}$ we may consider a superposition of such states
\be
\label{superposdistinct}
|\psisup \rangle
\equiv \left(\sum_{i=1}^m \coeff[i]|\psidual{{g_i}} \rangle  \right),
\ee
If the geometries above are reasonably distinct, then the states are almost orthogonal. This is also the case if we pick two generic microstates corresponding to the {\em same geometry}. As we will see below we expect that 
\be
\label{semiclassorth}
\begin{split}
&\langle \psidual{{g_1}} | \psidual{{g_2}} \rangle = \Or[e^{-\nc}],\\
&\langle \psidual{{g_1}} | \cop{g}_{\mu_1 \nu_1}(\vect{x}_{1}) \ldots \cop{g}_{\mu_m \nu_m}(\vect{x}_m) \cop{\phi}(\vect{x}_{m+1}) \ldots \cop{\phi}(\vect{x}_{m+p}) |\psidual{{g_2}} \rangle = \Or[e^{-\nc}],
\end{split}
\ee
both for states corresponding to distinct geometries, and for generic microstates corresponding to the same geometry.  Therefore, we require $\sum_i |\alpha_i|^2 = 1 + \Or[e^{-\nc}]$ in this situation. The important point is as follows. The smallness of the off-diagonal matrix elements above implies that a quantum superposition of a small number of geometries, or a small number of microstates corresponding to the same geometry, corresponds in effect to a classical probability distribution over these states.  On the other hand, it is clear that if we take $m = \Or[e^{\nc}]$ in the superposition above, then this intuition breaks down, and the cross terms become important.  
\item{\bf Excitations of semi-classical States \\ }
Furthermore, given a state $|\Psi_g \rangle$, which we have identified with a metric $g_{\mu \nu}$, one can consider ``excitations'' of this state. For example, one may ``act'' on this state using some of the operators corresponding to the metric or other light fields.  These new states correspond to excitations of the original state
\be
\label{excitation}
|\psidualexc{g} \rangle =  \cop{g}_{\mu_1 \nu_1}(\vect{x}_{1})  \ldots \cop{g}_{\mu_m \nu_m}(\vect{x}_m) \ldots  \cop{\phi}(\vect{x}_{m+1}) \ldots \cop{\phi}(\vect{x}_{m+n}) | \Psi_g \rangle.
\ee
In the large $\cn$ limit, after subtracting off the contribution of the background metric,  this state should be interpreted as an excitation with $n+m \ll \cn$ quanta on a background with metric $g$. 
Although these excited states occupy a very small fraction of the volume of the Hilbert space at any energy, they are important because there are several interesting physical questions about the response of equilibrium states to excitations.
\end{enumerate}
\paragraph{Coherent states vs metric eigenstates\\}
Although we have taken a CFT perspective on the states above in principle, we could also have viewed these states as solutions of the Wheeler de Witt equation that live in a Hilbert space obtained by quantizing gravity and the other light fields. From this perspective we  should emphasize, to avoid any confusion, that the semi-classical states $|\Psi_g \rangle$ that we refer to here are ``coherent states'', which correspond to an entire semi-classical spacetime; these states are {\em distinct} from ``metric eigenstates'' that are sometimes considered in conventional analyses of canonical gravity.\footnote{Strictly speaking, if we think of the degrees of freedom in gravity as being obtained from tracing out stringy and other heavy degrees of freedom, then we would expect a generic CFT state to correspond to a density matrix for the gravitational degrees of freedom, and not a pure state at all. However, because off-diagonal matrix elements of light operators between different coherent states are very small, a sum of coherent states effectively behaves like a classical superposition. Therefore we can neglect this complication here. Indeed, it is because of this fact that canonical gravity --- where the entanglement with these heavier degrees of freedom is ignored even in excited background geometries like the black hole  ---- makes sense at all.}

Let us make this more precise. We start by performing a $d+1$ split of the geometry
\be
d s^2 = -N^2 dt^2 + \gamma_{i j} (d x^i + N^i d t) (d x^j + N^j d t),
\ee
and promote the $d$-metric $\gamma_{i j}$ to an operator. The canonically conjugate momentum is 
\be
\pi^{i j} = -\gamma^{1 \over 2} \left(K^{i j} - \gamma^{i j} K \right),
\ee
where $K^{i j}$ is the extrinsic curvature \cite{DeWitt:1967yk}. (See \eqref{extrinsiccurv} for an explicit expression.)
Given a CFT operator $\cop{g}_{\mu \nu}$ we can therefore define two related CFT operators $\cop{\gamma}_{i j}$ and $\cop{\pi}^{i j}$. Now the key point is that the semi-classical/coherent states that we are discussing satisfy
\be
\label{coherentstatevar}
\begin{split}
&\langle \Psi_g | \cop{\gamma}_{i_1 j_1}(\vect{x}_{1}) \cop{\gamma}_{i_2 j_2} (\vect{x}_{2})  | \Psi_g \rangle - \langle \Psi_g | \cop{\gamma}_{i_1 j_1}(\vect{x}_{1}) |\Psi_g \rangle \langle \Psi_g | \cop{\gamma}_{i_2 j_2} (\vect{x}_{2})  | \Psi_g \rangle = \Or[{1 \over \nc}], \\
&\langle \Psi_g | \cop{\pi}^{i_1 j_1}(\vect{x}_{1}) \cop{\pi}^{i_2 j_2} (\vect{x}_{2})  | \Psi_g \rangle - \langle \Psi_g | \cop{\pi}^{i_1 j_1}(\vect{x}_{1}) |\Psi_g \rangle \langle \Psi_g | \cop{\pi}^{i_2 j_2} (\vect{x}_{2})  | \Psi_g \rangle = \Or[{1 \over \nc}]. 
\end{split}
\ee
 We can specify the $\Or[{1 \over \nc}]$ terms precisely, as we do in the next section. But for now we emphasize that these states have a small but finite uncertainty for both the three-metric and its canonically conjugate variable. Therefore they are distinct from ``metric eigenstates'' which would have satisfied 
\be
\cop{\gamma}_{i j} (\vect{x})  | \gamma \rangle = \gamma_{i j}(\vect{x}) |\gamma \rangle, \quad \text{metric~eigenstate.}
\ee
Such metric eigenstates would, on the other hand, have a large variance for $\cop{\pi}^{i j}$.

It is these coherent states that have a natural semi-classical interpretation. Metric eigenstates, on the other hand have maximum uncertainty in the value of $\cop{\pi}^{i j}$ and therefore, under time evolution, they quickly disperse into a superposition of several different eigenstates.  

\paragraph{Expectations from effective field theory\\}
We now turn to the other term used in the definition above: the expectations from effective field theory for correlators of these operators. 

Let us assume that we are given a state $|\Psi_g \rangle$ which is believed to be dual to a geometry  by the relation \eqref{statetometric}. Then, the most basic expectation from a putative CFT operator that could yield the metric in the bulk is that
\be
\label{linearopmetric}
\langle \Psi_g | \cop{g}_{\mu \nu}(\vect{x}) | \Psi_g \rangle = g_{\mu \nu}(\vect{x}).
\ee
Further, we demand that the $n$-point correlators of these operators have the property that
\be
\label{npointcorr}
\begin{split}
\langle \Psi_g | \cop{g}_{\mu_1 \nu_1}(\vect{x}_{1}) \ldots \cop{g}_{\mu_n \nu_n}(\vect{x}_{n}) | \Psi_g \rangle &= g_{\mu_1 \nu_1}(\vect{x}_{1}) g_{\mu_2 \nu_2}(\vect{x}_{2}) \ldots g_{\mu_n \nu_n}(\vect{x}_{n}) \\ &+  G_{\mu_1 \nu_1 \mu_2 \nu_2}(\vect{x}_{1}, \vect{x}_{2})  g_{\mu_3 \nu_3}(\vect{x}_{3}) \ldots g_{\mu_n \nu_n}(\vect{x}_{n})  + \text{perm.} \\ &+  G_{\mu_1 \nu_1 \mu_2 \nu_2 \mu_3 \nu_3}(\vect{x}_{1}, \vect{x}_{2}, \vect{x}_{3}) g_{\mu_4 \nu_4}(\vect{x}_{4}) \ldots g_{\mu_n \nu_n}(\vect{x}_{n})  + \text{perm.} \\ &+ \ldots.
\end{split}
\ee
where $G_{\mu_1 \nu_1 \ldots \mu_j \nu_j}(\vect{x}_{1}, \ldots \vect{x}_{j})$ are the {\em connected j-point correlators} as calculated by perturbatively quantizing metric fluctuations on the background of the metric $g_{\mu \nu}$ and $\ldots$ are the higher point functions which we have not shown explicitly. Note that this also fixes the ${1 \over \nc}$ corrections that appeared in \eqref{coherentstatevar}, because the connected correlators are subleading in ${1 \over \nc}$.

Similarly, 
we will declare that other bulk excitations are realized by state-independent operators, if there exist operators $\cop{\phi}(\vect{x})$ in the CFT, with the property that $n$-point correlators of these operators have an expansion
\be
\label{quasilocalphicond}
\begin{split}
\langle \Psi_g | \cop{\phi}(\vect{x}_{1}) \cop{\phi}(\vect{x}_{2}) \ldots \cop{\phi}(\vect{x}_n) | \Psi_g \rangle = G^{ }(\vect{x}_1, \vect{x}_2) G^{ }(\vect{x}_3, \vect{x}_4) \ldots G^{ }(\vect{x}_{n-1}, \vect{x}_n) &+ \text{perm.}  \\ + G^{  }(\vect{x}_1, \vect{x}_2, \vect{x}_3) G^{ }(\vect{x}_4, \vect{x}_5, \vect{x}_6) G^{ }(\vect{x}_7, \vect{x}_8) \ldots G^{ }(\vect{x}_{n-1}, \vect{x}_n) &+ \text{perm.} \\ &+ \ldots,
\end{split}
\ee
where the functions $G$ are the perturbative $j$-point connected correlation functions as obtained by quantizing the field $\phi$ about the metric $g$. 

 In this expansion, we emphasize that we are not interested in gravitational loop corrections at the moment, but would be satisfied
if the $n$-point correlators of the CFT operators have an expansion that agrees with that obtained from perturbative quantum field theory carried
out at tree-level. This tree-level contribution is already enough to fix the leading ${1 \over \nc}$ terms. It is also important to note that even the two-point function already knows about the background metric. This is simply because the graviton and matter propagators depend on the metric background. Therefore, in a sense, in the expansions \eqref{npointcorr} and \eqref{quasilocalphicond} we have already re-summed the ${1 \over \nc}$ series. It is in this re-summed series that we are only interested in
tree-level correlators.

Second, let us make a comment about superpositions of distinct geometries as in \eqref{superposdistinct}. Then we expect that
\be
\label{supasclass}
\begin{split}
&\langle \psisup  | \cop{g}_{\mu \nu}(\vect{x}) |  \psisup  \rangle  =  \sum_{i=1}^m |\alpha_i|^2 \langle \psidual{{g_i}} | g_{\mu \nu}(\vect{x}) | \psidual{{g_i}} \rangle  + \Or[e^{-\nc}].
\end{split}
\ee
A similar relation holds for $n$-point correlators, provided that $n \ll \nc$. 
This is the statement that cross-terms between macroscopically distinct geometries are very small. So, a superposition of the form above essentially behaves like a classical mixture for our purposes. 

This is an important point since there is no canonical way to speak of the ``same point'' in different macroscopic geometries. Stated precisely, this is the statement that quantum field theory in curved spacetime does not lead to any 
prediction for cross-correlators
\be
\langle \Psi_g | \cop{g}_{\mu \nu}(\vect{x}_{1}) \cop{g}_{\mu \nu}(\vect{x}_{2}) |\Psi_{g'} \rangle,
\ee
where $g_{\mu \nu}(\vect{x})$ and $g'_{\mu \nu}(\vect{x})$ are metrics corresponding to macroscopically different geometries.\footnote{For the case where these metrics are so close that one can be considered to be a coherent excitation of gravitons on the other, we refer the reader to Appendix \ref{appcoherentgravity}.}
 However, \eqref{superposdistinct} tells us that we {\em never} need to consider such cross-terms in correlators of the metric, which are exponentially suppressed and do not have any semi-classical interpretation. 

Finally, let us point out that if we declare that we do have a construction of state-independent local operators, then we should take it seriously. Therefore, if we find a state $|\psig\rangle$, in which $n$-point correlators of the operator $\cop{\phi}(\vect{x})$ cannot be reorganized
as perturbative correlators about any metric, then we must declare that the state $|\psig\rangle$ does not correspond to a semi-classical geometry. 

\paragraph{Gauge invariance and coordinates \\}
We now turn to the last remaining point in our definition of state-independent operators. The $d+1$ real parameters that parameterize CFT operators and are to be interpreted as coordinates in AdS. This is a tractable issue but two points sometimes lead to confusion: the fact that the metric and other local observables are not gauge invariant, and the fact that we are using a uniform coordinate system to represent all metrics. Both of these issues can be resolved simultaneously by an appropriate gauge fixing, as we now describe.

First, as we have already noted, given a family of  CFT operators labelled by coordinates $\vect{x}$, so that the family of operators satisfies \eqref{linearopmetric} and \eqref{npointcorr} we can clearly simply consider another family of CFT operators, which is related to the previous one by diffeomorphisms. 
\be
\label{trivialdiff}
\begin{split}
&\cop{\bar{g}}^{\mu \nu}(\vect{x}) = {\partial \diff^{\mu} \over \partial x^{\rho}} {\partial \diff^{\nu} \over \partial x^{\sigma}} \cop{g}^{\rho \sigma}(\vect{\diff}^{-1}(\vect{x})), \\
&\bar{\cop{\phi}}(\vect{x}) = \cop{\phi}(\diff^{-1}(\vect{x})).
\end{split}
\ee
The operators on the left hand side of \eqref{trivialdiff} are distinct CFT operators, but they obviously encode the same bulk physics.  We can choose to simply live with this lack of uniqueness, while keeping in mind that to extract any physics from the operator \eqref{linearopmetric} we need to form gauge-invariant quantities. But from a physical point of view, it is more convenient to pick a gauge so that the CFT operators that we are discussing become unambiguous. 

A related problem has to do with the the ``range'' of the real numbers in $\vect{x}$. Usually, we tailor the coordinate system to the metric. So it is often the case that the AdS Schwarzschild metric and the empty AdS metric are written in terms of coordinates that have different ranges.

In addressing these two issues, it is useful to recognize that they also arises in numerical general relativity. There we are given a grid of points, drawn from $R^{d,1}$, with a fixed range and we would like to place different metrics on this grid so that the resultant spacetime describes an entire range of physics, from empty AdS to black holes. 

To make this more precise, note that the empty AdS metric is given by
\be
d s_{\text{ads}}^2 
= -(1 + r^2) d t^2 + {d r^2 \over 1 + r^2} + r^2 d \Omega_{d-1}^2.
\ee
By a coordinate transformation, $r = {\rho \over 1 - \rho}$, we can bring the boundary to a finite coordinate distance 
\be
\label{rhometric}
d s_{\text{ads}}^2 
={1 \over (1 - \rho)^2} \left(-\hat{f}(\rho) d t^2 + {1 \over \hat{f}(\rho)} d \rho^2 + \rho^2 d \Omega_{d-1}^2 \right),
\ee
with $\hat{f}(\rho) = (1 - \rho)^2 + \rho^2$. The boundary is at $\rho = 1$, and manifold  in \eqref{rhometric} is $[0,1) \times R \times S^{d-1}$. In this paper we will only be interested in different metrics placed on this manifold  that asymptotically tend to the metric in \eqref{rhometric}, although they may differ in the bulk. Even if black holes are present, we simply consider nice slices that are parameterized by the coordinates $[0,1) \times S^{d-1}$ and then consider their evolution in time for a finite range of time. Note that by this finite-time restriction, we also avoid questions of ``topology changes.''
\begin{figure}[!h]
\begin{center}
\resizebox{0.3\textwidth}{!}{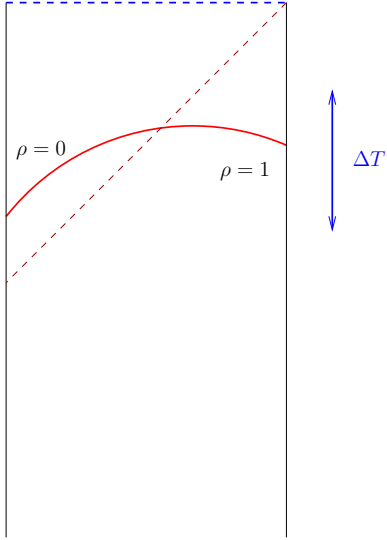}
\caption{Even in the presence of a black hole, nice slices can be parameterized by coordinates on $[0,1) \times S^{d-1}$. We examine physics for a finite interval $\Delta T$ so that the future singularity is irrelevant.}
\end{center}
\end{figure}

Having chosen a uniform coordinate system to describe the metrics that
we are interested in, we can further choose a gauge, to unambiguously specify
the CFT operators we are interested in. A convenient choice of gauge is given by the ``generalized harmonic gauges.'' In these gauges, we set
\be
\label{harmonicgauge}
\Box \vect{x}^{\mu} = H^{\mu}(\vect{x}).
\ee
A choice of the ``source functions'' $H^{\mu}(\vect{x})$ gives a choice of gauge. 

Note that once \eqref{harmonicgauge} is imposed as an additional {\em operator equation} that must be satisfied by the CFT operators that appear in \eqref{linearopmetric} and \eqref{npointcorr}, then this removes the redundancy \eqref{trivialdiff} in the identification of these operators in the CFT. So, if such operators exist then \eqref{harmonicgauge} picks out a specific family of them.

For the specific case of AdS, an appropriate choice of source functions 
is discussed in detail in \cite{Bantilan:2012vu}.  These details are not important here. 
The point that we can take away from the numerical analysis of \cite{Bantilan:2012vu} is that it is possible to describe a very broad range of metrics in AdS, including empty AdS and excited black holes that are dual to fluid dynamical situations on the boundary with a uniform choice of coordinate system and gauge.

\subsubsection{Relational observables }\label{subsecrelational}
There is another class of coordinate systems, which is particularly convenient in AdS. This is the class of coordinate systems that is defined
relationally with respect to the boundary. Here, we assume that we are already given the metric in some coordinate system, such as the ones above. We then describe a  coordinate transformation to a more {\em convenient} relational coordinate system. 

Intuitively, we would like to consider an experiment where an
observer jumps from the boundary, with no initial velocity along the $S^{d-1}$, falls for a given amount of proper time, and then makes a measurement. In fact this notion is a little hard to make concrete in this form because if we drop the observer from a point that is infinitesimally close to the boundary, he very rapidly approaches the speed of light. This problem cannot be solved by using an affine parameterization of null geodesics either, since any affine parameter that is finite in the bulk goes to infinity as we reach the boundary.

So, it is convenient to use the following slightly more complicated construction.  We start from a given point on the boundary, which we label by $(t_1, \Omega_1)$.  We know that the metric is of the asymptotically AdS form given by \eqref{rhometric}. We now consider a null geodesic, parameterized by ordinary asymptotic AdS time, that extends into the bulk, with no velocity along the $S^{d-1}$. More precisely, let us consider a null geodesic trajectory given by 
\be
\label{geodesic1}
\vect{x}_1(t) \equiv (t,\rho_1(t), \Omega_1(t)), \quad \rho_1(t_1) =  1, \quad \Omega_1(t_1) = \Omega_1, \quad \dot{\Omega}_1(t_1) = 0, \quad \dot{\rho}(t_1) = -1,
\ee
where by a slight abuse of notation we have used $\Omega_1$ both for the solution to the geodesic equation, and for the initial value of the solution. 
Note that initial ``velocity'' in the radial direction is fixed since the geodesic is null and the sign indicates that the geodesic is ingoing and moves into the bulk as time advances. This geodesic  reaches a finite coordinate distance in the bulk in finite time. Second, note that while we are starting with no angular momentum, intrinsic properties of the geometry may cause the geodesic to start moving on the sphere as  well after it departs from the boundary.

We now consider a {\em second} null geodesic that intersects the boundary at a {\em later} point $(t_1 + \tau, \Omega_2)$ and also has $\dot{\Omega}_2 = 0$ at its final point. This is the geodesic trajectory
\be
\label{geodesic2}
\vect{x}_2(t) = (t, \Omega_2(t), \rho_2(t)), \quad \rho_2(t_1 + \tau) = 1,  \quad \Omega_2(t_1 + \tau) = \Omega_2, \quad \dot{\Omega}_2(t_1 + \tau) = 0 \quad \dot{\rho}_2(t_1 + \tau) = 1,
\ee
and the sign of the radial derivative indicates that the geodesic is outgoing at the time $t_1 + \tau$.
Now given a particular value of $t_1, \Omega_1(t_1)$, we vary $\Omega_2(t_1 + \tau)$ so that the geodesics intersect. We expect that 
\be
\exists~\Omega_2~\text{and}~\exists~t_{i}, t_1 <  t_{i} < t_1 + \tau \quad \text{such~that}~ \rho_2(t_i) = \rho_1(t_i); \quad \Omega_2(t_i) = \Omega_1(t_i).
\ee
Intuitively, the existence of a such a solution seems clear. For example, in the case where the geometry has no angular momentum at all, we can solve the equation above simply by setting $\Omega_2 = \Omega_1$. If we start deforming the geometry so that it is rotating,  we should still be able to tune $\Omega_2$ so that the two geodesics intersect. Even for other, more complicated geometries, we expect that the intersection point should be well defined at least as long as we are close enough to the boundary and we will see below that this is all that we need.

We will denote the point of intersection by 
\be
\label{outsideparam}
P_{i}(t_1,  \Omega_1, \tau) \equiv (t_i,  \Omega_1(t_i), \rho_1(t_i)).
\ee
This is a bulk point that is parameterized by the starting point of the first geodesic and the time difference to the ending point of the second geodesic.

Note that by means of such a process we cannot reach behind the black hole horizon. However, once we have a parameterization of points in the exterior, it is simple to extend them behind the horizon. We once again consider geodesics that start from a point $(t_1, \vect{\Omega}_1)$ on the boundary but this time we parameterize them using an affine parameter so that the geodesic satisfies the equation
\be
\label{geodeqn}
{d^2 x_1^{\mu}(\lambda) \over d \lambda^2} + \Gamma^{\mu}_{\nu \sigma} {d {x}_1^{\nu} (\lambda) \over d \lambda} {d {x}_1^{\sigma} (\lambda) \over d \lambda} = 0.
\ee
This is just a reparameterization of the geodesic in \eqref{geodesic1}, and so we have denoted it with the same symbol $\vect{x}_1(\lambda)$.

The key point is that we can use our previous parameterization \eqref{outsideparam} to normalize the affine parameter. We set
\be
\vect{x}_1(0) = P_i(t_1, \Omega_1,  \tau_1), \quad \vect{x}_1(1) = P_i(t_1,  \Omega_1, \tau_2).
\ee
A choice of the intervals $\tau_1, \tau_2$  gives a specific normalization of the affine parameter. The reader can, for her convenience, think of any concrete value: say $\tau_1 = \ell_{\text{ads}}, \tau_2 = 2 \ell_{\text{ads}}$. 

\begin{figure}[!h]
\begin{center}
\resizebox{0.3\textwidth}{!}{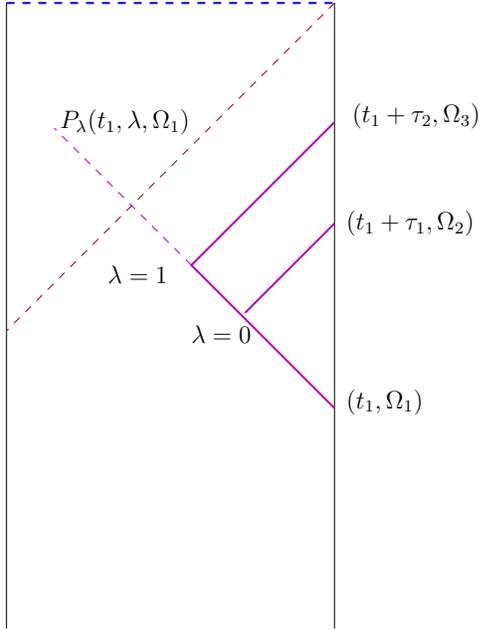}
\caption{The relational gauge fixing proceeds in two steps: first we use intersecting  geodesics to parameterize points outside the horizon. Then we use this set of points to normalize the affine parameter and follow null geodesics into the horizon.}\label{relational}
\end{center}
\end{figure}

Once this normalization has been fixed we now obtain the set of points
\be
\label{affinepoints}
P_{\lambda}(t_1, \Omega_1, \lambda) = (t_1(\lambda),  \Omega_1(\lambda), \rho_1(\lambda)).
\ee
The difference between \eqref{affinepoints} and \eqref{outsideparam} is that the points in \eqref{affinepoints} can also reach inside the horizon.  The entire process above is summarized in Fig \ref{relational}.

The advantage of this prescription is that classically, measurements of a scalar field defined in such a relational manner are {\em gauge-invariant.} We recall that when we define quantum gravity in anti-de Sitter space, we have to consider the set of all field configurations modulo trivial diffeomorphisms. The trivial diffeomorphisms are those that vanish at the boundary of anti-de Sitter space. Large gauge transformations --- which leave the boundary in asymptotically AdS form, but yet move points on the boundary --- correspond to symmetries in the boundary theory, and induce a change of the physical state. 

So, {\em gauge invariant} observables are those that are invariant under trivial diffeomorphisms. In the relational observables described above,  we start with a point on the boundary --- which is left fixed because the diffeomorphism vanishes there --- and then follow a gauge-invariant prescription to reach a point in the interior. Evidently, scalar fields evaluated at this point are themselves gauge invariant. 

There is an important stronger statement that we can make. Consider a large diffeomorphism that induces a conformal transformation on the boundary $(t, \Omega) \rightarrow {\cal C}^{-1}(t, \Omega)$, where ${\cal C}$ denotes an element of the conformal group. Geometrically, under the diffeomorphism the geodesic trajectories in \eqref{geodesic1} and \eqref{geodesic2} get mapped to new geodesic trajectories. Therefore we expect that the relationally define points in \eqref{affinepoints} will transform under the diffeomorphism as 
\be
P_{\lambda}(t, \Omega, \lambda) \rightarrow P_{\lambda}({\cal C}^{-1}(t, \Omega), \lambda).
\ee
The important point is that this transformation of the relational points does {\em not} depend on the details of the diffeomorphism in the bulk, but merely on how it acts on the boundary.

 Now consider a scalar field operator $\cop{\phi}(P_{\lambda}(t, \Omega), \lambda)$ with the bulk point defined as in \eqref{affinepoints}.  Corresponding to the conformal transformation ${\cal C}$, there is a unitary operator $U_{\cal C}$ on the boundary. Then, in order to be consistent with the geometric intuition, we expect that the CFT operator $\cop{\phi}$ will satisfy
\be
U_{\cal C}^{\dagger} \cop{\phi}^i(P_{\lambda}(t,  \Omega, \lambda)) U_{\cal C} = \cop{\phi}^i(P_{\lambda}({\cal C}(t, \Omega), \lambda).
\ee 
We will use this relation several times to obtain the commutator of bulk
operators with the Hamiltonian which arises from the special case where ${\cal C}$ is just taken to be time-translation above. In section \ref{seceternal}, we will apply this analysis in a more general setting where there are two boundaries.

The disadvantage of the relational prescription is that it is harder to make this precise at subleading order in ${1 \over \nc}$. Clearly, the affine parameter along a geodesic from the boundary to another point may itself be expected to fluctuate at order ${1 \over \nc}$. In this paper, these subtleties will not be important.

\subsection{The alternative: state-dependent bulk-boundary maps}
An alternative to the state-independent possibility above is that geometric quantities like the metric do not arise by evaluating a Hermitian operator, but is a more general ``measurable.''  More precisely, we would be led to state-dependence if there are no globally defined Hermitian operators $\cop{g}_{\mu \nu}(\vect{x})$ and $\cop{\phi}(\vect{x})$. Rather, about a given state $|\psidual{g}\rangle$ we would have operators $\cop{g}^{\{\Psi\}}_{\mu \nu}(\vect{x})$ and $\cop{\phi}^{\{\Psi\}}(\vect{x})$  so that the correlators
\be
\label{statedepcorrs}
C_{\Psi}(\vect{x}_1, \ldots \vect{x}_{m+p}) = \langle \psig| \cop{g}^{\{\Psi\}}_{\mu_1 \nu_1}(\vect{x}_{1}) \ldots \cop{g}^{\{\Psi\}}_{\mu_m \nu_m}(\vect{x}_m) \cop{\phi}^{\{\Psi\}}(\vect{x}_{m+1}) \ldots \cop{\phi}^{\{\Psi\}}(\vect{x}_{m+p}) |\psig\rangle,
\ee
reproduce the predictions of effective field theory that we outlined above. This definition is identical to the  definition \eqref{stateindcorrs} in terms of the semi-classical states $|\psig\rangle$ that appear here and the expectations we have for the values of the correlators. The difference is in the nature of the operators $\cop{g}^{\{\psi\}}_{\mu \nu}$ which now depend on the state.

One possible way to think about \eqref{statedepcorrs} is that the geometry emerges as a ``function of correlation functions''\footnote{We thank Nima Lashkari for this phrase.} and not by measuring linear operators. However, we have some additional structure in \eqref{statedepcorrs}. Since the bulk observer must see {\em quantum effective field theory}, it must be the case that to an excellent approximation the operators  $\cop{g}^{\{\Psi\}}_{ \mu \nu}(\vect{x})$ and  $\cop{\phi}_{\Psi}(\vect{x})$ act as linear operators. In terms of the classes of states that we have defined above, this can be turned into a sharp restriction: the {\em same} operators that represent the metric and other excitations in a state $|\psidual{g}\rangle$ must also represent these excitations in superpositions \eqref{superposdistinct} and \eqref{excitation}. We will show below that, in our construction, this is indeed the case. 

To lighten the notation we will no usually omit the superscript $\Psi$ in $g^{\{\Psi\}}_{\mu \nu}$ even when we are considering state-dependent operators. Although, in several cases we will discuss explicitly whether a given operator is state-dependent or state-independent, in others it should be clear from the context.

We now point out that many of the  existing methods of associating a geometry to a state as in \eqref{statetometric} are state-dependent in practice.\footnote{We cannot help making the curious observation that, within the string theory literature, this fact hardly attracted any attention or controversy until the recent discussions on the black hole interior.} We hasten to add that this, by itself, does not mean that the map \eqref{statetometric} can only be realized in a state-dependent manner. Our discussion in this subsection does {\em not} rule out the possibility that there may be an overarching state-independent prescription which encapsulates all of these state-dependent approaches in some approximation. Our purpose in this subsection is only to use these examples to explain the distinction between state-dependent and state-independent realizations of the maps.

We now proceed to discuss the Ryu-Takayanagi formula, the procedure for extracting the Einstein equations from the first law of entanglement, and the smearing function construction of operators  outside the  black hole. 

\subsubsection{State-dependence in geometry from entanglement }\label{secgeomfromentang}
The Ryu-Takayanagi (RT) formula \cite{Ryu:2006bv,Ryu:2006ef} and its generalization \cite{Hubeny:2007xt} by Hubeny, Rangamani and Takayanagi  provides a method of reading off geometric quantities from a state. We review the formula, and show how it is state-dependent. We also  show how to interpret it correctly and that this state-dependence does not imply any contradiction with quantum  mechanics.

In particular these formulas provide a relation 
between the entanglement entropy of a region on the boundary, and the area of an extremal surface in the bulk which is homologous to the boundary region.  So, given a region $R$ on the boundary and a semi-classical metric $g_{\mu \nu}$, we can calculate the area of this extremal area surface $A(g,R)$. The Ryu-Takayanagi formula now states
\be
\label{rtoriginal}
{1 \over 4 G_N} A(g,R) = S_R, \quad \text{Ryu-Takayanagi}
\ee
where $S_R$ is the entanglement entropy of the region $R$.

We  will now show the following
\begin{enumerate}
\item
The formula \eqref{rtoriginal} cannot be interpreted as an operator relation for the area, because there is no ``entanglement entropy'' operator.
\item
However, even though the entanglement entropy cannot, in general, be interpreted  as the expectation value of a Hermitian operator,  because of properties of the large-$\nc$ CFT Hilbert space, we expect to find a state-dependent operator $\cop{A}_R$ in the CFT which has the property that
\be
\langle \psig| \cop{A}_{R} |\psig\rangle = S_{R}(|\psig\rangle),
\ee
both in states \eqref{statetometric} and in superpositions of a small number of such states \eqref{superposdistinct}
\end{enumerate}

We start by noting that if the metric is a state-independent operator, then the area of the minimal area surface, which is a functional of the metric, is
also a state-independent operator. In fact, as we will see below, from the point of view of a semi-classical
quantization of gravity --- which is what yields the justification for expecting the metric to be an ordinary operator --- the area of the minimal
area surface should be as good an operator as the metric. Therefore, we might expect the existence of some operator $\cop{A}_R$, so that in the state dual to the geometry with metric $g_{\mu \nu}$, we have
\be
 \cop{A}_R | \Psi_ g \rangle = A(g,R) |\Psi_g \rangle + \Or[{1 \over \nc}].
\ee

However, on the other hand, the entanglement entropy is not a linear operator. The standard proof is as follows. Consider the division of the CFT Hilbert space
into that of the region and its complement: $H = \hilb[R] \otimes \hilb[\tilde{R}]$. Say that we want an operator $\cop{S}_{R}$ so that $\forall |\psig\rangle \in H$, we have 
$\langle \psi | \cop{S}_R |\psig\rangle = S(|\psig\rangle)$, where $S$ is the entanglement entropy between $R$ and its complement in that state.  Now, we note the following facts. Since $S(|\psig\rangle)$ is always non-negative, the expectation value of the putative $\cop{S}_R$ operator is non-negative in all states; therefore it can have no negative eigenvalues. Second, we can find a complete basis of unentangled states 
\be
\label{unentbasis}
|\Psi_{i j} \rangle = |R_{i} \rangle \otimes |\tilde{R}_{j} \rangle,
\ee
where $i \in [1, \ldots \text{dim}(\hilb[R])], j \in [1, \ldots \text{dim}(\hilb[\tilde{R}])]$. Clearly we expect $\langle \Psi_{i j}| \cop{S}_R |\Psi_{i j} \rangle = 0$. Moreover, since $|\Psi_{i j} \rangle$ is a basis, we also have $\tr(\cop{S}_R) = 0$. Since $\cop{S}_R$ has no negative eigenvalues, and its trace is zero, it must be the case that $\cop{S}_R = 0$ identically. This is absurd. Therefore, this is no operator $\cop{S}_R$ whose expectation value equals the entanglement entropy in general. 
A simple extension of this argument shows that this also true for the Renyi entropies $\tr(\cop{\rho}_{R}^n)$, where $\cop{\rho}_R$ is the reduced density matrix of the region. 

The fact that the entanglement entropy does not correspond to an ordinary linear operator may appear to be a formal statement, but it becomes acute in the following situation in the AdS/CFT correspondence. Consider a superposition
of two different classical geometries, as in \eqref{superposdistinct}. For simplicity, we can consider a pure state which is a superposition of a pure state corresponding to  a black hole at temperature $\beta$, with a corresponding metric $g_{\beta}$,  and another pure state corresponding to a black hole at a temperature $\beta'$, with a corresponding metric $g_{\beta'}$.  Provided that $\beta - \beta' \gg {1 \over \nc}$, we see that the corresponding pure states are almost orthogonal. We write the superposed state as
\be
|\psisup \rangle 
= \coeff[1] | \psidual{g_{\beta}} \rangle + \coeff[2] | \psidual{g_{\beta'}} \rangle,
\ee
and normalizability requires $|\coeff[1]|^2 + |\coeff[2]|^2 = 1 + \Or[e^{-\nc}]$. This is not a state that we usually consider, but it is certainly
possible to consider such superpositions in the CFT since distinct geometries do not belong to strict superselection sectors. 

  From the bulk point of view, quantum mechanics provides the following prediction. If one measures the ``area'' in this state, one expects to find the answer $A(g_{\beta}, R)$ with probability $|\coeff[1]|^2$ and $A(g_{\beta'}, R)$ with probability $|\coeff[2]|^2$. 

While the entanglement entropy cannot reproduce this probability distribution, with some work we can show that the entanglement entropy does correctly reproduce the expectation value of the area. The argument is as follows. Consider the reduced density matrix of the region $R$ in all three states
\be
\begin{split}
\cop{\rho}_R(\beta) &= \tr_{\tilde{R}}(|\psidual{g_{\beta}} \rangle \langle \psidual{g_{\beta}} | ),\\
\cop{\rho}_R(\beta') &= \tr_{\tilde{R}}(|\psidual{g_{\beta'}} \rangle \langle \psidual{g_{\beta'}} | ), \\ 
\cop{\rho}_R(\psisup) &= \tr_{\tilde{R}}(|\psisup \rangle \langle \psisup | ), \\ 
\end{split}
\ee
where $\tilde{R}$ is the complement of $R$. 

We can write both the states in terms of a Schmidt basis
\be
\label{schmidtbasis}
\begin{split}
|\psidual{g_{\beta}} \rangle &= \sum_i \gamma^{\beta}_i |R^{\beta}_i \rangle \otimes |\tilde{R}^{\beta}_i \rangle, \\
|\psidual{g_{\beta'}} \rangle &= \sum_i \gamma^{\beta'}_i |R^{\beta'}_i \rangle \otimes |\tilde{R}^{\beta'}_i \rangle, \\
\end{split}
\ee
where, by the definition of the Schmidt basis, we have
\be
\begin{split}
&\langle R^{\beta}_i | R^{\beta}_j \rangle = \delta_{i j}; \quad \langle \tilde{R}^{\beta}_i | \tilde{R}^{\beta}_j \rangle = \delta_{i j}; \\
&\langle R^{\beta'}_i | R^{\beta'}_j \rangle = \delta_{i j}; \quad  \langle \tilde{R}^{\beta'}_i | \tilde{R}^{\beta'}_j \rangle = \delta_{i j} \\
&\sum_i |\gamma^{\beta}_i|^2 = 1; \quad \sum_i |\gamma^{\beta'}_i|^2 = 1. \\
\end{split}
\ee
To simplify the analysis, without sacrificing anything of importance,  let us truncate the range of $i$ in \eqref{schmidtbasis} so that it runs over $\Or[e^{\nc}]$ states. In almost  any state, where the energy scales like $\nc$, it is in fact true that even if the exact expansion of the state involves an infinite number of eigenvectors, all but an $\Or[e^{\nc}]$ number of them are exponentially unimportant. 

Now, the key point is that in a very large Hilbert space we expect that the Schmidt basis decomposition for the
state $|\psidual{g_{\beta}} \rangle$ and the state $|\psidual{g_{\beta'}} \rangle$ is typically uncorrelated. This implies that
\be
\label{uncorrasumption}
|\langle R^{\beta}_i | R^{\beta'}_j \rangle|^2 = \Or[e^{-\nc}]; \quad |\langle \tilde{R}^{\beta}_i | \tilde{R}^{\beta'}_j \rangle|^2  = \Or[e^{-\nc}].
\ee
Strictly speaking \eqref{uncorrasumption} is valid if one takes a large Hilbert space and divides it into two parts. In a local quantum field theory, it is possible that the very short distance modes in the two regions are entangled in a universal manner. This will not affect any of our results since in considering the entanglement entropy we, in any case, must subtract off this universal part.

Now the first two reduced density matrices are given by
\be
\begin{split}
\cop{\rho}_R(\beta) &= \sum_i |\gamma^{\beta}_i|^2 |R^{\beta}_i \rangle \langle R^{\beta}_i |,\\
\cop{\rho}_R(\beta') &= \sum_i |\gamma^{\beta'}_i|^2 |R^{\beta'}_i \rangle \langle R^{\beta'}_i |. \\
\end{split}
\ee
The corresponding entanglement entropies are given by
\be
\begin{split}
S_{\beta} &= -\tr(\cop{\rho}_R(\beta) \ln \cop{\rho}_R(\beta)) = -2 \sum_i |\gamma^{\beta}_i|^2 \ln |\gamma^{\beta}_i|, \\
S_{\beta'} &= -\tr(\cop{\rho}_R(\beta') \ln \cop{\rho}_R(\beta')) = -2 \sum_i |\gamma^{\beta'}_i|^2 \ln |\gamma^{\beta'}_i|. 
\end{split}
\ee
Moreover, we see that
\be
\cop{\rho}_R(\psisup) = |\coeff[1]|^2 \cop{\rho}_R(\beta) + |\coeff[2]|^2 \cop{\rho}_R(\beta') + \cop{\rho}_{\text{cross}}.
\ee
where,  we see that the matrix involving the cross-terms is
\be
\cop{\rho}_{\text{cross}}  = \sum_{i, j} \Big[\coeff[1] \coeff[2]^* \gamma^{\beta}_i (\gamma^{\beta'}_j)^* \langle \tilde{R}^{\beta'}_j | \tilde{R}^{\beta'}_i \rangle \Big] |R^{\beta}_i \rangle \langle R^{\beta'}_j | + \text{h.c}.
\ee
Now even though this is an $e^{\nc} \times e^{\nc}$ sized matrix, we can check using \eqref{uncorrasumption} that $\tr(\cop{\rho}_{\text{cross}}) = \Or[e^{-\nc}]$ and also that $\tr(\cop{\rho}_{\text{cross}}^2) = \Or[e^{-\nc}]$. Therefore the from the cross terms will have an exponentially small effect in the computations below, and we will neglect it.

Now consider two positive integers $m_1, m_2$. We see that
\be
\tr(\cop{\rho}_R^{m_1}(\beta) \cop{\rho}_R^{m_2}(\beta')) = \sum_{i,j} |\gamma^{\beta}_i|^{2 m_1} |\gamma^{\beta'}_j|^{2 m_2}|\langle R^{\beta}_i | R^{\beta'}_j \rangle|^2.
\ee
Therefore, from \eqref{uncorrasumption}, we see that
\be
\tr(\cop{\rho}_R^{m_1}(\beta) \cop{\rho}_R^{m_2}(\beta')) = \Or[e^{-\nc}], \quad \text{if}~m_1, m_2 > 0.
\ee
This allows us to evaluate the entanglement entropy of the superposed state. In particular, using the result above, we see that m\textsuperscript{th} Renyi entropy for the superposed state is given by 
\be
\tr(\cop{\rho}_R(\psisup)^m) = |\coeff[1]|^{2 m} \tr(\cop{\rho}_{R}(\beta)^m) + |\coeff[2]|^{2 m} \tr(\cop{\rho}_{R}(\beta')^m) + \Or[e^{-\nc}].
\ee
Therefore the entanglement entropy is given by
\be
\begin{split}
S_R(\psisup) =  &-\lim_{m \rightarrow 1} {d \over d m}  \tr(\cop{\rho}_R(\psisup)^m) \\= &- |\coeff[1]|^2 \ln(|\coeff[1]|^2) \tr\big[\cop{\rho}_R(\beta)\big] -  |\coeff[2]|^2 \ln(|\coeff[2]|^2) \tr\big[\cop{\rho}_R(\beta')\big] \\ &- |\coeff[1]|^2 \tr\big[\cop{\rho}_R(\beta) \ln(\cop{\rho}_R(\beta)) \big] - |\coeff[2]|^2 \tr\big[\cop{\rho}_R(\beta') \ln(\cop{\rho}_R(\beta'))\big] \\
=  &- |\coeff[1]|^2 \ln(|\coeff[1]|^2) - |\coeff[2]|^2 \ln(|\coeff[2]|^2) + |\coeff[1]|^2 S_R(\beta) + |\coeff[2]|^2 S_R(\beta').
\end{split}
\ee
Therefore we see that
\be
S_R(\psisup) = {1 \over 4 G_N} \langle A(R) \rangle -  |\coeff[1]|^2 \ln(|\coeff[1]|^2) - |\coeff[2]|^2 \ln(|\coeff[2]|^2).
\ee
where $\langle A(R) \rangle = |\coeff[1]|^2 A(g_{\beta},R) + |\coeff[2]|^2 A(g_{\beta'},R)$ is the expectation value of the area obtained from a naive analysis. 

In fact the additional term that we have obtained is always subleading even if we take a superposition of a large number of states. This is because the the leading term is $\Or[\nc]$ as we can see from the explicit factor of $G_N$ in the formula above. Now, even if we superpose $m$-states in the form \eqref{superposdistinct} with coefficients $\sum_{i=1}^m |\coeff[i]|^2 = 1$, then the additional term is bounded by
\be
-\sum_{i=1}^m |\coeff[i]|^2 \ln(|\coeff[i]|^2) \leq \ln(m).
\ee
Therefore, unless we take a superposition of an $e^{\nc}$ number of states, we see that we can still consistently interpret the entanglement entropy as the expectation value of the operator, that classically, would correspond to the area.
\be
\label{rtasexp}
S_R=  {1 \over 4 G_N} \langle \cop{A}_R \rangle ,
\ee
If we do take a superposition of an exponentially large number of states, then the cross terms become important even for the area operator,
and we must re-evaluate the entire expression.

To summarize, we have concluded that once the original Ryu-Takayanagi formula is interpreted as a relation between an expectation value and the entanglement entropy as in  \eqref{rtasexp}, then it holds consistently even in states that are superpositions of classical geometries as advertised. Our analysis here does not rule out the existence of a state-independent ``area'' operator  $\cop{A}_R$ but such a state-independent operator cannot be dual to the entanglement entropy in general.

Before concluding, we should mention that there are several approaches that attempt to construct other bulk geometric quantities by massaging or refining the Ryu-Takayanagi formula. For example, the authors of \cite{Balasubramanian:2014sra,Balasubramanian:2013lsa} related the differential entropy --- obtained by considering the variation of the entanglement entropy as the interval on the boundary is altered --- to the area of a hole in the bulk. This can be used to read off the bulk metric more directly than the minimal area prescription. Of course, all of these approaches are also explicitly state-dependent. However, just as in our discussion above, we expect that when we interpret them appropriately they do not present any observable contradiction with
quantum mechanics in the bulk. 

\subsubsection{Equations of motion from the first law of entanglement}
Another approach of deriving the bulk from the boundary, which has attracted attention, is the program of deriving the bulk equations of motion
from the ``first law of entanglement'' \cite{Lashkari:2013koa,Faulkner:2013ica,VanRaamsdonk:2010pw,VanRaamsdonk:2009ar}. Consider, once again, a region $R$ on the boundary, and a CFT in the vacuum state. Then we may define
the modular Hamiltonian of the region by demanding that the reduced density matrix of $R$ have the form
\be
\cop{\rho}_R = {e^{-{\hmodr}} \over \tr_{\hilb[R]}(e^{-\hmodr})},
\ee
where the reader should note that the trace is in $\hilb[R]$ only.

In this case, if we consider the vacuum of the CFT and take the region $R$ to be a ball of radius $a$ centered around a point $\vect{y_0}$, then the modular Hamiltonian is given by \cite{Hislop:1981uh,Casini:2011kv}
\be
\label{linearizedmodh}
\hmodr = 2 \pi \int_R d^{d-1} \vect{y} {a^2 - |\vect{y} - \vect{y_0}|^2  \over 2 a} \cop{T}_{t t},
\ee
where $\cop{T}_{tt}$ is the time-time component of the stress-tensor. But this is a state-dependent formula that is obtained by defining the modular Hamiltonian about the vacuum. 

Using this formula it was shown \cite{Nozaki:2013vta,Blanco:2013joa,Lashkari:2013koa} that one can relate the linearized Einstein equations in the bulk to the ``1st law'' of entanglement
entropy under small changes of the state. By considering a generalization of the Ryu-Takayanagi conjecture, where the area is replaced by a Wald functional,
this was extended to higher derivative theories in \cite{Faulkner:2013ica} and ${1/\nc}$ interactions were included in \cite{Swingle:2014uza}.

However, although \eqref{linearizedmodh} looks like an operator equation, the modular Hamiltonian is also a state-dependent operator. There is no globally defined operator $\cop{H}^R_{\text{glob}}$ in the theory so that its action equals that of the modular Hamiltonian on every possible state. The proof is similar to the one above.  Let us say that we had an operator
\be
\label{putativehmod}
\cop{H}^R_{\text{glob}}|\psig\rangle = \hmodr |\psig\rangle,
\ee
so that its action on $\hilb[R]$ was that of the modular Hamiltonian and it acted as the identity on $\hilb[\tilde{R}]$. Considering again, the unentangled states in \eqref{unentbasis}.   
The density matrix of $R$ in this state is pure: $\cop{\rho}_{R}(|\Psi_{i j} \rangle) = |R_i \rangle \langle R_i|$. We can see that this implies that the putative modular Hamiltonian operator must have the action $\cop{H}^R_{\text{glob}} |\Psi_{i j} \rangle = 0$. However, if $\cop{H}^R_{\text{glob}}$ is a linear operator, then on any state $\cop{H}^R_{\text{glob}} \sum_{i j} \alpha_{i j} |\Psi_{i j} \rangle = 0$. This suggests that $\cop{H}^R_{\text{glob}} = 0$ as an operator, which is absurd. Therefore \eqref{putativehmod} cannot hold for any state-independent operator $\cop{H}^R_{\text{glob}}$.

Therefore, \eqref{linearizedmodh} must be interpreted as a relation that is true within expectation values taken in the vacuum or small fluctuations about the CFT vacuum. No operator generalization of this equation exists as we have shown above. 
Nevertheless, it should be possible to obtain similar formulae about different states by defining the action of the modular Hamiltonian relative to that state. Such formulae also work for superpositions of a small number of states, as we showed above in the case of the entanglement entropy, but this entire process is fundamentally state-dependent. 

The authors of \cite{Jafferis:2014lza} proposed that $\hmodr |\Psi \rangle = \cop{A}_R |\Psi \rangle$ should hold as an operator equation. However, as they noted explicitly this is a state-dependent relation which works in the neighbourhood of a given state. As we discussed above we would also expect it to work in superpositions of a small number of semi-classical states.

\subsubsection{Smearing function construction of local operators }\label{sectransferfunc}
Another commonly used method --- and one that we use in this paper --- of extracting local physics from a state uses a smearing function to represent bulk operators as smeared versions of boundary operators \cite{Banks:1998dd, Bena:1999jv, Hamilton:2006az, Hamilton:2005ju, Hamilton:2007wj,Kabat:2011rz}. We review this approach in greater detail in section \ref{subseclocalopcft}, where we will also derive the expressions below for some states. 
In this approach, given
a state $|\Psi_g \rangle$, we guess a smearing function and conjecture that local fields in the bulk have the form
\be
\label{transfunc}
\cop{\phi}(\vect{x}) = \int \op(\vectb{y}) K_g(\vect{y}^b, \vect{x}) d^d \vectb{y},
\ee
where $\vect{x}$ is a bulk point, $\vectb{y}$ is a boundary point, $\op$ is a single-trace operator on the boundary, and and $K_g$ is an appropriately chosen smearing function.
Strictly speaking, there are some difficulties in interpreting \eqref{transfunc} in position space, having to do with the convergence of the integral, which has led to some confusion in the literature \cite{Bousso:2012mh,Rey:2014dpa, Morrison:2014jha}. However, as we showed in \cite{Papadodimas:2012aq}, these difficulties go away if we work in momentum space and this subtlety is irrelevant for our present discussion.

One may object that one is putting in the answer by hand in \eqref{transfunc} in the Kernel $K_g$. However, it is a non-trivial fact that the
operators $\cop{\phi}(\vect{x})$ do obey \eqref{quasilocalphicond}, and also have the right boundary values (as one approaches the boundary of AdS) as CFT correlators. In particular for an operator $\op$ of dimension $\Delta$, we require that
\be
\label{alternateadscft}
\langle \op(\vectb{x}_1) \ldots \op(\vectb{x}_n) \rangle = Z^n \lim_{r \rightarrow \infty} r_1^{\Delta} \ldots r_n^{\Delta} \langle \cop{\phi}(\vectb{x}_1, r_1) \ldots \cop{\phi}(\vectb{x}_n, r_n) \rangle,
\ee
where $Z$ is a numerical wave-function renormalization factor, and we have written the bulk points as a boundary point combined with a radial coordinate $r$ which can be identified with the coordinate $r$ in \eqref{schwarzschild}. The fact that {\em both} \eqref{transfunc} and  \eqref{alternateadscft} hold  simultaneously  involves a delicate interplay between the kernel and the correlators of $\cop{\op}$ in the state $|\Psi_g \rangle$. 

As written,  the expression \eqref{transfunc} is explicitly state-dependent because the kernel $K_g$ depends on the metric, and is therefore different in different states $|\psidual{g} \rangle$.  So, for a given kernel $K_g$, this expression works only in a state that corresponds to this semi-classical geometry. 

In section \ref{secsmoothint}, we discuss whether it may be possible to lift \eqref{transfunc} to a state-independent prescription, at least, outside the horizon. While this is possible in a mini-superspace approximation as we show around \eqref{phistateind}, we are agnostic about whether this works in general, even outside the horizon. We will comment more on this issue in \cite{souvikprashant}.

\subsection{A  semi-classical obstruction to state-independence}
Given that all existing examples of extracting local physics from the boundary involve various measurables, which are nevertheless not linear operators, why should we expect that the metric is given by an ordinary operator in the CFT? More precisely, what is the basis for the the naive expectation that operators satisfying \eqref{linearopmetric} and \eqref{quasilocalphicond} should exist in the CFT? In this subsection, we try and explain the basis for this naive expectation, although,  as we will point out immediately, we believe that this intuition is flawed. 

For simplicity, we will consider whether one should expect a state-independent metric operator $\cop{g}_{\mu \nu}(\vect{x})$ to exist. A similar argument applies to other light fields in the theory.

The key point is that the classical metric $g_{\mu \nu}(\vect{x})$ is a  well defined functions on the classical phase space of the theory. Recall that the classical phase space can be put in 1--1 correspondence with the set of all classical solutions of the theory. Given initial data for the canonical variables, and their conjugate momenta, we can evolve it forward to generate the entire classical solution. Conversely, given a classical solution, we can take a ``section'' by evaluating the variables and their momenta at some point of time to obtain a point on the phase space. 

As we have explained above, once we go to a well defined gauge, the value of the metric $g_{\mu \nu}(\vect{x})$ is well defined in  any classical geometry. Therefore the metric is a well defined function on the phase space of the the theory.  Now, one usually expects that quantization takes functions on the phase space  to well defined operators in the Hilbert space. Therefore one might expect the metric $g_{\mu \nu}(\vect{x})$ in relational gauge to lift to a state-independent operator in the theory.

As we review in Appendix \ref{appcoherentgravity}, this is usually done as follows. In the quantum theory, we obtain coherent states, $|g \rangle$ corresponding to each semi-classical geometry. We then lift the classical function to an operator through
\be
\label{semiclasstateindg}
\cop{g}_{\mu \nu}(\vect{x}) \sim \sum_g g_{\mu \nu}(\vect{x}) |g \rangle \langle g|,
\ee
where the sum is over all metrics, discretized in some fashion.\footnote{For a concrete example of a formula of this sort, the reader may wish to look at \eqref{phistateind} although we caution the reader that \eqref{phistateind} sums only over spherically symmetric metrics and works only outside the horizon. In contrast, we would like \eqref{semiclasstateindg} to work for all kinds of metrics, and both inside and outside the horizon.}

Now, the analysis of Appendix \ref{appcoherentgravity} and section \ref{seceternal} shows that for such a construction to work, it is very important that if we consider the inner product of two distinct geometries, it dies off to arbitrarily small values 
\be
\langle g_1 | g_2 \rangle  =  e^{-\nc \distphase(g_1, g_2)}.
\ee
We can compute the function $\distphase$ on the right hand side in linearized gravity but in order for \eqref{semiclasstateindg} to converge we require that  for sufficiently ``distinct'' $g_1, g_2$, we can have $v \gg 1$. 

On the other hand, in the CFT, as we have discussed coherent states of the metric $|g \rangle$ correspond to CFT states $|\psidual{g} \rangle $. However, for generic states at the same energy $E \propto \nc$, we have
\be
\langle \psidual{g_1} | \psidual{g_2} \rangle = \Or[e^{-{S \over 2}}],
\ee
where $S \propto \nc$ is the thermodynamic entropy of the CFT at the energy $E$. 

This ``fat tail'' in the inner product of coherent states in the CFT subtly violates the expectation from a semi-classical quantization of gravity.\footnote{This is reminiscent of the fact \cite{Maldacena:2001kr} that thermal correlators in the CFT decay down to $\Or[e^{-{S}}]$, in contrast to the naive expectation from semi-classical gravity that the exponential decay in time should continue forever.} As a result of this tail, we cannot write down an expression of the form \eqref{semiclasstateindg} with the putative coherent states replaced by $|\psidual{g} \rangle$ because interference from ``distant'' microstates implies that the operator $\cop{g}_{\mu \nu}(\vect{x})$ on the left of \eqref{semiclasstateindg}  does not behave like  the classical function $g_{\mu \nu}(\vect{x})$. 

We direct the reader to section \ref{seceternal} for an example where this can be seen very clearly. In Appendix \ref{appcoherentgravity} we discuss the single-sided case in more detail and describe why we believe that the same obstruction prevents one from writing down state-independent operators for well defined classical geometric quantities.


\section{Local bulk operators in AdS/CFT: Conditions for a smooth interior }\label{secsmoothint}
In this section, we review the conditions that are required to obtain 
a smooth exterior and interior geometry for a black hole in AdS/CFT. 
The central point that we would like to emphasize in this section is that a smooth interior requires the existence of operators
in the CFT, with specific properties that we enumerate below. We have dealt with this question in our previous papers \cite{Papadodimas:2012aq,Papadodimas:2013jku,Papadodimas:2013wnh}, but we present
a slightly new perspective here to buttress the same conclusion. 

Before we proceed to the analysis, we briefly state our result and emphasize
the difference with previous derivations. Consider a black hole horizon, which may have been formed due to gravitational collapse or may be part of an eternal black hole. If we quantize a field on both sides of the horizon, we find that while the Schwarzschild left movers cross the horizon smoothly, the Schwarzschild right movers do not. The claim is that to obtain a smooth horizon, we must find new operators, which play the role of right movers behind the horizon, and are appropriately entangled with the right movers in front of the horizon.

There are various ways to reach this conclusion. These right movers were identified in Hawking's original analysis of this question  as modes from past null infinity that are concentrated in the time, just after the last null ray to escape the horizon. In Hawking's geometric analysis, these modes ``bounce'' from $r = 0$ to play the role of right movers behind the horizon. 
One can also argue for the existence of these right movers  and the appropriate entanglement --- as we did in \cite{Papadodimas:2012aq} --- by using the semi-classical intuition that, at late times, the collapsing geometry approaches the eternal black hole where these right movers originate from a left asymptotic region, which we call ``region III''.
\begin{figure}
\begin{center}
\begin{subfigure}[t]{0.2\textwidth}
\includegraphics[height=0.3\textheight]{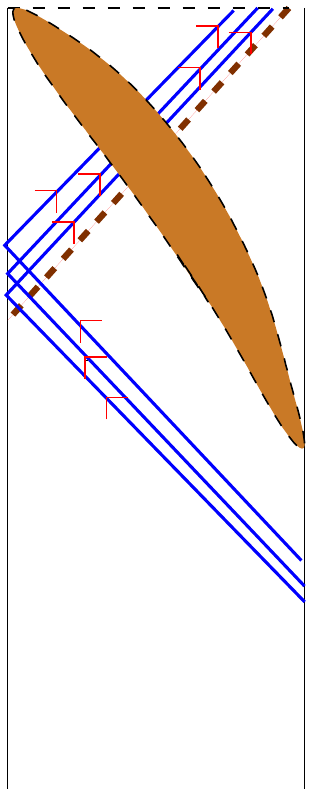}
\caption{Hawking's derivation}
\end{subfigure}
\qquad \qquad \qquad
\begin{subfigure}[t]{0.3\textwidth}
\includegraphics[height=0.3\textheight]{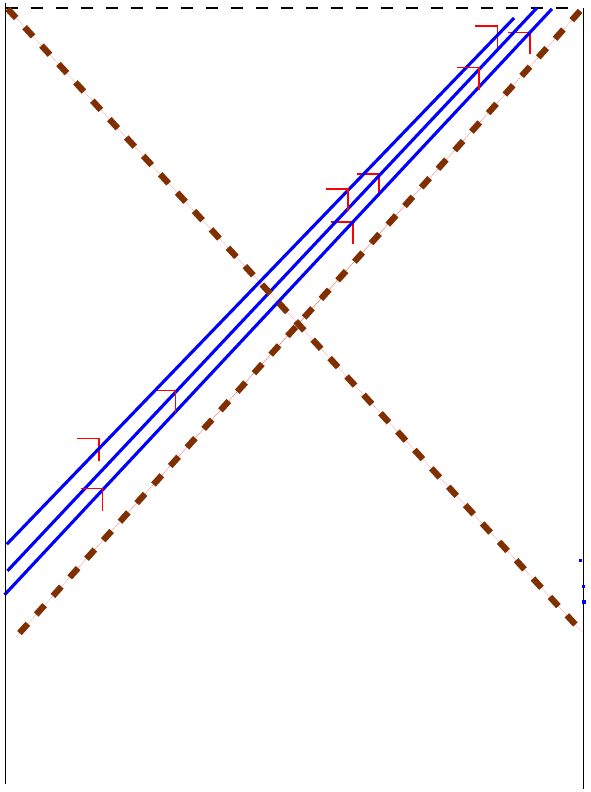}
\caption{The analogy to  the Eternal Black hole}
\end{subfigure}
\caption{Two ways of arguing that new right movers are necessary behind a black hole horizon. Hawking's original derivation on the left, where the right movers are modes that have ``bounced'' off $r=0$ and propagated through the infalling matter. The analogy to the eternal black hole on the right, where the right movers come from a left asymptotic region. Both of these suffer from difficulties, and so we perform a purely local derivation leading to the same result. \label{figmirrorglobal} }
\end{center}
\end{figure}

These derivations suffer from certain difficulties. Hawking's original work 
has a trans-Planckian problem because tracing these modes back to past null infinity boosts them to very high energies. Similarly, the intuition that
these modes come from an effective ``region ${\rm III}$'' is somewhat confusing because
we do not expect any such region to exist for a collapsing geometry.

\begin{figure}[!h]
\begin{center}
\resizebox{0.4\textwidth}{!}{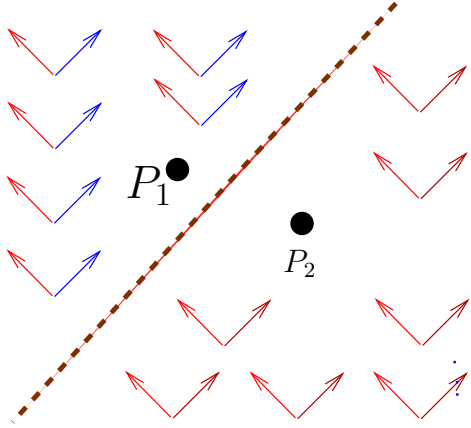}
\caption{We derive  the necessity of  new modes just  by demanding a regular two point function for points $P_1, P_2$ across  the horizon without invoking another asymptotic region or tracing  these modes  back into the past.}\label{figmirrorlocal}
\end{center}
\end{figure}

To solve these problems, in this section, we will perform a purely {\em local derivation} that reveals the necessity of the existence of appropriate entangled right-moving modes behind the horizon. Our  picture in  this paper is  shown in Figure \ref{figmirrorlocal}. We start with the sole assumption that the field in the near-horizon region outside and inside the horizon has an  effective perturbative description.
This assumption implies the universality of a certain two point function. By Fourier transforming this universal two point function, we infer that the
right movers behind the horizon must exist, and also infer their two point functions with modes in front of the horizon. 
We start by performing this analysis in the bulk, and then discuss the implications in the CFT.

\subsection{Bulk analysis of the mirror operators }\label{subsecbulkmirror}
Let us start from the bulk perspective. We will then examine how this must
be translated to the boundary. For simplicity, let us consider a massless 
scalar field in the bulk. This analysis carries over, almost entirely unchanged to the case of the graviton and other fields. 

\begin{figure}[!h]
\begin{center}
\includegraphics[height=0.3\textheight]{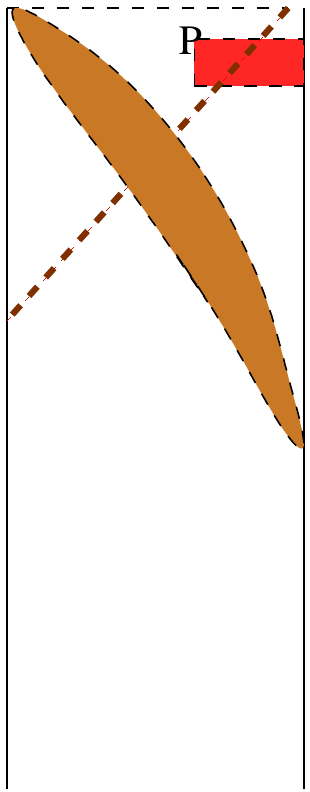}
\caption{We are interested in the late-time physics of the black hole geometry, schematically denoted by the rectangular patch $P$ above. }\label{figinteresting}
\end{center}
\end{figure}

Consider a big black hole in AdS. In the past this black hole could have been formed from the collapse of a star or some other physical process. However, we are interested in the late time region shown schematically as the rectangular patch  $P$ in Figure \ref{figinteresting}. This patch of spacetime overlaps with the region both in front of, and behind, the horizon. Classically, we expect that the initial collapsing matter, and any perturbations have died away and are irrelevant for physics in this region. In the analysis below, we will assume the validity of this classical expectation and derive various results for correlators of fields. Later we will need to check the consistency of these results by ensuring that it is possible to construct a bulk to boundary map that reproduces these correlators.

\paragraph{Geometry \\}
The metric, at late times, outside the horizon is given by
\begin{equation}
 \label{schwarzschild}
 ds^2 = -f(r) dt^2 + {1\over f(r)}dr^2 + r^2 d\Omega_{d-1}^2,
\end{equation}
where 
\be
\begin{split}
f(r) &= {r^2 } + 1 -c_d{ GM \over \,\,r^{d-2}},\\
c_d &= {8 \pi^{{2-d}\over 2} \Gamma(d/2)\over d-1}.
\end{split}
\ee
The numerical constant, $c_d$, arises from the volume of the $d-1$ dimensional sphere. and we have set the radius of AdS to $1$. 

The horizon is defined implicitly, by the equation $f(\rhor) = 0$. As usual, it is convenient to introduce tortoise coordinates by 
${d r_* \over d r} = f^{-1}(r).$ Unlike in the case of the Schwarzschild black hole in flat space, we cannot usually express the tortoise coordinates in terms of the original coordinates using elementary functions.  But we can choose the differential equation to satisfy
\be
r_* =0, ~\text{at}~r =\infty.
\ee
As $r \rightarrow \rhor$, we see that $f^{-1}(r)$ diverges and $r_* \rightarrow -\infty$. In order to approach the future
horizon we have to take the limit $r_* \rightarrow -\infty$ and at the same time $t \rightarrow +\infty$. 

We introduce the following coordinates
\be
\label{kruskaldef}
U = -e^{{2 \pi\over \beta} (\rtor - t) }; \quad V = e^{{2 \pi \over \beta}(\rtor + t) }.
\ee
The horizon is given by $U = 0$, but with $V$ finite. We can check that
with the factors of ${2 \pi \over \beta}$, the horizon is smooth in the $U,V$ coordinate system. Near the horizon, with $(r - \rhor) \ll 1$, we have $f(r) = \kappa(r - \rhor)$.  The constant $\kappa$ is related to the temperature. A shortcut to determine the relation is to continue to Euclidean time, $t \rightarrow i \tau$, identify $\tau \sim \tau + \beta$ and make the change of variables $x = 2 \sqrt{r - \rhor \over \kappa}$. Near the horizon, the analytically continued metric then takes the form 
\be
d s_E^2 \underset{x\rightarrow0}{\longrightarrow} d x^2 + {\kappa^2 \over 4} x^2 d \tau^2 + \rhor^2 d \Omega_{d-1}^2.
\ee
For the Euclidean circle $\tau$ to smoothly cap off at $x = 0$, we require ${\kappa^2 \beta^2 \over 4} = (2 \pi)^2$ or $\kappa = {4 \pi \over \beta}$. 

In the near horizon region, we now have the following relations
\be
f(r) = {4 \pi \over \beta} (r - \rhor), \quad \Rightarrow  r_* = {\beta \over 4 \pi}  \ln({r - \rhor \over \rhor}) + \text{const}.
\ee
From here, it follows that $f(r) \underset{r_* \rightarrow \infty}{\longrightarrow} \kappa' \left({2 \pi \over \beta}\right)^2 e^{{4 \pi r_* \over \beta}}$, where $\kappa'$ is another (irrelevant) constant. 

In Kruskal coordinates the metric takes the form
\be
d s^2 = \left({\beta \over 2 \pi} \right)^2 {f(r) \over U V} d U d V + r^2 d \Omega_{d-1}^2,
\ee
and we see that the factor of ${1 \over U V}$ precisely cancels off the growing exponential in $f(r)$ near the horizon to ensure that the metric is regular. 
\be
g_{\mu \nu} \underset{U \rightarrow 0}{\longrightarrow} -\kappa' d U d V + \rhor^2 d \Omega_{d-1}^2.
\ee

After we cross the horizon, we can introduce a second Schwarzschild patch. Since $U > 0$ in the region inside the black hole (which we sometimes also call region II), we write
\be
U = e^{{2 \pi\over \beta} (\rtor - t) }; \quad V = e^{{2 \pi \over \beta}(\rtor + t) }, \quad \text{in~region~II.}
\ee
Inside the horizon, the tortoise coordinate, $\rtor$, rises from its value of $-\infty$ at the horizon, while the Schwarzschild time decreases from its values of $\infty$ as one goes from right to left.

\paragraph{Two-point scalar correlators\\}
Now, we will consider a massless scalar field propagating in this background. We will define this field using the relational prescription of section \ref{subsecrelational}. We  derive various consequences of the fact that the horizon is smooth, simply by demanding that the two point function both outside and inside the horizon be smooth.

We expect that the two-point scalar function has the form
\be
\langle \phi({\vect{x}_1}) \phi({\vect{x}_2}) \rangle = G(\vect{x}_1, \vect{x}_2) + \Or[{1 \over \nc}].
\ee
We will be interested in the regime where $\vect{x}_1$ and $\vect{x}_2$ approach the light cone, but always remain spacelike with respect to each other. In this regime the Wightman and the time-ordered Green functions coincide and so we will not have to keep track of factors of $i \epsilon$.
In the expression above, we have also used the fact that corrections to this expression come from interactions that are suppressed by $1/\nc$. However, we will not need need the full form of the propagator. For a large black hole, provided that the geodesic distance $\ell_{12}$ between $\vect{x_1}$ and $\vect{x_2}$ is small in comparison to the scale of curvature $\ell_{12} \ll {1 \over \beta}$, we expect that
\be
\label{shortdistancetwopt}
\langle \phi({\vect{x_1}}) \phi({\vect{x_2}}) \rangle \approx {1 \over \Big[g^{\mu \nu}(x_1 - x_2)_{\mu} (x_1 - x_2)_{\nu} \Big]^{d-1 \over 2}}, \quad |\ell_{1 2}| \ll \beta^{-1}.
\ee
Recall that the dimension of the bulk theory is $d+1$.  The exponent above is the engineering dimension of the field, which is  ${(d+1) - 2 \over 2}$. The relation \eqref{shortdistancetwopt} above is a powerful constraint, which holds in the short distance limit for any field theory in the bulk that is controlled by a free ultra-violet fixed point\footnote{Of course here we are talking about the intermediate regime, where $\ell_{12}\ll \beta^{-1}$ but at the same time $\ell_{12}\gg l_p, l_s$ where the latter are the Planck and string scales in the bulk.}.

Now we consider the correlation function as one point approaches the light cone of the other in the U-V plane.\footnote{As we see below, to take this limit for correlators of the scalar itself is delicate, as a result of the usual complications of dealing with a massless scalar in  two  dimensions. This is the reason for taking correlators of its derivatives instead.} We will work in the regime where the two points are separated on this plane so that $-(U_1 - U_2)(V_1 - V_2) > 0$.

\be
\begin{split}
&\langle \partial_{V_1} \phi(U_1, V_1, \Omega_1) \partial_{V_2} \phi(U_2,V_2, \Omega_2) \rangle = \partial_{V_1} \partial_{V_2} {1 \over \left(- \kappa' (U_1 - U_2) (V_1 - V_2) + \Omega_{12}^2 \right)^{d-1 \over 2}} \\ &= {(d + 1) (d-1) \over 4} (\kappa')^2{(U_1 - U_2)^2 \over \left( -\kappa' (U_1 - U_2) (V_1 - V_2) + \Omega_{12}^2 \right)^{d+3 \over 2}},
\end{split}
\ee
where $\Omega_{12}^2$ is defined as the distance between the points $\Omega_1$ and $\Omega_2$ on the sphere of radius $\rhor$.
We will argue that this two-point function is actually proportional to a delta function in the coordinates on the sphere, as we take $U_1, U_2 \rightarrow 0$. If the transverse space had been planar, this would have been a planar delta function. 

First note that we clearly have that
\be
\lim_{U_1, U_2 \rightarrow 0} { (U_1 - U_2)^2 \over \left(- (U_1 - U_2) (V_1 - V_2) + \Omega_{12}^2 \right)^{d + 3 \over 2}} = 0, \quad \text{for~} \Omega_1 \neq \Omega_2.
\ee
But on the other hand, let us consider
\be
I(U_1 - U_2, V_1 - V_2) = \int d^{d-1} \Omega_2 {  (U_1 - U_2)^2 \over \left(  -\kappa' (U_1 - U_2) (V_1 - V_2) + \Omega_{12}^2 \right)^{d + 3 \over 2}}.
\ee
The integral above is on a sphere of radius $\rhor$, but we can rescale the sphere by introducing a new variable
$\Omega_2' = {\Omega_2 \over (\kappa' \delta)^{1 \over 2}}$ with $\delta   \equiv -(U_1 - U_2)(V_1 - V_2)$.
\be
\begin{split}
I(U_1 - U_2, V_1 - V_2) &= \int \left[{(\kappa' \delta)^{d-1 \over 2} (U_1 - U_2)^2 \over (\kappa' \delta)^{d+3 \over 2} \left( 1  + {\Omega_{12}^2 \over \kappa' \delta } \right)^{d + 3 \over 2}} d^{d-1} \Omega_2' \right] \\ &= {1 \over (\kappa')^2 (V_1 - V_2)^2} \int {d^{d-1} \Omega_{2}'  \over (1 + (\Omega_{12}')^2)^{d+3 \over 2}}. 
\end{split}
\ee
The final integral is clearly a constant independent of $\Omega_1$. 
This leads to the conclusion that
\be
\lim_{U_1 - U_2 \rightarrow 0} \langle \partial_{V_1} \phi(U_1, V_1, \Omega_1) \partial_{V_2} \phi(U_2,V_2, \Omega_2) = \normdelta {1 \over (V_1 - V_2)^2} \delta^{d-1}(\Omega_1 - \Omega_2),
\ee
where $\normdelta$ is a normalization constant that we will not fix here.
In the same way, we also have
\be
\label{universalshortdistance}
\lim_{V_1 - V_2 \rightarrow 0} \langle \partial_{U_1} \phi(U_1, V_1, \Omega_1) \partial_{U_2} \phi(U_2,V_2, \Omega_2) = \normdelta {1 \over (U_1 - U_2)^2} \delta^{d-1}(\Omega_1 - \Omega_2).
\ee
This is a powerful and broadly applicable result. The ultra-locality that we see in the transverse directions was also noted and used in the papers \cite{Wall:2010cj,Bousso:2014sda}.

Now, let us see what this result implies for the correlation functions of the Schwarzschild creation and annihilation operators. Consider again the region near the horizon of a black hole, but this time in the original time and tortoise coordinates. Outside the horizon, we have the expansion
\be
\label{phiexpansionoutside}
\phi(t,\rtor, \Omega) \underset{U \rightarrow 0^-}{\longrightarrow}  \sum_{\ang} \int_0^{\infty} {d \omega \over \sqrt{\omega}} a_{\omega, \ang} e^{-i \omega t} Y_{\ang}(\Omega)\left( e^{i \delta} e^{i \omega \rtor} + e^{-i \delta} e^{-i \omega \rtor} \right) + \text{h.c},
\ee
where $Y_{\ang}(\Omega)$ are spherical harmonics that we normalize below. 
The left and right movers get related to each other, and the phases $\delta$ depend on scattering in the black hole geometry \cite{Papadodimas:2012aq}. As we noted above, and will see again below, we can only use \eqref{phiexpansionoutside} for correlators of derivatives of the field.

Note that the canonical conjugate to the field outside the horizon is 
\be
\pi(t, \rtor, \Omega) = g^{tt} \sqrt{-g} {\partial \over \partial t} \phi(t, \rtor, \Omega) = r^{d-1}  {\partial \over \partial t} \phi(t, \rtor, \Omega).
\ee
We must impose the canonical commutation relations 
\be
[\phi(t, \rtors[1], \Omega_1), {\partial \over \partial t} \phi(t, \rtors[2], \Omega_2)] = {i \over r^{d-1}} \delta(\rtors[1] - \rtors[2]) \delta^{d-1}(\Omega_1 - \Omega_2).
\ee
Since the modes take this plane wave form in the near horizon region, as $r \rightarrow \rhor$, by imposing these commutation relation we find
that they are satisfied only if
\be
[a_{\omega, \ang}, a_{\omega', \ang'}^{\dagger}] = \delta(\omega - \omega') \delta_{\ang \ang'},
\ee
provided that we normalize the spherical harmonics by
\be
\sum_{\ang} Y_{\ang}(\Omega) Y_{\ang}^*(\Omega') = {1 \over 4 \pi \rhor^{d-1}} \delta^{d-1}(\Omega-\Omega').
\ee
Now the two point function, with both points outside the horizon but close to it, is given by
\be
\label{twoptfnexpand}
\begin{split}
&\langle \partial_{U_1} \phi(U_1, V_1,  \Omega_1) \partial_{U_2} \phi(U_2, V_2, \Omega_2) \rangle  = {\beta^2 \over 4 \pi^2 U_1 U_2}  \\ &\times \sum_{\ang} \int_0^{\infty} { \omega d \omega } \Bigg[(N_{\omega, \ang} +1)Y_{\ang}(\Omega_1) Y_{\ang}^*(\Omega_2) \left({U_1 \over U_2} \right)^{{i \beta \omega \over 2 \pi}}+ 
 N_{\omega, \ang}  Y_{\ang}(\Omega_1)^* Y_{\ang}(\Omega_2) \left({U_2 \over U_1} \right)^{{i \beta \omega \over 2 \pi}} \Bigg].
\end{split}
\ee
Here we have defined the two point expectation value 
\be
\langle a^{\dagger}_{\omega, \ang} a_{\omega' \ang'} \rangle = N_{\omega, \ang} \delta(\omega - \omega') \delta_{\ang, \ang'},
\ee
in the black hole state and assumed that it is proportional to a delta function which is reasonable at late times when nothing depends on the time or the angular position. 

Note that the expansion in two point function \eqref{twoptfnexpand} would not have converged without the derivatives on $U_1, U_2$. These derivatives pull down two factors of $\omega$ and ensure that the integrand is well behaved at $\omega = 0$. Now we will show that we must have
\be
N_{\omega, \ang} = {e^{-\beta \omega} \over 1 - e^{-\beta \omega}}.
\ee
To see this, note that
\be
\int_0^{\infty} {\omega d \omega} \left({e^{-\beta \omega}  \over 1 - e^{-\beta \omega}} \left({U_2 \over U_1} \right)^{{i \beta \omega \over 2 \pi}}  + {1  \over 1 - e^{-\beta \omega}}  \left({U_1 \over U_2} \right)^{i \beta \omega \over 2 \pi} \right) = \int_{-\infty}^{\infty}  {\omega d \omega} {e^{-\beta \omega}  \over 1 - e^{-\beta \omega}} \left({U_2 \over U_1} \right)^{i \beta \omega \over 2 \pi}. 
\ee
This integral can be completed in the lower half plane if $|U_1| > |U_2|$ and in the upper half plane otherwise. Picking up the poles at $\omega = {2 \pi i n \over \beta}$, we find that this integral evaluates to
\be
\int_{-\infty}^{\infty}  {\omega d \omega} {e^{-\beta \omega}  \over 1 - e^{-\beta \omega}} \left({U_2 \over U_1} \right)^{i \beta \omega \over 2 \pi} = 
 -{1 \over \beta} \sum_n n \left({U_2 \over U_1} \right)^{n} = -{U_1 U_2 \over \beta (U_1 - U_2)^2}.
\ee
Second note that the sum over $\ang$ in \eqref{twoptfnexpand} automatically leads to a delta function proportional to $\delta^{d-1}(\Omega_1 - \Omega_2)$. From the results above, we therefore find that \eqref{twoptfnexpand} and \eqref{universalshortdistance} coincide provided that
\be
\label{twoptfncondone}
\begin{split}
&\langle a_{\omega, \ang} a^{\dagger}_{\omega', \ang'} \rangle = {1 \over 1 - e^{-\beta \omega}} \delta(\omega - \omega') \delta_{m m'}, \\
&\langle a^{\dagger}_{\omega, \ang} a_{\omega', \ang'} \rangle = {e^{-\beta \omega} \over 1 - e^{-\beta \omega}} \delta(\omega - \omega') \delta_{m m'}.
\end{split}
\ee
Two caveats are in order. Note that \eqref{universalshortdistance} was derived in the near-horizon limit where $U_1, U_2 \rightarrow 0$ and therefore our derivation above for the value of $N_{\omega, \ang}$ is not valid for low frequencies $\omega \ll {1 \over \beta}$. It is also not valid for Planckian frequencies $\omega = \Or[\nc]$, where we do not expect effective field theory to give reliable results.

We now turn to the expansion behind the horizon. Here, as we quantize the field in region II, and approach the horizon from inside, we find an expansion. 
\be
\label{phiexpansioninside}
\phi(t,\rtor, \Omega) \underset{U \rightarrow 0^+}{\longrightarrow}  \sum_{\ang} \int_0^{\infty} {d \omega \over \sqrt{\omega}} \left(a_{\omega, \ang} e^{-i \delta} e^{-i \omega (t+\rtor)} Y_{\ang}(\Omega) + \widetilde{a}_{\omega, \ang} e^{-i \delta} e^{i \omega (t - \rtor)} Y_{\ang}^*(\Omega) \right) + \text{h.c}.
\ee
Several points are worth noting in \eqref{phiexpansioninside}. 
\begin{enumerate}
\item
By continuity of the mode $e^{i \omega(t + \rtor)} = V^{i \beta \omega \over 2 \pi}$, the operators $a$ in region II must be the same as the operators in region I. 
\item
Second we need some operators to multiply the right moving modes that vary as $e^{i \omega (t -\rtor)}$. In \eqref{phiexpansionoutside} we identified these modes with $a_{\omega, \ang}$, but we will find that this cannot be correct here. We will call the $\widetilde{a}_{\omega, \ang}$ operators the {\em mirror operators.}
\item
Note that the timelike coordinate inside the black hole is $\rtor$. Therefore, the operator multiplying $e^{i \omega(t - \rtor)}$ is classified as an ``annihilation'' operator. This is in spite of the fact that it has positive frequency with respect to $t$; the relevant point is that
it has negative frequency with respect to $\rtor$. 
\item
Note that we have also conjugated the spherical harmonic $Y_m$ for this mode. This is just a matter of choosing a convenient convention. 
\end{enumerate}
Inside the horizon, the canonical conjugate to the field is given by
\be
\pi(t, \rtor, \Omega) = g^{\rtor \rtor} \sqrt{-g} {\partial \over \partial \rtor} \phi(t, \rtor, \Omega) = r^{d-1}  {\partial \over \partial \rtor} \phi(t, \rtor, \Omega).
\ee
The canonical commutation relations are
\be
[\phi(t_1, \rtor, \Omega_1), {\partial \over \partial \rtor} \phi(t_2, \rtor, \Omega_2)] = {i \over r^{d-1}} \delta(t_1 - t_2) \delta^{d-1}(\Omega_1 - \Omega_2).
\ee
By repeating the analysis of the canonical commutation relations we find that
\be
[\widetilde{a}_{\omega, \ang}, \widetilde{a}^{\dagger}_{\omega', \ang'}] = \delta(\omega - \omega')\delta_{\ang \ang'},
\ee
where we have tacitly assumed that the possible mixed commutator $[\widetilde{a}_{\omega, \ang}, a^{\dagger}_{\omega', \ang'}]$ vanishes. The mirror annihilation operator $\widetilde{a}_{\omega, \ang}$ and the ordinary creation operator $a^{\dagger}_{\omega, \ang}$ have the same energy under the CFT Hamiltonian as we show in \eqref{commuthamilt}. So in a state that is time-translationally invariant, we do not expect this commutator to have a non-zero expectation value.\footnote{This assumption of time-translational invariance on the boundary is not true in some cases, like in the geon geometry considered in \cite{Guica:2014dfa} where the mirror operators can be identified with the ordinary ones.}

We now consider a two point function with one point in front of the horizon, and another point behind the horizon. This calls into play both the expansions \eqref{phiexpansionoutside} and \eqref{phiexpansioninside}. Recalling the fact that, the relation between the Kruskal and Schwarzschild coordinates inside and outside the horizon differs by a minus sign, and repeating the derivation above for this case, we find that
\be
\label{mixedexpand}
\langle \partial_{U_1} \phi(U_1, V_1,  \Omega_1) \partial_{U_2} \phi(U_2, V_2, \Omega_2) \rangle  = {\beta^2 \over 4 \pi^2 U_1 U_2} \sum_{\ang, \ang'} \int_0^{\infty} {\omega^{1 \over 2} d \omega}{(\omega')^{1 \over 2} d \omega'}  {\cal I}_{\omega, \omega', \ang, \ang'} ,
\ee
with
\be
\label{mixedexpandintegrand}
\begin{split}
{\cal I}_{\omega, \omega', \ang, \ang'}  &\equiv \langle a_{\omega, \ang} \widetilde{a}_{\omega', \ang'} \rangle Y_{\ang}(\Omega_1) Y_{\ang'}^*(\Omega_2) (-U_1)^{i \beta \omega \over 2 \pi} (U_2)^{-i \beta \omega' \over 2 \pi}  \\ &+  \langle a_{\omega, \ang} \widetilde{a}^{\dagger}_{\omega', \ang'} \rangle  (-U_1)^{i \beta \omega \over 2 \pi} (U_2)^{i \beta \omega' \over 2 \pi}  Y_{\ang}(\Omega_1) Y_{\ang'}(\Omega_2)  + \text{h.c.}
\end{split}
\ee
Note that the result \eqref{universalshortdistance} is valid regardless of whether the points are on opposite sides, or the same side of the horizon. Now we find, repeating the contour integral argument above that \eqref{mixedexpand} agrees with \eqref{universalshortdistance} only if the 
two point function between the two annihilation operators (and the two creation operators) is non-zero, whereas the mixed two-point function vanishes.
\be
\label{twoptfncondtwo}
\begin{split}
&\langle a_{\omega, \ang} \widetilde{a}_{\omega', \ang'} \rangle = {e^{-{\beta \omega \over 2}} \over 1 - e^{-\beta \omega}} \delta(\omega - \omega') \delta_{m m'}; \quad \langle a_{\omega, \ang} \widetilde{a}^{\dagger}_{\omega', \ang'} \rangle = 0, \\
&\langle a^{\dagger}_{\omega, \ang} \widetilde{a}^{\dagger}_{\omega', \ang'} \rangle = {e^{-{\beta \omega \over 2}} \over 1 - e^{-\beta \omega}} \delta(\omega - \omega') \delta_{m m'}; \quad \langle a^{\dagger}_{\omega, \ang} \widetilde{a}_{\omega', \ang'} \rangle = 0. \\
\end{split}
\ee
The additional factor of $e^{-{\beta \omega \over 2}}$ arises because of the relative minus sign between $U_1$ and  $U_2$ in \eqref{mixedexpandintegrand}.

We can also consider the case where both points are inside the black hole. This is very similar to the cases above, so we will just state the result. The smoothness of the two point function of $\phi$ requires
\be
\label{twoptfncondthree}
\begin{split}
&\langle \widetilde{a}_{\omega, \ang} \widetilde{a}^{\dagger}_{\omega', \ang'} \rangle = {1 \over 1 - e^{-\beta \omega}} \delta(\omega - \omega') \delta_{m m'}, \\
&\langle \widetilde{a}^{\dagger}_{\omega, \ang} \widetilde{a}_{\omega', \ang'} \rangle = {e^{-\beta \omega} \over 1 - e^{-\beta \omega}} \delta(\omega - \omega') \delta_{m m'}.
\end{split}
\ee

Finally, recall from the discussion of section \ref{subsecrelational} that relationally defined observables in the bulk must obey the Heisenberg equations of motion. Consider a bulk point obtained considering a geodesic that originates on the boundary at point $(t_b, \Omega_b)$, with no initial velocity along the sphere, and following it for an affine parameter $\lambda$. In \eqref{affinepoints}, this  this point was denoted by $P_{\lambda}(t_b, \Omega_b, \lambda)$. By solving the geodesic equation in the metric given by \eqref{schwarzschild}, we can trade these coordinates for Schwarzschild coordinates.
\be
P_{\lambda}(t_b, \Omega_b, \lambda) = (t, \Omega, \rtor).
\ee
Then it is easy to check that the isometry of the metric under time-translations implies that if we follow another geodesic that originates at $t_b + \capt$, then
\be
\label{shiftedtimerel}
P_{\lambda}(t_b + \capt, \Omega_b, \lambda) = (t + \capt, \Omega, \rtor).
\ee
The relation \eqref{shiftedtimerel} holds for points both outside and inside the horizon. In terms of the field this means that for the field written in Schwarzschild coordinates,
\be
\label{timetranslationlocal}
e^{i H \capt} \phi(t, \rtor, \Omega ) e^{-i H \capt}  = \phi(t + T,  \rtor, \Omega),
\ee
where $H$ is the boundary Hamiltonian that translates times on the boundary.
This translates into the following commutation relations for the modes introduced above
\be
\label{commuthamilt}
\begin{split}
&[H, a_{\omega, \ang}] = -\omega\, a_{\omega, \ang}; \quad [H, a^{\dagger}_{\omega, \ang}] = \omega \,a^{\dagger}_{\omega, \ang}, \\
&[H, \widetilde{a}_{\omega, \ang}] = \omega \,\widetilde{a}_{\omega, \ang}; \quad [H, \widetilde{a}^{\dagger}_{\omega, \ang}] = -\omega\, \widetilde{a}^{\dagger}_{\omega, \ang}. \\
\end{split}
\ee
Note the opposite signs in the two lines of \eqref{commuthamilt}. This is a result of the fact that we mentioned above --- the operator $\widetilde{a}_{\omega, \ang}$ multiplies a mode that is positive frequency with respect to the Schwarzschild time.

\paragraph{Summary\\}
In this section we considered a scalar field propagating in the geometry of a Schwarzschild black hole. By simply imposing the requirement
that the two point function had the correct short distance behaviour we were able to derive necessary conditions on the two point functions of the modes of the field in the black hole state. These conditions are given by \eqref{twoptfncondone}, \eqref{twoptfncondtwo} and \eqref{twoptfncondthree}. If the field is defined relationally with respect to the boundary, then the modes must also have the Hamiltonian commutators \eqref{commuthamilt}. 

In the CFT we must find operators that satisfy these conditions in any state that is dual to a smooth geometry.

\subsection{Local operators in the CFT }\label{subseclocalopcft}
Let us now understand what the analysis above implies for the CFT. As discussed in section \ref{secstatedepvsindep}, we would like
a family of operators in the CFT, parameterized by a set of real numbers, $\cop{\phi}(U, V, \Omega)$, so that the correlation functions of these operators reproduce the correlators of a perturbative field in AdS. In this subsection, we discuss how to find such correlators outside the horizon. We turn to the issue of the nature of these operators inside the horizon in section \ref{secparadoxes}. 

\subsubsection{Local operators outside the horizon}
For the CFT to successfully reproduce effective field theory correlators outside the horizon, it must  have operators which play the role of the modes $a_{\omega, \ang}$ that we encountered in \eqref{phiexpansionoutside}. If we allow ourselves to use state-dependent operators, then this can be done in a straightforward way, as we show below. 

Dual to each propagating field in the bulk, we have a generalized free field (GFF), $\op$ on the boundary --- usually it is a single trace operator in a gauge theory. The fact that bulk correlators factorize because the bulk theory is perturbative is reflected in the large-N factorization of boundary correlators.
 When evaluated in the vacuum,
\be
\begin{split}
\label{factorization}
&\lvac {\op}(t_1, \Omega_1) \ldots  {\op}(t_{2 n}, \Omega_{2 n}) \rvac \\ &= {1 \over 2^n} \sum_{\pi} \lvac{\op}(t_{\pi_1}, \Omega_{\pi_1}) {\op}(t_{\pi_2}, \Omega_{\pi_2}) \rvac \ldots  \lvac {\op}(t_{\pi_{2 n - 1}}, \Omega_{\pi_{2 n - 1}})  {\op}(t_{\pi_{2 n}}, \Omega_{\pi_{2 n}}) \rvac \\ &+ \Or[{1 \over \nc}],
\end{split}
\ee
where $\pi$ sums over all possible permutations.
A similar relation holds for thermal correlators.  
\be
\begin{split}
\label{factorizationthermal}
{1\over Z(\beta)}\tr\Big[e^{-\beta H} &{\op}(t_1, \Omega_1) \ldots  {\op}(t_{2 n}, \Omega_{2 n}) \Big] \\ = {1 \over 2^n} \sum_{\pi} \Bigg(&{1\over Z(\beta)}
\tr \left[e^{-\beta H} {\op}(t_{\pi_1}, \Omega_{\pi_1}) {\op}(t_{\pi_2}, \Omega_{\pi_2}) \right] \ldots \\ \times  &{1\over Z(\beta)}\tr \left[e^{-\beta H}  {\op}(t_{\pi_{2 n - 1}}, \Omega_{\pi_{2 n - 1}})  {\op}(t_{\pi_{2 n}}, \Omega_{\pi_{2 n}}) \right] \Bigg) \\ &+ \Or[{1 \over \nc}],
\end{split}
\ee
Note that \eqref{factorizationthermal} is subtly different from \eqref{factorization} and does not follow from it directly. In particular, in \eqref{factorizationthermal}, the thermal two point functions have {\em already} re-summed the ${1 \over \nc}$ series about the vacuum that appears in \eqref{factorization} into a {\em different} ${1 \over \nc}$ series. In particular, the thermal two point function
\be
\label{thermalo}
G_{\beta}(t_1, \Omega_1, t_2, \Omega_2) = {1 \over Z(\beta) } \tr \left[e^{-\beta H} {\op}(t_{1}, \Omega_{1}) {\op}(t_{2}, \Omega_{2}) \right],
\ee
where $Z(\beta)$ is the partition function, 
is very different from the vacuum two point function 
\be
G_{\text{vac}}(t_1, \Omega, t_2, \Omega_2) = \lvac  {\op}(t_{1}, \Omega_{1}) {\op}(t_{2}, \Omega_{2})   \rvac.
\ee
Also, note that the large $\nc$ factorization of the thermal correlators \eqref{factorizationthermal} may break down if the operators are separated by large distances in time. 

Finally, by the usual equivalence of ensembles, and the
eigenstate thermalization hypothesis \cite{Deutsch,srednicki1999approach,srednicki1994chaos}, a similar
statement holds when the thermal correlators on both sides of \eqref{factorizationthermal} are replaced by expectation values in typical energy eigenstate of the CFT. Explicitly, this is the statement that in a a typical eigenstate of the CFT $|E \rangle$ with energy $E \gg \nc$, we again have
\be
\langle E | {\op}(t_1, \Omega_1) \ldots  {\op}(t_{2 n}, \Omega_{2 n}) | E \rangle = {1 \over Z(\beta)} \tr\left[e^{-\beta H} {\op}(t_1, \Omega_1) \ldots  {\op}(t_{2 n}, \Omega_{2 n})  \right] + \Or[{1 \over \nc}],
\ee
where $\beta$ is the temperature corresponding to the energy $E$. At high temperatures in the CFT we expect that this is given by
\be
\label{entemprel}
\beta = f_{\beta}({E \over \nc}),
\ee
where $f_{\beta}$ is a smooth function. For example, in the ${\cal N}=4$ super Yang-Mills with $SU(N)$ gauge group at high temperature and at strong coupling on a sphere of volume $V$, we have
\be
\beta = \left({8 E \over 3 \pi^2 N^2 V}\right)^{-{1 \over 4}},
\ee
Therefore, in particular, correlators in an energy eigenstate also factorize, and the eigenstate two point function is close to the thermal one. We will use this important fact to switch freely between thermal and pure state expectations below.

Now consider the modes of these generalized free fields.
 \be
\label{cftmodedef}
 \op_{\omega_n, \ang} = {1 \over \tband^{1 \over 2}} \int_{-\tband}^{\tband} \op(t,\Omega) e^{i \omega_n t} Y_{\ang}^*(\Omega) \, d t \, d^{d-1} \Omega.
 \ee
Here we have discretized the modes by introducing a time band $[-\tband, \tband]$, and correspondingly we have introduced a discrete frequency $\omega_n = {n \over \tband}$. This is necessary because if we consider the strict Fourier modes of the CFT operators, they do not have the behaviour that we need below. In \cite{Papadodimas:2013jku,Papadodimas:2013wnh}, we performed this discretization by ``clubbing together'' these Fourier modes, whereas here we have reverted to a time band that has some other advantages. We also need a UV cutoff on $n$ because if we consider very high energy modes then the ${1 \over \nc}$ corrections that we have neglected above become important.

Now we find that in eigenstates
\be
 \label{godef}
\langle E | [\op_{\omega_n, \ang}, \op_{\omega_{n'},m'}^\dagger] | E \rangle = \comm_{\beta}(\omega_n, \ang) \delta_{\omega_n \omega_n'} \delta_{\ang \ang'} + \Or[{\nc}^{-1}].
\ee
On the right hand side the delta functions follow from the fact that both sides have the same CFT energy and CFT angular momentum. The non-trivial coefficient 
$\comm_{\beta}(\omega_n, \ang)$ is a function of the temperature $\beta$ corresponding to $E$ by \eqref{entemprel}.
Now we define the operators
\be
\label{adefo}
 \an[\omega_n,\ang] = {\op_{\omega_n, \ang} \over \sqrt{\comm_{\beta}(\omega_n, \ang})}+ \Or[{\nc}^{-1}].
\ee

These operators are the natural candidates for creation and annihilation operators in the bulk. By construction we have that up to $\nc^{-1}$ corrections
\be
[H, \an[\omega_n, \ang]] = -\omega_n \an[\omega_n, \ang], \quad [\an[\omega_n, \ang], \an[\omega_n', \ang']^{\dagger}] = \delta_{\omega_n, \omega_n'} \delta_{\ang, \ang'}.
\ee
It is not difficult to check that they have the right thermal two point function.
\be
\begin{split}
{1\over Z(\beta)}\tr \left(e^{-\beta \hcft} \an[\omega_n, \ang] \an[\omega_n, \ang]^{\dagger} \right) &= {1\over Z(\beta)}\tr \left(\an[\omega_n, \ang]^{\dagger} e^{-\beta \hcft} \an[\omega_n, \ang]  \right) = e^{\beta \omega_n} {1\over Z(\beta)}\tr \left(e^{-\beta \hcft} \an[\omega_n, \ang]^{\dagger} \an[\omega_n, \ang] \right) \\
&= e^{\beta \omega_n}  {1\over Z(\beta)}\tr \left(e^{-\beta \hcft} \an[\omega_n, \ang]  \an[\omega_n, \ang]^{\dagger} \right)  - e^{\beta \omega_n} {1\over Z(\beta)}\tr \left(e^{-\beta \hcft} \right),
\end{split}
\ee
where we have used the cyclicity of the trace and the commutation relations above.
A little algebra now shows that 
\be
{1 \over Z(\beta)} \tr \left(e^{-\beta \hcft} \an[\omega_n, \ang] \an[\omega_n, \ang]^{\dagger} \right) = \langle E | \an[\omega_n, \ang] \an[\omega_n, \ang]^{\dagger} | E \rangle = {1 \over 1 - e^{-\beta \omega_n}},
\ee
where we have used the equivalence of ensembles and the relations above hold only up to ${1 \over \nc}$ and other corrections from discretizations.

Now consider the CFT operator
\be
\label{phicftexpand}
\cop{\phi}(t, \rtor, \Omega) = \sum_{\omega_n, \ang} {1 \over \sqrt{\omega_n}} \an[\omega_n, \ang] f_{\omega_n, \ang}(t, \rtor) Y_{\ang}(\Omega) + \text{h.c.}
\ee
where $f_{\omega_n, \ang}$ is a solution of the Klein Gordon equation in the metric \eqref{schwarzschild} with the boundary condition at the horizon
\be
f_{\omega_n, \ang} \underset{r \rightarrow \rhor}{\longrightarrow} \left( e^{i \delta} e^{i \omega_n \rtor} + e^{-i \delta} e^{-i \omega_n \rtor} \right),
\ee
and normalizable boundary conditions at infinity. 
The expansion \eqref{phicftexpand} not only fulfills the necessary near-horizon conditions that we derived above, it also correctly reproduces the behaviour of a bulk field propagating in a smooth spacetime in the rest of AdS. 
This completes our construction of local operators in a high energy eigenstate. As we mentioned in section \ref{sectransferfunc}, we obtain a bonus, and a consistency check, from AdS/CFT. The fields constructed in \eqref{phicftexpand}, with the aid of \eqref{adefo} automatically satisfy
 \be
 \lim_{r \rightarrow \infty} r^{2\Delta} Z^2 \langle E| \cop{\phi}(t_1, \rtor, \Omega_1)  \cop{\phi}(t_2, \rtor, \Omega_2) | E \rangle = W_{\beta}(t_1 - t_2, \Omega_1, \Omega_2).
 \ee
 where $Z$ is a numerical factor and $W_{\beta}$ is defined in \eqref{thermalo}. Note that we did not put this relation in by hand. It follows from, and is a prediction of the claim that the eigenstate is dual to the black-hole geometry.

\subsubsection{A state-independent mini-superspace bulk-boundary map outside the horizon }\label{secstateindoutside}
In \eqref{adefo}, we explicitly put in the commutator in the energy-eigenstate. The modes in \eqref{phicftexpand} also contain information about the state. Therefore, as written the expression \eqref{phicftexpand} is state-dependent and will not correctly reproduce local correlation functions in states corresponding
to black holes with macroscopically different properties.

Now we consider whether it is possible to write down an expansion that will work outside the horizon in a larger class of states. The basic idea is to use projectors to try and ``detect'' the state. We will show how one can generalize \eqref{phicftexpand} so that it works in all high energy spherically symmetric eigenstates. 

Given a spherically symmetric energy eigenstate $|E \rangle$,  we can associate a temperature to the energy eigenstate by means of \eqref{entemprel}, and also an associated metric via \eqref{schwarzschild}. We denote this metric as $g_{E, \mu \nu}$. We also consider modes $f_{E,\omega, \ang}$; these are the same as the modes $f_{\omega, \ang}$ in \eqref{phicftexpand}, except that we have displayed their energy dependence explicitly.
Now, consider
\be
\label{phistateind}
\cop{\phi}_{\text{state-ind}}(t, \rtor, \Omega) = \sum_{E} \sum_{\omega, \ang} {1 \over \sqrt{\omega_n}} \left( {1 \over \sqrt{\comm_{\beta}(\omega_n, \ang)}} \op_{\omega_n, \ang} | E \rangle \langle E | \right) f_{E,\omega_n, \ang}(t, \rtor) Y_{\ang}(\Omega) + \text{h.c}, 
\ee
where, as we emphasized above, the expectation of the commutator that we have used to normalize the mode also depends on the energy eigenstate.
The claim is that this generalizes the construction \eqref{phicftexpand} so that, as long as we stay away from the horizon, it works in spherically symmetric states of the CFT corresponding to an arbitrary temperature. 

To verify this, note that the expression \eqref{phistateind} is designed so that when it acts directly on an energy eigenstate its action
reduces to that of \eqref{phicftexpand}. Now consider an excitation of an energy eigenstate by a polynomial in the modes \eqref{cftmodedef}
\be
\label{excitatede}
\op_{\omega_1, \ang_1} \ldots \op_{\omega_n, \ang_n} | E \rangle = \sum_i \coeff[i] |E_i \rangle.
\ee
If $\sum n  \ll \nc$ and $\sum n \omega_n \ll \nc$, then  all states $|E_i \rangle$ that appear above have ${E-E_i \over \nc} = 0 + \Or[{1 \over \nc}]$ and therefore, from  \eqref{entemprel},  the coefficients $\coeff[i]$ are restricted in support to states that have the same
macroscopic temperature and correspond to the same macroscopic metric. Therefore, \eqref{phistateind} again acts on this superposition as \eqref{phicftexpand} away from the horizon. This is the expected behaviour since we do not expect these excitations to have any significant back-reaction on the geometry.  

It is easy to verify that the action of \eqref{phistateind} is also consistent with the fact that we expect state of the form  \eqref{superposdistinct} to behave like classical superpositions of different geometries.

If we approach too close to the horizon, then not all quantities of physical interest are smooth functions of the energy. For example, there has been some debate in the literature on highly spacelike modes \cite{Bousso:2012mh,Rey:2014dpa} where the ratio of value of the mode function near the horizon to its value at the boundary can vary exponentially with temperature.  Although we showed in \cite{Papadodimas:2012aq} that these modes do not present an obstruction to reconstructing the field near the horizon in the thermal state, it is less clear how to deal with this difficulty in the putative state-independent expression \eqref{phistateind}. It is also not clear whether \eqref{phistateind} can be refined to work in all non-spherically symmetric situations.


\section{Arguments against state-independent operators }\label{secparadoxes}
In the previous section we explicitly found operators $\an[\omega_n, \ang]$ in the CFT that were dual to propagating modes in the bulk. However, if we want to describe local operators behind the horizon, then we also need to locate the operator $\ta_{\omega_n, \ang}$ in the CFT. Alternately, we could find operators $\tO_{\omega_n, \ang}$ related to $\ta_{\omega_n,\ang}$ by a relation analogous to \eqref{adefo}. At this order in ${1 \over \nc}$, we do not have to consider corrections to \eqref{adefo} and we will switch freely between $\tO_{\omega, \ang}$ and $\ta_{\omega, \ang}$. 

In this section, we will review and refine some of the arguments that suggest that these operators cannot be state-independent in the CFT.   In \cite{Almheiri:2012rt,Almheiri:2013hfa,Marolf:2013dba}, these arguments were used to argue that the CFT could not look past the black hole horizon, or even more dramatically that the horizon was just a cloak for a ``firewall''. Our interpretation is, instead, that these arguments tell us that the bulk to boundary map is state-dependent. From this point of view, the objective of this section is to prove that one must either accept state-dependence or firewalls.

\subsection{Some general results regarding projectors  \\}
\label{subsecprojector}
Before we continue with this analysis, let us make an elementary observation about matrix elements of projection operators. Eigenvalues of projection operators are either $1$ or $0$, so the operator norm of a projection operator is $||P||=1$. As a result projectors are bounded operators and this implies that the map from state vectors $|\Psi\rangle$ into expectation values $\langle \Psi| P |\Psi\rangle$ is a continuous map. 

Hence, to the extent that we can characterize the physical properties of a state by evaluating expectation values of projectors,  {\it nearby state vectors must have nearby physical properties.}

Let us try to make this a bit more precise. Suppose that we have two unit-normalized states $|\Psi_1\rangle$ and $|\Psi_2\rangle$ in the Hilbert space and we
denote their difference as $|\delta \Psi\rangle = |\Psi_1\rangle - |\Psi_2\rangle$. We define $\delta = ||\, |\delta \Psi\rangle\,||$. We consider a projector and  estimate the difference of its expectation value on the two nearby states
\be
\begin{split}
|\langle\Psi_1| P |\Psi_1 \rangle - \langle\Psi_2| P |\Psi_2 \rangle| &= 
|\langle\delta\Psi| P |\Psi_2 \rangle+\langle\Psi_2| P |\delta\Psi \rangle+\langle \delta \Psi| P |\delta\Psi \rangle |\\
& \leq |\langle \delta \Psi|P |\Psi_2\rangle| + |\langle\Psi_2|P |\delta\Psi\rangle|  + |\langle \delta \Psi| P |\delta\Psi \rangle |\\
& \leq 2 \delta + \delta^2.
\end{split}
\ee
Notice that it may also be useful to think of two nearby states as those obeying 
\be
\label{innerpclose}
|\langle \Psi_1|\Psi_2\rangle| = 1 - {\epsilon^2\over 2},
\ee
 with small positive $\epsilon$. Since physical states are represented by rays on the Hilbert space, we are free to chose the phase of the vectors as we like. It is easy to check that there is a choice where $\epsilon = \delta$ and the same result as before follows i.e. for any two vectors obeying \eqref{innerpclose}, we have
\be
\label{nearbystates}
|\langle\Psi_1| P |\Psi_1 \rangle - \langle\Psi_2| P |\Psi_2 \rangle| \leq  2 \epsilon + \epsilon^2.
\ee
We will use these results below. 

\subsection{$N_a \neq 0$ Argument }\label{secnazero}
First, let us consider the $N_a \neq 0$ argument \cite{Marolf:2013dba}. The essence of this argument is as follows. We would like the set of states in the CFT to obey two conditions, both of which seem motivated
on physical grounds.
\begin{enumerate}
\item
Typical superpositions of energy eigenstates are not excited states from the point of view of the infalling observer.
\item
If we consider states that are eigenstates of a Schwarzschild number operator $\nb \equiv \an[\omega_n, \ang]^{\dagger} \an[\omega_n, \ang]$, for the modes introduced in \eqref{cftmodedef} and \eqref{adefo}, then these are excited states from the point of view of the infalling observer.\end{enumerate}

To phrase the first condition more precisely consider the following set of energy eigenstates
\be
\enrange[E, \Delta] \equiv \{ |E_i \rangle: \quad E - \Delta \leq E_i \leq E + \Delta \},
\ee
where $E$ is some mean energy and $\Delta$ is a spread. We will use the same symbol $\enrange[E, \Delta]$ to denote the Hilbert space spanned by these states and the meaning should be clear from the context. We also denote
\be
\dimrange[E, \Delta] \equiv \text{dim}(\enrange[E, \Delta]).
\ee
Finally we introduce
\be
\projrange[E, \Delta] \equiv \text{projector~onto~}\enrange[E, \Delta].
\ee
Now consider a projection operator $\projf$  corresponding to the measurement of the infalling observer, defined so that $\projf=0$ corresponds to a smooth and empty interior. This projector can be constructed as an ordinary projector in the CFT Hilbert space if the bulk to boundary map is state-independent. The authors of \cite{Marolf:2013dba} used the number operator, as measured by the infalling observer, to detect whether the horizon was smooth but it is possible to use other operators and therefore we keep the analysis here general. 

From the first physical assumption mentioned above, we expect that for typical states in $\enrange[E, \Delta]$
the expectation value of $\projf$ should be small. Hence we expect
\be
\label{expempty}
{1 \over \dimrange[E, \Delta]} \tr_{\enrange[E, \Delta]} \left( \projf \right) = 0 + \Or[{1 \over \nc}],
\ee
The second condition means that for eigenstates $|N_i\rangle$ of the Schwarzschild number operator $\nb$ we have
\be
\label{eigenexc}
\langle N_i | \projf | N_i \rangle = \Or[1].
\ee
In the large $\nc$ limit we have $[\hcft,\nb]= 0 + O(\nc^{-1})$, so we intuitively expect
that we can find a basis of the Hilbert space $\enrange[E, \Delta]$  spanned by number operator eigenstates $|N_i\rangle$. 
The trace of an operator can be evaluated in any basis, so we can evaluate the trace \eqref{expempty} in the $|N_i\rangle$
basis. For each of the basis vectors \eqref{eigenexc} gives a significant contribution. Then it seems that we get
\be
\label{nanzarg}
{1\over \dimrange[E, \Delta]}\tr_{\enrange[E, \Delta]}\,\left( \projf \right)  = \Or[1] + {\rm small \,\, error.}
\ee
and that hence typical states are not smooth, in contradiction to the first assumption  above. This concludes the $N_a\neq 0$ argument of \cite{Marolf:2013dba}. The result
was interpreted by \cite{Marolf:2013dba} as an indication that typical pure states do not have a smooth interior. The small error above is due to the fact that the operators $\hcft$ and $\nb$
can be simultaneously diagonalized within $\enrange[E, \Delta]$ only in an approximate sense, in the large $N$ limit.

One might attempt to find a loophole in this argument by looking more carefully at the error terms mentioned above. Could it be that, contrary to what was
assumed in \cite{Marolf:2013dba}, these 
error terms are significant enough to make the RHS of equation \eqref{nanzarg} close to zero? In the following subsection, we perform a systematic analysis of
the error terms and 
exclude the possibility that they can  invalidate the $N_a \neq 0$ argument.

\subsubsection{Bounding errors in the $N_a \neq 0$ argument }\label{secnabounderror}
The linear algebra literature contains several results on ``almost commuting matrices'' \cite{lin1997almost,friis1996almost}, which could be used to make the argument above rigorous. Here, rather than
taking this path, we will follow an approach motivated by perturbation theory to make the $N_a \neq 0$ paradox sharper. 

We will assume that
\be
\label{onebyndecomp}
\hcft = \hinf + {1 \over \nc} V,
\ee
where the ``infinite $\nc$'' Hamiltonian, $\hinf$ has the property that $[\hinf, \nb] = 0$ and $V$ is a ``perturbation'', whose matrix elements have the property that ${\langle E | V | E \rangle \over E} = \Or[1]$ for high energy eigenstates.   Note that \eqref{onebyndecomp} is somewhat stronger than 
our original starting point --- which was simply that $\langle E| [\hcft, \nb] | E \rangle = \Or[{1 \over \nc}]$.\footnote{It is subtle to consider perturbations of the Hilbert space at high energies in ${1 \over \nc}$ because the Hilbert space changes discontinuously with $\nc$ and its dimension goes off to $\infty$ as $\nc \rightarrow \infty$. So we are assuming that \eqref{onebyndecomp} holds at each $\nc$ and some properties of these operators, such as the {\em ratio} of the dimensions of different sets below have a well defined large $\nc$ limit.}

\begin{figure}[!h]
\begin{center}
\includegraphics[width=0.3\textwidth]{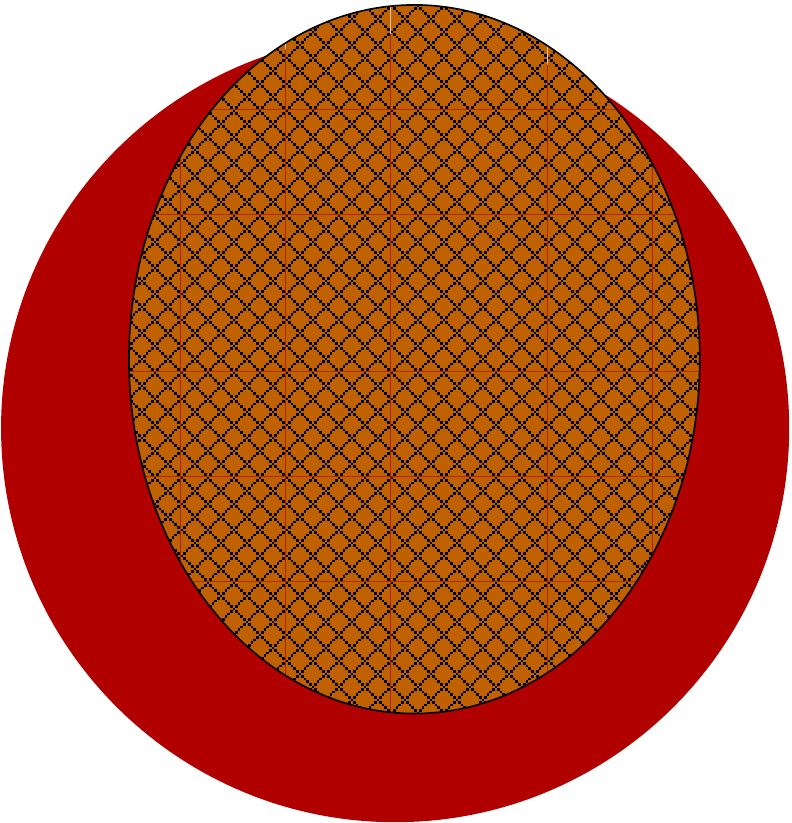}
\caption{The schematic structure of the two relevant sets. The solid circular set is the set of energy eigenstates. The smaller set of number eigenstates, shown as an elliptical patterned set, is almost completely contained inside the set of energy eigenstates.}\label{fignumberandenergyeigen}
\end{center}
\end{figure}

If \eqref{onebyndecomp} is correct, then by standard arguments from perturbation theory we expect that groups of eigenstates of $\hcft$ can 
be reorganized into eigenstates of $\nb$ and vice versa. Now consider the set of all number eigenstates that can be accurately approximated by energy eigenstates in $\enrange[E,\Delta]$. We will call this set of $\nb$ eigenstates  $\numinus$ and denote its dimension by $\diminus$. The projector onto $\numinus$ will be denoted by $\projminus$. 
By definition,
\be
\langle N_i | \projrange[E, \Delta] | N_i \rangle = 1-\Or[{1 \over \nc}], \quad \forall |N_i \rangle \in \numinus.
\ee
The structure of these two sets is shown in Figure \ref{fignumberandenergyeigen}.

The key physical consequence of \eqref{onebyndecomp} is that to form eigenstates of $\hinf$ with an eigenvalue $E$, we have take eigenstates of $\hcft$ with $\hcft$-eigenvalues $E \pm \Delta$, where $\Delta = \Or[E \over \nc] = \Or[1]$. 
Therefore if we take the original spread of energies $\Delta$ in $\enrange[E, \Delta]$ to be large, $\Delta \gg \Or[1]$, then we have
\be
\label{diffdimen}
{\dimrange[E, \Delta] - \diminus \over \dimrange[E, \Delta]} \ll 1.
\ee

If we accept these statements, then it is easy to produce a contradiction. 
From the assumptions above, given a $|N_i \rangle \in \numinus$, we have
\be
|N_i \rangle = \sum_m U^*_{m i} |E_m \rangle = \sum_{m \in \enrange[E, \Delta]} U_{m  i}^* |E_m \rangle + \sum_{m \notin \enrange[E, \Delta]} U_{m i}^* |E_m \rangle \equiv |M_i \rangle + |R_i \rangle,
\ee
where $U^*_{m i}$ is some matrix that implements the change in the two eigenvalue bases and where $\langle R_i | R_i \rangle = \Or[{1 \over \nc}]$.  Here, we have divided the sum into two parts and used the definition of $\numinus$ which is precisely that its elements can be re-expressed as elements in $\enrange[E, \Delta]$.  Moreover, using \eqref{nearbystates} we find that $\langle M_i | \projf | M_i \rangle = \langle N_i | \projf | N_i \rangle + \Or[{1 \over \nc}]$. But this implies that
\be
\label{manipulationone}
{1 \over \dimrange[E, \Delta]} \tr\big(\projminus \projrange[E, \Delta] \projf \projrange[E, \Delta] \projminus \big) = {1 \over \dimrange[E, \Delta]}  \tr\big(\projminus \projf \projminus\big) = {1 \over \dimrange[E, \Delta]}  \tr\big( \projf \projminus\big)  =  \kappa {\diminus \over \dimrange[E, \Delta]},
\ee
where $\kappa$ is some constant of $\Or[1]$ which determines the probability for an infalling observer to see an excitation in a number eigenstate and which
follows from \eqref{eigenexc}. Here  we have neglected $\Or[{1 \over \nc}]$ corrections.

Second notice that the original trace in the microcanonical ensemble can be transformed by a sequence of elementary manipulations to
\be
\begin{split}
&\tr\big(\projf \projrange[E, \Delta] \big) = \tr\big(\projrange[E, \Delta] \projf \projrange[E, \Delta] \big) 
= \tr\big( (1 - \projminus + \projminus) \projrange[E, \Delta] \projf \projrange[E, \Delta] \big) \\
&=  \tr\big( (1 - \projminus) \projrange[E, \Delta] \projf \projrange[E, \Delta] \big)  +  \tr\big( \projminus \projrange[E, \Delta] \projf \projrange[E, \Delta] \big) \\
&=  \tr\big( (1 - \projminus) \projrange[E, \Delta] \projf \projrange[E, \Delta] (1 - \projminus) \big)  +  \tr\big( \projminus \projrange[E, \Delta] \projf \projrange[E, \Delta] \projminus \big).
\end{split}
\ee
Here we have repeatedly used the cyclicity of the trace, and the fact that projectors square to themselves.
Now notice that given any product of projectors $X = P_1 \ldots P_n$, we find that $\tr(X) = \tr(X^{\dagger} X) \geq 0$. Therefore the first term in the last line above is positive and we find
\be
\label{manipulationthree}
\tr\big(\projf \projrange[E, \Delta] \big) \geq \tr\big( \projminus \projrange[E, \Delta] \projf \projrange[E, \Delta] \projminus \big) = \kappa {\diminus \over \dimrange[E, \Delta]}.
\ee

Combing the result of \eqref{manipulationthree} and the physical assumption \eqref{expempty}, we seem to find
\be
\label{contradictionna}
0 = \tr(\projf \projrange[E, \Delta]) \geq \kappa {\diminus \over \dimrange[E, \Delta]}.
\ee
This is clearly a contradiction, if we recall \eqref{diffdimen}. Note that the difference between the left and right sides of \eqref{contradictionna} is $\Or[1]$, and so the errors, which we have bounded to be $\Or[{1 \over \nc}]$ using the construction above cannot affect this result.

This was used by \cite{Marolf:2013dba} to suggest that \eqref{expempty} should be abandoned. We will show below how a more plausible explanation is that $\projf$ does not exist as a fixed (state-independent) linear projector; rather the question of whether a firewall exists or not depends on a state-dependent measurable.

\subsection{Negative occupancy argument }\label{secnegoccup}
We now present an argument that is closely related to the ``counting argument'' (or the lack of a left-inverse argument). As originally stated in \cite{Almheiri:2013hfa}, the counting argument is as follows. 
First, we consider a mode behind the horizon with creation and annihilation operators obeying the  algebra
\be
\label{fakealg}
[\ta_{\omega_n, \ang}, \ta_{\omega_n, \ang}^{\dagger}] = 1.
\ee
Notice that this equation unambiguously selects $\ta_{\omega_n,\ang}^\dagger$ as the ``creation operator'', since we can rewrite
it as $\left[(1 + \ta_{\omega_n,\ang}^{\dagger} \ta_{\omega_n,\ang})^{-1} \ta_{\omega_n,\ang} \right] \ta_{\omega_n,\ang}^{\dagger} = 1$, which means that the operator $\ta_{\omega_n,\ang}^\dagger$ has a left inverse and hence it does not annihilate any state.

Then we notice that, as explained in section \eqref{secsmoothint}, modes behind the horizon obey ``inverted'' commutators with the CFT Hamiltonian 
\be
\label{fakealgb}
[\hcft, \ta_{\omega_n,\ang}^{\dagger}] = -\omega_n \ta_{\omega_n,\ang}^{\dagger}.
\ee
This means that the operator $\ta_{\omega_n,\ang}^\dagger$, despite being a creation operator, lowers the energy of the CFT. Hence, it maps the space of states 
of energy $E$ into that of energy $E-\omega_n$.  However,  the density of states in the CFT increases monotonically with energy. This implies that the operator $\ta_{\omega_n, \ang}^\dagger$
maps the larger Hilbert space of energy $E$ into a smaller one of energy $E-\omega_n$. The {\it linear} operator $\ta_{\omega_n, \ang}^\dagger$ can do this only if it annihilates a fraction of the states of energy $E$. But this is in contradiction with the prediction of \eqref{fakealg} that $\ta_{\omega_n, \ang}^\dagger$ has a left inverse.

Hence it seems that imposing the algebra \eqref{fakealg}, \eqref{fakealgb} for {\it state-independent} linear operators is inconsistent with the growth of entropy in the CFT. This concludes the ``counting argument`` of \cite{Almheiri:2013hfa}.

One apparent difficulty with this argument is that it is phrased in terms of operator relations \eqref{fakealg}, \eqref{fakealgb}. One might wonder whether it
is possible to 
satisfy these relations, not as operator equations, but only {\em within simple correlation functions}. We now present a closely related argument, that is phrased entirely within
the context of low point correlation functions.

Let $\projrangenew{E}{\Delta}$ be the projector onto a narrow band of energy states. Define $\dimrangenew{E}{\Delta} = \tr(\projrangenew{E}{\Delta})$, which counts the number of states in this band. We consider the 
expectation value of the occupation level of the mode in this ensemble of states
\be
\label{occarg}
\begin{split}
\langle \cop{\widetilde{N}}_{\omega_n} \rangle & =  \dimrangenew{E}{\Delta}^{-1} \tr\left(\projrangenew{E}{\Delta} \,\ta_{\omega_n,\ang}^{\dagger} \ta_{\omega_n,\ang}\right) = \dimrangenew{E}{\Delta}^{-1} \tr \left(\ta_{\omega_n,\ang} \projrangenew{E}{\Delta} \ta_{\omega_n,\ang}^{\dagger} \right)\\
& =\dimrangenew{E}{\Delta}^{-1} \tr\left(\projrangenew{E+\omega_n}{\Delta}\, \ta_{\omega_n,\ang} \ta_{\omega_n,\ang}^{\dagger} \right) + \delta_1 \\ &=
e^{\beta \omega_n} + \dimrangenew{E}{\Delta}^{-1} \tr\left(\projrangenew{E+\omega_n}{\Delta}\, \ta_{\omega_n,\ang}^{\dagger} \ta_{\omega_n,\ang} \right) + \delta_1 + \delta_2.
\end{split}
\ee
In the first line we used the cyclicity of the trace. In the second line we used that \eqref{fakealgb} should hold inside simple correlators, which implies $\ta_{\omega_n,\ang} \projrangenew{E}{\Delta} = \projrangenew{E+\omega_n}{\Delta} \ta_{\omega_n, \ang}$ up to some small error $\delta_1$. In the last line we used that \eqref{fakealg} should hold in
simple correlators, up to some small error $\delta_2$.
Since the trace above consists just of a sum of low point correlators  we expect
that $\delta_1,\delta_2 \sim \Or[{1 \over \nc}]$.
This assumptions allows us to ignore these errors in deriving the contradiction that follows.
The factor  outside the trace of $e^{\beta \omega_n}$ arises because
\be
\dimrangenew{E}{\Delta}^{-1} \tr(\projrangenew{E+\omega_n}{\Delta}) = {\dimrangenew{E+\omega_n}{\Delta} \over \dimrangenew{E}{\Delta}} = e^{\beta \omega_n}.
\ee 
We also use the fact that for a reasonably smooth operator $\cop{\tN}_{\omega_n}$, we have
\be
 \dimrangenew{E}{\Delta}^{-1}\tr\left(\projrangenew{E+\omega_n}{\Delta} \, \ta_{\omega_n,\ang}^{\dagger} \ta_{\omega_n,\ang} \right) = e^{\beta \omega_n} \langle \cop{\tN}_{\omega_n} \rangle + \Or[\nc^{-1}].
\ee
Replacing this in \eqref{occarg} and dropping all subleading error terms we arrive at our final relation 
\be
\langle \cop{\tN}_{\omega_n} \rangle = e^{\beta \omega_n} +  e^{\beta \omega_n} \langle \cop{\tN}_{\omega_n}\rangle \Rightarrow \langle \cop{\tN}_{\omega_n}\rangle =  - {1 \over 1 - e^{-\beta \omega_n}},
\ee
which is negative! In some sense, this unphysical result is not surprising, because $\ta_{\omega_n,\ang}$ is an annihilation operator with positive energy, and the thermal properties of such an operator seem to be ill-defined. 

To summarize, the argument above demonstrates that there cannot exist
{\it linear, state-independent} operators in the CFT which approximately satisfy the relations \eqref{fakealg}, \eqref{fakealgb} inside simple correlation functions. One might conclude from this that the black hole does not have an interior that the CFT can describe. Instead, we advocate \cite{Papadodimas:2012aq, Papadodimas:2013jku, Papadodimas:2013wnh} that the desired relations \eqref{fakealg}, \eqref{fakealgb} can be consistently realized by allowing the operators $\ta_{\omega_n, \ang}, \ta_{\omega_n, \ang}^\dagger$ to depend on the state. For state-dependent operators the counting argument does not apply \cite{Papadodimas:2013jku} and the negative occupancy argument presented above does not apply since it is meaningless to evaluate the trace, if the 
operators vary as a function of the state in the ensemble.

\subsection{The generic commutator}\label{seccommutarg}
Now we consider the fact that there is not enough ``space'' in the CFT
Hilbert space to accommodate the commutant of the ordinary operators if they are finely spaced enough. There are two ways in which this argument can be phrased. One point, which was originally made in \cite{Almheiri:2013hfa} is as follows. If we assume that the algebra of the mirror operators is given by some ``scrambling`` unitary transform of the ordinary operators so that we have
\be
\ta_{\omega_n,\ang}^{\dagger} = \cop{U} \anc_{\omega_n, \ang}^{\dagger} \cop{U}^{\dagger},
\ee
then we find that, for a {\em generic} unitary operator $\cop{U}$, we have
\be
|\,[\ta_{\omega_n,\ang}^{\dagger}, \anc_{\omega_n, \ang}]\,|^2 \sim \Or[1].
\ee
This by itself is not a proof of the lack of existence of the commutant. In particular, if the Hilbert space has a factorization into coarse and fine pieces, as was discussed originally in \cite{Papadodimas:2012aq}, then this would break down. 

In what follows, we will discuss how finely an observer has to measure generalized free fields on the boundary, in order to exhaust the space of the CFT. However, first, we turn to two toy models: the spin chain and a set of decoupled harmonic oscillators.

Consider a chain of spins. We denote the operators acting on this chain by  $\paul^i_a$ as in \cite{Papadodimas:2013jku}. We assume that the spins are all decoupled. The index $i = 1 \ldots N$, where $N$ is the length of the spin chain, and $a = x,y,z$ as usual.  We normalize them to satisfy $[\paul^i_a, \paul^j_b] = {i \over 2} \delta^{i j} \epsilon_{a b c} \paul^i_c$.  A complete set of operators for the Hilbert space is obtained by taking arbitrary products of these single-spin operators. Nevertheless, even if we consider the significantly smaller set of just the $N$ single-spin operators, the commutant of this smaller set is trivial and consists only of the identity operator. 

One might hope that there exist (state-independent) operators $\pault$, apart from the identity, which {\it approximately} commute with all single-spin operators. We now demonstrate
that this is not possible: if $\pault$ has small commutators with all single-spin operators, then $\pault$ is small as an operator. To show this, we consider an arbitrary operator  $\pault$ acting on the spin chain. In order to factor out the identity operator, which is trivially in the commutant, we assume that $\pault$ is traceless, which means that 
we can represent it as a polynomial in the atomic spin operators
\be
\pault = \sum_{i_m, a_m, n}  c_{i_1 \ldots i_n}^{a_1 \ldots a_n}\, \paul^{i_1 \ldots i_n}_{a_1 \ldots a_n},
\ee
where $\paul^{i_1 \ldots i_n}_{a_1 \ldots a_n} \equiv \paul^{i_1}_{a_1} \ldots \paul^{i_n}_{a_n}$, and we impose the constraint that $i_1 < i_2 < \ldots i_n$ to avoid overcounting. 

We find that we have the following relation
\be
[\pault, \paul^j_b] = {i \over 2} \sum c_{i_1 \ldots i_n}^{a_1 \ldots a_n} \left(\delta_{i_1}^{j} \epsilon_{a_1 b c} \paul_c^{i_1} \paul^{i_2 \ldots i_n}_{a_2 \ldots a_n} + \delta_{i_2}^{j} \epsilon_{a_2 b c} \paul_c^{i_2} \paul^{i_1 i_3 \ldots i_n}_{a_1 a_3 \ldots a_n} + \ldots \right).
\ee
While we have written a sum of delta functions on the right, note that at most one of them is non-vanishing. A natural norm of an operator to consider in this space is $|X|^2 = {1 \over 2^n} \tr(X^{\dagger} X)$. With this definition
\be
|[\pault, \paul^j_b]|^2 = {1 \over 4} \sum |c_{i_1 \ldots i_n}^{a_1 \ldots a_n} \delta_{i_1}^{j} \epsilon_{a_1 b c}|^2 + |c_{i_1 \ldots i_n}^{a_1 \ldots a_n} \delta_{i_2}^{j} \epsilon_{a_2 b c}|^2 + \ldots
\ee
Note that there is no interference between the different terms in the sum due to the observation above. However, when we sum over $b$ we find that there are two values for which the completely anti-symmetric tensor is non-zero. This leads to
\be
\sum_{j, b} |[\pault, \paul^j_b]|^2 = {1 \over 2} \sum |c_{i_1 \ldots i_n}^{a_1 \ldots a_n}|^2  = {1 \over 2} |\pault|^2.
\ee
The physical implication of this is as follows. If an observer can measure the various single spin operators, then given any operator $\pault$, the observer can detect that it fails to commute with these ``ordinary'' operators. In particular, it is {\em not necessary} for the observer to measure very complicated observables. Even if the observer does not have access to more complicated products of these spin operators, she can determine that the commutant is trivial.

The argument presented above shows that an operator of unit-norm, $|\pault|^2$ must have an order $1$ commutator with at least one single-spin operator, or alternatively it could have $\Or[{1 \over S}]$ commutators with all the single-spin operators. In either case, the important point is that it cannot simultaneously have smaller commutators with all the $\paul^i_a$.

Now, we consider a similar argument for the case of decoupled harmonic oscillators. The setup was described in more detail in \cite{Papadodimas:2013jku}. We have unbounded creation and annihilation operators. The frequencies of the oscillators are given by $\omega_1 \ldots \omega_N$ and their respective creation and annihilation operators are specified by $a_1 \ldots a_N$. The only non-zero commutators are $[a_i, a^{\dagger}_j] = \delta_{i j}$. The Hilbert space is a Fock space indexed by the eigenvalues of the number operators $N_i = a^{\dagger}_i a_i$. 

We can still write any operator of interest as
\be
\tashm = \sum_{p_j, q_j} A(p_1, q_1 \ldots p_n, q_n) a_{1}^{p_1} (a^{\dagger}_{1})^{q_1} \ldots a_{N}^{p_N} (a^{\dagger}_{N})^{q_N}.
\ee
 Once again we factor out factors of $N_i$ from each monomial in the polynomial above so that either $p_i=0$ or $q_i =0$ for all $i$. the most general operator then lives in the direct product of the vector space of polynomials of $N_i$ and the space of operators above. But note that the sum above can also accommodate operators where a particular frequency, say $\omega_{i}$,  does not appear simply by 
setting $p_i = q_i = 0$.

Now in a typical equilibrium state, we see that the only non-zero expectation values are products of $N_i$. This implies that
\be
\langle \tashm^{\dagger} \tashm \rangle = \sum |A(p_1 , q_1 \ldots p_n, q_n)|^2 \langle  a_{1}^{q_1} (a^{\dagger}_{1})^{p_1}  a_{1}^{p_1}  (a^{\dagger}_{1})^{q_1} \ldots a_{N}^{q_N} (a^{\dagger}_{N})^{p_N}  a_{N}^{p_N}  (a^{\dagger}_{N})^{q_N} \rangle,
\ee
where the $\ldots$ indicate similar terms for all the other frequencies and cross terms vanish.

Evaluating the expectation value above in a state $|N_1 \ldots N_N \rangle$ we find that
\be
\langle \tashm^{\dagger} \tashm \rangle = \sum_{p_j, q_j}  |A({p_1, q_1 \ldots p_n, q_n})|^2 (N_{1} + 1)_{q_{1}} (N_{1} + q_{1} - p_{1} + 1)_{p_{1}} \ldots (N_{N} + 1)_{q_{N}} (N_{N} + q_{N} - p_{N} + 1)_{p_{N}} ,
\ee
where the Pochhammer symbol is $(x)_{n} \equiv x (x+1) \ldots (x+n-1).$

Next we notice that
\be
\begin{split}
&[\tashm, a_{j}] =  -\sum A(p_1, q_1, \ldots p_n, q_n) q_j a_{1}^{p_1} (a^{\dagger}_{1})^{q_1} \ldots a_j^{p_j} (a^{\dagger}_j)^{q_j - 1} \ldots a_{N}^{p_N} (a^{\dagger}_{N})^{q_N}.\\
&[\tashm, a_{j}^{\dagger}] =  \sum A(p_1, q_1, \ldots p_n, q_n) p_j a_{1}^{p_1} (a^{\dagger}_{1})^{q_1} \ldots a_j^{(p_j-1)} (a^{\dagger}_j)^{q_j} a_{N}^{p_N} \ldots (a^{\dagger}_{N})^{q_N}.\\
\end{split}
\ee
Defining a new function, by the recursion relations
\be
\begin{split}
&B(p_1, q_1 \ldots p_j, q_j, \ldots p_n, q_n) = (p_j + 1) A (p_1, q_1, \ldots p_j + 1, q_j, \ldots p_n, q_n), \\
&B(p_1, q_1 \ldots p_j, q_j, \ldots p_n, q_n) = (q_j + 1) A (p_1, q_1, \ldots p_j , q_j + 1, \ldots p_n, q_n), \\
\end{split}
\ee
we see that we have
\be
\begin{split}
\sum_j \langle |[\tashm, a_{j}]|^2 \rangle + \langle |[\tashm, a_{j}^{\dagger}]|^2 \rangle =  \sum\Big[ |B({p_1, q_1 \ldots p_n, q_n})|^2 &(N_1 + q_1 - p_1 + 1)_{p_1} (N_1 + 1)_{q_1} \ldots \\ &\times (N_N + q_N - p_N + 1)_{p_N} (N_N + 1)_{q_N} \Big].
\end{split}
\ee

In this case, we do not have a simple result like that of the simple harmonic oscillator. Indeed for some operators $\tashm$ that are comprised of 
creation and annihilation operators, which have a very high occupancy in the state, it seems possible to make $\langle \tashm^{\dagger} \tashm \rangle \gg \langle \sum_j \langle |[\tashm, a_j]|^2 \rangle + \langle |[\tashm, a_j^{\dagger}]|^2 \rangle$. However, in most configurations and for almost all operators $\tashm$, these two terms are comparable.

Note that in order to build an entire effectively isomorphic commuting algebra, we need a $\tashm$ operator for each ordinary operator. Therefore even if, in some states, some of these operators have a small commutator with the ordinary operators, it is clear that there is not enough space in this chain of simple harmonic oscillators to accommodate mirror operators for each oscillator.

It is this intuition that carries over to the CFT. Consider the set of modes of generalized free fields. For simplicity, imagine separating them in frequency by $\omega_{0}$, so that these  modes all appear to be $\op_{n \omega_{0}, \ang}$. As usual, there could be other GFFs, while we are displaying only one of them. 
The main observation is the following. By putting a cutoff at the stretched horizon, we can limit the maximum angular momentum $\ang$ that can appear for a given $\omega_n = n \omega_0$. Second, as we take $\omega_0 \propto {1 \over \nc^{\alpha}}$, where the precise power $\alpha$ depends on how we impose the cutoff above, then we find that these modes are already enough to account for the entropy of the CFT. (This is similar to the ``brick wall'' explanation of the black hole entropy in flat space \cite{'tHooft:1984re}.) Dimension counting, and the intuition from the simple harmonic oscillator above would then suggest that there are no operators $\tO_{\omega_n, \ang}$ that commute with all these modes.  

While this commutator argument is a powerful constraint in practice, and was an important guiding principle in our construction \cite{Papadodimas:2013jku,Papadodimas:2013wnh}, as the reader will notice it is hard to make it rigorous beyond this level. Moreover, power law suppressed commutators may be justified and even needed on physical grounds since the fields in the bulk are not strictly local. If we are willing to accept these small commutators, then the ``commutator argument'' above loses its power somewhat. For example, the reader can consult the talk \cite{suvratkitptalk} for an example that predates \cite{Papadodimas:2013wnh,Papadodimas:2013jku} and explores a model with such commutators.

This concludes our summary of the arguments that suggest that $\tO_{\omega_n, \ang}$ cannot be found as state-independent operators in the CFT. A logical possibility is to accept that black holes have no interior. However, we believe that a more compelling alternative is that the black hole interior is described by state-dependent operators in the CFT.


\section{Paradoxes for the eternal black hole}
\label{seceternal}
In this section, we show how versions of the paradoxes discussed in section \ref{secparadoxes} also appear in the thermofield double state.  It is sometimes believed, even by those who advocate that the single sided black hole does  not have an interior, that the thermofield double state nevertheless does correspond to an eternal black hole with a smooth horizon. For example, see \cite{Marolf:2013dba}.

We will now show that this position is inconsistent. If we  assume that the thermofield double state is dual to the eternal black hole, and demand only that the
bulk theory respects diffeomorphism invariance  --- which is a minimal requirement in a theory of quantum gravity --- then we can set up a large new class of states, all of which are dual to smooth black holes.  
This new class of states is obtained by performing one-sided diffeomorphisms on the geometry. We argue that diffeomorphisms that die off at the right boundary (but not, possibly, on the left boundary) should not affect the value of observables defined relationally from the right. This is a robust statement, and relies only on the fact that the gravity dual is diffeomorphism-invariant --- and not, in any way, on the equations of motion. 

We then show that demanding that we find operators that behave correctly in {\em all} the states above leads to the same paradoxes that one finds in the single-sided case. Therefore a map between the bulk and the boundary, which can successfully describe the black hole interior in all these states, must be state-dependent.

Our analysis is also useful because it indicates what state-dependence really means. To obtain the paradoxes above, we have to perform ``extremely large'' diffeomorphisms on one side --- shifting the left boundary by timescales of order $e^{\nc} \times \ell_{\text{AdS}}$ before gluing it back to the geometry. What the analysis below shows is that it is not possible to use the same operator in the original state, and in all states that are obtained by deforming it with diffeomorphisms that could be exponentially large. 

We start by reviewing the thermofield double state, and the geometry of the eternal black hole. Then we examine a class of ``phase shifted'' states, which are natural to consider from the point of view of the CFT, and show that they are also smooth because they are related to the original geometry by diffeomorphisms. We then set up analogues of the single-sided paradoxes. We defer the construction of {\em state-dependent} operators to section \ref{secdefnmirror}.

A shorter version of the  arguments of this section was also presented in \cite{Papadodimas:2015xma}. In this section we elaborate on the arguments there and fill some gaps. For some previous discussion of the eternal black hole see \cite{Marolf:2012xe}.

\subsection{Review of the eternal black hole and the thermofield double}
\label{secreview}
We start by reviewing the eternal black hole geometry and the duality proposed in \cite{Maldacena:2001kr}.  The important point that we want to emphasize is the ``time reversal'' that is involved in gluing the geometry to the CFT, which is sometimes under-emphasized.

\begin{figure}[!h]
\begin{center}
\includegraphics[width=5cm]{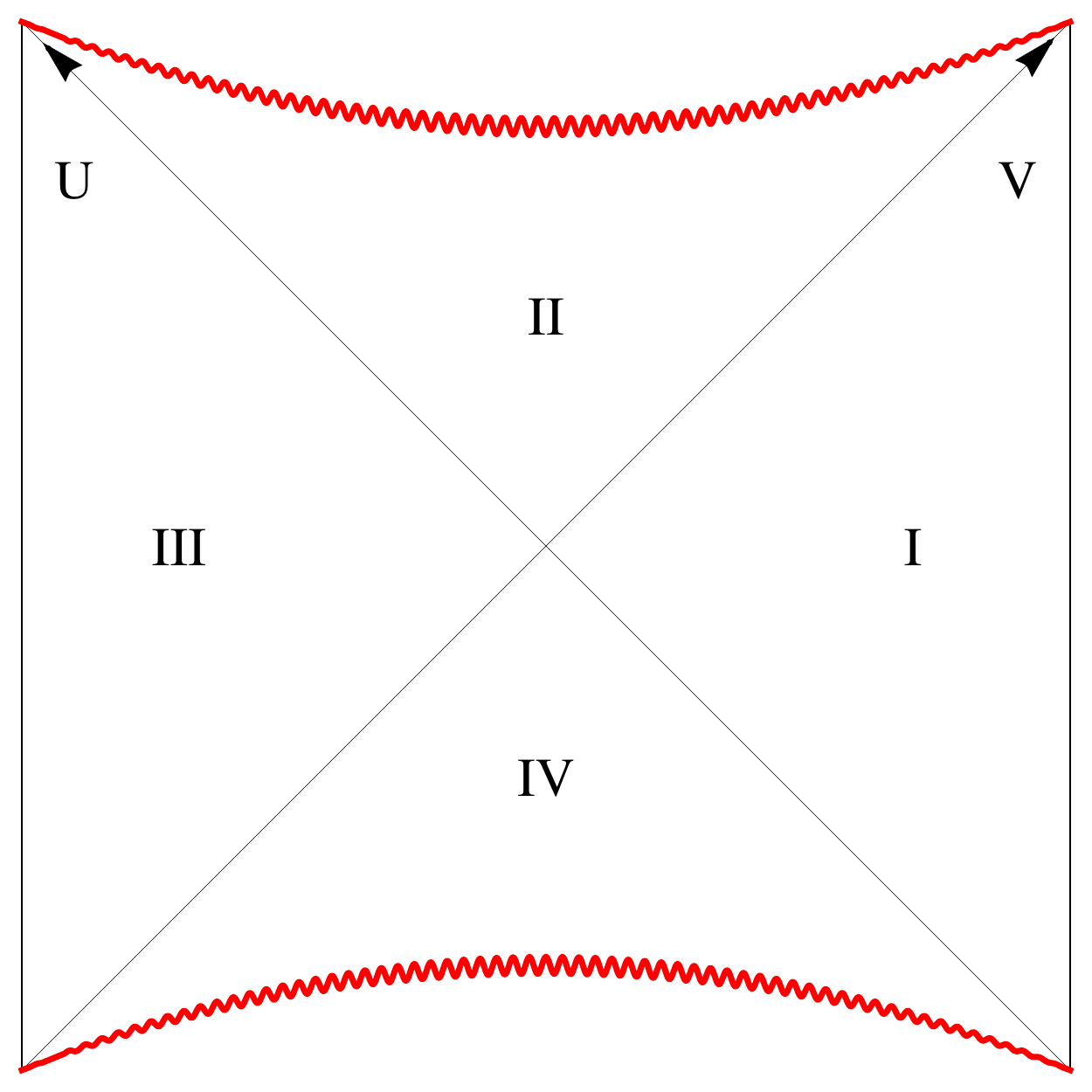}
\caption{\label{ethschematic} Eternal Black Hole in AdS.}
\label{figeternal}
\end{center}
\end{figure}

A schematic figure of the eternal black hole is shown in Figure \ref{ethschematic}.
For the eternal black hole, the metric is again given by 
\eqref{schwarzschild} outside the horizon. Just as in \ref{subsecbulkmirror} we introduce tortoise coordinates with the property that $r_* \rightarrow -\infty$ at the future horizon. The difference with the discussion in \ref{subsecbulkmirror} is that after introducing the Kruskal coordinates, and extending the geometry inside the black hole we now extend the metric in a maximal way while assuming that there is no matter anywhere. This leads to the eternal black hole shown in figure \ref{ethschematic}, which also contains regions III and regions IV as shown in the figure. We can introduce Schwarzschild coordinates in all regions, and the relationship between the Kruskal and Schwarzschild coordinates is given below.
\be
\label{regiondef}
\begin{array}{l|l|l}
\text{Region}&\text{signs~of~}(U,V)&\text{Relationship~to~}(t,\rtor)\\ \hline
\text{I}& U < 0, V > 0 & U = -e^{{2 \pi\over \beta} (r_* - t)},~V=e^{{2\pi \over \beta} (r_* + t)} \\
\text{II} & U > 0, V > 0 & U = e^{{2 \pi\over \beta} (r_* - t)},~V=e^{{2\pi \over \beta} (r_* + t)} \\
\text{III} & U > 0, V < 0 & U = e^{{2 \pi\over \beta} (r_* - t)},~V=-e^{{2\pi \over \beta} (r_* + t)} \\
\text{IV} & U < 0, V < 0 & U = -e^{{2 \pi\over \beta} (r_* - t)},~V=-e^{{2\pi \over \beta} (r_* + t)} \\
\end{array}
\ee

The boundary, in these coordinates, is determined by the hyperbola $U V = -1$. On the other hand, the singularity lives at another hyperbola $U V = \text{positive~constant}$. The two null rays $U = 0, V = 0$ determine all four horizons. The horizon between region I and region II, which would be the ``future horizon'' for the right infalling observer is at $U = 0$. This same null ray also demarcates the boundary between regions IV and III and is therefore the ``past horizon'' for the left observer.  The ray $V=0$ is the ``future horizon'' for the left infalling observer, and the past horizon for the right observer.

The advantage of the choice of coordinates in \eqref{regiondef} is that, in the U-V plane, surfaces of $t = \text{const}$ are simply straight lines running through the origin. This includes the horizons, which are $t = \infty$ and $t = -\infty$ respectively. Therefore, in these coordinates, geometrically we can think of time-translations as ``rotations'' of the Kruskal diagram about the bifurcation point. Of course, we caution the reader that no finite rotation can rotate a line past the horizons.  On the other hand, surfaces of constant $r_*$ are hyperboloids that always stay within a single region. 

Now, we mention an important point. When we associate the Schwarzschild time with the CFT time, we must ``glue'' the geometry to the left CFT with a flip in the time coordinate in region III. Therefore, denoting the time in CFT$_R$ by $\rt$ and the time in CFT$_L$ by $\lt$ we have the identifications
\be
\label{gluetimerev}
\lt = -t, \quad \rt = t,
\ee
where $t$ is the Schwarzschild time. 
 An alert reader might ask that, given that there is no natural choice of the origin of time, why one should not glue the geometry on the left as $\lt = -t + T$, where $T$ is some constant. This is indeed possible, and will be a central point in our discussion below. 

We now turn to a description of the thermofield double state of the CFT. Maldacena conjectured \cite{Maldacena:2001kr} that the geometry we have described above is dual to an entangled state of {\em two} identical, non-interacting CFTs
\be
\label{tfdfull}
|\tfd \rangle = {1 \over \sqrt{Z(\beta)}} \sum_E e^{-{\beta E \over 2}} \timerev | E, E \rangle,
\ee
Here $Z(\beta)$ is the partition function of a {\em single} CFT at the inverse temperature $\beta$ and $|E, E \rangle \equiv |E \rangle_{L} \otimes |E \rangle_{R}$ is a tensor-product state of two energy eigenstates. Although the CFTs are entangled, they are non-interacting, and $\timerev$ is the time-reversal operator, which acts on left energy eigenstates.\footnote{For simplicity, we assume that the CFT under consideration is invariant under time-reversal and direct the reader to \cite{Andrade:2013rra} for comments about the more general case.} The formula \eqref{tfdfull} is usually written with a  tacit choice of the time-reversal operator 
\be
\label{timereversechoice}
\timerev | E \rangle = | E \rangle,
\ee
in which case \eqref{tfdfull} reduces to the standard form
\be
\label{tfd}
|\tfd \rangle = {1 \over \sqrt{Z(\beta)}} \sum_E e^{-{\beta E \over 2}} | E , E \rangle,
\ee
We denote the Hamiltonian of the ``left CFT'' by $\cop{H}_L$ while that of the ``right CFT'' by $\cop{H}$\footnote{We use the notation $(\cop{H}_L, \cop{H})$ instead
of what would be the more symmetric $(\cop{H}_L, \cop{H}_R)$ in order to keep the notation consistent with section \ref{secentangled} and also because we try to define ``right-relational'' observables, thus
breaking the symmetry between the two CFTs.}. 

We immediately see that $|\tfd \rangle$ has a symmetry
\be
\label{tfdisom}
\begin{split}
&(\hleft - \hcft) | \tfd \rangle= 0 \\
&\Rightarrow e^{i (\hleft - \hcft) T} | \tfd \rangle = | \tfd \rangle.
\end{split}
\ee
This symmetry of the thermofield double state corresponds to the isometry of the bulk geometry under $t \rightarrow t + T$. However, as is clear from the equation above, this symmetry corresponds to a shift in the CFT time in opposite directions in the two CFTs.
\be
t \rightarrow t + T\quad \Rightarrow \quad \rt \rightarrow \rt  + T; \quad \lt \rightarrow \lt - T.
\ee

Now, let us examine why the eternal black hole, glued to the boundary as described above, is dual to the thermofield state $|\tfd\rangle$, which involves a time-reversal on the left rather than a time-reversal combined with a time-translation.  Consider {\em mixed} correlators of a single trace operator in the thermofield state with one point, $(t_1, r_1, \Omega_1)$ in region III and the other point $(t_2, r_2, \Omega_2)$ in region I. We would like to ensure that the bulk two-point function in this geometry has a limit that leads to these correlators.
\be
\label{tfdaslimitofeth}
Z^2 \lim_{r_1, r_2 \rightarrow \infty}(r_1)^{\Delta} (r_2)^{\Delta}  \langle \phi(t_1, r_1, \Omega_1) \phi(t_2, r_2, \Omega_2) \rangle_{\text{EBH}} = \tfdbra \op_1(-t_{1},{\Omega}_{1})  \op_R(t_2, {\Omega}_2) |\tfd\rangle,
\ee
where the left hand side is computed using bulk effective field theory in a metric that behaves asymptotically on both the right and the left as \eqref{schwarzschild}, and the right hand side is computed as an expectation value in the thermofield state.

To compute the bulk two point function in the eternal black hole metric is non-trivial, but we can do it patch-wise as follows. We write down expansions for the field in regions I, II, and III of the eternal black hole geometry. Only the near-horizon expansions are relevant and, with a short
extension of the analysis of section \ref{secsmoothint} these expansions can be written as follows.
\begin{align}
\label{futhorone}
&\phi(t, \rtor, \Omega) \xrightarrow[U \rightarrow 0^-]{ V > 0} \sum_{\ang} \int_0^{\infty} {d \omega \over \sqrt{\omega}} a_{\omega, \ang} e^{-i \omega t} Y_{\ang}(\Omega)\left( e^{i \delta} e^{i \omega \rtor} + e^{-i \delta} e^{-i \omega \rtor} \right) + \text{h.c}\\
\label{futhortwo}
&\phi(t,\rtor, \Omega) \xrightarrow[U \rightarrow 0^+]{V > 0}  \sum_{\ang} \int_0^{\infty} {d \omega  e^{-i \delta} \over \sqrt{\omega}}   \left(a_{\omega, \ang} e^{-i \omega (t+\rtor)} Y_{\ang}(\Omega) + \widetilde{a}_{\omega, \ang}  e^{i \omega (t - \rtor)} Y_{\ang}^*(\Omega) \right) + \text{h.c} \quad \\
\label{pasthortwo}
&\phi(t,\rtor, \Omega) \xrightarrow[V \rightarrow 0^+]{U > 0}  \sum_{\ang} \int_0^{\infty} {d \omega  e^{-i \delta} \over \sqrt{\omega}}  \left(\widetilde{a}_{L,\omega, \ang} e^{-i \omega (t+\rtor)} Y_{\ang}(\Omega) + a_{L,\omega, \ang}  e^{i \omega (t - \rtor)} Y_{\ang}^*(\Omega) \right) + \text{h.c} \hphantom{\text{du}}\\
\label{pasthorthree}
&\phi(t,\rtor, \Omega) \xrightarrow[V \rightarrow 0^-]{U > 0}  \sum_{\ang} \int_0^{\infty} {d \omega \over \sqrt{\omega}} a_{L, \omega, \ang} e^{i \omega t} Y_{\ang}(\Omega) \left(e^{i \delta} e^{i \omega \rtor}  +  e^{-i \delta} e^{-i \omega \rtor} \right) + \text{h.c} \quad \\
\end{align}
Here we have introduced two new operators $a_{L, \omega \ang}$ and its mirror $\widetilde{a}_{L, \omega, \ang}$. At the horizon between region III and region II, the field is defined using a left relational coordinate system using the techniques of \eqref{subsecrelational} and at the horizon between region I and II, it is defined using a right relational coordinate system as usual.

The phase factors of $e^{i \delta}$ in the expansion above are slightly subtle. In \eqref{futhorone} the two phase factors are fixed by the behaviour of the mode at infinity by demanding \eqref{tfdaslimitofeth} and by scattering in the bulk. In \eqref{futhortwo} the factor of $e^{-i \delta}$ multiplying the left mover is fixed but we have a choice of convention for the right movers. In region IV we have the same geometry but time-reversed and this fixes the phase factors in \eqref{pasthorthree} once again. We once again have some freedom in \eqref{pasthortwo} for left relational mirror.

Now notice that \eqref{futhortwo} and \eqref{pasthortwo} have an overlapping regime of validity near the bifurcation point. Imposing the condition for the regularity of the two point function that was discussed in section \ref{secsmoothint} we find that we must have
\be
\langle a_{\omega, \ang} a_{L, \omega', \ang'} \rangle = {e^{-{\beta \omega \over 2}} \over 1 - e^{-\beta \omega}} \delta(\omega - \omega') \delta_{\ang \ang'}.
\ee
Since the two point function of the generalized free fields is the same in both CFTs, we can assume that \eqref{adefo} holds on both sides after we appropriate discretize the CFT modes. Therefore, from the bulk geometry and from \eqref{tfdaslimitofeth} and after taking \eqref{gluetimerev} into account we find that from the bulk we obtain the prediction for the boundary two-point function
\be
\label{twoptfnneed}
\langle \tfd |  \Oright[\omega_n, \ang] \Oleft[\omega_n, \ang] | \tfd \rangle = e^{-{\beta \omega_n \over 2}} G_{\beta}(\omega_n, \ang).
\ee
Note that here we have used a relationship between the boundary two point function $G_{\beta}(\omega_n, \ang)$ and the boundary commutator $\comm_{\beta}(\omega_n, \ang)$ that appears in \eqref{adefo}. This follows from the KMS condition and is reviewed in \cite{Papadodimas:2012aq}.

To prove this we allow the matrix elements of these operators to be $c_{j i}$ so that
\be
\label{opactionright}
\Oright[\omega_n, \ang] \sum_i e^{-{\beta E_i \over 2}} |E_i,E_i \rangle = \sum_{i,j} e^{-{\beta E_i \over 2}} c_{j i} |E_i, E_j \rangle .
\ee
If the time reversal symmetry acts as $\timerev | E \rangle = | E \rangle$ then using the fact that $\timerev \Oright[\omega_n, \ang] \timerev = \Oright[\omega_n, \ang]$, it follows that the $c_{j i}$ must be real. Therefore 
\be
\label{opactionleft}
\Oleft[\omega_n, \ang] \sum e^{-{\beta E_j \over 2}} |E_j, E_j \rangle = \sum e^{-{\beta E_j \over 2}} c_{i j} |E_i, E_j \rangle.
\ee
Since the matrix elements of $c_{j i}$ are concentrated around $E_i - E_j = \omega_n$ we see that 
This is indeed true in the CFT because we can show that
\be
\Oleft[\omega_n, \ang] |\tfd \rangle = e^{-{\beta \omega_n \over 2}} \Oright[\omega_n, \ang]^{\dagger} |\tfd \rangle.
\ee
From here \eqref{twoptfnneed} follows automatically.

We have therefore shown that the thermofield double state corresponds to the eternal black hole geometry glued with the specific identification \eqref{gluetimerev}. We return to this question below. We will see that states with different correlators between the left and right boundary can also correspond to smooth geometries, albeit ones which are ``glued'' differently to the boundary. 

\subsection{Time-evolved thermofield states}

We start by examining the effect of time evolution on the thermofield state. We consider the state
\be
\label{tfdtdef}
| \tfdT \rangle =e^{i (\hleft + \hcft) {T \over 2} } |\tfd \rangle = e^{i \hleft T} |\tfd \rangle.
\ee
This is obtained by performing Hamiltonian evolution on the base thermofield state. We now perform both a geometric and a CFT analysis of these states. Our main results about these states come from understanding their geometry, as we do in the next subsection. However, we then provide some supporting arguments for these conclusions directly from the CFT.

\subsubsection{Geometric analysis of time shifted states}
The action of the global symmetry group of the theory (which includes the Hamiltonian, of course) has been the subject of significant analysis in the general relativity literature \cite{Regge:1974zd}. The reader may find it useful to recall the analysis of Brown and Henneaux \cite{Brown:1986nw} who used such diffeomorphisms to analyze the action of the conformal group on the AdS$_3$ vacuum. For some more recent applications see \cite{Guica:2008mu}. The point is that Hamiltonian evolution ---  or evolution by some other global charge --- corresponds to {\em large diffeomorphisms.} These operations may change the  state of the theory.

\begin{figure}[!h]
\begin{center}
\includegraphics[height=0.3\textheight]{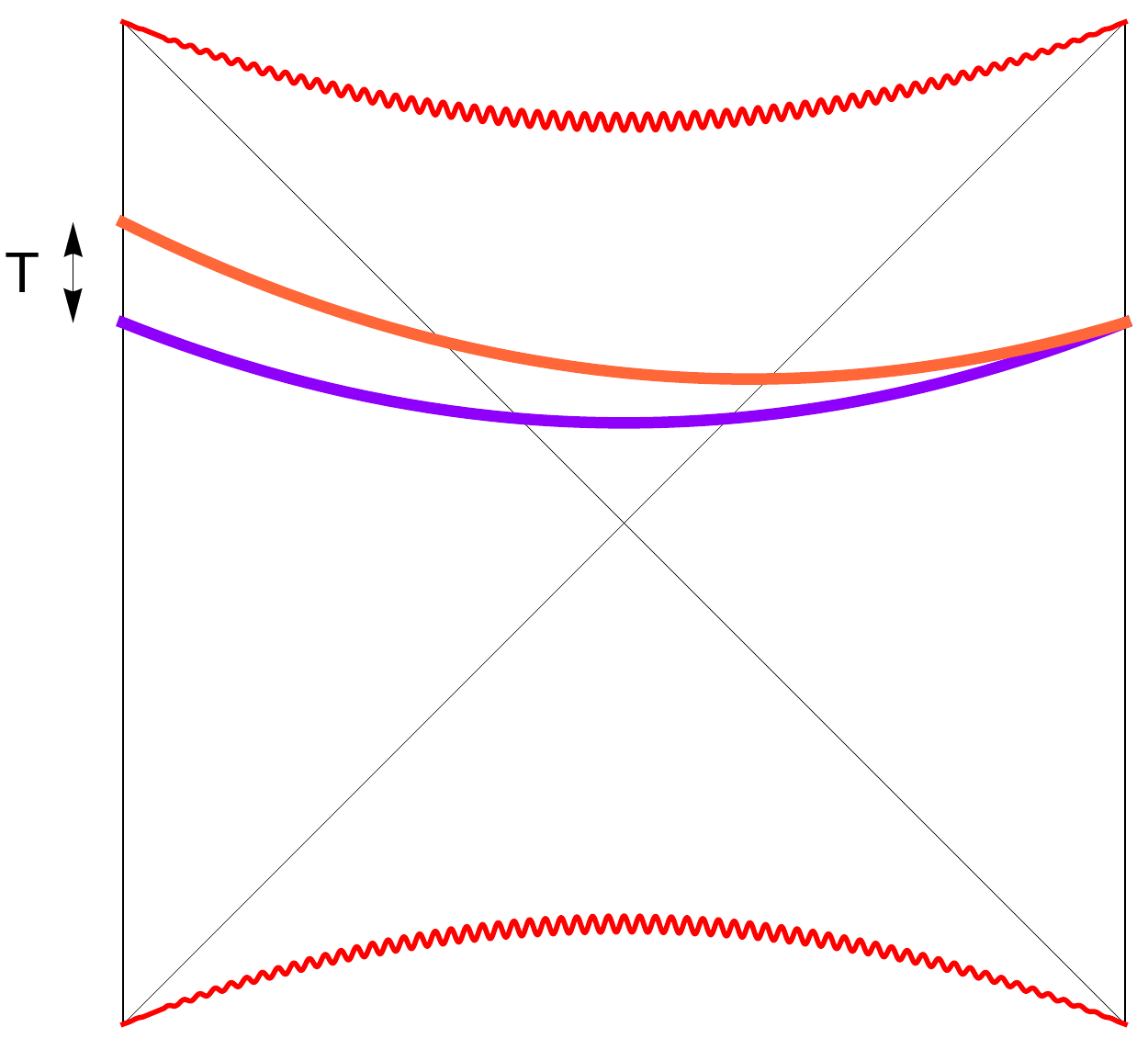}
\caption{The action of $e^{i \hleft \capt}$ is a large diffeomorphism that does not vanish on the left boundary. Its action on one nice slice is shown above.}\label{figleftdiff}
\end{center}
\end{figure}

A quick way to see this is as follows. Consider a nice slice that that runs through the interior of the black hole and is anchored at the points $(\lt, \rt)$. 
According to the standard analysis of the  Hamiltonian constraint \cite{DeWitt:1967yk}, the bulk Hamiltonian (including that of gravity and the other matter fields) must satisfy  $H_{\text{bulk}} |\tfd \rangle = 0$. Therefore, time evolution of this slice is generated only by the boundary Hamiltonians $\hcft$ and $\hleft$.  The action of $e^{i \hleft \capt}$ evolves this slice to another slice that is anchored at $(\lt + \capt, \rt)$. This is shown in Figure \ref{figleftdiff}. 

To summarize the geometric action of the left and right Hamiltonians is as follows.
\begin{enumerate}
\item
$e^{i \hleft \capt} \leftrightarrow $ large diffeomorphisms that die off at the right boundary, but not at the left boundary. On the left boundary, these diffeomorphisms shift points  by $(\lt, \Omega_L) \rightarrow (\lt + \capt,  \Omega_L)$.
\item
$e^{i \hcft \capt} \leftrightarrow $ large diffeomorphisms that die off at the left boundary, but not on the right boundary. On the right boundary, these diffeomorphisms shift points by  $(\rt, \Omega_R) \rightarrow (\rt + \capt,  \Omega_R)$.
\end{enumerate}

We emphasize two important points. First, note that the operation $e^{i \hleft T}$ does not correspond to a unique diffeomorphism. Rather there is an {\em equivalence class} of diffeomorphisms, all of which have the property outlined above. All diffeomorphisms in this equivalence class differ by {\em trivial diffeomorphisms}, which are those that die off at both boundaries. In terms of the nice slice picture of Figure \ref{figleftdiff}, this corresponds to the fact that we can choose to extend the nice slice in any way we like in the bulk, and a particular choice of nice slices is related to a choice of gauge. The left Hamiltonian must nevertheless evolve these slices forward in time. It achieves this because its Dirac brackets with operators in the interior depend on the choice of gauge. Therefore gauge invariant statements about the diffeomorphism can only make reference to its action on the boundary and not in the interior.

Second, from the CFT we can see that while $e^{i \hleft T}$ and $e^{i \hcft T}$ change the state, an operation by $e^{i (\hleft - \hcft) T}$ leaves the thermofield state invariance, since it satisfies $(\hleft - \hcft) | \tfd \rangle= 0$. Geometrically, this has the following meaning. Apart from the form of the metric itself, the thermofield state also has an additional piece of information that specifies the {\em relative placement} of the two boundaries. More specifically, there is an entire class of states --- all of which correspond to the same gauge invariant geometric quantities --- which differ in how the left boundary is glued to the geometry.

\begin{figure}[!h]
\begin{center}
\includegraphics[height=0.3\textheight]{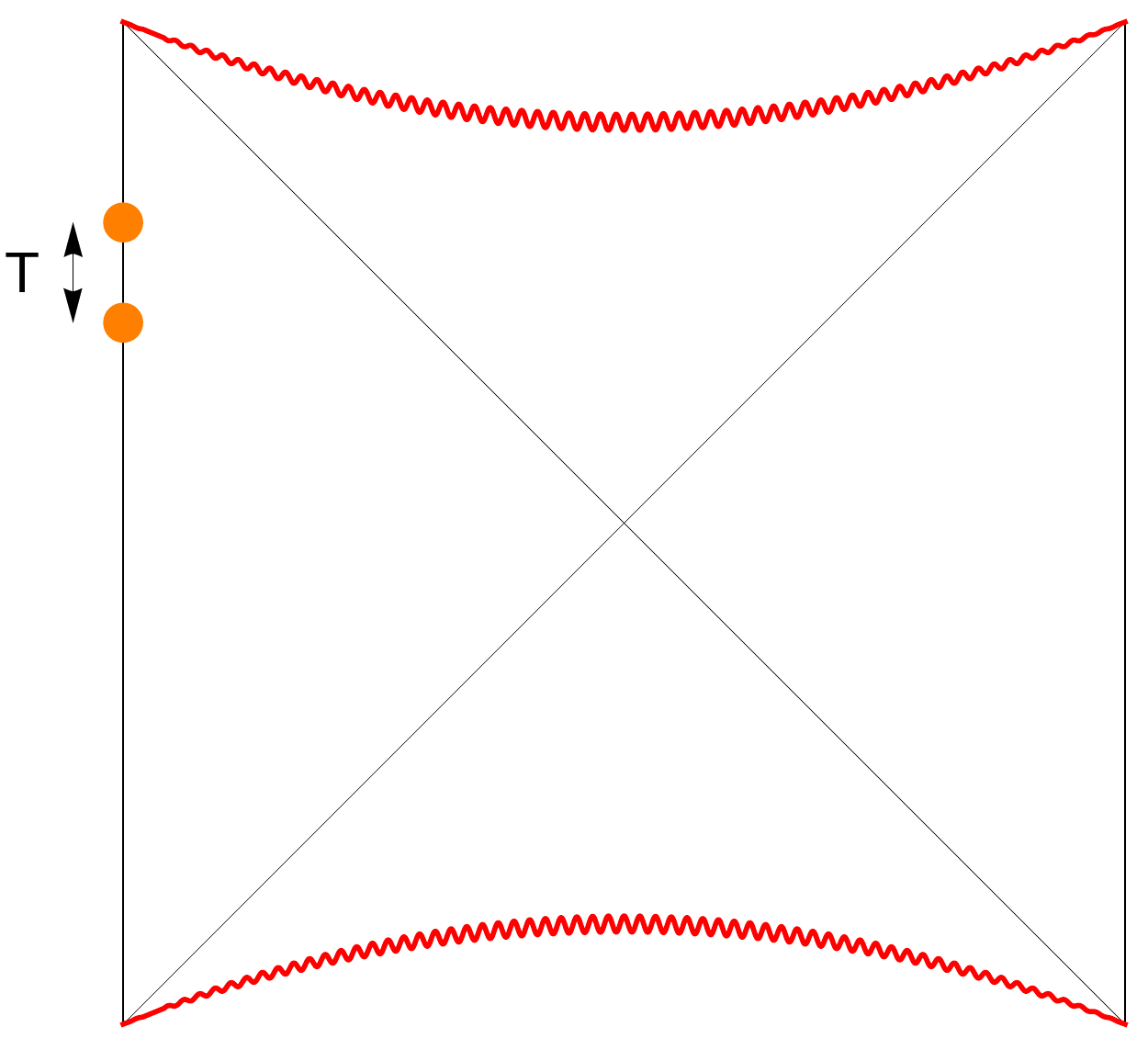}
\caption{Another diffeomorphism in the equivalence class of the diffeomorphism of Figure \ref{figleftdiff}: it slides points on the boundary but acts trivially in the bulk. This can be achieved by composing the diffeomorphism of Figure \ref{figleftdiff} with a trivial diffeomorphism that cancels its action everywhere except for a region that is infinitesimally close to the boundary.}\label{figleftgaugeinv}
\end{center}
\end{figure}

To make this more precise, we describe a specific element of the class of diffeomorphisms that induce the action of $e^{i \hleft T}$. In the Kruskal coordinates $U,V$ described  above, we consider the following  diffeomorphism $U \rightarrow U_T, V \rightarrow V_T$, where $U_T, V_T$ are defined by
\be
\begin{split}
&U_T = U \left(e^{2 \pi \capt \over \beta } \hat{\theta}(U - V) + \hat{\theta}(V-U) \right), \\
&V_T =  V \left(e^{-{2 \pi T \over \beta}} \hat{\theta}(U - V) + \hat{\theta}(V-U) \right), \\
\end{split}
\ee
where $\hat{\theta}(x)$ is an infinitely differentiable version of the theta function with the property that
\be
\hat{\theta}(x) = \left\{\begin{array}{cc} 1 & x  > \epsilon \\ 
0 & x < -\epsilon  \end{array} \right.
\ee
In the intermediate region $-\epsilon \leq x \leq  \epsilon$ we can take $f$ to be any smooth interpolating function between $0$ and $1$. For example, a function that satisfies all these criterion is given  by
\be
\hat{\theta}(x) = {\theta(x + \epsilon) \over 1 + \theta(\epsilon-x) e^{{\epsilon \over \epsilon + x} + {\epsilon \over x - \epsilon}}}.
\ee
Since this is just a diffeomorphism, it does not actually change any gauge invariant  quantity that we can calculate in the bulk geometry. The correct way to picture the gauge-invariant effects of this diffeomorphism is to think of it as one  that  {\em slides} the left boundary by an amount $T$. The figure \ref{figleftgaugeinv} may help the reader think of  the effect of this diffeomorphism which, as we emphasized above, just changes the relation between the bulk and the boundary.

 It is clear from the  analysis above that the states $| \tfdT \rangle$ are also smooth states. This is an exact  statement that does not rely on the bulk equations of motion and should be respected in any theory of quantum gravity that is diffeomorphism-invariant. In particular, this implies that even for very large $\capt$, such as $\capt = e^{\nc}$, the geometry remains smooth. 

\paragraph{Time-shifted states for an infalling observer}
Consider the experience of an infalling observer in the time shifted thermofield state. This observer starts from region I, and falls towards the singularity.  For example, such an observer could measure CFT correlators
\be
\langle \tfdT| \cop{\phi}(t_1, r_1, \Omega_1) \ldots \cop{\phi}(t_n, r_n, \Omega_n) |\tfdT \rangle,
\ee
where all the points along his trajectory are defined relationally  with respect to the right boundary as in section \ref{subsecrelational}.

We consider the relational observables, and the mirror creation and annihilation operators a little more carefully in the next subsection. However, for now we note 
an important property of the unshifted, standard thermofield state $|\tfd\rangle$: if the observer jumps ``earlier'' or ``later'' in $|\tfd\rangle$,  according to the classical geometry, he will measure the same correlators. As the reader can verify, using classical geometry and quantum field theory quantized around this geometry we have
\be
\langle \tfd | \cop{\phi}(t_1, r_1, \Omega_1) \ldots \cop{\phi}(t_n, r_n, \Omega_n) |\tfd \rangle = \langle \tfd | \cop{\phi}(t_1 + T, r_1, \Omega_1) \ldots \cop{\phi}(t_n + T, r_n, \Omega_n) |\tfd \rangle.
\ee

Next, we note that
\be
|\tfdT\rangle= e^{i \hleft T} | \tfd \rangle= e^{i \hcft T}| \tfd\rangle.
\ee
This results from the isometry \eqref{tfdisom} of the eternal black hole. So
\be
\begin{split}
&\langle \tfd | e^{-i \hleft T} \cop{\phi}(t_1, r_1, \Omega_1) \ldots \cop{\phi}(t_n, r_n, \Omega_n) e^{i \hleft T} |\tfd \rangle \\
&= \langle \tfd | e^{-i \hcft T} \cop{\phi}(t_1, r_1, \Omega_1) \ldots \cop{\phi}(t_n, r_n, \Omega_n) e^{i \hcft T} |\tfd \rangle \\
&=\langle \tfd | \cop{\phi}(t_1-T, r_1, \Omega_1) \ldots \cop{\phi}(t_n-T, r_n, \Omega_n)|\tfd \rangle.
\end{split}
\ee
Therefore, if we combine the isometry of the eternal black hole with the fact that an infalling observer from the right observes the same geometry whenever he jumps in, then we obtain the same conclusion: the states $| \tfdT \rangle$ are smooth for all times. This is a second method to reach the conclusion that we already reached above. We now discuss these states from the perspective of the CFT.

\subsubsection{CFT analysis of time-shifted states}
We emphasize that the statement that we have made above --- namely that the eternal black hole geometry should appear to be smooth
under arbitrarily large diffeomorphisms --- could be considered to be rather strong.   Since, we do not usually make statements about quantities that are exponentially large, using the geometry, let us understand these time shifted states directly from the CFT. 

The point we are making above is equivalent to the assertion that there is {\em no natural common origin of time} for the two CFTs. Usually, the origin of time is not relevant to any experiment. On the right CFT, for example,  we declare some point in time to be $t= 0$, pick some basis of operators that we can measure at that time, which we denote by $\op(0,\Omega)$ and declare that these are the Schr\"{o}dinger operators. We can then classify states, using the eigenstates of these operators. 

In our case, we have two CFTs. Roughly speaking, the original thermofield state involves entanglement between $\op(0, \Omega)$ and $\Oleft[](0,\Omega)$. The relation
\be
\tfdbra \op(0,{\Omega}) \Oleft[](0, {\Omega}') | \tfd \rangle= \langle \tfdT | \op(0,{\Omega}) \Oleft[](\capt, {\Omega'}) | \tfdT \rangle,
\ee
tells us that the shifted states involve entanglement between $\op(0, {\Omega})$ and $\Oleft[](\capt, \Omega)$.
We can make an even stronger statement, as follows. Let us consider eigenstates of  the Schr\"{o}dinger picture
operators which satisfy
\be
\begin{split}
&\op(0,{\Omega}) \,|\eigens_L(\Omega), \eigens(\Omega) \rangle =  \eigens(\Omega)\,|\eigens_L(\Omega), \eigens(\Omega) \rangle,
\\
&\Oleft[](0,{\Omega}) \,|\eigens_L(\Omega), \eigens(\Omega) \rangle = \eigens_L(\Omega)\, |\eigens_L(\Omega), \eigens(\Omega) \rangle,
\end{split}
\ee
where $\eigens_L(\Omega), \eigens(\Omega)$ are c-number functions that specify the eigenstate. We have a corresponding basis of eigenstates for the time-shifted Schr\"{o}dinger basis operators, which are given by
\be
\begin{split}
&\op(0,{\Omega}) \,|\eigens_L(\Omega), \eigens(\Omega) \rangle_T =  \eigens(\Omega)\,|\eigens_L(\Omega), \eigens(\Omega) \rangle_T, \\
&\Oleft[](T,{\Omega})\, |\eigens_L(\Omega), \eigens(\Omega) \rangle_T = \eigens_L(\Omega) \,|\eigens_L(\Omega), \eigens(\Omega) \rangle_T.
\end{split}
\ee
Then the thermofield state and the time-shifted thermofield state are identical when considered as wave-functions on these states
\be
\tfdbra \eigens_L(\Omega), \eigens(\Omega) \rangle = \langle \tfdT | \eigens_L(\Omega), \eigens(\Omega) \rangle_T.
\ee
So, unless we have some means of preferentially choosing the states $|\eigens_L(\Omega), \eigens(\Omega) \rangle$ over the states $|\eigens_L(\Omega), \eigens(\Omega) \rangle_T$, we must treat both the thermofield  state and the time-shifted thermofield state on the same footing.

One distinguishing principle that is sometimes invoked in problems of this kind is to appeal to the ``environment.'' We could state that
the environment picks out the operators $\op_L(0,\Omega)$ and distinguishes them from the operators $\op_L(T,\Omega)$. However, this would tacitly
break the time-translational invariance on the boundary. Moreover, from the point of view of gravity this would be very unusual; we would like
the two coupled CFTs to autonomously describe the bulk geometry, and it would be unusual if some tacit reference to an external environment was
important for deciding whether the geometry was smooth or not.

Let us consider some other methods that appear to uniquely pick the thermofield state but, on closer inspection, do not actually do so. 

\paragraph{Euclidean path integral\\}
The thermofield state can be defined by a Euclidean path integral on an interval of length $\beta$. More precisely we specify
\be
\tfdbra \eigens_L(\Omega), \eigens(\Omega) \rangle = \int_{O(0,\Omega) = \eigens_L(\Omega)}^{O(\beta, \Omega) = \eigens(\Omega)} e^{-S}  [{\cal D}O],
\ee
where we have used $[{\cal D} O]$ to schematically represent the measure over fields in the theory, and placed boundary conditions so that, at time $0$, the field is in the state specified by $\eigens_L(\Omega)$ and at Euclidean time $\beta$ it is in the state $\eigens(\Omega)$. 
However, we see immediately that while the path integral on the right side has an unambiguous value, the interpretation of the path
integral as a wave-function on the left requires us to choose an origin of time. We could as well, write 
\be
\langle \tfdT | \eigens_L(\Omega), \eigens(\Omega) \rangle_T = \int_{O(0,\Omega) = \eigens_L(\Omega)}^{O(\beta, \Omega) = \eigens(\Omega)} e^{-S}  [{\cal D}O].
\ee
So, using the Euclidean path integral to define the wave-function begs the question of whether we should privilege $|\eigens_L(\Omega), \eigens(\Omega) \rangle_T$ versus the states $|\eigens_L(\Omega), \eigens(\Omega) \rangle$.

\paragraph{Time-reversal invariance \\}
Another ostensible method of choosing the phases is to use invariance under the time-reversal operation. If we define the time-reversal operator in the left CFT as $\timerev | E \rangle = |E \rangle$, then the thermofield state is the only one of the family of time-shifted states that satisfies
\be
\timerev | \tfd \rangle= |\tfd \rangle.
\ee
For the other states, recalling that the time-reversal operator acts anti-linearly, we have
\be
\timerev | \tfdT \rangle = \tfdmT.
\ee
However, it is clear that this time-reversal operator itself involves the choice of an origin of time. We could just as well define
a new time-reversal operation by a shift of the time-reversal above and a time-translation. On the basis of energy eigenstates, we define
\be
\timerev^T  |E \rangle = e^{2 i E T} | E \rangle,
\ee
and extend this operation anti-linearly on linear combinations of energy eigenstates. It is clear that
\be
\timerev^T |\tfdT \rangle = | \tfdT \rangle.
\ee
The new operator ${\timerev^T}$ is as valid a time-reversal operator as the operator $\timerev$. Therefore, the idea that time-reversal invariance picks a particular origin of time is also specious; it can only do so, if the origin of time is built into the time-reversal operator.

\paragraph{Time-shifted states as phase-modified states \\}
We  now turn to another property of the time-shifted states.  This property is again suggestive of the fact that nothing very special happens if we take a long time limit of the time-translation. 
Note that we can write the time-shifted states as
\be
\label{phaseshifted}
| \tfdT \rangle = e^{i \hleft T} | \tfd \rangle= {1 \over \sqrt{Z(\beta)}} \sum_E e^{-{\beta E \over 2}}  e^{i \phi_E} |E, E \rangle,
\ee
where $\phi_E$ are real phases. Since we expect the spectrum of the CFT to be chaotic at the high energies that dominate the state \eqref{phaseshifted}, we can obtain {\em almost any} choice of phases $\phi_E$ by choosing a suitable time translation. The relevant equation that we need to satisfy is
\be
E \, T ~\text{mod} ~ 2 \pi = \phi_E,
\ee
and we can satisfy this to arbitrary accuracy for a chaotic collection of energies, if we are allowed to choose $T$ from a large enough range.

There are some exceptions to the kinds of phases we can generate. For example, the energies of supersymmetric states are quantized integrally, and therefore we cannot choose their phases all independently. However, the set of supersymmetric states constitute an {\em exponentially unimportant} subset in the thermofield state $|\tfd \rangle$. More importantly, the energies within a conformal representation are integrally quantized. Therefore by time evolution with the Hamiltonian,\footnote{The reader might notice that we can generate a slightly more general class of phases using other diffeomorphisms, such as those that rotate the $S^{d-1}$, but this is not relevant to our discussion.}
we can only generate phases that satisfy 
\be
\phi[E] - \phi[E+1] = \phi[E+1] - \phi[E+2] \quad \text{mod} ~2 \pi.
\ee

The statement that there is no natural common origin of time translates, in this language, to the statement that there is no natural choice
of phases for the energy eigenstates on both sides.  (This is, subject, of course, to the relations above.) The advantage of thinking in this
language is that it is clear that the phases do not have any special behaviour at late times. Therefore if we accept the standard interpretation that $e^{i \hleft T}$ acts as a large diffeomorphism in the bulk, for $\Or[1]$ times, and preserves a smooth geometry, then it is natural to expect that this also happens for arbitrarily long $\capt$.

We caution the reader however that the argument above is a ``naturalness'' argument. It is predicated on the assumption that a ``natural'' bulk to boundary map should not privilege one pattern of random phases (obtained by translations of $\Or[1]$) from another pattern of random phases (obtained by translations of $\Or[e^{\nc}]$). So it is suggestive and not a proof.

\subsection{Relational observables in time shifted states}
\label{subsec:reltss}
We now turn to a detailed discussion of relational observables in time-shifted states. 
These operators are particularly important in our discussion of the eternal black hole. 

We have already carefully defined relational observables in section \ref{subsecrelational}. Now, the key point is as follows. These observables are defined relationally with respect to the right boundary. Therefore, if we consider diffeomorphisms that die off at the right boundary, then right-relational observables are invariant under such diffeomorphisms, {\em even} if the diffeomorphisms do not die off at the left boundary.

This point may be slightly confusing if one thinks of diffeomorphisms that shift the left boundary as acting everywhere in the spacetime. However, as we pointed out,   these diffeomorphisms belong to an equivalence class, and a limiting element of the class is the diffeomorphism that simply ``slides'' the left boundary up and down while leaving the rest of the geometry invariant. If we consider this element of the class, it is clear
that right relational observables are left invariant.

Let us check this more explicitly by carefully repeating the derivation of \ref{subsecrelational}. We start by defining points in the bulk
as intersection points of null geodesics which end on the boundary. We introduce asymptotically AdS coordinates, so the near the boundary the metric coincides with \eqref{rhometric}. These coordinates are $(t, \rho, \Omega)$ and the boundary is at $\rho = 1$. 
We now consider two solutions to the geodesic differential equation parameterized by  ordinary AdS time (not necessarily an affine parameter) with the property that
\be
\label{rightboundary}
\begin{split}
&\vect{x}_1(t_1) = (t_1, \rho=1,{\Omega_1}); \quad \dot{\vect{x}}_1(0) = (1, -1, 0), \\
&\vect{x}_2(t_1 + \tau) = (t_1+\tau, \rho=1,{\Omega_1}); \quad \dot{\vect{x}}_2(t_1 + \tau) = (1, 1, 0). \\
\end{split}
\ee
We then tune $\Omega_1$ so that the geodesics meet.
Given a particular value of $t_1, \Omega_1(t_1)$, we vary $\Omega_2(t_1 + \tau)$ so that the geodesics intersect at some $t_i$ with $t_1 < t_i < t_1 + \tau$,
\be
\rho_2(t_i) = \rho_1(t_i); \quad \Omega_2(t_i) = \Omega_1(t_i),
\ee
and we denote the intersection point by $\vect{P}_i(t_1,\Omega_1, \tau)$ as in section \ref{subsecrelational}.

Let us now make a large diffeomorphism that dies off at the right boundary: 
\be
\label{largediffleft}
\vect{x} \rightarrow \vect{\diff}(\vect{x}).
\ee 
To implement this diffeomorphism in a quantum field theory, we can act on all {\em fields} (including the metric), rather than points,  with the inverse transformation. The new scalar fields $\bar{\phi}(\vect{x})$ are given by
\be
\bar{\phi}(\vect{x}) = \phi(\vect{\diff}^{-1}(\vect{x})).
\ee
The action of the diffeomorphism on the metric is
\be
\label{newmetric}
g_{\bar{\mu} \bar{\nu}}(\vect{x}) \rightarrow {\partial x^{\mu} \over \partial \diff^{\bar{\mu}}} {\partial x^{\nu} \over \partial \diff^{\bar{\nu}}} g_{\mu \nu} (\vect{\diff}^{-1}(\vect{x})).
\ee
Now if we transform the entire entire geodesic trajectory specified by the solution to the geodesic equation with initial conditions \eqref{rightboundary} by means of the diffeomorphism  \eqref{largediffleft}, then we get a new trajectory that is a geodesic with respect to the new metric \eqref{newmetric}.

The boundary conditions \eqref{rightboundary} remain invariant under the diffeomorphism since, by assumption, $\diff$ turns into the identity at the boundary. 
Moreover, if the original geodesics intersected, then the new geodesics also intersect. In particular the new intersection point, $\bar{P}_i$ is just given by the transform of the original intersection point
\be
\vect{\bar{P}}_{i}(t_1, \Omega_1, \tau)=\vect{\diff}(\vect{P}_i(t_1, \Omega_1, \tau)),
\ee
where we are using the same notation as \eqref{outsideparam}.

Now consider evaluating a scalar field at this intersection point. Clearly we have
\be
\bar{\phi}(\vect{\bar{P}}_i) =  \phi(\vect{\diff}^{-1}(\vect{\diff}(\vect{P}_i)))  = \phi(\vect{P}_i),
\ee
which is the same value as it had before the diffeomorphism. Therefore, scalar observables defined at points which are related relationally to the right boundary are invariant under left diffeomorphisms.

This logic extends to points behind the horizon. Recall that these points were defined by solutions to the geodesic equation, where the affine parameter was normalized by using the points outside the horizon already defined above. Clearly, in the new metric the new geodesics are again given by $\vect{\diff}(\vect{x}(\lambda))$, and by the same logic scalar variables evaluated inside the horizon are invariant under any diffeomorphism that dies off at the right boundary.

\subsubsection{Commutator of mirror operators }\label{subsec:commutatormirror}
Note that, in the analysis above, it was important that the boundary conditions \eqref{rightboundary} were not altered by the diffeomorphisms. If we consider diffeomorphisms that do not die off at the right boundary, then the right relational observables do transform, but in a simple manner.
Under a diffeomorphism that shifts points on the right boundary by $\rt \rightarrow \rt + \capt$, we have
\be
\vect{\bar{P}}_i(t, \Omega, \tau) = \vect{\diff}(\vect{P}_i(t-\capt, \Omega, \tau)).
\ee
For the field operators, defined relationally with respect to the right boundary, this leads to
\be
\label{commutH}
\begin{split}
&e^{i H_L \capt} \phirelr(\rt,\Omega, \lambda) e^{-i H_L \capt} = \phirelr(\rt,\Omega,\lambda), \\
&e^{i H \capt} \phirelr(\rt,\Omega, \lambda) e^{-i H \capt} = \phirelr(\rt + \capt, \Omega, \lambda),
\end{split}
\ee
where $H_L$ and $H$ are the left and right boundary Hamiltonians respectively.

We now write down a mode expansion for the fields in front of and behind the horizon, as in \eqref{futhorone} and \eqref{futhortwo}. 
The conditions \eqref{commutH} imply that when we try and find CFT operators that can play the role of these mirrors
then they must have the CFT commutation relations
\be
\label{commutmodes}
\begin{split}
&[\hcft \,,\, \anc_{\omega, \ang}] =\, - \omega \,\anc_{\omega, \ang}, \quad [\hleft \,, \,\anc_{\omega, \ang}]  = 0, \\
&[\hcft \,,\, \ta_{\omega, \ang}] = \omega \,\ta_{\omega, \ang}, \quad [\hleft \,,\, \ta_{\omega, \ang}]= 0 .
\end{split}
\ee
We remind the reader that the asymmetry above arises because these are right relational modes. The relation \eqref{commutmodes} must hold approximately within low point correlation functions, and not necessarily as operators. However, within correlators they are crucial to ensure that the field operators transform correctly under large diffeomorphisms.

We will proceed to now argue that it is impossible to find state-independent operators $\widetilde{a}_{\omega, \ang}$ that have the right properties to play the role
of mirror operators behind the horizon in the entire family of time-shifted states.

\subsection{Naive construction of local operators in the thermofield double}\label{subsecnaiveeternal}
We start by considering the naive construction of local operators in the thermofield double. We will show that this does not satisfy the conditions above and, therefore, cannot be correct. In particular we would like to identify CFT operators $\ta_{\omega_n, \ang}$ with the properties that we derived from the bulk above.

The naive construction of local operators proceeds by simply identifying discretized mirror modes with modes on the left CFT 
\be
\ta_{\omega_n, \ang} \underset{\text{naive}}{\longrightarrow} \cop{a}_{L \omega_n, \ang}.
\ee
 However, this is clearly wrong as a computation of the two point function across the horizon shows.  
If we now compute this two point correlator in the time-shifted state, we find that
\be
\begin{split}
&\langle \tfdT | \an[L \omega_n, \ang] \an[\omega_n, \ang] | \tfdT \rangle = e^{i \omega_n T}{ e^{-{\beta \omega_n \over 2}} \over 1 - e^{-\beta \omega_n}}, \\
&\langle \tfdT | \an[L \omega_n, \ang]^{\dagger} \an[\omega_n, \ang]^{\dagger} | \tfdT \rangle =e^{-i \omega_n T} { e^{-{\beta \omega_n \over 2}} \over 1 - e^{-\beta \omega_n}}.\\
\end{split}
\ee
Let us call the CFT operator obtained by using this ``naive'' mode $\cop{\phi}^{\text{n}}$. Now, repeating the computation of the two point function that we performed in section \ref{secsmoothint}, with point $1$ outside the horizon and point $2$ behind the horizon we find that
\be
\label{lightconeresulttfdt}
\begin{split}
&\lim_{V_1 - V_2 \rightarrow 0} \langle \tfdT | \partial_U \cop{\phi}^{\text{n}}(U_1, V_1, \Omega_1) \partial_U \cop{\phi}^{\text{n}}(U_2,V_2, \Omega_2)| \tfdT \rangle = c {\delta^{d-1}(\Omega_1 - \Omega_2) \over (U_1 - U_2 e^{-{2 \pi \capt \over \beta}})^2} , \\
&\lim_{U_1 - U_2 \rightarrow 0} \langle \tfdT | \partial_V \cop{\phi}^{\text{n}}(U_1, V_1, \Omega_1) \partial_V \cop{\phi}^{\text{n}}(U_2,V_2, \Omega_2) | \tfdT \rangle  = c {\delta^{d-1}(\Omega_1 - \Omega_2) \over (V_1 - V_2 )^2}, 
\end{split}
\ee
where $c$ is a normalization constant. 
Clearly this is not the correct result. In particular, the first line of \eqref{lightconeresulttfdt} does not have the right behaviour when $U_1 \rightarrow U_2$.  We obtain a similar pathology by considering the boundary  between region II and region III. 

This was only to be expected since the operators $\cop{a}_{L \omega}$ clearly do not obey the correct commutators with the Hamiltonian that we demanded above. Therefore, it is incorrect to identify $\ta_{\omega_n, \ang}$ with $\an[L,\omega_n, \ang]$ as has been done commonly in the literature. As we will discuss below, this led to some errors in the analysis of \cite{Harlow:2014yoa}.

\subsection{Paradoxes for the eternal black hole}
We now set out various paradoxes, similar to the ones outlined by \cite{Almheiri:2012rt,Marolf:2013dba,Almheiri:2013hfa}
 which show that the relational observable defined above, {\em cannot} be realized by a linear operator. These paradoxes were already outlined concisely in \cite{Papadodimas:2015xma}, and we suggest that the reader consult that paper in parallel with this section. Our arguments here are more detailed variants of the arguments there. 

Let us assume that some state-independent operators $\ta_{\omega_n, \ang}$ exist with the properties that we derived earlier. If so we can multiply them with the appropriate modes and  construct state-independent operators $\cop{\phi}(U,V, \Omega)$ in the thermofield double state and in a right relational gauge. Then, consider 
\be
 C(U_1, V_1, \Omega_1, \ldots U_n, V_n, \Omega_n) =  \langle \tfdT | \cop{\phi}(U_1, V_1, \Omega_1) \ldots \cop{\phi}(U_n, V_n, \Omega_n) | \tfdT \rangle.
\ee
From the arguments above we have
\be
{d \over d \capt} C(U_1, V_1, \Omega_1, \ldots U_n, V_n, \Omega_n) = 0.
\ee
Second, from the discussion in section \ref{secstatedepvsindep}, we expect this $\capt$-independent answer to correspond to the 
correlators as computed by effective field theory in the eternal black hole. This expectation is indicated in \ref{quasilocalphicond}. 
Now, for any operator $\al_{\alpha}$ we have
\be
\langle \tfdT |  \al_{\alpha} | \tfdT \rangle = {1 \over Z(\beta)} \Big[\sum_{E} e^{-\beta E} \langle E, E | \al_{\alpha} | E, E \rangle + \sum_{E' \neq E} e^{-\beta (E + E') \over 2} e^{i (E - E') T} \langle E', E' | \al_{\alpha} | E, E \rangle \Big].
\ee
Even if we know that this expectation value is $\capt$-independent, we must be careful not to immediately discard the second term above. This is because, if $\al_{\alpha}$ happens to be an operator with support on narrowly separated eigenstates $E - E' = \Or[e^{-{S \over 2}}]$, then the time-variation of the second term will be negligible and so it may appear to be time-independent for short times. However, if we demand
\be
\langle \tfdT | \al_{\alpha} | \tfdT \rangle = \langle \tfd | \al_{\alpha} | \tfd \rangle,
\ee
even for exponentially long times, then the contribution to the expectation value can only come from diagonal terms. 

In the case of the correlator under consideration this implies that
\be
{1 \over Z(\beta)} \sum_{E} e^{-\beta E} \langle E,E | \cop{\phi}(U_1, V_1, \Omega_1) \ldots \cop{\phi}(U_n, V_n, \Omega_n) | E,E \rangle = C(U_1, V_1, \Omega_1, \ldots U_n, V_n, \Omega_n).
\ee
Using the standard arguments from the equivalence of the canonical and the microcanonical ensemble this means that for a typical eigenstate pair  $|E,E \rangle$ at the energy relevant to the eternal black hole
\be
\langle E, E | \cop{\phi}(U_1, V_1, \Omega_1) \ldots \cop{\phi}(U_n, V_n, \Omega_n)| E, E \rangle =  C(U_1, V_1, \Omega_1, \ldots U_n, V_n, \Omega_n).
\ee

At an intuitive level this is already a strange conclusion because the energy-eigenstate pair that appears above has {\em no entanglement.}  We have shown above that no state-independent operators $\cop{\phi}(U, V, \Omega)$ can reproduce the effective field theory correlators in arbitrary {\em single sided} energy eigenstates. How can such operators correctly reproduce this answer in {\em two sided} eigenstate pairs?

We can turn this into a sharp contradiction as follows. In the eigenstate pair $|E, E \rangle$ with no entanglement, we expect that there is no geometric wormhole. Therefore no excitation generated by the left observer can affect the correlators observed by the right infalling observer. In particular, if the left observer decides to act with an arbitrary unitary, $\cop{U}_L$ we should have
\be
\label{nowormholecond}
\begin{split}
&\langle E, E |\cop{U}_L^{\dagger} \cop{\phi}(U_1, V_1, \Omega_1) \ldots \cop{\phi}(U_n, V_n, \Omega_n) \cop{U}_L | E, E \rangle \\ &=  \langle E, E |\cop{\phi}(U_1, V_1, \Omega_1) \ldots \cop{\phi}(U_n, V_n, \Omega_n)  | E, E \rangle.
\end{split}
\ee
We can use this freedom to map the left energy eigenstate to some fixed state --- $\cop{U}_L |E, E \rangle = |F,E \rangle$, where $F$ could even correspond to the left CFT vacuum. This means that the operators $\cop{\phi}(U,V, \Omega)$ must reproduce the correct correlators in all states $|F, E \rangle$ and must be independent of $F$. This can only be if they are ordinary operators in the right CFT. But we have already proved that there are no state-independent operators in the right CFT. Therefore our starting assumption --- that such operators exist in the doubled CFT --- must be wrong.

The reader may consult \cite{Papadodimas:2015xma} for concrete versions of the $N_a \neq 0$ argument, and the negative occupancy argument phrased directly in the doubled CFT. Here, we will conclude by briefly re-emphasizing the importance of the \eqref{nowormholecond}, which states that there is no wormhole in eigenstate pairs. 

In section \ref{secdefnmirror} we will review the construction of {\em state-dependent} operators in a single CFT that can correctly reproduce effective field theory correlators about a black hole. This construction was first described in \cite{Papadodimas:2013jku,Papadodimas:2013wnh}. Let us denote such operators acting only in the original (right) CFT, and defined about an energy eigenstate $|E \rangle$ by $\cop{\phi}^{\{E\}}(U,V, \Omega)$. The superscript $E$ indicates that they reproduce the expected effective field theory answers when evaluated in correlators about $|E \rangle$ and  reasonable excitations of this state. Now, consider the following {\em state-independent} operator, which acts in the Hilbert space of two CFTs
\be
\cop{\Theta}(U,V,\Omega) = \sum_{E} P_{E_L} \otimes \cop{\phi}^{\{E\}}(U,V,\Omega),
\ee
where $P_{E_L}$ is the projector onto the energy eigenstate on the left: $P_{E_L} \equiv |E_L \rangle \langle E_L|$, and the sum is over all energy eigenstates. 

Now $\cop{\Theta}(U,V,\Omega)$ has some interesting properties. When evaluated in the thermofield double, we find
\be
\label{thetaintfd}
\begin{split}
&\langle \tfd | \cop{\Theta}(U_1, V_1, \Omega_1) \ldots \cop{\Theta}(U_n, V_n, \Omega_n) | \tfd \rangle \\ &= {1 \over Z(\beta)} \sum_{E} e^{-\beta E} \langle E |  \cop{\phi}^{\{E\}}(U_1,V_1,\Omega_1) \ldots \cop{\phi}^{\{E\}}(U_n,V_n,\Omega_n) | E \rangle.
\end{split}
\ee
Note that the sum on the right is in a single CFT since the $P_{E_L}$ term simply makes cross terms vanish and gives $1$ for the diagonal terms. 

Since $\cop{\phi}^{\{E\}}(U,V,\Omega)$ is only evaluated in the state $|E \rangle$ and its excitations, the expression above does yield the answer expected from effective field theory. Note that $\cop{\Theta}(U,V,\Omega)$ also produces the following correlators about eigenstate-pairs.
\be
\langle E, E | \cop{\Theta}(U_1, V_1, \Omega_1) \ldots \cop{\Theta}(U_n, V_n, \Omega_n) | E, E \rangle = \langle E |  \cop{\phi}^{\{E\}}(U_1,V_1,\Omega_1) \ldots \cop{\phi}^{\{E\}}(U_n,V_n,\Omega_n) | E \rangle.
\ee
Using the equivalence between the canonical and microcanonical ensemble, these correlators are approximately the same as the thermofield correlators in \eqref{thetaintfd}. These correlators would suggest that the geometry in eigenstate pairs, as seen by the right infalling observer is almost the same in eigenstate pairs as in the thermofield. While this conclusion is correct, as we will see below, the operator $\cop{\Theta}(U,V,\Omega)$ cannot be the correct CFT operator dual to a local bulk fields.

This is because  $\cop{\Theta}(U,V,\Omega)$ violates the no wormhole condition and keeps the wormhole open even when there is no entanglement. In particular, using a left unitary that acts as $\cop{U}_L |E,E \rangle = |F,E \rangle$ we find that 
\be
\begin{split}
&\langle E, E | \cop{U}_L^{\dagger} \cop{\Theta}(U_1, V_1, \Omega_1) \ldots \cop{\Theta}(U_n, V_n, \Omega_n) \cop{U}_L| E, E \rangle \\ &= \langle E |  \cop{\phi}^{\{F\}}(U_1,V_1,\Omega_1) \ldots \cop{\phi}^{\{F\}}(U_n,V_n,\Omega_n) | E \rangle.
\end{split}
\ee
But these are correlators of $\cop{\phi}^{\{F\}}(U,V,\Omega)$ evaluated about a different eigenstate and, in general, these lead to exponentially small answers. Therefore, $\cop{\Theta}(U,V,\Omega)$ cannot be the correct field operators in the eternal black hole because they would predict that even in eigenstate pairs, by performing the unitary transformation discussed above a left observer could alter the correlators of a right infalling observer. So we see that the condition \eqref{nowormholecond} is important in ruling out such putative state-independent operators.  In the next section, we will show how the interior of the eternal black hole can be correctly constructed using state-dependent bulk to boundary maps.

Before concluding this section, we should mention that our arguments should be distinguished from those of \cite{Avery:2013bea,Mathur:2014dia}, who suggested that the duality between the eternal black hole the thermofield double does not hold. Although we will not engage with this in detail, we briefly indicate our point of disagreement. The authors of \cite{Avery:2013bea} suggested that there was an ambiguity in the duality between the thermofield double and the eternal black hole. In particular, they argued that the CFT cannot distinguish between this case and another bulk geometry where the bulk Hamiltonian has been modified by removing the ``interaction'' between the left and the right at the bifurcation point. Alternately, this corresponds to adding a delta-function source there in a manner that appears to be hidden from both CFTs. They argued that this leads to an ambiguity that invalidates the duality.

While this argument may have been plausible if the bulk theory had been an ordinary quantum field theory, it is inapplicable to a theory of quantum gravity. The Hamiltonian constraint rules out the alternate bulk Hamiltonian considered above. It is this crucial feature of the bulk that allows
the boundary to know the ``details'' of the bulk Hamiltonian and allows the duality to be consistent.


\section{Definition of the mirror operators }\label{secdefnmirror}
In the past sections, we have set up paradoxes that show that no state-independent operator can correctly satisfy the conditions outlined in section \ref{secsmoothint}. We have shown that these paradoxes apply to both the single-sided CFT and the thermofield double.

We now review and extend the definition of the mirror operators provided in \cite{Papadodimas:2013wnh,Papadodimas:2013jku}. These operators are state-dependent. What this means, in our context is as follows. Say that we are computing expectation values of a mirror operator within a correlation function
\be
\langle \Psi | \op_{\omega_1, \ang_1} \ldots  \tO_{\omega_p, \ang_p} \ldots \op_{\omega_n, \ang_n}    |\Psi \rangle,
\ee
where $|\Psi \rangle$ is an equilibrium state.
Then, the statement is that the operator $\tO_{\omega, \ang}$ depends, in a subtle manner on the sandwiching state $|\Psi \rangle$.

This would imply that when one speaks of local operators in gravity, or of their modes, then at least behind the horizon of a black hole it is important to specify the state that one is referring to.  
A given local operator is good to describe physics in a given state and in small excitations about that state. If we consider another microstate which is ``far away'', in the sense that it cannot be obtained from the original microstate by the action of a small number of single-trace operators, then we must use a different operator to describe the ``same physical quantity.''

In this section we will first review the construction that we presented in \cite{Papadodimas:2013wnh,Papadodimas:2013jku} both for equilibrium and near-equilibrium states. We show how this completely resolves
all the paradoxes of \cite{Almheiri:2012rt,Almheiri:2013hfa,Marolf:2013dba}. Our review will be brief, and we direct the reader to those papers for a more detailed exposition. 

A significant new element in this paper is that we will discuss the action of our operators on {\em superpositions} of states. This is important, because we show that even though our operators are state-dependent, the infalling observer will not observe any deviations from linearity for small superpositions of equilibrium or near-equilibrium states. 

Next, we also describe the construction of mirror operators for the thermofield double and its time-shifted cousins. This construction can be obtained as a special case of our construction, as applied to an entangled state. However, in this section we also show how one could guess this solution independently. The analysis of \eqref{secexpliciteternal} is useful because it helps to elucidate the nature of state-dependence.

\subsection{The set of natural observables and the little Hilbert space about a state}
Consider the modes of the generalized free field operators that were defined in \eqref{cftmodedef}. 
As we explained there, we have discretized these modes $\op_{\omega_n, \ang}$ 
both by selecting some discrete set of frequencies, and also by choosing a time-band on the boundary that we integrate over to transform to frequency space. 

We now consider the set of polynomials in these modes that we denote by
\be
\label{alsetgff}
\alsetgff = \text{span}\{{\op}_{\omega_1,\ang_1},\,\, {\op}_{\omega_{1}, \ang_1} {\op}_{\omega_{2}, \ang_2}, \ldots, {\op}_{\omega_{1}, \ang_1} {\op}_{\omega_{2}, \ang_2} \ldots {\op}_{\omega_{K}, \ang_K}\}.
\ee
This means that this set comprises all monomials of the form displayed above, and arbitrary linear combinations of these monomials. In addition, we consider the set of polynomials --- limited to small orders --- in the CFT Hamiltonian.\footnote{For a more careful treatment of other conserved charges, including in cases where the CFT has a non-Abelian symmetry we refer the reader to section 3.2.4 of \cite{Papadodimas:2013jku}.}
\be
\label{alsetH}
\alseth = \text{span}\{\hcft, \hcft^2 \ldots \hcft^n\}.
\ee
We then consider the set of observables involving insertions of both the generalized free fields and the CFT Hamiltonian
\be
\label{alsetdef}
\alset = \alset_{\text{gff}} \otimes \alset_{\hcft}.
\ee
The dimension of this set is denoted by
\be
\label{dimadef}
\dima = \text{dim}(\alset).
\ee
We will often refer to arbitrary elements of this set, comprising generalized free fields by 
\be
\label{aldef}
\al_{\alpha} \in \alsetgff.
\ee
We emphasize by default the notation $\al_{\alpha}$ {\em does not}  include the CFT Hamiltonian. If we want to consider an element from $\alset$ that might include $\hcft$, we will state this explicitly.

We want to restrict $\alset$ to be the set of ``reasonable'' experiments
that one can perform in the bulk, and still expect to observe effective field
theory about a given background. This excludes any monomial in \eqref{alsetdef} that has a very high
total energy
\be
\sum \omega_{i} \ll \Or[\nc].
\ee
Similarly, this also excludes any monomial that has a very large number of insertions. So
\be
K \ll \Or[\nc],
\ee
for all monomials displayed in \eqref{alsetdef}. These restrictions imply, as a consequence that
\be
\dima \ll \Or[e^{\nc}].
\ee
The set $\alset$ is approximately an algebra because we can usually multiply
two of its element to obtain another element. However, this is not always the case because of edge effects --- where such a multiplication may take us beyond
the cutoff we have imposed. In this paper we will usually not keep track of these ``edge effects'.'

The set of ``reasonable operators'' can be used to excite a state. This leads us to consider the space
\be
\hilb[\Psi] = \alset |\Psi \rangle \equiv \text{span}\{\sum \coeff[p] \al_{p} |\Psi \rangle \},
\ee
where $\al_{\alpha}$ may include $\hcft$.
We will denote the projector on this subspace by $\projh[\Psi]$.
The fact that $\alset$ is approximately an algebra implies that we can consider the action of its elements as $\al_{\alpha}: \hilb[\Psi] \rightarrow \hilb[\Psi]$. This is subject to the same edge-effect caveat above. 

\begin{figure}[!h]
\begin{center}
\includegraphics[width=0.4\textwidth]{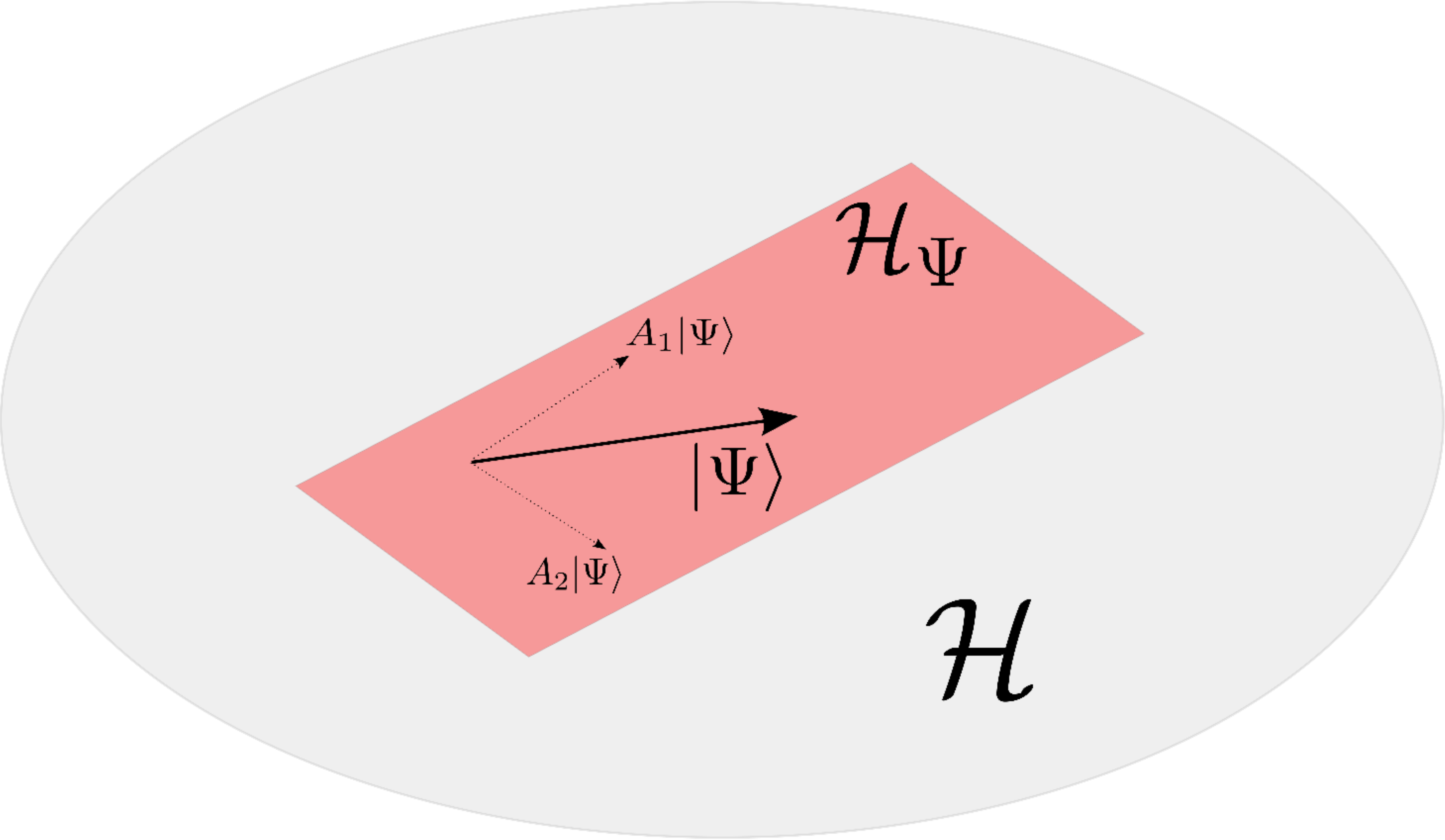}
\caption{A cartoon of the little Hilbert space $\hilb[\Psi]$ as the relevant subspace in the full Hilbert space.}\label{fighpsi}
\end{center}
\end{figure}

We will sometimes call the space $\hilb[\Psi]$ the ``little Hilbert space'' about the space $|\Psi \rangle$, since it contains the part of the Hilbert space 
that is accessible within effective field theory. Conceptually, this little Hilbert space is very important.  We show a schematic figure of this set in Figure \ref{fighpsi}.

\subsection{Equilibrium and near-equilibrium states }\label{seceqandneareq}
The next ingredient in our construction is the classification of states. First we would like to consider equilibrium states. Intuitively, these are states where a black hole in the bulk has not been disturbed for a long time. 
We then expect that all excitations both outside and inside the horizon have died off, leaving behind a smooth horizon and an empty interior. We now want to make this precise in the CFT.

Let us review some {\em necessary} conditions for us to classify a state as being in equilibrium. (As we will discuss in section \ref{secremoveamb} these conditions are not quite sufficient.) The first  is that correlation functions in an equilibrium state should be invariant under time-translation. 

We consider the expectation value of an element of the set of observables $\al_p \in \alset$, as a function of time. This is defined as
\be
\label{deviationdefone}
\chi_p(t) = \langle \Psi| e^{i H t} \al_p e^{-i H t} | \Psi\rangle,
\ee
where it is important that $\al_p$ may include $\hcft$.
Intuitively, while  there may be small fluctuations in this expectation value, we expect that in an equilibrium state, these fluctuations are extremely unlikely. The size of the fluctuations is measured by
\be
\label{deviationdeftwo}
{\nu}_p = {1 \over \tband}   \int_0^{\tband}   |(\chi_p(t) - \chi_p(0))| d t. 
\ee
An estimate of these fluctuations \cite{Papadodimas:2013jku} suggests
that a state should be classified as being in equilibrium if
\be
\label{equilibriumcrit}
\nu_p = \Or[e^{-{S \over 2}}], ~ \forall p.
\ee
Note that the definition requires this to hold for all observables in $\alset$. 

The condition for time-independence of correlators can be imposed
very accurately. However, this condition is necessary but not sufficient in order for us to apply our definition of the mirror operators.  In particular, to apply our definition,  we would also like the state to correspond to a state at a single temperature. For example, consider the state ${1 \over \sqrt{2}} \left(|E_1 \rangle + | E_2 \rangle \right)$ where $E_1, E_2$ are two distinct energy eigenstates at substantially separated energies. For example, we could take $E_2 \approx 10 E_1$.  It is easy to verify, using the eigenstate thermalization hypothesis,  that this state meets the criterion \eqref{equilibriumcrit} above. However we would like to think of this as a sum of two separate equilibrium states.

Now we describe near-equilibrium states. Near-equilibrium states are simply obtained by exciting an equilibrium state with an exponentiated Hermitian element of the set of observables ${\cal A}$. 
\be
\label{neareq}
|\psine \rangle = \cop{U} |\Psi \rangle, \quad \cop{U} = e^{i \al_p}, \al_p^{\dagger} = \al_p.
\ee
In \cite{Papadodimas:2013jku,Papadodimas:2013wnh}, we showed that given a state $| \psine \rangle$ of this kind, the decomposition into a unitary $\cop{U}$ and a base-equilibrium state $|\Psi \rangle$ was essentially unique. The reason for this is very simple. Given an equilibrium state $|\Psi \rangle$, if we excite it with a unitary we necessarily spoil the time-translational invariance criterion of \eqref{equilibriumcrit}. 
Therefore, given a state $|\psine \rangle$, once we have found a decomposition \eqref{neareq} that works to make all correlators time-translationally invariant in the base state $|\Psi \rangle$, we know that it must be the right one.

\subsection{Mirrors for equilibrium and near-equilibrium states }\label{subsecmirrordef}
We now consider the definition of mirror operators for the states considered above.  We start with an equilibrium state $|\Psi \rangle$ with inverse temperature $\beta$. First we consider excitations of this state with $\al_{\alpha} \in \alset_{\text{gff}}$. This set was defined in \eqref{alsetgff} and excludes the Hamiltonian.  We now {\em define} mirror operators on this subspace of $\hilb[\Psi]$ through the linear equations
\be
\label{todef}
\tO_{\omega_n, \ang} \al_{\alpha} |\Psi \rangle = e^{-{\beta \omega_n \over 2}} \al_{\alpha} \op_{\omega_n, \ang}^{\dagger} |\Psi \rangle.
\ee 
We can use this definition recursively to define the mirrors of products of operators as well
\be
\tal_{\alpha} \al_{\beta} |\Psi \rangle = \al_{\beta} e^{-{\beta H \over 2}} \al_{\alpha}^{\dagger} e^{\beta H \over 2} |\Psi \rangle.
\ee
These relations specify the action of $\tO_{\omega_n, \ang}$ on $\hilb[\Psi]$. The action of this operator outside this space is irrelevant for questions within effective field theory.  We expect \eqref{todef} to hold at leading order in ${1 \over \nc}$. 

However, we do specify its commutator with the Hamiltonian and this fixes some ${1 \over \nc}$ corrections.
\be
\label{tohamiltcommut}
[\tO_{\omega_n, \ang}, \hcft] \al_{\alpha} |\Psi \rangle = -\omega_n \tO_{\omega_n, \ang} \al_{\alpha} |\Psi \rangle.
\ee
Note that this means that $\tO_{\omega_n, \ang}$ has ``positive energy''. It is possible to check that \eqref{tohamiltcommut} implies certain corrections to \eqref{todef} at $\Or[{1 \over \nc}]$. 

It is easy to check that \eqref{tohamiltcommut} is equivalent to 
\be
\label{todefwithH}
\tO_{\omega_n, \ang} \al_{\alpha} \hcft |\Psi \rangle = \al_{\alpha} e^{-{\beta \omega_n \over 2}} \op_{\omega_n, \ang}^{\dagger}  \hcft |\Psi \rangle.
\ee
This equation is  equivalent to \eqref{todef} when $|\Psi \rangle$ is an energy eigenstate satisfying $H |\Psi \rangle = E |\Psi \rangle$. In other situations $H|\Psi \rangle$ is an independent descendant and \eqref{todefwithH} gives an independent set of constraints on the definition of $\tO_{\omega_n, \ang}$. 

We pause to make a slightly subtle point related to a discussion in \cite{Harlow:2014yoa}. The operator product expansion in the CFT implies that the stress tensor always appears in the OPE of two local generalized free fields. The Hamiltonian is the zero mode of the stress-tensor. Nevertheless, it is consistent for the mirrors to effectively commute with the modes of these operators, but not with the Hamiltonian. 
This is because if we 
attempt to express the CFT Hamiltonian in terms of the modes of the GFFs we expect to get an expression involving not just quadratic but also higher order terms.
\be
\label{hcftinmodes}
\hcft \doteq \sum_n \omega_n \an[\omega_n, \ang]^{\dagger} \an[\omega_n, \ang] + \ldots + \Or[{1 \over \nc}],
\ee
where the $\ldots$ are similar quadratic terms from other fields and the $\Or[{1 \over \nc}]$ terms can be obtained from bulk interactions. As usual, the $\doteq$ in the equation above indicates that this holds within low point correlators. The form of \eqref{hcftinmodes} is dictated by bulk effective field theory, but a similar expression arises from a careful analysis of boundary correlators. 

Now, due to the cutoffs on the set $\alset$ above, there is no strict relation between $\hcft$ and other elements $\al_{\alpha} \in \alset$. Therefore it is {\em mathematically consistent} to define the mirrors to have a zero commutator to very high order with ordinary operators but have a non-zero commutator with the Hamiltonian. 

However, we must mention another physical point. The $\tO_{\omega_n, \ang}$ operators that we have defined above are auxiliary variables, which do not have any direct physical significance. This is  because there is no left asymptotic region in the geometry.  It is the $\ta_{\omega_n, \ang}$ operators that appear in right relational observables. Since these observables are defined relationally, they are not strictly local. Therefore, depending on the precise choice of gauge, it is possible --- without any loss of locality in the bulk ---  to consider operators that have a non-zero commutator with $\an[\omega_n, \ang]$ at subleading
$\Or[{1 \over \nc}]$. This may even be convenient from some perspectives. We will comment more on this issue in forthcoming work.

We now return to the definition of the mirror operators. The equations \eqref{todef} can be considered to be {\em linear equations}
that define the operator $\tO_{\omega_n, \ang}$. We now explain why these
equations are consistent. 

First, note that if $\al_p \in \alsetgff$ then,  in general, we cannot annihilate an equilibrium state by its action.  
\be
\label{noannihil}
\al_p  | \Psi \rangle \neq 0\quad ,\quad \forall \al_p \in \alsetgff.
\ee
This is simply a consequence of the fact that $\text{dim}(\alsetgff) \ll e^{\nc}$ and therefore the space of states annihilated by an element of $\alsetgff$ is of a very high co-dimension. 

For physical reasons we would like to consider energy eigenstates, which can be annihilated by elements of $\alseth$. In such cases, we might have $(\hcft - E) |\Psi \rangle = 0$ for some eigenvalue $E$. However, as we noted above, in such cases \eqref{todefwithH} reduces to \eqref{todef}, and therefore does not lead to an inconsistency.\footnote{Here we have been careful to consider these special states where some descendants obtained by the action of conserved charges are null. In the rest of the paper, when we consider the action of the mirror operators in other settings, we will not always consider this case separately. However, our construction can smoothly accommodate charge or energy eigenstates in all cases.}

To summarize \eqref{todef} and \eqref{todefwithH} specify the action of the mirror operator, $\tO_{\omega_n, \ang}$ on a set of linearly independent vectors. This guarantees that we can find a linear operator with the desired action. We can even write down an explicit solution for these linear equations as follows.

We consider a basis of $\hilb[\Psi]$ given by
\be
\al_{1} |\Psi \rangle \ldots \al_{\dima} |\Psi \rangle,
\ee
and denote an element of this basis by $|v_p \rangle$, where the corresponding $\al_{p}$ may include $\hcft$.
The linear equations \eqref{todef} and \eqref{todefwithH} specify the action of the operator $\tO_{\omega_n, \ang}$ on this basis as
\be
\tO_{\omega_n, \ang} |v_p \rangle = |u_p \rangle,
\ee
where $|u_p \rangle$ can be read off from the right hand side of \eqref{todef} and \eqref{todefwithH}. With $g_{p q
} = \langle v_p | v_q \rangle$, we can simply
define
\be
\label{explicitsol}
\tO_{\omega_n, \ang} = \sum_{p,q} g^{p q} |u_q \rangle \langle v_p|,
\ee
where $g^{p q}$ is the inverse of $g_{p q}$.  The solution \eqref{explicitsol} has the property that it acts only within $\hilb[\Psi]$. If $\projh[\Psi] |w \rangle = 0$ for a state $|w \rangle$, then $\tO_{\omega_n, \ang} |w \rangle = 0$. 

This definition directly extends to near-equilibrium states. Given a state of the form \eqref{neareq}, we define the action of the mirrors by
\be
\label{neareqmirror}
\tO_{\omega_n, \ang} \al_{\alpha} |\psine \rangle = e^{-{\beta \omega_n \over 2}} \al_{\alpha} \cop{U} \op_{\omega_n, \ang}^{\dagger} \cop{U}^{-1} |\psine \rangle.
\ee
The commutator with the Hamiltonian is unchanged.
\be
\tO_{\omega_n, \ang} \hcft \al_{\alpha} |\psine \rangle = \hcft \tO_{\omega_n, \ang} \al_{\alpha} |\psine \rangle - \omega_n \tO_{\omega_n, \ang} \al_{\alpha} |\psine \rangle,
\ee
where all elements on the right hand side can be computed using \eqref{neareqmirror}.

\subsection{Resolution of paradoxes}
We emphasize that our construction above resolves {\em all} of the paradoxes set out by AMPSS in \cite{Almheiri:2012rt,Almheiri:2013hfa,Marolf:2013dba}. We reviewed and sharpened these paradoxes in section \ref{secparadoxes} but none
of these arguments apply to state-dependent operators.

Our construction resolves the $N_a \neq 0$ argument as follows. It is true that typical energy
eigenstates are smooth, whereas number eigenstates may not be smooth. However,
as we saw in \eqref{secnazero} to obtain a contradiction we have to 
perform a {\em basis change} to go from \eqref{expempty} where the trace is evaluated in the energy-eigenbasis to \eqref{nanzarg} where the trace is evaluated in the number eigenbasis. If the operator $\projf$ that appears there is state-dependent, then this change of basis is impermissible because it is a different
operator in each eigenstate. We can see this immediately if we make
the state-dependence explicit by adding a small superscript
\be
{1 \over \dimicro} \sum_{\microen} \langle E |  \projf^{\{E\}} | E \rangle \neq 
{1 \over \dimicro} \sum_{\microen} \langle N_i |  \projf^{\{N_i\}} | N_i \rangle,
\ee
even if these two sets of eigenstates span the same space $\microen$.

In \eqref{secnegoccup} we refined the original ``lack of a left inverse paradox'' of \cite{Almheiri:2013hfa} to argue that no state-independent operator
could have the commutator required of $\ta_{\omega_n, \ang}$ with its adjoint and with the CFT Hamiltonian. However, the argument breaks down if we attempt
to apply it to state-dependent operators. In \eqref{occarg} we had to use
the cyclicity of the trace. But if the operator $\ta_{\omega_n, \ang}$ that appears varies as we vary the energy eigenstate then we cannot use this.

As we explained in section \ref{seccommutarg}, the commutator argument is
not really a paradox but more of a ``genericity argument.'' Our construction
sidesteps this because our mirrors are designed to explicitly commute
with the ordinary operators within correlation functions as \eqref{todef} shows.

Finally, consider the strong-subadditivity paradox of \cite{Mathur:2009hf,Almheiri:2012rt}. Our construction resolves this through a  version of black hole complementarity \cite{'tHooft:1984re,Susskind:1993if}. The statement is that it is {\em impossible} to define mirror operators so that they exactly commute with all CFT operators in any finite time band. From the CFT this is clear from general
principles of local quantum field theory. Therefore the mirror operators that
describe the interior of the black hole must appear to commute with simple
observables within correlation functions but cannot do so exactly. This is a precise version of the colloquial statement that the ``interior is a scrambled version of the exterior.''  The strong subadditivity paradox assumes that the Hilbert space of gravity factorizes exactly into parts that can be associated with the outside and inside of the black hole. 
If complementarity is correct, then this assumption is wrong and the strong subadditivity paradox vanishes.

We direct the reader to \cite{Papadodimas:2013jku,Papadodimas:2013wnh} for further discussion of the resolution of these paradoxes.

\subsection{Small superpositions of equilibrium and near-equilibrium states }\label{secsmallsuperpositions}
We now describe how our construction extends to small superpositions of states.  Such superpositions will be important, and will obtain a direct observational significance, when we consider entangled states of the CFT with an external system of qubits in section \ref{secentqub}.  For now we are interested in the following abstract question. 
\begin{quote}
{\bf Question:} {\em Is exciting a superposition of states by a mirror operator the same as superposing the excited states.}
\end{quote}
We show that the answer to this question is affirmative. This follows almost trivially from the definition above and ensures that the infalling observer
does not observe any departures from linearity.

\subsubsection{Superpositions of equilibrium states}
Consider a superposition of equilibrium states $|\Psi_k \rangle$,
\be
\label{superposeq}
|\psisup \rangle = \sum_{k=1}^M |\Psi_k \rangle,
\ee
where, $M$ is an $\Or[1]$ number and we assume that $\langle \Psi_k | \Psi_p \rangle = 0$ for $k \neq p$ and also that $\sum_k |\langle \Psi_k | \Psi_k \rangle|^2 = 1$ so that the state \eqref{superposeq} is normalized.

We first show that for generic $|\Psi_k \rangle$, the superposition \eqref{superposeq} is also in equilibrium. Let us assume that 
each equilibrium state can be expanded $|\Psi_k \rangle = \sum_{i} \coeff[k,i] |E_i \rangle$, so that the entire superposition is
\be
|\psisup \rangle = \sum_{i,k} \coeff[k,i] |E_i \rangle,
\ee
We now consider $\al_p \in \alset$ and assume that it obeys the eigenstate thermalization hypothesis  \cite{Deutsch, srednicki1999approach, srednicki1994chaos}.
\be
\label{eth}
\langle E_i | \al_p | E_j \rangle = A(E_i) \delta_{i j} + e^{-{1 \over 2} S\left({E_i + E_j \over 2} \right)}  B(E_i, E_j) R_{i j},
\ee
Here, the quantity $S\left({E_i + E_j \over 2}\right)$ is the log of the density of states at the mean energy, for which we just write  $S$. The function $A, B$ are ``smooth'' functions, and $R_{i j}$ is a matrix with erratically varying phases in its entries but with magnitudes of order $1$. 

We see now that
\be
\langle \psisup | \al_p | \psisup \rangle = \sum_{i,k,n} \coeff[k,i]^* \coeff[n,i] A(E_i) + \sum_{i\neq j,k,n} e^{-{1 \over 2} S\left({E_i + E_j \over 2} \right)} B(E_i, E_j) R_{i j} \coeff[k,i]^* \coeff[n,j].
\ee
Consider the first term in the sum above. This involves a sum over $\Or[e^{S}]$ energy eigenstates, but for $k \neq n$ the terms in this sum are erratic. Since each $\coeff[k,i] = \Or[e^{-{S \over 2}}]$, this turns into an erratic sum over $e^{S}$ terms over size $e^{-S}$. We expect it to typically be only of size $\Or[e^{-{S \over 2}}]$.  The same argument applies to the second term in the sum, involving $R$. This term --- irrespective of whether $n=k$ or $n \neq k$ ---
 turns into an erratic sum over $e^{2 S}$ terms, each of size $e^{-{3 S \over 2}}$. This is again expected to typically only
be of size $e^{-{S \over 2}}$. This leads to the conclusion that
\be
\langle \psisup | \al_p | \psisup \rangle =  \sum_{k=1}^M \langle \Psi_k | \al_p | \Psi_k \rangle + \Or[e^{-{S \over 2}}].
\ee
Therefore if the equilibrium criterion \eqref{seceqandneareq} applies to each state $|\Psi_k \rangle$ it also applies to the superposition $|\psisup \rangle$, as long as $M = \Or[1]$. Therefore the superposition is also in equilibrium. 

 The interesting case is where the $|\Psi_i \rangle$ are microstates corresponding to the same black hole.\footnote{The case where they correspond to different geometries simply corresponds leads to a classical probability distribution over the various possibilities as we described around \eqref{semiclassorth}. This situation is not of significant physical interest but, in any case, it can be dealt with easily by extending the results obtained here.} We can now define the mirrors independently for $|\psisup \rangle$ and each of the $|\Psi_i \rangle$. We display this state-dependence explicitly with a superscript below. 

We now notice the following simple fact. 
\be
\tO_{\omega_n, \ang}^{\{\text{sup}\}} \al_{\alpha} |\psisup \rangle = e^{-{\beta \omega_n \over 2}} \al_{\alpha} \op_{\omega_n, \ang}^{\dagger} |\psisup \rangle.
\ee
This follows because $|\psisup \rangle$ is also in equilibrium and at the temperature $\beta^{-1}$.
On the other hand
\be
\tO_{\omega_n, \ang}^{\{k\}} \al_{\alpha} |\Psi_k \rangle = e^{-{\beta \omega_n \over 2}} \al_{\alpha} \op_{\omega_n, \ang}^{\dagger} |\Psi_k \rangle.
\ee
Therefore we find that
\be
\label{linearitysmallsup}
\tO_{\omega_n, \ang}^{\{\text{sup}\}} \al_{\alpha} |\psisup \rangle = \sum_{k=1}^M \tO_{\omega_n, \ang}^{\{k\}} \al_{\alpha} |\Psi_k \rangle.
\ee
This equation shows that the mirror operators act consistently with the superposition principle, as long as we are looking at small superpositions of equilibrium states. As we will see later, this is important in order for the infalling observer not to be able to detect any violations of quantum mechanics.

\subsubsection{Superpositions of near-equilibrium states}
\label{subsecneareqsuper}

Now, we consider an 
$\Or[1]$ superposition of near-equilibrium states
\be
\label{neareqdecomp}
|\psisupne \rangle = \sum_{k=1}^M \cop{U}_k |\Psi_k \rangle,
\ee
where $|\Psi_k \rangle$ are orthogonal equilibrium states, as previously,
and we again assume that the sum in \eqref{neareqdecomp} is normalized to $1$. Here, as in \eqref{neareq},  $\cop{U}_k = e^{i \al_{k}}$, where $\al_{k}$ are Hermitian elements of $\alsetgff$.  

We now define the action of the tildes via
\be
\label{tildesumneareq}
\tO_{\omega_n, \ang} \al_{\alpha} |\psisupne \rangle= \sum_{k=1}^M \al_{\alpha} \cop{U}_k e^{-{\beta \omega_n \over 2}} \op_{\omega_n, \ang}^{\dagger} |\Psi_k \rangle.
\ee
Note that, strictly speaking, \eqref{tildesumneareq} is an extension of our definition of mirror operators since a superposition of near-equilibrium states is not itself a near-equilibrium state by the definition of such states in \eqref{neareq}. 

We also note that in this case the action of $\tO_{\omega_n, \ang}$ is not closed within  the span of $\alset |\psisupne\rangle$. This can be seen from \eqref{tildesumneareq} where the right hand side is not just an ordinary operator acting on $|\psisupne \rangle$. It is convenient to imagine that we expand the little Hilbert space to the direct sum of the little Hilbert spaces produced by acting on the equilibrium states in \eqref{neareqdecomp}
\be
\hilb[\psisupne] = \bigoplus_k \hilb[\Psi_k].
\ee
This may be used as a general rule when the space obtained by acting with $\alset$ does not contain any equilibrium state at all.

Let us check that \eqref{tildesumneareq} immediately passes a consistency check. The decomposition of a state in the form \eqref{neareqdecomp} is not unique. As we explained above, almost all sums of $\Or[1]$ equilibrium states are also equilibrium states.  Correspondingly $\hilb[\psisupne]$ contains many equilibrium states.

This implies that we can just as well write \eqref{neareqdecomp} as
\be
|\psisupne \rangle = \sum_{k,q,p=1}^M  \cop{U}_k Q^{-1}_{k q} Q_{q p} |\Psi_p \rangle = \sum_{q=1}^{M} \cop{V}_q |\Psi'_q \rangle,
\ee
with
\be
\cop{V}_q = \sum_{k=1}^M \cop{U}_k Q^{-1}_{k q}; \quad |\Psi'_q \rangle = \sum_{p=1}^M Q_{q p} |\Psi_p \rangle.
\ee
Here $Q$ is any invertible $M \times M$ matrix and $Q^{-1}$ is its inverse: $\sum_q Q^{-1}_{k q} Q_{q p} = \delta_{k p}$. It is important to us that the matrices $\cop{V}_q$ also be invertible. This is true for generic choices of the $\cop{U}_k$ and we will only consider cases of this sort.

Now, since the state $|\Psi'_q \rangle$ will also typically be in equilibrium,
it is equally natural to demand that
\be
\label{nearalternate}
\tO_{\omega_n, \ang} \al_{\alpha} |\psisupne\rangle = e^{-{\beta \omega_n \over 2}} \al_{\alpha} \sum_{q=1}^M  \cop{V}_q  \op_{\omega_n, \ang}^{\dagger} |\Psi_q' \rangle.
\ee
We would like to ensure that \eqref{nearalternate} is consistent with \eqref{tildesumneareq}. But this follows immediately by inserting the definitions of $V$ and $|\Psi_q' \rangle$ above. 

We can also repeat the check we performed for equilibrium states above. Using the definition \eqref{tildesumneareq} of mirror operators on superpositions of near-equilibrium states on the left hand side of the equation below, we have
\be
\label{linearityneareq}
\tO_{\omega_n, \ang}^{\{\psisupne\}} |\psisupne \rangle = \sum_{k=1}^M   \tO_{\omega_n, \ang}^{\{k\}} \al_{\alpha} \cop{U}_k  |\Psi_k \rangle,
\ee 
where on the right hand side we use the standard definition of the mirrors on non-equilibrium states given in \eqref{neareqmirror}, and we have again indicated the state-dependence explicitly by means of the superscript.

The result \eqref{linearityneareq} shows that the infalling observer does not observe any violation of linearity even for superpositions of near-equilibrium states. This includes, as a special case, a superposition of an equilibrium and a near-equilibrium state, and thereby answers a question about superposition
raised in \cite{polchinskistrings2014}.

\subsection{The interior of the eternal black hole  }\label{secexpliciteternal}
We conclude this section by constructing state-dependent local operators 
in the eternal black hole. We already showed in \eqref{subsecnaiveeternal}
that the naive state-independent construction of local operators where we identify $\tO_{\omega_n, \ang} = \Oleft[\omega_n, \ang]$ does not work correctly
in the states $|\tfdT \rangle$ defined in \eqref{tfdtdef}. 

We will proceed as follows. We start by reviewing the conditions that we need from the mirrors in the eternal black hole. Based on these, we guess an appropriate solution. We then verify that it meets the conditions that we outlined. We hasten to add that the formulas we present here can be derived in a completely systematic fashion using the formalism for entangled states that we present in section \ref{secentangled}. We present this alternate method of obtaining the answer only because it provides some additional insight into the nature of state-dependence. 

We would like to suggest that the reader also consult \cite{Papadodimas:2015xma} --- where this result is stated concisely --- before examining the detailed calculation below.
\paragraph{Constraints on $\tO_{\omega_n, \ang}$ \\}
The precise conditions that $\tO_{\omega_n, \ang}$ need to satisfy are given in section \ref{seceternal}. These modes need to be correctly entangled with $\op_{\omega_n. \ang}$ in all states $| \tfdT \rangle$, they need to commute with the $\op_{\omega_n, \ang}$ within correlators, and also have the commutator with the Hamiltonians given in \eqref{commutH}. 

In fact all of these conditions would be met if
\be
\label{imitateolt}
\langle \tfdT | \al_{\alpha} \tO_{\omega_n, \ang} \al_{\beta} |\tfdT \rangle = \langle \tfdT | \al_{\alpha} \Oleft[\omega_n, \ang](\capt) \al_{\beta} |\tfdT \rangle + \Or[{1 \over \nc}],
\ee
where 
\be
\label{olmodet}
\Oleft[\omega_n, \ang](\capt) \equiv {1 \over \tband^{1 \over 2}} \int_{-\tband}^{\tband} \op_L(t+T,\Omega) e^{i \omega_n t} Y_{\ang}^*(\Omega) \, d t \, d^{d-1} \Omega.
\ee
Note that for small $\capt$ we have $\Oleft[\omega_n, \ang](\capt) = \Oleft[\omega_n, \ang] e^{-i \omega T}$. However, this is no longer true when $\capt \gg \tband$. Since we allow exponentially large $\capt$ in the states $|\tfdT \rangle$, we must adopt the more careful definition \eqref{olmodet}. 

We can try and achieve \eqref{imitateolt} through the use of projectors as in section \ref{secstateindoutside}. In particular, we would like to use a projector to ``detect'' the state as an excitation of $|\tfdT \rangle$ and then modulate $\tO_{\omega_n, \ang}$ accordingly.  We caution the reader that this program will be only partly successful. But to this end, we investigate these projectors in some detail below. We have to construct these projectors and then in order to put them together correctly, we also need to examine their overlaps.

\paragraph{Projectors on $\hilb[\tfdT]$\\}
We define the projector $\projh[\tfdT]$ as follows
\be
\begin{split} 
&\projh[\tfdT] \al_{\alpha} | \tfdT \rangle  = \al_{\alpha} |\tfdT \rangle, \\
&\text{if}\,~\forall \al_{\alpha}\,,\, \langle v | \al_{\alpha} |\tfdT \rangle = 0  \Rightarrow \projh[\tfdT]  | v \rangle = 0.
\end{split}
\ee
In these equations we restrict $\al_{\alpha} \in \alsetgff$ and do not allow it to include $\hcft$. 

We can construct the projector explicitly. Define
\be
g_{\alpha \beta} = \langle \tfdT | \al_{\beta}^{\dagger} \al_{\alpha} | \tfdT \rangle.
\ee
Note that $g_{\alpha \beta}$ is actually independent of $\capt$ because the operators above come from the right CFT
and commute with the left Hamiltonian that is used to evolve $|\tfd \rangle$ to $|\tfdT \rangle$.
Then the projector can be written as
\be
\projh[\tfdT] = \sum_{\alpha \beta} g^{\alpha \beta} \al_{\alpha} |\tfdT \rangle \langle \tfdT | \al_{\beta}^{\dagger},
\ee
where $g^{\alpha \beta}$ is the inverse of $g_{\alpha \beta}$. We can check that
\be
\projh[\tfdT] \al_{\gamma} |\tfdT \rangle = \sum_{\alpha \beta} g^{\alpha \beta} \al_{\alpha} |\tfdT \rangle g_{\beta \gamma} = \al_{\gamma} |\tfdT \rangle.
\ee
Obviously, in the orthogonal subspace, $\projh[\tfdT]$ gives $0$. 

\paragraph{Overlaps of the projectors  $\projh[\tfdT]$ \\ }
Next we have to account for the fact that the different projectors $\projh[\tfdT]$ are not  quite orthogonal for different values of $\capt$. We can calculate the overlap between the states $|\tfdT\rangle$ and their descendants as follows. 
We have
\be
\label{overlap}
\langle \tfd |   \al_{\alpha} | \tfdT \rangle = {1 \over Z(\beta)} \sum_{E} e^{-\beta E}  \langle E| \al_{\alpha}| E \rangle e^{i E \capt},
\ee
where all cross terms have dropped out because the operator $\al_{\alpha}$ acts only within the right CFT, and we can use the eigenstates in the left CFT to impose a delta function in energy.

First, let us consider this this quantity for $\capt \ll 1$.  In this situation we can approximate \eqref{overlap}  by
\be
\langle \tfd |  \al_{\alpha} | \tfdT \rangle = {1 \over Z(\beta)} \int e^{-\beta E} e^{S(E)} A(E) e^{i E T},
\ee
where we have indicated the diagonal element of $\al_{\alpha}$ by $A(E)$ as in \eqref{eth}.  

We can compute this integral using a saddle point approximation. We write the exponent as 
\be
-\beta E + S(E) = -\beta E_0 + S(E_0) + {1 \over 2} (E-E_0)^2 \left. {\partial^2 S \over \partial^2 E} \right|_{E=E_0},
\ee
where $E_0$ satisfies 
\be
\left. {\partial S \over \partial E}\right|_{E=E_0} = \beta. 
\ee
Consider the second derivative term. We write the  temperature as a function of energy $\tau(E)$, and then this is just
\be
{\partial {1 \over \tau(E)} \over \partial E} = -{1 \over \tau^2(E)} \partial {\tau(E) \over \partial E} = -{1 \over \tau^2(E) C},
\ee
where $C$ is the specific heat. Note that $C \propto \nc$. Evaluated at $E = E_0$, we find
\be
\left. {\partial^2 S \over \partial^2 E} \right|_{E=E_0} = -{\beta^2 \over C}.
\ee

Therefore the integral above can be written
\be
{1 \over Z(\beta)}  \int \exp\left[-{\beta^2 \over C} {(E - E_0)^2 \over 2} + i E \capt\right] A(E) d E.
\ee
Now notice that if $A(E)$ is a smooth function of ${E \over \nc}$ it varies slowly over the energy scales $\sqrt{C}$  that are relevant here, since ${E \over \nc}$ changes only by ${1 \over \sqrt{C}}$ over this scale. Second since we have assumed that $\capt \ll 1$, we conclude that
\be
\label{tfdtaoverlap}
\langle \tfd |   \al_{\alpha} | \tfdT \rangle  =\left( \langle \al_{\alpha} \rangle  + \Or[{1 \over \nc}] \right) e^{- {C \capt^2 \over 2 \beta^2}} e^{i E_0 \capt},
\ee
where the expectation value on the right is the normal expectation value taken in $|\tfd \rangle$. Note that we can actually get the pre-factor right, and it precisely cancels the factor of  ${1 \over Z(\beta)}$ in the integral. In particular note that \eqref{tfdtaoverlap} also has the correct limit at $\capt = 0$.
Below, we will use
\be
f(\capt) =  e^{- {C \capt^2 \over 2 \beta^2}} e^{i E_0 \capt}.
\ee
We caution the reader that the estimates for the overlap between different projectors are
no longer valid for $\capt \sim \Or[1]$. We will consider this case separately below. 

\paragraph{Guess for $\tO_{\omega_n, \ang}$ \\}
We can now use these projectors and the idea explained above to write down
a guess for the $\tO_{\omega_n, \ang}$ that will reproduce \eqref{imitateolt}. 
We consider
\be
\label{statedependental}
\tO_{\omega_n \ang} = {\sqrt{ C \over \pi \beta^2}} \int_{-\cutoffT}^{\cutoffT } \Oleft[\omega_n \ang](\intT) \projh[\Psi_{\intT}] d \intT,
\ee
where $\cutoffT$ is a cutoff that we explore further below. 
The idea of \eqref{statedependental} is that the projector $\projh[\Psi_{\intT}]$ detects the state it is acting on as an excitation of $|\Psi_{\intT} \rangle$, and therefore the insertion of $\tO_{\omega_n, \ang}$ effectively turns into an insertion of $\Oleft[\omega_n \ang](\intT)$ as required in \eqref{imitateolt}.

We now verify in detail that the guess \eqref{statedependental} does satisfy all the conditions that we need in the state $|\tfd \rangle$ and in states $| \tfdT \rangle$ for $|\capt| < \cutoffT$. For states where $\capt$ does not satisfy this condition we will need to change the operator \eqref{statedependental} as we describe below.

\paragraph{Correlators of  $\tO_{\omega_n \ang}$\\}
We are interested in inserting the proposed mirror defined in \eqref{statedependental} in correlators. We find that
\be
\langle \tfdT| \al_{\alpha} \tO_{\omega_n \ang} \al_{\beta} |\tfdT \rangle =  {\sqrt{ C \over \pi \beta^2}} \int_{-\cutoffT}^{\cutoffT } d \intT \langle \tfd | e^{-i \cop{H} \capt} \al_{\alpha} \Oleft[\omega_n \ang](\intT) \projh[\Psi_{\intT}] \al_{\beta} e^{i \cop{H} \capt} |\tfd \rangle.
\ee
To evaluate the integral on the right hand side we consider the integrand 
\be
\begin{split}
\langle \tfd | e^{-i \cop{H} \capt} \al_{\alpha} \projh[\Psi_{\intT}] \al_{\beta} e^{i \cop{H} \capt} |\tfd \rangle &= \langle \tfd |  \al_{\alpha} \projh[\tfdvar{\capt-\intT}] \al_{\beta} | \tfd \rangle \\ &= \sum_{\gamma \delta} \langle \tfd | \al_{\alpha} g^{\gamma \delta} \al_{\gamma} |\tfdvar{\capt-\intT} \rangle \langle \tfdvar{\capt-\intT} | \al_{\delta}^{\dagger} \al_{\beta} |\tfd \rangle,
\end{split}  
\ee
where we have first used the factors of $e^{i \cop{H} \capt}$ to convert the projector to $\projh[\tfdvar{\capt-\intT}]$ and then we have inserted the explicit expression for the projector derived above. This quantity can be further be simplified to
\be
\begin{split}
 \langle  \tfd | \al_{\alpha} \projh[\tfdvar{\capt-\intT}] \al_{\beta} | \tfd \rangle &= |f(\capt-\intT)|^2  \sum_{\gamma \delta} \langle \tfd | \al_{\alpha} g^{\gamma \delta} \al_{\gamma} |\tfd \rangle \langle \tfd | \al_{\delta}^{\dagger} \al_{\beta} |\tfd \rangle  \\
&= |f(\capt-\intT)|^2 \langle \tfd | \al_{\alpha}\projh[\tfd] \al_{\beta} |\tfd \rangle  \\
&= |f(\capt-\intT)|^2  \langle \tfd | \al_{\alpha} \al_{\beta} |\tfd \rangle,
\end{split}
\ee
where we have used the expression for mixed correlators in \eqref{tfdtaoverlap},then re-absorbed the sum over $\gamma, \delta$ into another projector, and recognized that the projector acts as the identity on descendants of $|\tfd \rangle$.

Plugging this into the original integral we find that
\be
\begin{split}
\langle \tfdT| \al_{\alpha} \tO_{\omega_n \ang} \al_{\beta} |\tfdT \rangle &= 
 {\sqrt{ C \over \pi \beta^2}} \int_{-\cutoffT}^{\cutoffT } d \intT |f(T-\intT)|^2 \langle \tfd | e^{-i \cop{H} \capt} \al_{\alpha} \Oleft[\omega_n \ang](\intT) \al_{\beta} e^{i \cop{H} \capt} |\tfd \rangle\\  &= \langle \tfdT | \al_{\alpha} \Oleft[\omega_n \ang](\capt) \al_{\beta} |\tfdT \rangle + \Or[{1 \over \nc}].
\end{split}
\ee
Here we have used the fact that $\Oleft[\omega_n \ang](\intT)$ varies very slowly with respect to the function $f(T-\intT)$, provided $\omega_n \ll \nc$ since $C \sim \Or[\nc]$. Therefore, to leading order in ${1 \over \nc}$ we can simply evaluate this integral in the saddle point approximation which leads to the result above.  This result is, of course, valid provided that $|\capt| < \cutoffT$ and it agrees with what was required in \eqref{imitateolt}.

Note that this immediately leads to the right two point and higher point functions. For example, 
\be
\langle \tfdT |  \Oleft[\omega_n \ang](\capt) \op_{\omega_n \ang} |\tfdT \rangle = \langle \tfd | \Oleft[\omega_n \ang] \op_{\omega_n \ang} |\tfd \rangle  = {e^{-\beta \omega_n \over 2} G_{\beta}(\omega_n, \ang)},
\ee
which is precisely what is required.

\paragraph{Commutator with Hamiltonians \\}
Finally we check the behaviour of the proposed $\tO_{\omega_n \ang}$ under
time evolution with the left and right Hamiltonians. 
Notice that
\be
\projh[\tfdvar{T_i}] e^{-i \cop{H} T} = e^{-i \cop{H} T} \projh[\tfdvar{T_i + T}].
\ee
Therefore,
\be
\begin{split}
e^{i \hcft \capt} \tO_{\omega_n \ang} e^{-i \hcft \capt} &=   {\sqrt{C \over \pi \beta^2}} \int_{-\cutoffT}^{\cutoffT} \Oleft[\omega_n \ang](\intT) \projh[\tfdvar{\intT + \capt}] d \intT  \\ &=  {\sqrt{C \over \pi \beta^2}} \int_{\capt-\cutoffT}^{\cutoffT+\capt} \Oleft[\omega_n \ang](\intT - \capt) \projh[\tfdvar{\intT}] d \intT,
\end{split}
\ee
where the  last equality comes from a change of variables inside the integral. Note that 
\be
\Oleft[\omega_n \ang](\intT - \capt) = e^{i \omega_n \capt} \Oleft[\omega_n \ang],
\ee
for $\capt \sim \Or[1]$, and as long as $\capt \ll \tband$.
Now, when inserted into correlation functions, the cutoffs are exponentially irrelevant as the analysis above shows. The dominant contribution when $\tO_{\omega_n \ang}$ is inserted into a correlator always comes from a saddle point in the interior of the integral. Therefore we find that within correlation functions
\be
e^{i \hcft \capt} \tO_{\omega_n \ang} e^{-i \hcft \capt} \doteq e^{i \omega_n \capt} \tO_{\omega_n \ang},
\ee
which is precisely what is required as long as we do not evolve for a very long time.

A very similar analysis shows that conjugation by $e^{i \hleft \capt}$ leaves $\tO_{\omega_n \ang}$ invariant within correlators because of the transformation of $\Oleft[\omega_n \ang]$ in the integral above. This completes our verification of \eqref{commutH}.

\subsubsection{Analysis  of state-dependence in the eternal black hole}

The reader should note that our construction is explicitly state-dependent. The operators \eqref{statedependental} fail to click correctly when they are inserted in states $|\tfdT \rangle
$ with $\capt \gg \cutoffT$. It is easy to verify this by repeating the exercise above. The reader will find
that when $\tO_{\omega_n, \ang}$ is inserted into a correlator, the saddle point of the integral over $\intT$ occurs outside the range of integration, and therefore the correlator is exponentially suppressed.  

Now, we might naively believe that this can be fixed simply by taking $\cutoffT$ to infinity. However, we will show below that if we do this, then instead of behaving correctly in every state, the integral \eqref{statedependental} would fail to behave correctly in any state. To see this we need to reconsider the overlap estimate of \eqref{tfdtaoverlap}. The expression in \eqref{tfdtaoverlap} is not the correct
answer for $\capt \gg 1$ since our saddle-point technique of evaluating the thermal correlator breaks down if the phase factor that arises from the term involving $\capt$ varies too rapidly. 

At large $\capt$, we simply note that the overlap is a sum over approximately $\Or[e^S]$ uncorrelated complex numbers of $\Or[1]$.
\be
\label{fattail}
\langle \tfd | \al_{\alpha} |\tfdT \rangle = {1 \over Z(\beta)} \sum e^{-\beta E} e^{i E \capt} A(E) = \Or[e^{-{S \over 2}}], \quad \capt \gg 1.
\ee
In particular for $\capt \gg 1$, this overlap is much larger than the overlap predicted by \eqref{tfdtaoverlap}. It has a ``fat tail.''

Therefore if we take $\cutoffT$ to be exponentially large, $\cutoffT \gg \Or[e^{S}]$ and insert \eqref{statedependental} into a correlator, then the contributions from this fat tail will overwhelm the contribution of the dominant saddle. This is the reason that we are forced to use state-dependence. 

For the states $|\tfdT \rangle$ with $\capt \gg e^{S}$, we can still write down interior operators. These operators are given by
\be
\op_{\omega_n, \ang}^{\{T\}} = {\sqrt{ C \over \pi \beta^2}} \int_{\capt-\cutoffT}^{\capt+\cutoffT } \Oleft[\omega_n \ang](\intT) \projh[\Psi_{\intT}] d \intT,
\ee
where we have explicitly moved the range of integration.

This discussion helps to shed light on the nature of state-dependence. By performing these large diffeomorphisms we have, in a sense, ``geometrized'' the microstates of the black hole. The states $|\tfdT \rangle$ are all identical states from the perspective of the right infalling observer,
but the left and right modes are entangled differently in each of them. The novel part of this situation is that these are also distinct and well separated solutions from the point of view of the semi-classical theory if we keep track of how the solution is ``glued'' to the boundary.  

Now, classically the right-relational observables are well defined objects on each of these geometries. Often, in such situations, it is possible to lift such classical observables to operators as we describe in more detail in Appendix \ref{appcoherentgravity}. This is usually done by identifying classical solutions as coherent states in the Hilbert space, and using projectors to map classical functions to operators. (See, for instance, \eqref{projectorrep}.) However, if we consider the states $|\tfdT \rangle$ for exponentially large ranges of $\capt$, then \eqref{fattail} tells us they are ``overcomplete''. This ``overcompleteness'' goes beyond the usual overcompleteness of coherent states. In fact, we believe that
a computation using coherent states to represent the different states $|\tfdT \rangle$ in canonical gravity should yield the overlap \eqref{tfdtaoverlap} but at large $\capt$ this is very different from \eqref{fattail}. This forces us to use state-dependent operators for the black hole interior, even in this one-parameter class of states.

By considering time-shifted versions of the geon solution analyzed in \cite{Guica:2014dfa}, we believe that it should not be difficult
to find a similar one-parameter set in a single CFT where state-dependence can be analyzed in detail.


\section{Removing ambiguities in the construction}\label{secremoveamb}
We now turn to the issue of some ambiguities in our construction. There are two sorts of ambiguities that have been described in the literature. The first is
related to an observation about the eternal black hole by Marolf and Wall \cite{Marolf:2012xe} and a similar observation by van Raamsdonk \cite{VanRaamsdonk:2013sza} which was framed more directly in terms of our construction. We show here how this ambiguity should be resolved.

The second ambiguity was discussed by the authors of \cite{Marolf:2013dba} and some of these objections were expanded in a paper by Harlow \cite{Harlow:2014yoa}. However, Harlow's construction attempted to add to this ambiguity by adopting a modified definition of the mirror operators, which had a different commutator with the Hamiltonian from the one in our construction. We will show  that this alternate definition of the mirror operators of \cite{Harlow:2014yoa} suffers
from certain inconsistencies which we point out below.

As a consequence of this, the alternate mirror operators described by Harlow do not themselves have direct physical significance. However, it is true that there is an interesting class of excited states that we will consider in section \ref{seccanonicalambiguity}; these are related to the analysis of \cite{Harlow:2014yoa} but we will consider them independently so as to separate them from the main claims of that paper.

We should mention that an additional class of ambiguities, involving only ordinary operators was described in \cite{Almheiri:2013hfa}. The authors of \cite{Almheiri:2013hfa} suggested that one could act with the Schwarzschild number operator $e^{i \theta {\cop{N}}_{\omega}} |\Psi \rangle$ on an equilibrium state to obtain another state that was approximately time-translationally invariant. We have addressed this issue previously. (See page 46 of  \cite{Papadodimas:2013jku}.) If we use a finite time-band to extract the modes of the CFT generalized free-fields, and then combine them into a number operator then such an operator does not commute exactly with the CFT Hamiltonian. One may attempt to improve this construction by considering an extremely slow acting source, which inserts only a finite amount of energy into the system over an extremely long time scale. The action of such a source might be consistent with our equilibrium condition but this would not be a contradiction since the infalling observer would also not see any excitation in this case.

\subsection{Mirror unitary behind the horizon}\label{mirrorbehind}
Consider an equilibrium state $|\Psi \rangle$ and perform the construction described in section \ref{secdefnmirror}, leading to the mirror operators. Now, consider the state
\be
\label{unitbehind}
|\psiexc \rangle = e^{i \alpha \tal_p} |\Psi \rangle \equiv \cop{\widetilde{U}} |\Psi \rangle.
\ee
Here $\tal_p$ is the mirror of a Hermitian operator satisfying $(\al_p)^{\dagger} = \al_p$. The parameter $\alpha$ is a real number that will be useful below.
\begin{figure}[!h]
\begin{center}
\includegraphics[width=3.5cm]{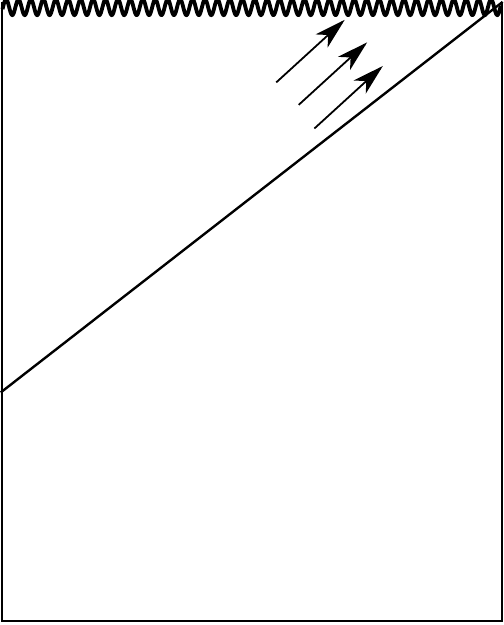}
\caption{A state $|\psiexc \rangle = \cop{\widetilde{U}} |\Psi\rangle$ corresponding to an equilibrium state $|\Psi\rangle$ excited with a mirror 
unitary behind the horizon $\cop{\widetilde{U}}$. \label{figunitbehind} }
\end{center}
\end{figure}

In our construction above, we have not really defined the exponentiated version of the mirror operators.  To exponentiate the mirror we need to be able to evaluate 
\be
e^{i \alpha \tal_p} |\Psi \rangle = \sum_{n=0}^{\infty} {(i \alpha)^n \over n!} (\tal_p)^n |\Psi\rangle,
\ee
which involves arbitrarily high products of the mirror operator and necessarily takes us outside the space $\hilb[\Psi]$. To be precise, beyond some cutoff $K$, we  expect $\langle \Psi | [(\tal_p)^K, \al_s] | \Psi \rangle \neq 0$. The precise value of $K$ depends on the precise definition of $\tal_p$. We return to this ``edge effect'' below in the discussion of Harlow's ambiguity.

The first putative ambiguity mentioned in the beginning of section \ref{secremoveamb} is the following: if we assume that the state $|\Psi\rangle$ is a black hole in an equilibrium state, then the state $|\psiexc \rangle$ should be an excited state. Intuitively we expect $|\psiexc \rangle$ to be a state with an excitation behind the horizon as shown in
figure \ref{figunitbehind}. In particular, an observer crossing the horizon in the state $|\psiexc \rangle$, within a suitable time-range, should detect this excitation. Now,
the question is, suppose we are given the state $|\psiexc \rangle$ without the additional information that it came by acting with $e^{i \alpha \tal_p}$ on some equilibrium state $|\Psi\rangle$. How can we directly detect that the state $|\psiexc \rangle$ is a non-equilibrium state? The difficulty comes from the fact that since $\cop{\widetilde{U}}$ approximately commutes with elements of the small algebra, so we have
\be 
\langle \Psi | \cop{\widetilde{U}^{\dagger}} \op_{\omega_1, \ang_1} \ldots \op_{\omega_n,  \ang_n} \cop{\widetilde{U}} |\Psi \rangle = \langle \Psi | \op_{\omega_1, \ang_1} \ldots \op_{\omega_n, \ang_n} |\Psi \rangle + R,
\ee
where $R$ is the small remainder that we discussed above. We will neglect this remainder in what follows.  Hence, simple correlators of the small algebra on the state $|\psiexc \rangle$ seem to be almost the same as those in the state $|\Psi\rangle$. This might lead to the erroneous conclusion that $|\psiexc \rangle$  is an equilibrium state. This mistake
would lead to the definition of mirror operators as if $|\psiexc \rangle$ were equilibrium, and using these wrong mirror operators would lead to the incorrect prediction that the infalling observer will not detect any
excitation behind the horizon. In order to avoid this ambiguity in the mirror operator construction, we need to find a way to detect
from the CFT that $|\psiexc \rangle$ is an excited state.

The key point is that we have also included the Hamiltonian in our set of observables. The Hamiltonian does {\em not} commute with the mirror operators. Hence, correlators
of operators in the small algebra, together with insertions of the Hamiltonian will differ between typical equilibrium states and states which have
been excited by mirror unitary operators $|\psiexc \rangle = \widetilde{U}|\Psi\rangle$. We can use these differences as a diagnostic of the non-equilibrium nature of these states. This resolves the ambiguity of the mirror unitaries behind the horizon.

To make this more clear, let us consider the state $|\psiexc \rangle$ in \eqref{unitbehind} and let us define
\be
\label{defcomh}
\tal_s \equiv [\cop{H}, \tal_p].
\ee
We can detect the non-equilibrium nature of the state $|\psiexc \rangle$ by considering the correlation function with $\cop{H}$ and  the corresponding $\al_s$ operator
\be
\label{markdetector}
\begin{split}
&\langle \psiexc | \cop{H} \al_s |\psiexc \rangle = \langle \Psi | \cop{\widetilde{U}^{\dagger}} \cop{H} \al_s \cop{\widetilde{U}} |\Psi \rangle = \langle \Psi | (\cop{1} - i \alpha \tal_p) \cop{H} \al_s (\cop{1} + i \alpha \tal_p) |\Psi \rangle + \Or[\alpha^2] \\ &= \langle \Psi | \cop{H} \al_s |\Psi \rangle + i \alpha \langle \Psi |  \tal_s \al_s | \Psi \rangle + \Or[\alpha^2] \\
&= \Or[e^{-{S \over 2}}] + i \alpha \langle \Psi | \al_s e^{-{\beta \cop{H} \over 2}} (\al_s)^{\dagger} e^{\beta \cop{H} \over 2} |\Psi \rangle + \Or[\alpha^2].
\end{split}
\ee
Here we have used the fact the equilibrium expectation value of the operator $\cop{H} \al_s$ is exponentially small, if $\al_s$ has non-zero ``energy''. On the other hand, we expect that the expectation value in the second term of the last line above to be $\Or[1]$. So, we see that for the observable in \eqref{markdetector}, we discern a substantial deviation from its equilibrium value. This allows us to classify the state $|\psiexc \rangle$ as an ``excited state'', as expected
intuitively.

For a concrete example, let us take $\tal_p$ in \eqref{unitbehind} to be $\tal_p = \tO_{\omega, \ang} + \tO_{\omega, \ang}^{\dagger}$.\footnote{In this section and in section \ref{secentangled}, to lighten the notation, instead of $\omega_n$ for the discretized frequencies, we drop the subscripts and simply write $\omega$.}
 We consider
\eqref{defcomh} for this case, to find $\tal_s = \omega(\tO_{\omega, \ang} - \tO_{\omega, \ang}^{\dagger})$. In an equilibrium state we have
\be
\label{eqvancond}
\omega \langle \Psi| \cop{H}(\op_{\omega, \ang} - \op_{\omega, \ang}^{\dagger})|\Psi\rangle = 0,
\ee
up to exponentially small corrections. On the other hand, for the state $e^{i\alpha(\tO_{\omega, \ang} + \tO_{\omega, \ang}^\dagger )}|\Psi \rangle$ we find
to linear order in $\alpha$ and up to exponentially small corrections that
\be
\begin{split}
&\omega \langle \Psi |\cop{\widetilde{U}^{\dagger}} \cop{H} (\op_{\omega, \ang} - \op_{\omega, \ang}^{\dagger})  \cop{\widetilde{U}} | \Psi \rangle =
\omega \langle \Psi | e^{-i\alpha (\tO_{\omega, \ang} + (\tO_{\omega, \ang})^{\dagger}) } \cop{H} (\op_{\omega, \ang} - \op_{\omega, \ang}^{\dagger}) 
e^{i\alpha(\tO_{\omega, \ang} + \tO_{\omega, \ang}^\dagger )}|\Psi \rangle \\
 &= i\alpha \omega^2 \langle \Psi |(\tO_{\omega, \ang} - \tO_{\omega, \ang}^{\dagger}) (\op_{\omega, \ang} - \op_{\omega, \ang}^{\dagger})   | \Psi \rangle + \Or[\alpha^2]\\
 &=i\alpha  \omega^2 \langle \Psi | (\op_{\omega, \ang} - \op_{\omega, \ang}^{\dagger}) \left( e^{-{\beta \omega \over 2}} \op_{\omega, \ang}^{\dagger} -  e^{\beta \omega \over 2} \op_{\omega, \ang} \right) | \Psi \rangle  + \Or[\alpha^2]\\
&= 2i \alpha \omega^2 {e^{-{\beta \omega \over 2}}} G_{\beta}(\omega,\ang)+ \Or[\alpha^2],\end{split}
\ee
which is $\Or[1]$. So this correlator is different on $|\psiexc \rangle$ from that on the equilibrium state \eqref{eqvancond} and by measuring this correlator we can detect the excitation by the mirror unitary behind the horizon.

\paragraph{Uniqueness of the behind-horizon unitaries \\}
We note that given a state $|\psiexc \rangle$ of the form \eqref{unitbehind} it has an essentially unique decomposition into an equilibrium state and a unitary behind the horizon. The reason is as follows. First, it is clear that we cannot have such a decomposition with two different basis states, since in that case we would have
\be
\cop{\widetilde{U}_1} |\Psi_1 \rangle = \cop{\widetilde{U}_2} |\Psi_2 \rangle \Rightarrow |\Psi_1 \rangle = \cop{\widetilde{U}_1^{\dagger}} \cop{\widetilde{U}_2} |\Psi_2 \rangle.
\ee
As we have shown above, if $|\Psi_2 \rangle$ is in equilibrium a relation of the sort above implies that $|\Psi_1 \rangle$ cannot be in equilibrium, and vice versa. 

Furthermore, with $\cop{\widetilde{U}_1} = e^{i \tal_1}$, and  $\cop{\widetilde{U}_2} = e^{i \tal_2}$, it is clear from a chain of reasoning that
\be
\begin{split}
&\cop{\widetilde{U}_1} |\Psi \rangle = \cop{\widetilde{U}_2} |\Psi \rangle \Rightarrow \left(\cop{\widetilde{U}_1^{\dagger} }\cop{\widetilde{U}_2} \right) |\Psi \rangle = |\Psi \rangle\\ &\Rightarrow (\tal_1 - \tal_2) |\Psi \rangle = 0  \\
&\Rightarrow (\al_1^{\dagger} - \al_2^{\dagger}) |\Psi \rangle = 0,
\end{split}
\ee
which is prohibited by \eqref{noannihil} unless $\al_1 = \al_2$, and so $\cop{\widetilde{U}_1} = \cop{\widetilde{U}_2}$. 
This concludes our proof of the uniqueness of the decomposition.

Therefore, to summarize, given a state of the form \eqref{unitbehind} we can not only detect that it is out of equilibrium, but even detect the operator with which it has been excited. 

\subsection{Comments on the Harlow unitaries }\label{seccommentsharlow}
Now, let us turn to a second set of unitaries described by Harlow \cite{Harlow:2014yoa}, who attempted to define a new set of mirror operators $\ohar_{\omega,\ang}$ which act on an equilibrium state as follows 
\begin{align}
\label{harloword} 
&\ohar_{\omega, \ang} \al_{\beta} |\Psi \rangle = \al_{\beta} e^{-{\beta \omega \over 2}} (\op_{\omega, \ang})^{\dagger} |\Psi \rangle, \\
\label{harlowhamiltonian}
&[\ohar_{\omega, \ang}, \cop{H}] \al_{\beta} |\Psi \rangle \stackrel{?}{=} 0.
\end{align}
Notice that the first equation, \eqref{harloword}, is the same as the one in our definition, \eqref{todef}, but the commutator with the Hamiltonian given in \eqref{harlowhamiltonian} differs from ours, which is specified by \eqref{tohamiltcommut}.

We will now show that the definition of mirror operators given by Harlow is inconsistent, and runs into difficulties in several physical situations. We discuss an energy eigenstate, and then a state drawn from the microcanonical ensemble. We then discuss a more serious problem --- definition  \eqref{harlowhamiltonian} leads to operators that do not satisfy the Heisenberg equations of motion. Therefore, these operators $\ohar_{\omega, \ang}$ cannot be used to build up gauge invariant relational observables.

\subsubsection{Inconsistency of $\ohar_{\omega, \ang}$ mirrors in energy eigenstates}
First, we point out that the second line above, \eqref{harlowhamiltonian} does not have any solutions at all, when defined about energy eigenstates. We find that
\be
\label{xhe}
\ohar_{\omega, \ang} \cop{H} | E \rangle = E \ohar_{\omega, \ang} | E \rangle =  e^{-{\beta \omega \over 2}}E\op_{\omega, \ang}^{\dagger} | E \rangle.
\ee
But\footnote{Note that these results are unaffected by a possible small correction to the commutator between the Hamiltonian and the ordinary operator:  $\cop{{\cal R}_C} = [\cop{H}, \op_{\omega, \ang}^{\dagger}] - \omega \op^{\dagger}_{\omega, \ang}$. This may arise because we define the ``modes'' by considering only a finite time interval as we discussed above. However, we expect that $|| \cop{{\cal R}_C} | E \rangle ||^2 \ll 1$, and particularly that $\langle E | \op_{\omega, \ang}  \cop{{\cal R}_C}  | E \rangle  = \Or[{1 \over \nc}]$. These statements just point out that the ``remainder'' is small and, in particular, it does not have an overlap with $\op_{\omega, \ang}^{\dagger} | E \rangle$ at $\Or[1]$.}
\be
\label{hxe}
\begin{split}
\cop{H} \ohar_{\omega, \ang} | E \rangle &\stackrel{?}{=}  e^{-{\beta \omega \over 2}}\cop{H} (\op_{\omega, \ang})^{\dagger} | E \rangle =  e^{-{\beta \omega \over 2}}[\cop{H}, (\op_{\omega, \ang})^{\dagger}] | E \rangle +   e^{-{\beta \omega \over 2}}(\op_{\omega, \ang})^{\dagger} \cop{H} | E \rangle \\
&= e^{-{\beta \omega \over 2}} \omega \op_{\omega, \ang}^{\dagger} | E \rangle +   e^{-{\beta \omega \over 2}}E  \op_{\omega, \ang}^{\dagger}| E \rangle\\
&=  e^{-{\beta \omega \over 2}}(E + \omega)  \op_{\omega, \ang}^{\dagger}| E \rangle.
\end{split}
\ee

To understand the inconsistency of Harlow's definition for eigenstates, we consider the correlator
$\langle E|\op_{\omega,\ang}  [\cop{H}, \ohar_{\omega, \ang}] | E \rangle$. We can compute it in two ways. The first is
to subtract  \eqref{xhe} from \eqref{hxe}  and multiply the resulting state with the bra $\langle E | \op_{\omega, \ang}$. This leads to the prediction
\be
\label{straightcalc}
\begin{split}
\langle E | \op_{\omega, \ang} [\cop{H}, \ohar_{\omega, \ang}] | E \rangle &=  e^{-{\beta \omega \over 2}}\langle E | \op_{\omega, \ang} \omega \op^{\dagger}_{\omega, \ang} | E \rangle  \\ &= \omega e^{-{\beta \omega \over 2}} G_{\beta}(\omega, \ang).
\end{split}
\ee
On the other hand, using directly \eqref{harlowhamiltonian}, we find that
\be
\label{harlowcalc}
\langle E | \op_{\omega, \ang} [\cop{H}, \ohar_{\omega, \ang}] | E \rangle \stackrel{?}{=} 0.
\ee
Clearly \eqref{harlowcalc} and \eqref{straightcalc} are in contradiction, and therefore the equation \eqref{harlowhamiltonian}, which was used by Harlow to define the mirrors, is actually inconsistent in an energy eigenstate. Moreover note that at this level the contradiction arises at $\Or[1]$ and cannot be resolved by ${1 \over \nc}$ corrections.

Now, we move away from a strict energy eigenstate and turn to a state with an $\Or[1]$ spread in energies. We show that even in such a state, the modified definition of the mirror operators in \cite{Harlow:2014yoa} cannot be used consistently. 

\subsubsection{Inconsistency of $\ohar_{\omega, \ang}$  in microcanonical states}
We now show that the inconsistency in Harlow's unitaries in not restricted to energy eigenstates. It persists in states that are drawn from a microcanonical ensemble with an $\Or[1]$ spread in energies. Consider a state of the following kind
\be
|\psimicro \rangle = \sum_i \coeff[i] |E_i \rangle,
\ee
where the coefficients $\coeff[i]$ have the property that they are peaked around a given energy, which we will call $E$, but the spread in energies is $\Or[1]$. More precisely, we demand
\be
\label{statemicro}
\begin{split}
&\langle \psimicro | \cop{H} | \psimicro \rangle = E, \\
&\langle \psimicro | \projrange[E, \Delta] | \psimicro \rangle = 1 - \Or[\nc^{-1}],
\end{split}
\ee
where 
\be
\label{projectepmd}
\projrange[E, \Delta]  = \sum_{i=E-\Delta}^{i= E+\Delta} |E_i \rangle \langle E_i |,
\ee
is the projector onto states in the range $E \pm \Delta$, and $\Delta \ll \nc$ is some $\Or[1]$ number.

Now, the key point is as follows. In \eqref{harlowhamiltonian} we have imposed the relation that the commutator of the operator $\ohar$ with the Hamiltonian annihilates the state. However, the projector onto a range of energies, like the one that appears in \eqref{projectepmd}, is also a good observable. In fact, physically we expect to be able to measure this observable rather easily both on the boundary, and in the bulk. On the boundary, this observable is completely determined by considering the zero mode of the stress tensor. In the bulk, it can be determined by considering the subleading falloff in the metric. This is in contrast to a projector onto a Schwarzschild number eigenstate which, as we reviewed in Appendix C of \cite{
Papadodimas:2013jku}, requires an extremely long time to measure and projects the final state onto a firewall. 

Now, consider again the relation \eqref{harlowhamiltonian}, but extended to products of the operator $\ohar_{\omega, \ang}$. As we discussed above, unless we can define such products consistently to a high order, it is not possible to consider ``unitaries'' made out of this operator, which are required to produce the ambiguity that was discussed in \cite{Harlow:2014yoa}.

However, for any $\Or[1]$ frequency $\omega$, we have an $\Or[1]$ number $n_{\DDelta}$, so that
\be
n_{\DDelta}\, \omega > 2 \Delta.
\ee
Now, following \eqref{harlowhamiltonian}, we would like to impose
\be
\label{haractonmicro}
(\ohar_{\omega, \ang})^{n_{\DDelta}} |\psimicro \rangle = e^{-{n_{\DDelta} \beta \omega \over 2}} (\op_{\omega, \ang}^{\dagger})^{n_{\DDelta}} |\psimicro \rangle + {1 \over \nc} |{\cal R}_C^{\text{micro}} \rangle,
\ee
where we have included a small possible ${1 \over \nc}$ correction with the property that 
\be
\langle {\cal R}_C^{\text{micro}} | {\cal R}_C^{\text{micro}} \rangle = \Or[1].
\ee
However, now we note that
\be
\label{directprojector}
e^{-n_{\DDelta} \beta \omega} \langle \psimicro| (\op_{\omega, \ang})^{n_{\DDelta}} \projrange[E, \Delta] (\op_{\omega, \ang}^{\dagger})^{n_{\DDelta}} |\psimicro \rangle \ll 1.
\ee
This is because the action of $n_{\DDelta}$ insertions of $\tO_{\omega, \ang}^{\dagger}$ raises the energy by the state by $n_{\DDelta} \,\omega$ and so necessarily takes it out of the band $E \pm \Delta$. On the other hand, if the operator $\ohar_{\omega, \ang}$ is defined to commute also with $\projrange[E, \Delta]$ then
we would expect
\be
\label{indirectprojector}
\begin{split}
\langle & \psimicro| \Big[(\ohar_{\omega, \ang})^{\dagger}\Big]^{n_{\DDelta}} \projrange[E, \Delta]  (\ohar_{\omega, \ang})^{n_{\DDelta}} |\psimicro \rangle \\ &\stackrel{?}{=} \langle \psimicro|  \projrange[E, \Delta] \Big[[(\ohar_{\omega, \ang})^{\dagger} \Big]^{n_{\DDelta}}  (\ohar_{\omega, \ang})^{n_{\DDelta}} |\psimicro \rangle + \Or[\nc^{-1}].\\
&= \langle \psimicro| \projrange[E, \Delta]  (\op_{\omega, \ang}^{\dagger})^{n_{\DDelta}} (\op_{\omega, \ang})^{n_{\DDelta}} |\psimicro \rangle  + \Or[\nc^{-1}]\\ &= \Or[1],
\end{split}
\ee
where in the final result we have noted that action of $(\op_{\omega, \ang})^{n_{\DDelta}}$ followed by the action of its adjoint maps us back to the same band of energies.
Clearly the results of \eqref{directprojector} and \eqref{indirectprojector} are in contradiction given the general results about the expectation value of projectors in states that are almost parallel, which we reviewed in section \ref{subsecprojector}. 

\subsubsection{Failure of $\ohar_{\omega, \ang}$ to satisfy the Heisenberg equations of motion}
Now we turn to an even more serious difficulty with the mirror operators defined by \eqref{harlowhamiltonian}: their failure to satisfy the Heisenberg equations of motion. This failure persists even in states with a canonical spread of energies. In such states, the fundamental relation \eqref{harlowhamiltonian} does not suffer from an obvious inconsistency, unlike in energy eigenstates or states with a microcanonical spread. However, as we show below these operators nevertheless do not have the correct geometric properties to play the role of interior mirror operators. 

In particular, as we described in detail in section \ref{subsec:commutatormirror}, if the bulk operators are defined relationally with respect to the boundary, in order to be gauge invariant, then they must satisfy
\be
e^{i \cop{H} T}  \cop{\phi} (t,r,\Omega) e^{-i \cop{H} T} = \cop{\phi} (t+T,r, \Omega).
\ee
It is clear that if we attempt to create these operators by means of the operators defined in \eqref{harlowhamiltonian}, then the local operators will not obey the Heisenberg equations of motion. Let us check this explicitly by computing a two point function across the horizon.

Outside the horizon we have the usual expansion of the field in terms of CFT modes
\be
\cop{\phi}^H(t,\rtor, \Omega) \underset{U \rightarrow 0^-}{\longrightarrow}  \sum_{\ang, \omega}  {1 \over \sqrt{\omega \comm_{\beta}(\omega, \ang)}} \op_{\omega, \ang} e^{-i \omega t} Y_{\ang}(\Omega)\left( e^{i \delta} e^{i \omega \rtor} + e^{-i \delta} e^{-i \omega \rtor} \right) + \text{h.c}.
\ee
This expansion does not depend on our definition of the mirror operators. Inside the horizon, however, using the Harlow mirror operators we find
\be
\cop{\phi}^H(t,\rtor, \Omega) \underset{U \rightarrow 0^+}{\longrightarrow}  \sum_{\ang, \omega} { e^{-i \delta} \over \sqrt{\omega \comm_{\beta}(\omega, \ang)}} \Bigg[ \op_{\omega, \ang}  e^{-i \omega (t+\rtor)} Y_{\ang}(\Omega)  + \ohar_{\omega, \ang} e^{i \omega (t - \rtor)} Y_{\ang}^*(\Omega) \Big) + \text{h.c}.  \Bigg].
\ee
Now, let us compute correlation functions with this operator in an {\em equilibrium state}, $|\Psi \rangle$. Moving to the usual Kruskal coordinates $U,V$, let us consider two points, so that one of them, $(U_1, V_1, \Omega_1)$, is just outside the horizon whereas the other $(U_2, V_2, \Omega_2)$ is just inside. Then we find
\be
\begin{split}
&\langle \Psi |e^{-i \hcft T} \cop{\phi}^H(U_1, V_1, \Omega_1) \cop{\phi}^H(U_2, V_2, \Omega_2) e^{i \hcft T} |\Psi \rangle = 
\\  &=\sum_{\ang, \omega}  {1 \over \omega C_{\beta}(\omega, \ang)}  \left[ \langle \op_{\omega, \ang} \op_{\omega, \ang}^{\dagger} \rangle \left({V_1 \over V_2} \right)^{i \omega}  + e^{i \omega T} \langle \op_{\omega, \ang} \ohar_{\omega, \ang} \rangle \left({-U_1 \over U_2} \right)^{i \omega} \right]Y_{m}(\Omega_1) Y^*_{m}(\Omega_2) + \text{h.c}.
\end{split}
\ee
Notice the extra factor of $e^{i \omega T}$ which appears in front of the ${U_1 \over U_2}$ factor. In particular, this implies that if we compute the derivative of the two point function and take the two points to be close then we find (using the techniques of section \ref{secsmoothint}), substituting the relevant two point functions and converting the sum to an integral that
\be
\lim_{V_1 - V_2 \rightarrow 0} \langle \Psi | e^{-i \hcft \capt} \partial_U \cop{\phi}^{\text{H}}(U_1, V_1, \Omega_1) \partial_U \cop{\phi}^{\text{H}}(U_2,V_2, \Omega_2) e^{i \hcft \capt} | \Psi \rangle = c {\delta^{d-1}(\Omega_1 - \Omega_2) \over (U_1 - U_2 e^{-{2 \pi \capt \over \beta}})^2}.
\ee
However, this is in explicit contradiction with the universal short distance form of the correlator that we derived in \eqref{universalshortdistance}. In fact, such a correlator would suggest the presence of a firewall. 

Therefore, we have reached the following conclusion. Even in an equilibrium state, where we expect correlation functions to be time-invariant, if one uses Harlow's definition of the mirror operators, this leads to the prediction that if one starts with a state with no firewall, a firewall appears immediately!

This is a straightforward consequence of the fact that these putative mirror operators do not obey the Heisenberg equations of motion. The commutator with the Hamiltonian \eqref{harlowhamiltonian} was not derived from a gauge fixing procedure, which we carried out carefully in \cite{Papadodimas:2013jku}, nor was it derived 
from a careful consideration of relational observables in the geometry, which we performed in subsection \ref{subsec:reltss}.

In fact, the source of this error is apparent. The motivation of \cite{Harlow:2014yoa} to propose the vanishing commutator of the interior operators with the Hamiltonian \eqref{harlowhamiltonian} was partly based on the analogy with the thermofield doubled state. In fact, it was argued in \cite{Harlow:2014yoa} that in some specific pure states, one may expect bulk correlators to approximate thermofield correlators to high orders in ${1 \over \nc}$. However, {\em even} in the thermofield state, as we showed in section \ref{seceternal}, when we carefully consider commutators of the right Hamiltonian with the mirrors that are relevant for the right relational observables, one finds non-zero commutators. It is only if one uses the naive but incorrect expansion \ref{subsecnaiveeternal}, that one obtains the incorrect expectation for the commutator used in \eqref{harlowhamiltonian}. 

\subsection{States in the ``canonical'' ensemble }\label{seccanonicalambiguity}
We now turn precisely to an interesting class of excitations of states in the canonical ensemble. The point is that we need to refine our notion of equilibrium, since the time-independence of correlators of single-trace operators may not be sufficient to classify these states into equilibrium and non-equilibrium. We do not explicitly perform this classification here, but we show that such a classification should exist. 
 
These states were also discussed in \cite{Harlow:2014yoa}, but we phrase the issue  independently of Harlow's mirror operators, since these do not have any geometric significance as we pointed out above. 
 
Consider a state $|\psican \rangle$ that satisfies the following condition. For any element $\al_p$ of the set of observables $\alset$, we have
\be
\label{staterhoequiv}
\langle \psican | \al_p | \psican \rangle = \tr(\cop{\rho} \al_p) + \Or[e^{-S}],
\ee
where $\cop{\rho}$ is an {\em invertible} matrix. Note that if the state $|\psican \rangle$ is in equilibrium then the density matrix $\cop{\rho}$ satisfies $[\cop{H}, \cop{\rho}] = 0$. This is important for correlation functions to be time-translationally invariant.

We pause to make two important points. Given a state $|\psican \rangle$ the density matrix that appears on the right of \eqref{staterhoequiv} is not unique. In fact, the possible solutions to this equation are the subject of entropy maximization \cite{jaynes1957information, jaynes1957informationII}. Second, both the energy eigenstate and the sharp microcanonical state that we considered above are not relevant here. We cannot find any {\em invertible} choice of  $\cop{\rho}$ to satisfy \eqref{staterhoequiv} for these states without making some matrix elements of the inverse arbitrarily large.

Now, given any Hermitian element of the set of observables $\al_p$, we consider the transformation
\be
\label{transformedstate}
|\Psi'_{\text{can}} \rangle =  \cop{\rho}^{1 \over 2} e^{i \al_p} \cop{\rho}^{-{1 \over 2}} |\psican \rangle.
\ee
We can check that correlators of elements of $\alset$  in the state $|\Psi'_{\text{can}} \rangle$ are the same as those in $|\psican \rangle$.  We see that
\begin{align}
&\langle \Psi'_{\text{can}} | \al_m |\Psi'_{\text{can}} \rangle =  \langle \psican| \cop{\rho}^{-{1 \over 2}} e^{-i \al_p} \cop{\rho}^{1 \over 2}
\,\al_m\, \cop{\rho}^{1 \over 2} e^{i \al_p} \cop{\rho}^{-{1 \over 2}} |\psican \rangle \\ \label{firstprimeequiv}
&= \tr\Big[\cop{\rho} \Big( \cop{\rho}^{-{1 \over 2}} e^{-i \al_p} \cop{\rho}^{1 \over 2}\al_m \cop{\rho}^{1 \over 2} e^{i \al_p} \cop{\rho}^{-{1 \over 2}} \Big) \Big]   + \Or[e^{-S}]\\ \label{finalprimeequiv}
&= \tr(\cop{\rho} \al_m) + \Or[e^{-S}] = \langle \psican | \al_m | \psican \rangle  + \Or[e^{-S}].
\end{align}
In obtaining \eqref{firstprimeequiv}, we simply used \eqref{staterhoequiv}, and then we use the cyclicity of the trace and \eqref{staterhoequiv} to obtain the final result in \eqref{finalprimeequiv}. The question now is as follows: is the state $|\Psi'_{\text{can}} \rangle$ in equilibrium or not?

Consider a concrete example. Take the state that was discussed in \cite{Harlow:2014yoa}
\be
\label{canonicalstate}
|\psican \rangle = {1 \over \sqrt{Z(\beta)}} \sum_i e^{-{\beta E_i \over 2}} e^{i \phi_i} |E_i \rangle,
\ee
where $\phi_i$ are arbitrary phases, the sum is over all energy eigenstates and $Z(\beta)$ is the partition function of the boundary theory. As discussed in 
\cite{Harlow:2014yoa} for simple correlators this state behaves like the canonical ensemble to exponential accuracy, and for this state we can take $\cop{\rho} =
{1\over Z(\beta)} e^{-\beta \cop{H}}$ and satisfy \eqref{staterhoequiv}.  

To see this, consider {\em any} operator, $\al_p$ obeying the eigenstate thermalization hypothesis \eqref{eth}. 
Adopting the notation of \eqref{eth}, we consider
\be
\langle \psican | \al_p|\psican \rangle = {1 \over Z(\beta)} \sum_i A(E_i) e^{-\beta E_i} + {1 \over Z(\beta)} \sum_{i j} e^{-\beta {E_i + E_j \over 2}} R_{i j} e^{-S} B(E_i, E_j) e^{i(\phi_j-\phi_i)}.
\ee
To convert the second term to a sum over $i$, we sum over all $j$ that can be connected by the cross-terms. We make the further reasonable assumption that the unitary links states that are separated only by a ``finite'' band, i.e. $B(E_i, E_j) \ll 1$ for $|E_i - E_j| \gg 1$.  Now, we see that for each value of $i$, the sum over $j$ runs over effectively $\Or[e^{S}]$ states. However, since these states contribute with varying phases the typical size of this sum over $j$ is suppressed by $e^{-{S \over 2}}$ compared to the first term involving $A(E_i)$. So we can estimate that
\be
\langle \psican | \al_p|\psican \rangle = {1 \over Z(\beta)} \sum_i A(E_i) e^{-\beta E_i}  + \Or[e^{-{S \over 2}}] = {1\over Z(\beta)}\tr(e^{-\beta H} \al_p)+ \Or[e^{-{S \over 2}}]. 
\ee
Now, we consider the group of transformations of the form \eqref{transformedstate} that we can make to this state, where now $\cop{\rho} = {1\over Z(\beta)}
\tr(e^{-\beta\cop{H}})$
\be
\label{transformedcanonical}
{\cop{M}} |\psican \rangle \equiv e^{-{\beta \cop{H} \over 2}} e^{i \al_p} e^{\beta \cop{H} \over 2} |\psican \rangle.
\ee
The question is, if $|\psican \rangle$ is an equilibrium state, then is $\cop{M} |\psican \rangle$ in equilibrium or not? 

We will work with this concrete example to consider this question. Of course, the reader can easily generalize this discussion to states that mimic a density matrix that is distinct from the thermal one. 

At first sight, this question is a little puzzling because of two seemingly contradictory facts. On the one hand, all correlators of elements of $\alset$ in this new state \eqref{transformedcanonical} are the same as in the canonical ensemble
\be
\langle \psican| \cop{M}^{\dagger} \al_m \cop{M} |\psican \rangle = {1\over Z(\beta)}\tr\left(e^{-\beta H} \al_m \right) + \Or[e^
{-{S \over 2}}].
\ee
On the other hand, it is easy to verify that
\be
\label{innercontra}
\langle \psican| \cop{M}^{\dagger}  e^{-i \tal_p} | \psican \rangle = 1 - \Or[\nc^{-1}],
\ee
where here $ e^{-i \tal_p} | \psican \rangle$ is an excited state, as discussed in subsection \ref{mirrorbehind}.
So if we declare the transformed state in \eqref{transformedcanonical} as an equilibrium state, then we would have the unusual situation of having equilibrium and excited states separated by a ``distance'' ${1 \over \nc}$ in the Hilbert space \eqref{innercontra}. This would not be a contradiction, since the operators $\tO$ are state-dependent, but it would be a rather striking departure from the behaviour of state-independent operators. 

Therefore, the better alternative is to enlarge the set of observables $\alset$ to include an operator that can distinguish between the states $\cop{M} |\psican \rangle$ and $|\psican \rangle$. There are many such operators because it is certainly not true that all physical properties of these states can be captured by the thermal density matrix. For example, if we take the boundary to be on $S^{d-1}$ and ask for the entanglement entropy of a subregion on this boundary, then in both states, this entanglement entropy starts to decrease after the volume of the subregion increases past half the volume of the $S^{d-1}$, which would not be the case for a truly thermal mixed state.

We will return to the discussion of the appropriate operators that can detect this excitation in future work. However, for now, we perform an important 
consistency check. Consider the set of states formed by the action of the group  of exponentiated unitaries
\be
\label{orbitcan}
\{|\psican \rangle, \cop{M}(\al_1) |\psican  \rangle, \cop{M}(\al_2) |\psican 
\rangle \ldots \cop{M}(\al_n) |\psican  \rangle \},
\ee
where $\al_1, \al_2, \ldots \al_n$ are elements of $\alset$ and $\cop{M}(\al_p) |\psican \rangle \equiv e^{-{\beta H \over 2}} e^{i \al_p} e^{\beta H \over 2} |\psican \rangle$ as above. We will show that it is consistent, in principle, to have sets of this form, where only one element of the set is an equilibrium state,
and all others are non-equilibrium states.  The consistency check that we need to perform is to ensure that such a classification will not violate the rule that ``most'' states in the Hilbert space must be equilibrium states. 

\subsubsection{Consistency condition for maps from equilibrium to non-equilibrium states}
\label{secconsistencyeq}
Let us state this consistency condition more precisely. It is applicable not only to this case, but to more general statistical mechanical questions of classifying equilibrium. Let us say that we have two regions of the Hilbert space, ${\cal D}$, and ${\cal I}$. We have a function on the Hilbert space $\Theta_E(\Psi)$, with the property that $\Theta_E(\Psi) = 0$ for equilibrium states and $\Theta_E(\Psi) = 1$ for non-equilibrium states. This function provides a classification of equilibrium.  Next, we have a measure on the Hilbert space $d \meas(\Psi)$, which has the property that by this measure ``most states'' in both ${\cal D}$ and ${\cal I}$ are in equilibrium.
\begin{align}
\label{neqissmallinD}
&{\int_{\cal D} d \meas (\Psi) \Theta_E(\Psi) \over \int_{\cal D} d \meas(\Psi)} \ll 1, \\
\label{neqissmallinI}
&{\int_{\cal I} d \meas (\Psi) \Theta_E(\Psi) \over \int_{\cal I} d \meas(\Psi)} \ll 1. \\
\end{align}
This means that the volume of non-equilibrium states as a fraction of the total volume is very small both in ${\cal D}$ and in ${\cal I}$. Finally, consider a map $M$
\be
M: {\cal D} \longrightarrow {\cal I},
\ee
which has the property that it maps equilibrium to non-equilibrium states. 

Let $M({\cal D})$ be the image of ${\cal D}$ under this map. Now, let 
 ${\cal I}_{\cal D}$ be the intersection of an $\epsilon$-ball about this image with the set ${\cal I}$.  More precisely, for $\epsilon \ll 1$,
\be
|\psiexc \rangle \in {\cal I}_{\cal D} \Leftrightarrow \exists |\Psi \rangle \in {\cal D},~\text{s.t.}~ |\langle \psiexc |M | \Psi \rangle|^2 \geq 1 - \epsilon^2.
\ee

Then we have the following important {\em consistency condition} on this map
\be
\label{consistencycond}
{\int_{{\cal I}_{\cal D}} d \meas(\Psi)  \over \int_{\cal I} d \meas(\Psi)} \ll 1.
\ee
We explain this condition in a little more detail below. Intuitively, it means that states that are close to the image of ${\cal D}$ under $M$ must have very small volume in ${\cal I}$.

From this condition it follows immediately that an invertible map ${\cal D} \longrightarrow {\cal D}$ {\em cannot} map equilibrium to non-equilibrium states consistently. For example, consider the microcanonical measure where we pick states 
 in an energy band. (We define this more precisely below.) We expect most such states to be in equilibrium. Now consider time-translations, which  map this region back to itself. Therefore, the image under time-translations of the original region is the region itself. Thus time-translations do not satisfy \eqref{consistencycond} and therefore cannot have the property 

\subsubsection{Microcanonical ensemble and unitaries}
To warm up for the problem of maps from ``canonical states'' back to themselves, we consider a similar problem for the microcanonical ensemble. We will define this ensemble, define an appropriate measure so that \eqref{neqissmallinD} and \eqref{neqissmallinI} are satisfied and show how unitaries of simple operators do satisfy \eqref{consistencycond}.

Consider the set of all states of the form
\be
\label{microensemble}
|\psimicro \rangle = \sum_{E_i = E - \Delta}^{E_i = E + \Delta} a_i |E_i \rangle,
\ee
where $\sum_i |a_i|^2 = 1$ for the state to be normalized. We now write down an invariant Haar measure on this set, $d \meas (\psimicro )$, with the property that for {\em any unitary} that
maps states of the form \eqref{microensemble} back to another state of the same form, $|\psimicro' \rangle = \cop{U}|\psimicro \rangle$, we have $d \meas(\psimicro') = d \meas(\psimicro)$.  Explicitly, to obtain the ``microcanonical ensemble'', we consider the
uniform probability measure
\be
\label{uniformmeasure}
d \meas(a_i) = N_{\meas} \delta(1 - \sum_i |a_i|^2) d^2 a_1 \ldots d^2 a_D,
\ee
where $D$ is the total number of energy eigenstates in this range, and $N_{\meas}$ is a normalization constant that we will fix below. In the measure above, note that we have not identified states that differ by a phase.

In terms of the objects introduced in section \ref{secconsistencyeq}, the set ${\cal D}$ is the set of all states of the form \eqref{microensemble}. We have not specified a precise equilibrium function. However, with almost any reasonable choice of $\Theta_E(\Psi)$ --- for example, we can choose this function so that it implements our equilibrium condition in \eqref{equilibriumcrit} ---  and with the measure \eqref{uniformmeasure}, we see that \eqref{neqissmallinD} is satisfied. 

We can take the map under consideration to be the unitary matrix, $\cop{U}_m = e^{i \al_m}$. 
Now one might naively imagine that there are ``as many'' states of the form $\cop{U}_m |\psimicro \rangle$ as of the form $|\psimicro \rangle$. The reason this is still consistent with the fact that most states are equilibrium states is that  $\cop{U}_m |\psimicro \rangle$ does {\em not} belong to the original microcanonical ensemble. Even if we consider $\al_m = \op_{\omega} + \op_{\omega}^{\dagger}$ where $\omega$ is a very low frequency we see that the new state $\cop{U}_m|\psimicro \rangle$ contains energy eigenstates of higher energies. The term ${\al_m^k \over k!}$ in the expansion of the unitary operator leads to a new ensemble with states
$E \pm \Delta \pm k \omega$. The point is that even a small increase in energy increases the ``volume'' of the ensemble by a huge amount, and therefore the state $\cop{U}(\al_m) |\psimicro \rangle$ come from a larger ensemble, where they are extremely atypical.  

Let us see this more precisely, let us define ${\cal I}$ to be the set of states that can be written in the form \eqref{microensemble}, but with a width $\Delta' > \Delta$. In the example above, if we take ${\Delta' - \Delta \over \omega} \gg 1$, then we can consistently think of the unitary as a map from ${\cal D}$ to ${\cal I}$.  Strictly speaking the image of the lower dimensional manifold in the higher dimensional manifold is measure 0. However, this does mean that non-equilibrium states are infinitely unlikely. To answer physical questions we must examine how many states in the higher dimensional manifold are within an $\epsilon$ distance of the states obtained by exciting the lower dimensional manifold with a unitary. The relevance of this condition is that by the arguments of section \ref{subsecprojector} the expectation value of any projector in states which have an almost unit inner product is almost identical and therefore such states have similar physical properties. 

To verify \eqref{consistencycond}, we consider the volume of the manifold of all states of the form \eqref{microensemble}. This is just that of a $2 D - 1$ dimensional sphere under the measure \eqref{uniformmeasure} and is given by
\be
V_{\text{micro}} = N_{\meas} {\pi^{D}  \over \Gamma(D)}.
\ee
So we should take $N_{\meas} = {\Gamma(D) \over \pi^D}$ for the distribution to be normalized. 

However, we may also ask the following useful question. Let us discretize this region into $\epsilon$-nets. Given a state $|\Psi \rangle$, we ask: what is the volume of the space in the neighbourhood, comprising states $|\Psi' \rangle$ so that $|\langle \Psi | \Psi' \rangle|^2 \geq 1 - \epsilon^2$? We can see that this is given by the volume of the $2 D - 2$ dimensional {\em ball} with small radius $\epsilon$, with an additional factor of $\pi$ that comes from the possible relative phase between the two vectors.\footnote{Choose a basis so that  $|\Psi \rangle = (1,0, \ldots 0)$. Then we must have $|\Psi' \rangle = (a_1, \ldots a_D)$ where $|a_1|^2 \geq 1 - \epsilon^2$ and so $|a_2|^2 + \ldots |a_D|^2 \leq \epsilon^2$.} This is 
\be
V_{\epsilon-{\text{ball}}} = N_{\meas} {\pi^{D - 1} \epsilon^{2 D - 2} \over \Gamma\left(D \right)}.
\ee
So the number of such $\epsilon$-balls in a space of dimension $D$ is given by
\be
N_D= {V_{\text{micro}} \over \pi V_{\epsilon-{\text{ball}}}}  = {1 \over \epsilon^{2 D - 2}}.
\ee
Now we consider the unitary map above, which takes us from the microcanonical space of dimensionality $D$ to a set of states of dimensionality $D(1 + \delta)$.  This map clearly sends $\epsilon$-balls in the lower dimensional space to a cross-section of an $\epsilon$-ball in the higher dimensional space. However, the higher dimensional space has far more $\epsilon$-balls, and therefore the image of the lower dimensional space only intersects a small fraction of these. The precise fraction is given by
\be
{N_D \over N_{D(1 + \delta)}} = \epsilon^{2 D \delta}
\ll 1.
\ee
Note that $D$ is exponentially large: $D = \Or[e^{S}]$. Therefore even if the unitary increases the dimension of the new ensemble by only a small fraction, it is completely consistent with thermodynamic expectations to classify almost all states both in the original ensemble, and in the new ensemble, as equilibrium states. 

\subsubsection{Excitations of canonical states}
Now we want to show that the same principle holds for the canonical states that we discussed above. More precisely, we consider some possible measures on a subset of the Hilbert space, so that typical states picked using this measure are of the form \eqref{canonicalstate}. Then the action of the operators $\cop{M}$ takes us to another subset of the Hilbert space where the image of the original subset occupies a vanishingly small volume. By the remark below \eqref{canonicalstate}, as a corollary, this provides some evidence for the claim that there is no subset of the CFT Hilbert space, with a nice measure satisfying \eqref{neqissmallinD} which has the property that it is left invariant by the action of $\cop{M}$. 

First, let us attempt to make precise what we mean by states ``of the form'' \eqref{canonicalstate}. In \eqref{canonicalstate} we ensured that each coefficient was precisely the Boltzmann factor. This is clearly a very special class of states and we would set ourselves too simple a problem by focussing on these states. So we can generalize this slightly to consider states of the form 
\be
\label{cangeneralized}
|\psican\rangle = \sum_{E_1}^{E_2} {1 \over \sqrt{Z(\beta)}} a_i e^{-{\beta E_i \over 2}} |E_i \rangle,
\ee
where the $a_i$ are complex numbers that are drawn from a distribution so that their norms can each independently fluctuate a little about $1$ but
\be 
\label{aifluctuation}
\langle |a_i|^2 \rangle = 1. 
\ee
We will comment more on the range of the sum $[E_1, E_2]$ below. It is easy to verify, by repeating the argument above, that even for the states \eqref{cangeneralized} we have
\be
\langle \psican | \al_p |\psican \rangle = {1\over Z(\beta)}\tr(e^{-\beta H} \al_p) + \Or[e^{-{S \over 2}}].
\ee
By the central limit theorem, since there are an exponentially large number of energy eigenstates in \eqref{cangeneralized} the the fact that the coefficients $a_i$ can fluctuate in magnitudes as well as phases is unimportant. To see this consider a range of energies of size $e^{-{\nc \over 2}}$. Even this tiny range of energies has an exponentially large number of eigenstates. In the notation of \eqref{eth}, the expectation value $A(E_i)$  is constant over this range, and therefore the fluctuations of $|a_i|^2$ average out. Therefore,  for any smooth function, it is only the mean magnitude of the $|a_i|^2$ that matters, which is what leads to the result above. 

Now consider the action of an element of $\cop{M}$ on the state \eqref{cangeneralized}. If we write $\cop{M} = e^{-{\beta \cop{H} \over 2}} \cop{U} e^{\beta \cop{H} \over 2}$. If the matrix elements of $\cop{U}$ are $\cop{U} |E_i \rangle = \sum_j U_{j i} |E_j \rangle$, then we reach the new state
\be
|\psican'\rangle \equiv N_M \cop{M} |\psican\rangle  = N_M \sum_{E_i = E_1}^{E_i = E_2} \sum_{E_j} {1 \over \sqrt{Z(\beta)}} e^{-\beta E_j} a_i U_{j i} |E_j \rangle.
\ee
where the factor
\be
N_M = \langle \psican | \cop{M}^{\dagger} \cop{M} |\psican\rangle^{-{1\over 2}},
\ee
is required to normalize the state.
If we neglect the ``edge effects'' for the moment (these will be important below), then we see that
we again have a state of the form \eqref{cangeneralized}, although with coefficients
\be
\label{newstate}
a_j' = N_M \sum_i U_{i j} a_i.
\ee
From the argument above we can check that $N_m = 1 + \Or[e^{-{S \over 2}}]$. Therefore the action of the group of transformations denoted by $\cop{M}$, is basically like that of a unitary transformation on the elements $a_i$.  

We now see the following
\begin{enumerate}
\item
Physically the range of energies that is relevant in \eqref{cangeneralized} is limited. So, we may truncate this range so that the lower bound is  $E_1 = E-\Delta$ and the upper bound is $E_2 = E + \Delta$.  In that case, by an extension of the arguments of the previous 
subsection we find that $\cop{M}$ maps us to a slightly larger band of energies. Under almost any reasonable measure, this larger band has a much larger volume and therefore 
\eqref{consistencycond} is met. The technical details of this argument are identical to the previous subsection since, as we noted, $\cop{M}$ acts precisely as a unitary transformation on the coefficients $a_i$.
\item
We may try and avoid this conclusion in the following artificial manner. We extend the band of energies $[E_1, E_2]$ in \eqref{cangeneralized} so that it spans a very large range. We now truncate the action of $\cop{M}$ so that it acts only within this large energy range. By construction, now $\cop{M}$ maps this set back to itself. This may suggest that \eqref{consistencycond} cannot be met. This conclusion is clearly physically incorrect since the higher energies in \eqref{cangeneralized} are physically unimportant and therefore artificially extending the band should have no effect. However, there is another important point. If we indeed take our original domain ${\cal D}$ to be the subspace of this large range of energies then, and attempt to define a measure that is left invariant by the action of $\cop{M}$ then as we show below we find that the states \eqref{cangeneralized} are extremely unlikely states and themselves occupy only a small volume of the space. 
\end{enumerate}

The point is that there is a {\em tension} between the requirement \eqref{aifluctuation} which mandates that all the $a_i$ must have equal and approximately unit magnitude and the fact that $\cop{M}$ acts as a ``unitary'' on this space. We now consider one particular example to bring out this tension. In an attempt to write down a measure that is invariant under the action of $\cop{M}$ we may try and write the ``uniform'' measure on the space $a_i$. More precisely, we consider the measure
\be
\label{canonicalmeasure}
\meas_{\text{can}}(a_i)  d^2 a_1 \ldots d^2 a_D = 2 \pi N_{\meas} \delta(Z(\beta) - \sum_i |a_i|^2 e^{-\beta E_i})  d^2 a_1 \ldots d^2 a_D.
\ee
Here, to make the measure well-defined we had to truncate the range of energies $[E_1, E_2]$ so that the total number of eigenstates that enter the range are $D$. If we take this range to be large enough so that $E_2 - E_1 \gg \sqrt{\nc}$ then, for the purposes of its action on states \eqref{cangeneralized}, the action of $\cop{M}$ can be consistently restricted to this range. Now, naively, one might believe that this leads to a contradiction with \eqref{consistencycond}. However, we find that under \eqref{canonicalmeasure} with a large range of energies the states \eqref{cangeneralized} are themselves very atypical. Therefore the fact that the truncated version of $\cop{M}$ maps the energy-range back to itself and also leaves the measure \eqref{canonicalmeasure} invariant still does not lead to a contradiction with \eqref{consistencycond}.

We now explicitly bring out the tension between measures like \eqref{canonicalmeasure} which are the natural guesses for measures invariant under $\cop{M}$ and the fact that we would like the magnitudes of the $a_i$ to be approximately constant in  \eqref{aifluctuation}. We compute the reduced probability distribution, $\mu_{\text{red}}$ for the coefficient $a_1$ by integrating out $a_2 \ldots a_D$.
We write the delta function as
\be
\delta(Z(\beta) - \sum_i |a_i|^2 e^{-\beta E_i}) = \lim_{\epsilon \rightarrow 0} \int {d l \over 2 \pi} e^{i l(Z(\beta) - \sum_i |a_i|^2 e^{-\beta E_i}) - \epsilon l^2},
\ee
where $\epsilon$ is a small regulator. We also add small regulators $\epsilon' e^{-\beta E_i} |a_i|^2$ to make the integrals over $a_2 \ldots a_{D}$ well defined. Then we find
\be
\begin{split}
& \meas_{\text{red}}(a_1) \equiv \int \meas_{\text{can}}(a_i)  d^2 a_2 \ldots d^2 a_D \\&= 
N_{\meas} \int  d^2 a_2 \ldots d^2 a_D \lim_{\epsilon, \epsilon' \rightarrow 0} \int dl \, e^{i l(Z(\beta) - \sum_i |a_i|^2 e^{-\beta E_i}) - \epsilon l^2} 
e^{-\epsilon'  \sum_i e^{-\beta E_i} |a_i|^2} \\
&= \Big[{N_{\meas} \pi^{D-1} \over e^{-\beta \sum_i E_i}} \Big] \lim_{\epsilon, \epsilon' \rightarrow 0}\int d l \, {e^{i l(Z(\beta) - |a_1|^2 e^{-\beta E_1}) - \epsilon l^2} \over (\epsilon' + i l )^{D-1}} \\
&= \Big[{N_{\meas} \pi^{D-1} \over  \Gamma(D-1) e^{-\beta \sum_i E_i}} \Big] \int d l \, d x \, x^{D-2} e^{-x(i l + \epsilon')} e^{i l(Z(\beta) - |a_1|^2 e^{-\beta E_1}) - \epsilon l^2}   \\
&= \Big[{N_{\meas} \pi^{D-1} \over  \Gamma(D-1) e^{-\beta \sum_i E_i}} { \sqrt{\pi \over \epsilon}}\Big] \int  d x \, x^{D-2}e^{-{(x + Z(\beta) - |a_1|^2 e^{-\beta E_1})^2 \over 4 \epsilon} - x \epsilon'} \\
&= \kappa (1 - {|a_1|^2 e^{-\beta E_1} \over Z(\beta)})^{D-2}.
\end{split}
\ee
In the last step here, we have absorbed all the normalization factors into an irrelevant constant $\kappa$ and taken all regulators to $0$ and kept the part that is non-vanishing in this limit.

Generalizing this computation to the other coefficients, we find that the reduced probability distribution for the coefficient $|a_i|^2$ can be written as 
\be
\label{probreduceda1}
\mu_{\text{red}}(a_i) = \kappa (1 - {|a_i|^2 e^{-\beta E_i} \over Z(\beta)})^{D-2} \approx \kappa \exp\Big[{- {D e^{-\beta E_i} |a_i|^2 \over Z(\beta)}}\Big].
\ee

Now, we see something interesting. If we take the range of energies $[E_1, E_2]$ that appeared in \eqref{cangeneralized} to be much larger than $\sqrt{\nc}$ as we would need to make $\cop{M}$ act effectively in this space then \eqref{probreduceda1} suggests that the different $a_i$ have very different typical magnitudes. To ensure that the typical magnitudes of the coefficients $a_i$ are the same in \eqref{probreduceda1}, we have to take the range of energies $E_1 - E_2 \ll 1$. However, in this case the ensemble is clearly not invariant under the action of $\cop{M}$.

\paragraph{Physical intuition \\}
Let us briefly summarize the physical intuition behind the analysis above. The action of $\cop{M}$ is like a unitary on the coefficients $a_i$. Therefore, just like unitaries in a microcanonical ensemble, $\cop{M}$ tends to ``move'' the coefficients slightly from lower to higher energies. From this point of view, in the states \eqref{cangeneralized}, as written, the high energy states are weighted with coefficients that are typically too small and the low energy states are weighted with coefficients that are typically too large.  If we truncate the coefficients $a_i$ to a small range of energies, then $\cop{M}$ simply moves us out of this range. This suggests  that it may be difficult to find a measure on the Hilbert space that satisfies \eqref{neqissmallinD} and \eqref{neqissmallinI} for which $\cop{M}$ does not meet \eqref{consistencycond}. 

So, in principle it is consistent to expect that there may exist further criteria, based on the magnitudes and the phases of \eqref{cangeneralized} which can be detected by various operators beyond the simple operators in our algebra, which will determine that in the set \eqref{orbitcan} at most one of the states is in equilibrium whereas the others are not. We will return to this issue in future work.

\subsection{Summary}
We now summarize the results of this section. 
\begin{enumerate}
\item
For ordinary excitations of an equilibrium state with unitary operators, we can detect them using ordinary correlators and modify the construction of our mirrors accordingly.
\item
For the van Raamsdonk type unitaries, which act behind the horizon, we can detect them by using correlators of the Hamiltonian.
\item
Harlow attempted to define new mirrors that could evade detection by the Hamiltonian. However, we have shown here that this was predicated on an error in the computation of the Hamiltonian with the mirror operators. Harlow's operators do not have the right geometric properties to play the role of mirror operators, and do not even obey the Heisenberg equations of motion.
\item
Nevertheless, for some states with a canonical spread, we can find a group of transformations as in \eqref{orbitcan} so that we can map one state to another where the correlators are almost the same. There is no strict ambiguity involved here, because none of these states coincide exactly with the states obtained by acting on an equilibrium state with a mirror operator.
\item
However, while it is true that at the moment we do not know how to classify the
states in the orbit \eqref{orbitcan}, we have further shown that it is  consistent with statistical mechanics expectations to classify one of these as equilibrium and the others as non-equilibrium. Although, it appears that all these states are ``equally'' generic, this is specious, and such a classification would be perfectly consistent with the notion that most states are equilibrium states.
\end{enumerate}
We will return to this issue of the classification in further work. However, we note that this is a broader question in AdS/CFT --- that of precursors. At the moment, we do not know how to write down the bulk to boundary map for all possible states but this is an issue that extends beyond our construction, and is independent of the recent discussions on the information paradox.
We emphasize again that, our results in this subsection show that, within the class of states we have considered --- equilibrium states, near-equilibrium states excited by the ordinary and mirror operators, and small superpositions of these --- there is no ambiguity in our construction.


\section{State-dependence in entangled systems and ER=EPR}\label{secentangled}

We now describe the construction of our operators in general entangled systems. In section \ref{secexpliciteternal}, we already examined the construction of the interior in a specific entangled state --- the eternal black hole. Here we generalize the construction to more general entangled states. We show, also, that the construction of section \ref{secexpliciteternal} follows automatically from our generalized definition here.

We first present a general construction of interior operators. This construction is a very natural generalization of the one-sided interior constructed in section \ref{secdefnmirror} and in fact the defining equations for the mirror are unchanged. The only difference is in the construction of the ``little Hilbert space'' $\hilb[\psient]$. This is because for entangled systems we have two sets of possible natural excitations: one, where we act with excitations in the original CFT, and the other where we act with excitations in the entangled system. 

We then examine the consequences of this construction. We divide this analysis into two parts. We first consider states where the CFT is entangled with another CFT in a maximal manner so that the entanglement entropy scales with $\nc$. Next we consider states where the CFT is entangled with a small ``pointer'', which could be a collection of a few qubits so that the entanglement entropy is $\Or[1]$. 

In both cases, we obtain interesting results. When the CFT is entangled with another CFT, our construction leads to a precise and natural formulation of the ER=EPR conjecture \cite{Maldacena:2013xja}. When light operators on the right are entangled with light operators on the left, we find that excitations on the left can affect the experience of the right infalling observer in precisely that manner predicted by a geometric wormhole. On the other hand, in a generic state where there is no such entanglement we find that an observer on the left CFT loses his power to affect the region behind the  right horizon by means of simple operations, although he could do possibly do so by using some very complicated operators. This is consistent with the heuristic notion that the wormhole becomes ``very long'' for these states.

On the other hand,  when the CFT is  entangled with a small system then no such geometric wormhole  appears for any state. However, for this case, there is another crucial question, which is as follows. As we show below, the important test of whether there are any observable violations of quantum mechanics for the infalling observer arises when the observer entangles the CFT with a small system, jumps into the black hole and observes whether the state-dependence leads to any deviations from linearity. {\em We show below that such an experiment does not lead to any observable departure from the predictions of quantum mechanics}. 

We wish to emphasize throughout this section that these predictions arise as a natural consequence of our construction and not because
we have tailored the definition of the interior operators to entangled systems. As we mentioned above, the only change in an entangled system is that we have additional ``coarse'' or ``light'' operators to excite the system from the left and therefore we must enlarge the space $\hilb[\psient]$. 

We should mention that, our emphasis and approach is complementary to the approach of directly studying density matrices that was adopted in  \cite{Verlinde:2013uja}. 

\paragraph{Notation and objective \\}
In this section, we will consider entangled states
\be
\label{entangledstate}
|\psient \rangle = \sum_{i} \alpha_i |\widetilde{\Psi}_i \rangle \otimes  |\Psi_i \rangle.
\ee
Here $\alpha_i$ are some coefficients,  $|\Psi_i \rangle$ are orthonormal states in the original CFT, and $|\widetilde{\Psi}_i \rangle$ are states in a second system that may be another CFT or a collection of qubits.  We will refer to this system as the ``{\em left}'' system. The sum may be over a small number of states, or an exponentially large number.  

In this section, our primary objective is to reconstruct the experience of the infalling observer from the original CFT, which we also call the ``{\em right}'' CFT. Our construction of the mirrors, and also the little Hilbert space is appropriate for right-relationally defined local observables. In many cases where the left system is also a CFT, we can perform an analogous construction to describe the experience of a left-infalling observer. But apart from indicating this briefly below,  we do not focus on this.

\subsection{Mirror operators for entangled systems}

\paragraph{Summary of the construction\\}

The construction can be summarized as follows. We call $\alset$ the small algebra of the right CFT and $\alset_L$ for the algebra of observables
of the left system. We also define the product of the two algebras $\alsetent = \alset_L \otimes \alset$. 

The ``little Hilbert space'' is defined as the span of states $\{\alsetent |\Psi_{\text{en}}\rangle\}$. In general this will be bigger than just the span
of states $\{\alset|\Psi\rangle\}$, but there are some cases (like the thermofield double state) where the two spaces are the same. In the general case, the Hilbert space $\hilb[\psient]$ can be decomposed into the direct sum of subspaces $\hilb[\psient]^j$, each of which is closed under the action of the right algebra $\alset$ 
\be
\hilb[\psient] = \bigoplus_j \hilb[\psient]^j.
\ee
For each $j$ we can identify
a unique state $|\Psi_{\text{en}}^j\rangle \in \hilb[\psient]^j$ which is an equilibrium state with respect to the right CFT.\footnote{As in section \ref{subsecneareqsuper} when considering superpositions, it may happen that there is no equilibrium state inside $\hilb[\psient]^j$. In this case we need to enlarge $\hilb[\psient]^j$ to the direct sum of little Hilbert spaces built on equilibrium states.} The rest of the subspace $\hilb[\psient]^j$ can be generated
by acting on this equilibrium vector with elements of the algebra $\alset$.

Hence, within each of these subspaces we have a representation of the algebra $\alset$ which obeys all the conditions that 
we encountered in the case of non-entangled systems. More precisely, no element of the algebra $\alset$ can annihilate the state $|\Psi_{\text{en}}^j\rangle$ and the entire Hilbert space $\hilb[\psient]^j$ can be generated by acting with $\alset$ on $|\Psi_{\text{en}}^j\rangle$. The first condition
follows from our assumption that right-CFT states in \eqref{entangledstate} are black hole states. 

We can now 
define the mirror operators acting within this subspace using exactly the same rules as in section \ref{secdefnmirror}. Finally, the mirror operators acting on the full ``little Hilbert space'' 
$\hilb[\psient]$ are just the  sums of the individual mirror operators on the subspaces $ \hilb[\psient]^j$.

We emphasize that this is the natural extension of our construction of the mirror operators for systems without entanglement. As we will see, this simple definition is able to reproduce the expected physics for ER=EPR and other types
of entangled states with or without wormholes. Below we describe this construction in more detail.

\subsubsection{Construction of the ``little Hilbert space'' for entangled systems}
We now discuss in detail how to construct the ``little Hilbert space'' about an entangled state $\hilb[\psient]$. We first discuss the set of allowed excitations. We then use this to discuss the notion of ``equilibrium'' in entangled systems. Finally we put these notions together to construct $\hilb[\psient]$. 

\paragraph{Allowed excitations of entangled systems\\}
There are  two differences from the single-sided construction. In an entangled system, we have first the operators from the original CFT, which are part of $\alset.$ 
Additionally, observers should also have the ability to excite the state by acting with operators in the left system as well. 
In the left system, we can again build up a set of operators, which we will denote by $\alset_L$. If the left system is a holographic CFT, we should restrict the set of allowed operators in the same way that we restrict them 
for the original CFT. On the other hand if the left system is a collection of qubits, then there is no notion of light and heavy operators, and we  can allow $\alset_L$ to include {\em all} operators in the left theory.
Since operators on the left commute with operators on the right the full set of allowed operators has the structure of a direct product 
\be
\alsetent = \alset_L \otimes \alset.
\ee
We will denote elements of the left algebra by $\al_{L, \alpha} \in \alset_L$, and elements of the original algebra by $\al_{\alpha} \in \alset$ as usual.

We will explore this in greater detail  below but we caution the reader that  unlike in the case of the single sided CFT the little Hilbert space $\hilb[\psient]$ is not isomorphic to $\alsetent$. 

\paragraph{Equilibrium in entangled systems \\}
We now turn to the notion of equilibrium in entangled systems.  Since we are now allowing excitations of the state by operators in $\alsetent$ it is natural to modify the notion of equilibrium as well. This is a natural generalization of the definition of equilibrium in section \ref{seceqandneareq} for the original CFT. We define the deviation from equilibrium on the right using the same parameters as in \eqref{deviationdefone} and \eqref{deviationdeftwo}
\be
\begin{split}
\chi_p(t) &= \langle \psient | e^{i \hcft t} \al_p e^{-i \hcft t} | \psient \rangle, \\
{\nu}_p &= \tband^{-1 \over 2}   \int_0^{\tband}  |(\chi_p(t) - \chi_p(0))| d t,
\end{split}
\ee
where $\hcft$ is the right Hamiltonian. 
In addition, we consider similar deviations from equilibrium in the left CFT.
\be
\begin{split}
\chi_{Lp}(t) &= \langle \psient | e^{i \hleft t} \alen[p] e^{-i \hleft t} | \psient \rangle, \\
{\nu}_{Lp} &= \tband^{-1 \over 2}  \int_0^{\tband}  |(\chi_{Lp}(t) - \chi_{Lp}(0))| d t,
\end{split}
\ee
A necessary condition for the system to be in equilibrium is then that both left and right correlators are time translationally invariant. 
\begin{align}
\label{righteq}
&\nu_p = \Or[e^{-{S \over 2}}], ~ \forall p, \\
\label{lefteq}
&\nu_{Lp} = \Or[e^{-{S \over 2}}], ~ \forall p.
\end{align}
As above this condition is necessary but not strictly sufficient because of the  class of excitations that we discussed in section \ref{seccanonicalambiguity}. We will also see below 
that \eqref{lefteq} is often superfluous and we can perform the construction of the mirrors provided that the state is in right equilibrium even if it is not in left equilibrium. 

\paragraph{$\hilb[\psient]$ for entangled states \\ }
We now turn to the construction of the little Hilbert space, which describes the space of simple excitations about the base state. 
The main difference compared to our discussion above is that in the presence of entanglement, it is not necessary that all operators in  $\alsetent$ will give rise to independent descendants of the state $|\psient \rangle$. In particular, it is possible that
\be
\left(\alen[p] - \al_{q} \right) |\psient\rangle = 0,
\ee
for some correlated choices of $\alen[p]$ and $\al_{q}$.
Let us consider two examples of this. 

In the thermofield state $\tfdk$, we have
\be
\label{examplenull1}
\left( \op_{L \omega} - e^{-{\beta \omega \over 2}} \op_{\omega}^{\dagger} \right) |\tfd \rangle = 0.
\ee
It is understood, above and in other equations below that when we write an operator purely from the left system, it can be lifted to
an operator on the product system through $\op_{L, \omega} \equiv \op_{L,\omega} \otimes \cop{1}_R$ and vice versa.

Next, consider the CFT entangled with a two qubit system. This system has four states, which we denote by $|1\rangle \ldots |4 \rangle$. Now we may have a state that is not maximally entangled
\be
|\psient \rangle = {1 \over \sqrt{3}} \left(|\Psi_1 \rangle \otimes |1 \rangle + |\Psi_2 \rangle \otimes |2 \rangle + |\Psi_3 \rangle \otimes |3 \rangle \right),
\ee
where $|\Psi_i \rangle$ are some orthogonal states in the original CFT. Denoting the projector onto state $|4 \rangle$ by $P_4 = |4 \rangle \langle 4 |$, we see clearly that
\be
\label{examplenull2}
P_4 |\psient \rangle = 0.
\ee
Note that both these kinds of states, where we obtain null relations, are very special. States where relations of the form \eqref{examplenull1} hold are special because the entanglement is between ``simple'' operators on both sides. As we see below, generic states do not have such relations. Similarly, when the left system is ``small'', relations of the form \eqref{examplenull2} also occur only when the entanglement is non-maximal. Nevertheless, our construction will be able to account for these null relations correctly. 

We now define $\hilb[\psient]$ as follows. Starting with the state $|\psient \rangle$, we act with all elements of $\alset$ to obtain the space
\be
\label{hpsientzero}
\hilb[\psient]^0 = \text{span~of}\{\al_1 |\psient \rangle, \ldots \al_{\cal D} |\psient \rangle  \},
\ee
where we remind the reader that the elements of $\alset$ displayed above form a complete basis for this linear set. As usual we assume that there are no null vectors in the set displayed in \eqref{hpsientzero}. We define $\cop{P_{\text{en}}^0}$ to be the projector onto this subspace. This means
that
\be
\begin{split}
|v \rangle \in \hilb[\psient]^0 \quad &\Rightarrow \quad \cop{P_{\text{en}}^0} |v \rangle  = |v \rangle, \\
\langle v | \al_{p} |\psient \rangle  = 0 \,,\,\forall p \quad &\Rightarrow\quad \cop{P_{\text{en}}^0} |v \rangle = 0.
\end{split}
\ee
Next we pick a Hermitian element, $\alen[1]$  of $\alset_L$ and construct
\be
\label{psientone}
|\psient^1 \rangle =  (1 - \cop{P_{\text{en}}^0}) \alen[1] |\psient \rangle.
\ee
We pick $\alen[1]$ so that $|\psient^1  \rangle$ is non-vanishing and in {\em right equilibrium}. Note that it is not necessary for $|\psient^1\rangle$
to be in left equilibrium. (The reason for the restriction that $\alen[1]$ be Hermitian is explained below.) We now construct the space
\be
\label{hpsientone}
\hilb[\psient]^1 = \text{span~of}\{\al_1 |\psient^1 \rangle, \ldots \al_{\cal D} |\psient^1 \rangle  \}.
\ee
Then we define $\cop{P_{\text{en}}^1}$ to be the projector on $\hilb[\psient]^1$. Similarly, we look for $\alen[2] \in \alset_L$ so that 
\be
|\psient^2 \rangle = (1 - \cop{P_{\text{en}}^0}) (1 -  \cop{P_{\text{en}}^1}) \alen[2] |\psient \rangle,
\ee
is non-vanishing and in right equilibrium. We then construct $\hilb[\psient]^2$ analogously to \eqref{hpsientzero} and \eqref{hpsientone} and continue recursively in this manner until it is no longer possible to find any elements of $\alset_L$  which can produce descendants of $|\psient \rangle$ that are orthogonal to all the previous subspaces.

To summarize this construction, we find elements $\alen[1] \ldots \alen[{\cal D}_{\text{max}}]$ (where ${\cal D}_{\text{max}}$ may be smaller than the dimension of the left algebra) with the property that
\be
\alen[1] |\psient \rangle \ldots \alen[{\cal D}_{\text{max}}] |\psient \rangle,
\ee
are all in right equilibrium and have the property that
\be
\langle \psient | \al_{p} \alen[j] |\psient \rangle = 0\quad,\quad \forall p, j.
\ee
On each of these we construct the space $\hilb[\psient]^m$ as shown in \eqref{hpsientzero} and \eqref{hpsientone}. The full space $\hilb[\psient]$ is then defined by
\be
\hilb[\psient] = \bigoplus_j \hilb[\psient]^j.
\ee

It is worth discussing the structure of the space $\hilb[\psient]$ that results from the construction above, and the examples that we consider below will elucidate this. In the thermofield state, an action by a ``simple'' operator in the left CFT corresponds to the action of a ``simple'' operator on the right CFT. Therefore in this case $\hilb[\psient]$ coincides with $\hilb[\psient]^0$. On the other hand, in a generic entangled state of two CFTs, there is no relation between the action of simple operators on the left and the right, and therefore $\hilb[\psient]$ is isomorphic to $\alset \otimes \alsetent$. In intermediate cases where there is some entanglement, but not maximal, we obtain an $\hilb[\psient]$ that is intermediate between these two cases: its dimension is larger than $\hilb[\psient]^0$ but not maximal. We describe this in detail in several cases below.

\begin{figure}
\begin{center}
\begin{subfigure}[t]{0.5\textwidth}
\includegraphics[width=\textwidth]{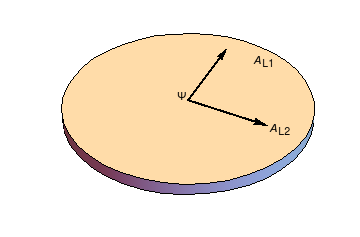}
\caption{$\hilb[\psient]$ where the action of the ``left algebra'' is entirely contained within the space  obtained by acting with the right algebra.}
\end{subfigure}
\qquad
\begin{subfigure}[t]{0.4\textwidth}
\includegraphics[width=\textwidth]{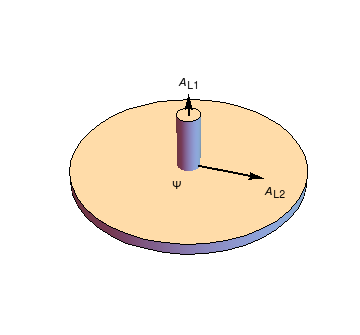}
\caption{$\hilb[\psient]$ in cases with less entanglement. Now the action of ``left  operators'' opens up new  directions.}
\end{subfigure}
\caption{The structure of the wormhole is directly linked to the structure  of $\hilb[\psient]$. In the case on the left above, where $\hilb[\psient]$ coincides with  $\hilb[\psient]^0$,  we  obtain a geometric wormhole. The case on the right can be  understood as an elongated wormhole. In the  extreme case  where $\hilb[\psient]$  becomes a  direct product  space, the geometric wormhole disappears. \label{figent} }
\end{center}
\end{figure}

The structure of $\hilb[\psient]$ is directly related to whether we obtain a wormhole on this. This is  shown schematically in Figure \ref{figent} and explained further below.

\paragraph{Definition of the mirror operators \\}
The mirror operators are now defined via precisely the  same linear equations as section \ref{subsecmirrordef}. Note that each vector in $\hilb[\psient]$ can be written as a linear combinations of vectors of the form $\al_{p} |\psient^j \rangle$ for some choice of $p$ and $j$.  We define 
\be
\label{entangledef}
\begin{split}
&\tO_{\omega,\ang} \,\al_{p} |\psient^j \rangle = \al_p \,e^{-{\beta \omega\over 2}} (\op_{\omega,\ang})^{\dagger} |\psient^j \rangle, \\
&[\tO_{\omega,\ang},\hcft]\, \al_p |\psient^j \rangle = - \omega \,\tO_{\omega,\ang} \al_p |\psient^j \rangle.
\end{split}
\ee
As usual, these equations have a solution because we have $\al_p |\psient^j \rangle \neq 0, \forall p,j$.  As the reader will note this is a direct extension of our definition of the mirrors for the original CFT. We now show how this simple extension
has remarkable properties and allows us to derive a precise version of the ER=EPR conjecture and also show that the infalling observer
will not observer any violations of quantum  mechanics.

\subsection{The wormhole in the thermofield double state }\label{secenttherm}
We now show how the construction above leads to a wormhole in the thermofield double state, where we take $|\psient \rangle = |\tfd\rangle$.
First, let us examine the construction of $\hilb[\psient]$. In the thermofield state we have the following relations
\be
\label{tfdrelations}\begin{split}
&\Oleft[\omega,\ang]| \tfd\rangle = e^{-{\beta \omega \over 2}} \Oright[\omega,\ang]^{\dagger} |\tfd\rangle ,\\
&\Oleft[\omega,\ang]^\dagger |\tfd\rangle = e^{{\beta \omega \over2}} \Oright[\omega,\ang] |\tfd\rangle .
\end{split}
\ee
Now consider an arbitrary polynomial in the $\Oleft[\omega,\ang]$, which we denote by $\alen[\alpha]$. In the thermofield state we have the relation
\be
\alen[\alpha]| \tfd\rangle = e^{-{\beta \hcft \over 2}} \al_{\alpha}^{\dagger} e^{\beta \hcft \over 2} |\tfd\rangle,
\ee
where, on the right of the equation above, we have an operator acting purely in the right CFT.  If $\alen[\alpha] \in \alset_L$ then, 
barring edge effects, we have $ e^{-{\beta \hcft \over 2}} \al_{\alpha}^{\dagger} e^{\beta \hcft \over 2} \in \alset$.  Therefore, in this case we start by constructing 
\be
\hilb[\tfd]^0 = \alset |\tfd\rangle,
\ee
and then we do not get any new states by acting with $\alset_L$. As a result, the full ``little Hilbert space'' is simply
\be
\hilb[\tfd] = \hilb[\tfd]^0.
\ee
Then the  construction of the mirror operators results in the same answer as  the construction in section \ref{secexpliciteternal} but we repeat it here from the general perspective of mirrors in entangled systems that we have presented above. The action of the mirror operators is   specified by the linear equations \eqref{entangledef}. Since in this case the structure of $\hilb[\tfd]$ is so simple, these equations reduce to
\begin{align}
\label{normalcommut} &\tO_{\omega,\ang} \al_{\alpha} |\tfd\rangle = \al_{\alpha} e^{-{\beta \omega \over 2}} \op_{\omega,\ang}^{\dagger} |\tfd\rangle, \\
\label{hrcommut} &[\tO_{\omega,\ang}, \hcft] \al_{\alpha}|\tfd\rangle = -\omega \al_{\alpha} e^{-{\beta \omega \over 2}} \op_{\omega,\ang}^{\dagger} |\tfd\rangle. 
\end{align}

Now the first point we note is that $\tO_{\omega,\ang}$ does not {\em commute} with elements of $\alset_L$, and moreover that this non-zero commutator is very special.  We can check this explicitly by considering the commutator of $[\tO_{\omega,\ang}, \Oleft[\omega',\ang']^{\dagger}]$. We have
\be
\tO_{\omega,\ang} \Oleft[\omega',\ang']^{\dagger} |\tfd\rangle = e^{\beta \omega' \over 2} \tO_{\omega,\ang} \Oright[\omega',\ang'] |\tfd\rangle = e^{\beta (\omega' - \omega) \over 2} \Oright[\omega',\ang'] \Oright[\omega,\ang]^{\dagger} |\tfd\rangle ,
\ee
where in the first equality we used \eqref{tfdrelations}. And also
\be
\begin{split}
\Oleft[\omega',\ang']^{\dagger} \tO_{\omega,\ang}  |\tfd\rangle & = e^{-{\beta \omega \over 2}} \Oleft[\omega',\ang']^{\dagger}  \Oright[\omega,\ang]^{\dagger} |\tfd\rangle  = e^{-{\beta \omega \over 2}} \Oright[\omega,\ang]^{\dagger} \Oleft[\omega',\ang']^{\dagger} |\tfd\rangle  \\ &= e^{\beta (\omega' - \omega) \over 2} \Oright[\omega,\ang]^{\dagger} \Oright[\omega',\ang']  |\tfd\rangle . 
\end{split}
\ee
This leads to an $\Or[1]$ effective commutator 
\be
\label{effectivetfdcommut}
[\tO_{\omega,\ang}, \Oleft[\omega',\ang']^{\dagger}] |\tfd\rangle = \comm_{\beta}(\omega, \ang) \delta_{\omega \omega'} \delta_{\ang \ang'} |\tfd\rangle.
\ee
These are very special commutators, and suggest that within correlators involving only elements of $\alsetgff$, it is possible to replace
$\tO_{\omega, \ang}$ with $\Oleft[\omega, \ang]$. However, as we have emphasized one cannot equate these operators. In particular,
to compute the commutator of the mirrors with the left Hamiltonian we consider
\be
\begin{split}
\tO_{\omega, \ang} \hleft |\tfd \rangle &= \tO_{\omega, \ang} \hcft | \tfd \rangle = e^{-{\beta \omega \over 2}} \op^{\dagger}_{\omega, \ang} \hcft |\tfd \rangle = e^{-{\beta \omega \over 2}} \op^{\dagger}_{\omega, \ang} \hleft |\tfd \rangle \\ &= \hleft e^{-{\beta \omega \over 2}} \op^{\dagger}_{\omega, \ang}  |\tfd \rangle = \hleft \tO_{\omega, \ang} |\tfd \rangle.
\end{split}
\ee
In this chain of equalities we have first used the isometry of the thermofield state, then used the definition \eqref{hrcommut} and then manipulated this expression by using the isometry again and the fact that $\hleft$ commutes with right operators.  So we find that 
within simple correlators
\be
\label{hlcommut}
[\tO_{\omega, \ang}, \hleft] |\tfd \rangle \doteq 0.
\ee
Therefore the mirror operators have a vanishing commutator with the left Hamiltonian. Note that this follows as a {\em consequence} of our defining relations and is not something that we have to put in by hand.

\begin{figure}[!h]
\begin{center}
\resizebox{0.4\textwidth}{!}{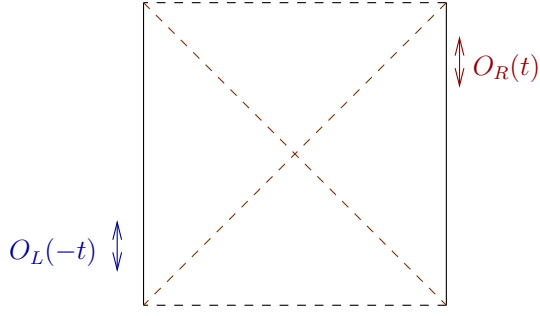}
\caption{The standard wormhole described in section \ref{secenttherm}: operators on the right $\op_R(t)$ are entangled with left operator $\op_L(-t)$ }\label{figstandworm}
\end{center}
\end{figure}

For the sake of completeness, we can also evaluate the two point function
\be
\label{tfdtwoptfun}
\langle \tfd | \tO_{\omega,\ang} \Oleft[\omega,\ang]^{\dagger} |\tfd \rangle = e^{\beta \omega \over 2} \langle \tfd | \tO_{\omega,\ang} \Oright[\omega,\ang] |\tfd \rangle = G_{\beta}(\omega, \ang).
\ee
We can proceed to evaluate other correlators along the lines of \eqref{effectivetfdcommut} and \eqref{tfdtwoptfun}. If we now try and reproduce these correlators from a geometry then the geometric picture that arises from this is that of the standard thermofield wormhole.  See Figure \ref{figstandworm}. Now we will show how, in a generic entangled state of the two CFTs, a very different geometric picture emerges.

\subsection{The generic entangled state of two CFTs }\label{secentgeneric}
We now show how our construction works in the ``generic'' entangled state of two CFTs. Consider scrambling the thermofield double state with a left unitary.  So we now consider
\be
\label{psigendef}
|\psigen\rangle = \ulgen |\tfd \rangle,
\ee
where the unitary is {\em not} an exponentiated element of the algebra of simple operator:  $\ulgen \neq e^{i \alen[\alpha]}$, but rather some ``generic unitary'' that changes the structure of entanglement of the two sides.  As a result,  as shown in \cite{Marolf:2013dba}, simple operators on the left and right are uncorrelated.
\be
\label{ethnoworm}
\langle \tfd | \ulgen^{\dagger} \alen[\alpha] \al_{\beta} \ulgen | \tfd \rangle = \Or[e^{-{S \over 2}}]\quad ,\quad \forall \alpha, \beta.
\ee

The construction of $\hilb[\psigen]$ proceeds according to the algorithm described in the beginning of this section. Notice that there is a 
qualitative difference from the thermofield double state, because we no longer have relations of the form \eqref{tfdrelations}. The relation \eqref{ethnoworm}
implies that  for an arbitrary element $\alen[1] \in \alset_L$, the left descendant constructed via \eqref{psientone} is non-null and in right equilibrium. Hence the ``little
Hilbert space'' $\hilb[\psigen]$ will have the direct sum decomposition as explained earlier. We select a set of operators  $\alen[1] \ldots \alen[{\cal D}_L]$ which
form a basis of $\alset_L$ and generate the equilibrium vector in each of these subspaces. Finally we find
\be
\hilb[\psigen] = \text{span~of}\{\al_{\beta} \alen[\alpha] |\psigen \rangle, \quad \beta= 1 \ldots {\cal D};~\alpha = 1 \ldots {\cal D}_L \}.
\ee
Now, the definition of the mirror operators above reads 
\be
\label{genericdefiningeq}
\tO_{\omega,\ang} \al_{\beta} \alen[\alpha] |\psigen\rangle = \al_{\beta} e^{-{\beta \omega \over 2}} \Oright[\omega,\ang]^{\dagger} \alen[\alpha] |\psigen\rangle.
\ee
But since operators in $\alset$ and $\alset_L$ commute this becomes
\be
\tO_{\omega,\ang} \al_{\beta} \alen[\alpha]  |\psigen\rangle  = e^{-{\beta \omega \over 2}} \al_{\beta}   \alen[\alpha] \Oright[\omega,\ang]^{\dagger}  |\psigen\rangle.
\ee
Therefore for the generic entangled state $|\psigen \rangle$, we have 
\be
\label{genericommutator}
[\tO_{\omega,\ang}, \alen[\alpha]] |\psigen \rangle = 0, \quad \text{generic~state}.
\ee

We can also compute the two point function
\be
\label{generictwopoint}
\langle \psigen | \tO_{\omega, \ang} \Oleft[\omega, \ang]^{\dagger} | \psigen \rangle = e^{-\beta \omega \over 2} \langle \psigen | \Oleft[\omega, \ang]^{\dagger} \op^{\dagger}_{\omega, \ang} | \psigen \rangle = \Or[e^{-S \over 2}].
\ee
Other two point functions of simple operators vanish in the same manner.
Therefore the mirrors not only effectively commute, they are also uncorrelated with the simple left operators.

Note that both \eqref{genericommutator} and \eqref{generictwopoint} --- just like \eqref{effectivetfdcommut} --- came automatically from our definition of the mirror operators for entangled systems and the different structure of $\hilb[\psigen]$ in these cases, without having to put anything in by hand. 

\begin{figure}[!h]
\begin{center}
\resizebox{0.4\textwidth}{!}{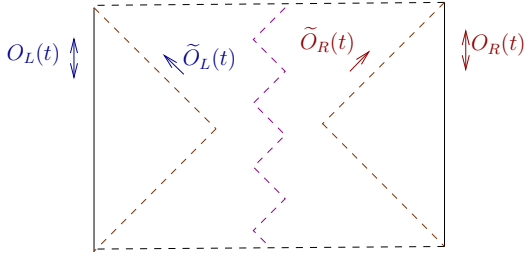}
\caption{The dual to the generic entangled state described in section \ref{secentgeneric}. Simple operators on the right $\op_R(t)$ and left are not correlated. This is indicated by the jagged broken line in the middle and there is no geometric wormhole. But both sides see a smooth horizon with the emergence of new mirror operators behind the horizon. }\label{fignoworm}
\end{center}
\end{figure}

Now, we may try and write down a geometry that reproduces \eqref{generictwopoint} and \eqref{genericommutator}. We remind the reader that correlators between the mirror operators and ordinary operators are unchanged showing that right-infalling observer still perceives a smooth horizon. However, the vanishing commutator \eqref{genericommutator} shows that in the generic state it is not possible
to affect the experience of the right-infalling observer by simple operators on the left. Hence the geometric wormhole has disappeared. Instead, geometrically we obtain the  Penrose diagram of Figure \ref{fignoworm}. This Penrose diagram was also conjectured in \cite{Shenker:2013yza}.

\subsubsection{Mirrors as scrambled left operators in the generic state}
We conclude with a further observation on the mirror operators in the generic state $|\psigen \rangle$. The relation \eqref{genericommutator} is somewhat deceptive. Our construction automatically leads to the conclusion that the commutator of the mirror operators for the right infalling observer  and simple left operators --- where simple is defined through membership in $\alset_L$ ---vanishes when inserted in low point correlation functions. However, another interesting consequence is that when we  have a  high  degree of entanglement of the  CFT with  another  system, then generically the  mirror  operators  act  on the  left system as well.  This follows as  an inevitable consequence of their  defining  equations.   It is easy to prove this as follows.

Let us write the generic entangled state in a Schmidt basis so that
\be
\label{entangstate}
|\psigen \rangle = \sum_i \kappa_i  |\tilde{v}_i \rangle \otimes |v_i \rangle ,
\ee
where the $\kappa_i$ are arbitrary coefficients and we have ``diagonalized'' the entanglement so that $|v_i \rangle$ are some 
orthonormal states in the right CFT and $|\tilde{v}_i \rangle$ are some states in the left CFT.
Now consider just one of the defining equation for $\tO_{\omega, \ang}$ 
\be
\label{entangledsyseqn}
\tO_{\omega,\ang} \al_{\alpha}  |\psigen \rangle = \al_{\alpha} e^{-{\beta \omega \over 2}}  \op_{\omega,\ang}^{\dagger} |\psigen \rangle,
\ee
and look for a solution to \eqref{entangledsyseqn} with the $\tO_{\omega, \ang}$ acting entirely within the Hilbert space of the right CFT. We emphasize that \eqref{entangledsyseqn} is just a special case of \eqref{genericdefiningeq} with the element of the left algebra that appears there set to the identity. Let us denote this putative solution by  $\cop{X} =\tO_{\omega, \ang}$.

We see that this demand that $\cop{X}$ is an operator in the right CFT means that for each $\alpha$, the single equation \eqref{entangledsyseqn} leads to a {\em system} of linear equations given by
\be
\label{entangledsyseqn_onlyCFT}
\cop{X} \al_{\alpha} |v_i \rangle= \al_{\alpha} e^{-\beta \omega \over 2} \op^{\dagger}_{\omega,\ang} |v_i \rangle, \quad \forall \alpha, i.
\ee
However, if the set $i$ in \eqref{entangstate} runs over a large enough range, then in general \eqref{entangledsyseqn_onlyCFT} has no solutions. For example, consider the situation where the states $|v_i\rangle$ provide a basis of the Hilbert space. Then, with $\al_{\alpha} \in \alsetgff$, the states 
\be
|w_{\alpha,i} \rangle = \al_{\alpha} |v_i \rangle,
\ee
provide an {\em overcomplete} basis for the space if we span over all $i$ and all $\alpha$. Therefore in \eqref{entangledsyseqn_onlyCFT} we are trying to specify the action of the putative purely right mirror operator on an overcomplete basis and this is not possible in general.

For example, we can find coefficients $z_{\alpha i}$ so that
\be
\sum_{\alpha, i} z_{\alpha i} \al_{\alpha} |v_i \rangle  = 0,
\ee
and in general it will not be the case that \eqref{entangledsyseqn_onlyCFT} map this
vector to 0. In particular on this vector we would find
\be
0 = \cop{X} \sum_{\alpha, i} z_{\alpha i} \al_{\alpha} |v_i \rangle = \sum_{\alpha, i} z_{\alpha i} e^{-{\beta \omega \over 2}} \al_{\alpha} \op_{\omega,\ang}^{\dagger}  |v_i \rangle  \neq 0 \, ?
\ee
Here we have used the fact that generically the right hand side of the relation above will not vanish with the same coefficients $z_{\alpha,i}$. 

So we have shown that in the situation with a high entanglement entropy the $\tO_{\omega, \ang}$ operators must act on the left as well and the operator $\cop{X}$ that acts only in the right CFT does not exist.

We conclude with some speculative comments on the possible physical implications of this fact. The authors of \cite{Maldacena:2013xja} suggested that the generic state $|\psigen \rangle$ may nevertheless be understood through a ``very long'' wormhole. Now note that our discussion of the {\em generic commutator} in section \ref{seccommutarg} suggests that if we take a {\em generic} operator in the left CFT, $\cop{Y}$ then we would find that
\be
\label{genstategencommut}
\langle \psigen | |[\cop{Y}, \tO_{\omega, \ang}]|^2 | \psigen \rangle = \Or[1].
\ee
We emphasize that $\cop{Y}$ is not one of the simple operators that are part of $\alset_L$ which commute with the mirrors within low point correlators. Now \eqref{genstategencommut} suggests that with a suitably complicated operation the left observer can affect the experience of the right infalling observer. This may be taken as some evidence of the existence of a long wormhole although it would be nice to make this more precise .

\subsection{A superposition of the thermofield and a generic state }\label{secentsup}
As a further example, we now show how our construction works in the superposition of the thermofield and a generic state. We consider
\be
\label{psisupent}
| \psisupent \rangle = \kappa \left( |\tfd \rangle + |\psigen \rangle \right).
\ee
For the generic left unitary of the sort discussed in \eqref{psigendef}, we have $\kappa = {1 \over \sqrt{2}} + \Or[e^{-S}]$. 

We start with 
\be
\hilb[{\psisupent}]^0 = \alset |\psisupent \rangle.
\ee
On the other hand, on acting with an element of $\alset_L$ we find that
\be
\label{psisupentdesc}
\begin{split}
|\psisupent^1 \rangle &= (1 - \cop{P}_s^0) \alen[1] |\psisupent\rangle = \kappa(1 - \cop{P}^0_{s}) \left(e^{-{\beta \cop{H} \over 2}} \al_1^{\dagger} e^{\beta  \cop{H} \over 2} |\tfd \rangle + \alen[1] |\psigen\rangle \right) \\ &=  \kappa\alen[1] |\psigen\rangle - {1 \over 2 } {\kappa \langle \alen[1] \rangle} \big(|\psigen \rangle + |\tfd \rangle \big)  + {\kappa \over 2}  e^{-{\beta \hcft \over 2}} \al_1^{\dagger} e^{\beta \hcft \over 2} \big(|\tfd \rangle  - |\psigen \rangle \big).
\end{split}
\ee
Here $\langle \alen[1] \rangle \equiv \langle \psigen | \alen[1] | \psigen \rangle$. In deriving this result, we have used two intermediate results.
\be
\begin{split}
&\cop{P}_s^0 \big(\alen[1] - \langle \alen[1] \rangle \big) |\psigen \rangle = 0,\\
&\cop{P}_s^0 \al_{m} |\psigen \rangle = \cop{P}_s^0 \al_{m} |\tfd \rangle = {1 \over 2} \Big(\al_{m} |\psigen \rangle + \al_{m}|\tfd \rangle \Big),
\end{split}
\ee
where $\al_m$ is any element of $\alset$.

In the final expression in \eqref{psisupentdesc}  we have, once again, a superposition of an equilibrium and a  near-equilibrium state from the point of view of observables in $\alset$. This is a special case of the superposition of near-equilibrium states that was considered in section \ref{secdefnmirror}. In such states, as explained there, we must enlarge the little Hilbert space slightly and upon doing that we find
\be
\hilb[{\psisupent}] = \hilb[\tfd] \oplus \hilb[\psigen].
\ee

The action of the mirror operators can be deduced in a straightforward way from the definition provided in \eqref{entangledef}.
\be
\tO_{\omega,\ang} \alen[\alpha] \al_{\beta} |\psisupent \rangle = \kappa \al_{\beta} e^{-{\beta \cop{H} \over 2}} \al_{\alpha}^{\dagger} e^{\beta \cop{H} \over 2} e^{-{\beta \omega \over 2}} \op_{\omega,\ang}^{\dagger} |\tfd \rangle + \kappa \al_{\beta} \alen[\alpha] e^{-{\beta \omega \over 2}} \tO_{\omega,\ang}^{\dagger} |\psigen \rangle. 
\ee
Consequently correlators involving mirrors and ordinary operators separate into
\be  
\langle \psisupent| \tal_{\alpha_3} \alen[\alpha_2] \al_{\alpha_1} |\psisupent \rangle  = |\kappa|^2 \left(\langle \tfd | \tal_{\alpha_3} \alen[\alpha_2] \al_{\alpha_1} | \tfd \rangle + \langle \psigen|  \tal_{\alpha_3} \alen[\alpha_2] \al_{\alpha_1} |\psigen \rangle \right).
\ee
Therefore the superposition of states \eqref{psisupent} acts like a classical mixture of a thermofield and a state with no wormhole. This is precisely what is expected. Note that standard Penrose diagrams cannot capture this superposition of two geometries, although the correlators are very simply related to the correlators in the two individual geometries.

\subsection{The microcanonical double state and a low-pass wormhole }\label{secentmicro}
We now consider a modification of the thermofield state: a ``microcanonical double state''. We will show that in the appropriate regime this leads to a new kind of wormhole with interesting properties.

Consider a range of energy $E \pm \Delta$ that contains ${\cal D}_{E, \Delta}$ states. Here $\Delta = \Or[1]$. It is also useful to consider energies that are high enough so that the associated temperature satisfies $\beta \Delta \ll 1$.  These are all hierarchies between $\Or[1]$ quantities and neither $\beta$ nor $\Delta$ scale with $\nc$. Now consider
\be
\label{microdouble}
|\psimd\rangle= {1 \over \sqrt{\dimrange[E, \Delta]}} \sum_{E_i =E - \Delta}^{E_i = E + \Delta} |E_i, E_i \rangle.
\ee
This state was also considered in \cite{Harlow:2014yoa} (see page 15), but we will reach a conclusion that is different from the conclusion reached there. In particular, the state  \eqref{microdouble} does  have a smooth interior and, contrary to the suggestion made in \cite{Harlow:2014yoa},  our construction generates it correctly. The error made in \cite{Harlow:2014yoa} follows from the error alluded to in section \ref{seccommentsharlow}: an incorrect expectation that the mirror operators must correspond to simple operators in the left CFT. 

Consider a frequency $\omega_l \ll \Delta$. The subscript indicates that this is a ``low'' frequency. For correlators involving such modes, the fact that the entanglement has been truncated is invisible. Let us denote the matrix elements of this operator in the energy eigenbasis by $c_{j i}$ as in \eqref{opactionright} so that we have
\be
\sum_{E_i=E - \Delta}^{E_i = E + \Delta}  \op_{\omega_l, \ang} |E_i,E_i \rangle = \sum_{E_i=E - \Delta}^{E_i = E + \Delta} \sum_{E_j} c_{j i} |E_i, E_j \rangle.
\ee
Note that, as we explained around \eqref{opactionright}, we can choose these matrix elements $c_{j i}$ to be real because of the T-invariance of the modes of local operators.  While the sum over $j$ above technically runs over all energies, since we know that the matrix elements $c_{j i}$ should be peaked around $E_i - E_j = \omega_l$, we can write
\be
 \op_{\omega_l} |\psimd \rangle = {1 \over \sqrt{\dimrange[E,\Delta]}} \sum_{E_i=E - \Delta}^{E_i= E + \Delta} \left[ \sum_{E_j = E - \Delta - \omega_l}^{E_j=E + \Delta - \omega_l} c_{j i} |E_i, E_j \rangle \right].
\ee

Now, notice that we also have
\be
\begin{split}
\sum_{E_i=E - \Delta}^{E_i = E + \Delta} \Oleft[\omega_l, \ang]^{\dagger} |E_i,E_i \rangle &= \sum_{E_i = E - \Delta}^{E_i = E + \Delta} \left[ \sum_{E_j = E - \Delta + \omega_l}^{E_j = E + \Delta + \omega_l}  c_{i j} |E_j, E_i \rangle \right]\\
&= \sum_{E_i = E - \Delta + \omega_l}^{E_i = E + \Delta + \omega_l} \sum_{E_j = E - \Delta}^{E_j = E + \Delta}  c_{j i} |E_i, E_j \rangle .
\end{split}
\ee
In the last step, we have interchanged $i$ and $j$ above to bring it into a form where we can compare it with the action of the right operator. However, the ranges of the sums over $i,j$ are different. In the case where $\omega_l \ll \Delta$ and $\beta \omega_l \ll 1$ we can approximately neglect this to obtain
\be
\label{microdoublelowfreq}
 \op_{\omega_l, \ang} |\psimd \rangle = \Oleft[\omega_l, \ang]^{\dagger} |\psimd \rangle + \Or[{\omega_l \over \Delta}] +  \Or[\beta \omega_l], \quad \omega_l \ll \Delta.
\ee

On the other hand, for large $\omega_h \gg \Delta$ we see that
\be
\label{microdoublehighfreq}
\langle \psimd | \Oleft[\omega_h]^{\dagger} \op_{\omega_h} | \psimd \rangle \ll 1, \quad \omega_h \gg \Delta.
\ee
Note that the result \eqref{microdoublehighfreq} holds even if $\beta \omega_h \ll 1$.

We can now perform the construction above to define the right-relational mirrors on this state.  The relations \eqref{microdoublehighfreq} and \eqref{microdoublelowfreq} then tell us that inside correlation functions evaluated on \eqref{microdouble} (except those involving the Hamiltonian, where ${1 \over \nc}$ corrections are important) we can approximately perform the replacement for low frequencies
\be
\widetilde{\op}_{\omega_l, \ang} \rightarrow \Oleft[\omega_l, \ang], \quad \omega_l \ll \Delta.
\ee
However, no such replacement is possible for high frequency modes $\widetilde{\op}_{\omega_h, \ang}$, which cannot be related to the action of simple left operators. These are independent operators that can be constructed using the algorithm that we have outlined. Using this we can compute correlators involving both ordinary operators on the left and the right, and the mirror operators precisely.  

It would be interesting to develop a more precise picture of the geometric dual to this state. However, some qualitative properties are clear.
The state \eqref{microdouble} is a ``low-pass wormhole'' --- where low frequency modes on the left and right are entangled, but the mirrors for high frequency modes on both sides are independent operators.   In this geometry both the left and the right infalling observer see smooth horizons. These observers can ``communicate'' using low frequencies but not using high frequencies.

It may also be possible to think of these wormholes as ``elongated wormholes''. It is interesting to notice that the geometries described in \cite{Bak:2011ga,Bak:2007qw,Bak:2007jm}, which were also considered in \cite{Maldacena:2013xja}  have somewhat similar properties.  However, these geometries involve infalling matter and  cannot be 
a precise dual to $|\psimd \rangle$, since the state $|\psimd \rangle$ is invariant under $e^{i (\hleft - \hcft) \capt} |\psimd \rangle = |\psimd \rangle$ and this isometry is not evident in these geometries.

\subsection{Entangled qubits and linearity}\label{secentqub}
We now consider a final case in some detail: the situation where the CFT is entangled with a few qubits. In this situation not only is there no geometric wormhole, but we find that it is possible to select the interior operators to strictly commute (as operators) with all operators in the qubit system.

For now we make no assumption about the Hamiltonian of the qubit system. However, the combined CFT and qubit system can be in equilibrium only in states of the form
\be
\label{qubent}
|\psiqub\rangle = \sum_i \alpha_i |E_{qi} \rangle \otimes|\Psi_i \rangle,
\ee
where $|E_{qi} \rangle$ are energy eigenstates in the qubit system and $|\Psi_i \rangle$ are equilibrium states in the CFT, and the coefficients $\alpha_i$ obey 
$\sum_i|\alpha_i|^2 =1$.

The reason that the entanglement structure has to be of this form in an equilibrium state is because in the qubit system, we assume that we have access to {\em all} operators. Therefore the only ``equilibrium'' states in this system are strict energy eigenstates which  remain invariant under time evolution. If, upon tracing out the CFT, we were to obtain any significant off-diagonal terms in the qubit density matrix, then it would be possible to find an appropriate operator whose expectation value would be time-dependent. These energy-eigenstates must be entangled with states that are independently in equilibrium in the CFT. This fixes equilibrium states to be of the form \eqref{qubent}. 

We  now find that
\be
\hilb[\psiqub]^0  =  \sum_i \coeff[i] |E_{q i} \rangle  \otimes \alset |\Psi_i \rangle.
\ee
We now act with an arbitrary operator from the qubit system $\alen[1]$ to obtain
\be
\label{qubdesc}
\alen[1] |\psiqub \rangle = 	 \sum_{i,j} \coeff[i] A_{L,1}^{j i} |E_{q j} \rangle \otimes |\Psi_i \rangle,
\ee
where $A_{L,1}^{j i}$ are the matrix elements in the qubit-energy eigenbasis of the left operator.
This state is not  in left-equilibrium but because a small superposition of equilibrium state is still an equilibrium state we see that \eqref{qubdesc} still represents a right equilibrium state  and does not  lie in $\hilb[\psiqub]^0$. 

Proceeding in this manner,  we find that the little Hilbert space has the form
\be
\hilb[\psiqub] = \bigoplus_{i,j} |E_i \rangle \otimes \alset |\Psi_j \rangle.
\ee
Now, using the prescription above, we find that  the action of the mirrors is given by
\be
\label{qubitmirror}
\tO_{\omega,\ang} \left(|E_i \rangle \otimes \al_{\alpha} |\Psi_j \rangle \right) = |E_i \rangle \otimes \al_{\alpha} e^{-{\beta \omega \over 2}}  \op_{\omega,\ang}^{\dagger} |\Psi_j \rangle.
\ee
Therefore in this situation the mirror operators are entirely operators within the right CFT and do not act in the qubit system at all. Moreover  the mirror operators above can be  understood as follows. We construct mirror operators on each of the equilibrium states $|\Psi_i \rangle$. We then take  the union of these operators and  this yields the  operators above.

\paragraph{Avoiding possible superluminality in the presence of state-dependence \\}
Let us briefly mention the significance of the observation above. Our state-dependent operators are  sometimes conflated with notions of ``non-linear'' quantum mechanics that have been proposed earlier.  Gisin \cite{gisin1990weinberg} and Polchinski \cite{PhysRevLett.66.397} pointed out sharp difficulties with one such idea that was advanced by Weinberg \cite{weinberg1989testing}. In particular,  Gisin noted that non-linear evolution in quantum mechanics could lead to superluminal communication.

We emphasize that in  our proposal we do not add any non-linear terms to the Hamiltonian, which is simply the CFT Hamiltonian. Nevertheless, one  may still be concerned about this issue of superluminality. We now show that this also  does not arise in  our construction.

Consider the following  experiment.  An experimenter entangles black hole
microstates  in the CFT  with states of a  ``small pointer'' comprising a few qubits. Then the qubits and the CFT are  separated by a large distance. An observer from the CFT now jumps into the black hole and makes a measurement. Physically,  we expect that such an observer  should not be able to send messages  to another observer who has access only to the qubits.

To make this more precise, consider a qubit system with $M+1$ states, that we denote by $|1 \rangle, | 2 \rangle,  \ldots | M + 1\rangle$, where $M \ll \cn$.  Now, we consider $M$ equilibrium states of the CFT,  $|\Psi_1 \rangle \ldots |\Psi_M \rangle$, and we take them to be orthogonal without loss of generality. Let us prepare the joint qubit-CFT system in the state
\be
\label{initialentstate}
|\psiqub \rangle = \sum_{i=1}^M \alpha_i |i \rangle \otimes |\Psi_i \rangle  + |M+1 \rangle \otimes \left(\sum_j \beta_j |\Psi_j \rangle \right).
\ee
In order for the state to be normalized correctly, we have the condition
\be
\sum_i |\alpha_i|^2 + |\beta_i|^2 = 1.
\ee

Now, we act with a {\em unitary} of the mirror operators on $|\psiqub \rangle$. Let us call this unitary $\widetilde{\cop{U}}$. We see that from \eqref{qubitmirror} we have
\be
\label{finalentstate}
\widetilde{\cop{U}} |\Psi \rangle = \sum_{i=1}^M \alpha_i |i \rangle \otimes \widetilde{\cop{U}} |\Psi_i \rangle  + |M+1 \rangle \otimes \widetilde{\cop{U}} \left(\sum_j \beta_j |\Psi_j \rangle \right).
\ee

The key physical requirement to ensure that no messages can be sent  from the black hole interior to the  qubit system is that this process should leave the density matrix of the pointer invariant. The density matrix of the pointer in
\eqref{initialentstate} has the following components
\be
\label{initdensity}
\begin{split}
&\langle M+1 | \cop{\rho}^{\text{init}} | M + 1 \rangle = \sum |\beta_i|^2, \\
&\langle i | \cop{\rho}^{\text{init}} | i \rangle = |\alpha_i|^2, \\
&\langle i | \cop{\rho}^{\text{init}}  | M + 1 \rangle = \alpha_i \beta_i^*, \\
&\langle M+1 | \cop{\rho}^{\text{init}}  | i \rangle = \alpha_i^* \beta_i.
\end{split}
\ee
For convenience, let us denote $|\chi \rangle = \widetilde{\cop{U}} \left(\sum_j \beta_j |\Psi_j \rangle \right)$.
Then the components of the  density matrix of the pointer in the final state \eqref{finalentstate} are
\be
\label{finaldensity}
\begin{split}
&\langle M+1 | \cop{\rho}^{\text{fin}} | M + 1 \rangle = \langle \chi | \chi \rangle, \\
&\langle i | \cop{\rho}^{\text{fin}} | i \rangle = |\alpha_i|^2, \\
&\langle i | \cop{\rho}^{\text{fin}}  | M + 1 \rangle = \alpha_i \langle \chi | \widetilde{\cop{U}} |\Psi_i \rangle, \\
&\langle M+1 | \cop{\rho}^{\text{fin}}  | i \rangle = \alpha_i^* \langle \Psi_i | \widetilde{\cop{U}}^{\dagger} |\chi \rangle.
\end{split}
\ee

Demanding that the infalling observer cannot send messages is equivalent to setting $\cop{\rho}^{\text{fin}} = \cop{\rho}^{\text{init}}.$ From \eqref{initdensity} and \eqref{finaldensity} we see that this implies
\be
\begin{split}
&\langle \chi | \chi \rangle = \sum |\beta_i|^2, \\
&\langle \chi | \widetilde{\cop{U}} |\Psi_i \rangle = \beta_i^*, \\ 
&\langle \Psi_i | \widetilde{\cop{U}}^{\dagger} |\chi \rangle = \beta_i.
\end{split}
\ee
In fact, since the states $\widetilde{\cop{U}} |\Psi_i \rangle$ also give an orthogonal set, we see that we are forced to the conclusion that
\be
| \chi \rangle = \beta_i \widetilde{\cop{U}} |\Psi_i \rangle.
\ee
This implies that the operator $\widetilde{\cop{U}}$ must act linearly on a superposition of a small number of states.

This is precisely what  is  ensured by the construction above. As we mentioned,  this construction  proceeds by constructing mirrors  for each of  the individual equilibrium states and then just taking the union of their actions, which ensures  that the  constraint above  is satisfied. The reader may recall the discussion of section \ref{secsmallsuperpositions} where we verified that our operators naturally respect linearity
 in their action on small superpositions.

This result is important because it shows that in the context of entanglement with pointers, and experiments of the kind considered above, the state-dependence of our operators is {\em completely transparent} to the infalling observer. Therefore, in no experiment, that can be described within effective field theory, does the observer detect a violation of linearity.
 
We conclude by remarking on a slightly subtle point. We have now described two situations  where there  is entanglement but no geometric wormhole between the CFT and the system that it is entangled  with. However, from the point of view of the microscopic operators, this is attained rather differently when the left  system is a CFT, and when it is just  a collection of qubits.  In the case where the left system is a CFT and the entanglement entropy is large, the right mirror operators commute with simple left operators but not with all operators on the left. On the other hand, in the case where the CFT  is entangled with a few qubits or with a system that does  not have $\Or[e^{\nc}]$ states, then we can indeed find mirrors entirely within the original CFT. As we saw above this  was important to ensure the  absence of  superluminal effects in such cases.

\subsection{Refining the notion of equilibrium for entangled states}
In some cases, the fact that our notion of ``equilibrium'' as time-independence of simple correlators is necessary but not sufficient --- as we discussed in section \ref{seccanonicalambiguity}  --- is also relevant to the discussion of entangled states. Consider the state 
\be
\label{tfdleftexcite}
\cop{M}(\al_{\alpha}) |\psient \rangle = e^{-{\beta \hcft \over 2}} \left(e^{i \al_{\alpha}}\right)^{\dagger} e^{\beta \hcft \over 2} |\psient\rangle.
\ee
In the thermofield state, correlation functions of this state are time-invariant on the right, but not on the left. This is because we have
\be
\label{tfddetectexcite}
\cop{M}(\al_{\alpha}) |\tfd \rangle = e^{i \alen[\alpha]} |\tfd \rangle.
\ee
Therefore, in this case, this lack of equilibrium can be detected by our left-equilibrium criterion.

On the other hand, in a generic entangled state there is no such relation between these states and left-excited states. Therefore, in such states the ambiguity from the single-sided case carries over. The reason we imposed the restriction that the left excitation in \eqref{psientone} be Hermitian was to prevent this ambiguity in descendants. Given the state in \eqref{psientone} we can dress it with a left unitary to obtain another valid descendant, which also appears to be in right equilibrium. With $\alen[1]^{U} = e^{i \alen[\alpha]} \alen[1]$, we could have considered
\be
|\psient^{1,U} \rangle =  (1 - \cop{P_{\text{en}}^0}) \alen[{1}]^U  |\psient \rangle,
\ee
in \eqref{hpsientone}.
However, when $\alen[\alpha]$ is entangled with a right operator, we want to ensure that we do not mistake $|\psient^{1,U} \rangle$ for an equilibrium descendant. However, the restriction that the left excitation be Hermitian excludes operators of the form $\alen[1]^U.$

As we explained in section \ref{seccanonicalambiguity}, even though all correlators on the right are left invariant under the excitation \eqref{tfdleftexcite}, it should still be possible to find ``measurables'' that can detect this excitation. Although we have not yet identified such measurables precisely, it is possible that the physical quantity that is capable of detecting the excitation in \eqref{orbitcan} in a single-sided CFT will also be able to detect the excitation \eqref{tfdleftexcite} in the two-sided case.


\section{Discussion}
In this paper we have presented strong evidence for the claim
that the black hole interior must be described using state-dependent bulk-boundary maps. We showed that a state-independent construction of the interior
was impossible, not only for single-sided AdS black holes, but even for
the eternal black hole. It is possible that this indicates that AdS/CFT
does not describe black hole interiors at all. However, this is in contradiction with many other calculations that suggest that the eternal black hole, at least, does have a smooth interior that can be probed by the CFT.

State-dependent bulk to boundary maps provide a solution to these versions
of the information paradox that preserves the predictions of effective field theory.  Our state-dependent construction of the black hole interior explicitly
identifies the duals of bulk local operators in the CFT. These bulk probes
do not see any sign of a pathology at the horizon, and so this should be 
taken as additional evidence that generic states do not correspond to firewalls.

In this paper, we demonstrated that our construction does not lead to any
violation of quantum mechanics or the ``Born rule.'' We also  
successfully resolved some of the ambiguities in our definition of an equilibrium state. 

Furthermore, we showed that our construction admitted a natural extension to entangled systems. This extension leads to a surprising bonus: a precise version of the ER=EPR conjecture emerges automatically from our construction without having to put anything in by hand. 

We have described our construction in significant detail and discussed how it works in equilibrium states --- which are generic at high energy. We have
also considered a large class of non-equilibrium states, including those that have been excited outside and inside the horizon. Although it is possible to 
consider other special classes of states in the CFT, we believe that our results 
provide persuasive evidence for the consistency of our construction.

There are several natural questions that arise from this analysis. It would be interesting to examine local operators outside the horizon in greater detail. Although we presented a state-independent description of such operators, in the mini-superspace approximation in section \ref{secstateindoutside}, the question of  whether state-dependence is also required outside the horizon is open. We will comment more on this in \cite{souvikprashant}. 

It would also be interesting to understand whether our construction can shed some light on the nature of the black hole singularity. So far we have used techniques from effective field theory to motivate the bulk to boundary map. Any investigation of the singularity will require new ideas.

Recent studies \cite{Shenker:2013pqa,Maldacena:2015waa} have shown that the naive ${1 \over \nc}$ expansion can often break down unexpectedly. We would like
to understand the implications of this breakdown for effective field theory on the nice slices and for the limitations of locality in quantum gravity.

Finally, as we have explained, while the use of state-dependent operators is perfectly consistent with quantum effective field theory, they are both unusual and interesting. It would be very useful to develop a more comprehensive measurement theory for these objects and understand whether they appear in other settings.


\section*{Acknowledgments}
We have discussed these ideas with a large number of people over the past year. We are particularly grateful to 
to all members of ICTS-TIFR, and the string theory groups at TIFR (Mumbai),  IISc (Bangalore), CERN, the University of Groningen and Harvard University.
We are  also grateful to the organizers and participants of the ``Bulk Microscopy from Holography and Quantum Information'' (PCTS, Princeton 2013), the summer workshop on ``Emergent Spacetime in String Theory'' (2014) at the Aspen Center for Physics (which was supported by NSF Grant No. PHYS-1066293 and the Simons Foundation), the Santa Barbara Gravity Workshop (2014), the CERN Winter School on Strings and Supergravity (2015), the QUC Autumn Symposium on String/M Theory (KIAS, Seoul, 2014), the IPM Winter School and Workshop (Tehran, 2014), the Asian Winter School (Puri, 2014), the discussion meeting on the Black Hole Information Paradox (HRI, Allahabad, 2014), the discussion meeting on Entanglement and Gravity (ICTS-TIFR, 2014), the Bangalore Area Discussion Meeting (ICTS-TIFR, 2015), the ``Workshop on holography, gauge theory and black holes'' (Southampton 2014), the ``Solvay Workshop on
Holography for Black Holes and Cosmology'' (Brussels 2014), the COST meeting ``String theory Universe'' (Mainz 2014), the ``Institut d'ete'' (ENS 2014), the
``Amsterdam Summer String Workshop'' (2014), the Nordic String Meeting 2015 (Groningen) and the Swiss String Meeting, GeNeZiSS 2015 (Bern). K.P. acknowledges the hospitality of the Institute for Advanced Study (Princeton) and of the Crete Center for Theoretical Physics. S.R. also acknowledges the hospitality of the Center for the Fundamental Laws of Nature at Harvard, Delhi University, and the Institute for Advanced Study (Princeton). S.R. is partially supported by a Ramanujan fellowship of the Department of Science and Technology (India). S.R. also acknowledges the partial support of the Center of Mathematical Sciences and Applications at Harvard University and the Cheng Yu-Tung Fund for Research at the Interface of Mathematics and Physics. K.P. would like to thank
the Royal Netherlands Academy of Sciences (KNAW).

\appendix

\section*{Appendix}

\section{State-dependence and semi-classical quantization }\label{appcoherentgravity}
In this appendix, we explore the semi-classical origins of state-dependence. Some of the ideas in this appendix were anticipated in \cite{luboscoherentblog}, although our analysis differs in some eventual details.  As we mentioned in  section \ref{secstatedepvsindep}, the belief that 
geometric quantities such as the metric should be represented by state-independent operators in the CFT is predicated on intuition
from geometric quantization. We elaborate on this intuition here. But we also explain why this intuition fails because of important ways in which the Hilbert space
of the CFT differs from what one might expect from a semi-classical linearized analysis of gravity. 

\subsection{Review of semi-classical quantization }\label{secsemireview}
We briefly remind the reader of the elementary concepts involved in quantizing the phase space of a system so as to make the classical
limit manifest. We will closely follow the excellent review by Yaffe \cite{Yaffe:1981vf}.

Before we proceed to the analysis for gravity, we briefly remind the reader 
of the elementary notions that are involved  in semi-classical quantization.  Consider a system with canonical variables $x_i, p_i$, with $i = 1 \ldots n$,  obeying the classical Poisson bracket relations $\{x_i, p_i \}_{\text{P.B.}} = 1$, and some classical functions on the phase space $f_m(\vect{x},\vect{p})$. We assume that all the first class constraints have been converted to second class constraints by gauge-fixing and that  all the second class constraints have been solved to eliminate the dependent variables. So the phase-space is unconstrained.

 Here we have denoted the coordinates on phase-space by two vectors $\vect{x}, \vect{p}$, with  $\vect{x} = (x_1, \ldots x_n)$ and $\vect{p} = (p_1, \ldots p_n)$. We also define $\vect{z} =  \left({1 \over \sqrt{2}} (x_1 + i p_1), \ldots{1\over \sqrt{2}} (x_n + i p_n)\right)$. Now, we want to show that in the quantum theory it is possible to find (a) an appropriate set of operators $\hat{f}_m$ and (b) a set of  semi-classical {\em coherent} states $|\vect{x}, \vect{p} \rangle$ in one to one correspondence with
the phase space so that, {\em when evaluated} on these states the operators $\hat{f}_m$ behave like the classical functions $f_m(x,p)$ as we discuss more precisely below. 

First, since we already have a simple and explicit description of the phase space and symplectic form in this setting, we quantize the system and define the canonical operators $\hat{x}_i, \hat{p}_i$ satisfying $[\hat{x}_i, \hat{p}_j] = i \delta_{i j}$.  This provides us with eigenstates of the operators $\hat{x}_i$ that satisfy $\hat{x}_i | \vect{x} \rangle = x_i |x_1, \ldots x_n \rangle$.   We also define $\hat{a}_i = {1 \over \sqrt{2}} (\hat{x}_i + i \hat{p}_i); \hat{a}_i^{\dagger} = {1 \over \sqrt{2}} (\hat{x}_i - i \hat{p}_i)$. 

With the vacuum $|\Omega \rangle$ defined as $a_i |\Omega \rangle = 0$, we consider the coherent states
\be
|\vect{z} \rangle =  e^{-{\sum_i |z_i|^2 \over 2}} e^{\sum_i a_i^{\dagger} z_i} |\Omega \rangle.
\ee
The wave-function of this state in the basis of eigenvectors of $\hat{x}_i$ can be calculated by noticing that $a_i |\vect{z} \rangle = z_i |\vect{z} \rangle$. With $\Psi_z(\vect{x_i}) = \langle \vect{x} | \vect{z} \rangle$, and using the fact that in the position eigenbasis $\hat{p}_i = -i {\partial \over \partial x_i}$, this turns into the differential equation
\be
\left(x_i + {\partial \over \partial x_i} \right) \Psi_{\vect{z}}(\vect{x}) = \left(z_{x i} + i z_{p i} \right) \Psi_{\vect{z}}(\vect{x}).
\ee
where we have written the components of $z_i$ as $z_i = z_{x i} + i z_{p i}$ to avoid confusion with the $x_i$ variable on the left. This is solved by the normalized position space wave-function for the coherent states. 
\be
\label{posspace}
\Psi_{\vect{z}}(\vect{x}) = \left({2 \over \pi}\right)^{n \over 4} \exp \Big\{-\sum_i \left[(x_i - z_{x i})^2 + i {z_{p i}} (x_i - z_{x i}) \right] \Big\}.
\ee
These states play the role of semi-classical states, and we can place them in a bijective correspondence with the phase space. 

These coherent states have several important properties. They are not orthonormal; in fact, it is important that they form an overcomplete basis
of the Hilbert space. We have
\be
\label{coherentinner}
\begin{split}
&\langle \vect{u} | \vect{z} \rangle = e^{-{|\vect{z}|^2 \over 2}} e^{-{|\vect{u}|^2 \over 2}} \langle \Omega | e^{\vect{a} \cdot \vect{\bar{u}}} e^{\vect{a}^{\dagger} \cdot  \vect{z}} |\Omega \rangle = e^{-{|\vect{z}|^2 \over 2} -{|\vect{u}|^2 \over 2} + {\vect{\bar{u}}} \cdot \vect{z} }, \\
&|\langle \vect{u} | \vect{z} \rangle|^2 = e^{-|\vect{z} - \vect{u}|^2}.
\end{split}
\ee
Nevertheless, we can partition the identity by using projectors onto these states. 
\be
\label{coherentcomplete}
1 = {1 \over (2 \pi)^n} \int d^2 \vect{z} P_{\vect{z}}; \quad P_{\vect{z}} = | \vect{z} \rangle \langle \vect{z} |.
\ee
This identity can be easily proved using, for example, the position space representation of the coherent states in \eqref{posspace}. 

Next, we need a way of lifting functions from the phase space to operators. Consider a function $f(\vect{z})$ on the phase space. (We have suppressed the dependence on $\bar{\vect{z}}$ simply to lighten the notation; we do not necessarily consider only holomorphic functions.) We now consider the operator defined by
\be
\label{projectorrep}
\hat{f} = \int f(\vect{z}) | \vect{z} \rangle \langle \vect{z} | {d^{2 n} \vect{z} \over (2 \pi)^n}.
\ee
This representation of operators is the so-called Sudarshan-Mehta P-representation \cite{sudarshan1963equivalence,mehta1967diagonal}. It differs from the more commonly used Weyl representation of operators, by operator ordering. The Weyl representation is sometimes favoured in the literature, since this map also allows one to represent the product of operators in the quantum theory by a Moyal star product of functions on the phase space.  However \eqref{projectorrep} yields more insight for our discussion, and has the same classical limit as the Weyl representation. 

Note that when this operator is inserted back into a coherent state we have
\be
\label{smearedresult}
\langle \vect{u} | \hat{f} | \vect{u} \rangle = \int f(\vect{z}) e^{-|\vect{z} - \vect{u}|^2} {d^{2 n} \vect{z} \over (2 \pi)^n}.
\ee
Therefore, the expectation value of the quantum operator is a {\em slightly smeared version} of the classical function. We have suppressed factors of $\hbar$ here, but if we consider classical functions that do not vary rapidly within a volume of $\hbar$ about a point in phase space, then the
expectation value of the  corresponding quantum operators faithfully reproduces their behaviour. 

Furthermore, if we consider the expectation value of the product of two operators then by using \eqref{coherentcomplete}
\be
\begin{split}
\langle \vect{y} | \,\hat{f} \,\hat{g} \,| \vect{y} \rangle  &= {1 \over (2 \pi)^2} \int f(\vect{z}) g(\vect{u}) \langle \vect{y} | \vect{u} \rangle \langle \vect{u} | \vect{z} \rangle \langle \vect{z} |  \vect{y} \rangle  d^2 \vect{z} d^2\vect{u} \\
&= {1 \over (2 \pi)^2} \int f(\vect{z}) g(\vect{u})  e^{-{|\vect{z}|^2} -{|\vect{u}|^2} - |\vect{y}|^2+ \vect{\bar{u}} \cdot \vect{z} + \vect{\bar{y}} \cdot \vect{u} + \vect{\bar{z}} \cdot \vect{y} }d^2 \vect{z} d^2\vect{u}.
\end{split}
\ee
We see that this integral is peaked around $z = u = y$ and expanding $g(\vect{u}) = g(\vect{y}) + (\vect{u} - \vect{y}) \cdot \partial_{\vect{y}} g(\vect{y}) + \ldots$, and similarly for $f$,  we see that the leading term is obtained by doing the Gaussian integral and we find
\be
\langle \vect{y} |\, \hat{f}\, \hat{g}\, |\vect{y} \rangle \approx f(\vect{y}) g(\vect{y}) .
\ee
On the other hand, we can also compute the commutator between two functions, in which case we need to keep the first subleading term to obtain a non-zero answer. Here, we find
\be
\langle \vect{y}  |  [\hat{f}, \hat{g}]\, | \vect{y}  \rangle  = i\,\{f, g\}_{\text{P.B.}} (\vect{y}).
\ee

\subsection{Geometrical quantities as classical functions on the phase space}
We now turn to the case of gravity where we first discuss the classical phase space and then describe coherent states in the linearized theory. In this subsection we are interested in establishing the following\\
{\bf Claim:} ``the metric $g_{\mu \nu}(\vect{x})$ is a well defined function on the classical phase space of gravity.''

The phase space of gravity is often discussed in canonical terms, where we specify the three-metric and the extrinsic curvature on a spacelike slice. This provides Cauchy data that we can evolve forward and backward in time. However, a covariant description of the phase space is given by considering the set of all classical solutions to gravity with asymptotic AdS boundary conditions \cite{dedecker1953cvf, goldschmidt1973hcf,Kijowski:1973gi,gawedzki1974cfl,szczyrba1976sss,garcia570rss,zuckerman1987apa,crnkovic1987cdc,Lee:1990nz}. The map between these two pictures is straightforward. 

Given a solution to the classical equations of motion, and a metric with a $d+1$ split,
\be
\label{threeonesplit}
d s^2 = -N^2 d t^2 + \gamma_{i j} \left(d x_i + N_i d t \right) \left(d x_j + N_j d t \right),
\ee
one may simply evaluate the fields at the spacelike slice $t = 0$. Then the variables
\be
\gamma_{i j}(\vect{x}, 0), \quad \pi^{i j}(\vect{x}, 0) = -\gamma^{1 \over 2} \left(K^{i j} - \gamma^{i j} K \right),
\ee
provide the standard parameterization of gravitational phase space. 
Here $K$ is the extrinsic curvature
\be
\label{extrinsiccurv}
K_{i j} = {1 \over 2} N^{-1} \left(\partial_j N_i + \partial_i N_j - \partial_t \gamma_{i j} \right),
\ee
and for the purposes of this $d+1$ split we have displayed the time coordinate separately in $(\vect{x}, t)$. 

Conversely, given the variables $\gamma_{i j}(\vect{x}, 0)$ and $\pi^{i j}(\vect{x},0)$, one may use the equations of motion to evolve them forward in time and generate the entire metric in the form \eqref{threeonesplit}. Of course, such a solution requires a choice of gauge, as we have already discussed. 

It is also possible to write down a symplectic form on the phase space described covariantly as the set of classical solutions, and this was done by \cite{crnkovic1987cdc}. 

For us the important point is that each point on the phase space corresponds to an entire spacetime. Now, evidently given the entire spacetime, classically, we may ask any question we wish; even one that involves global notions like an event horizon. For example, we may set up relational coordinates as in section \ref{subsecrelational} and just evaluate the metric at a point $g_{\mu \nu}(\vect{x}, t)$. The same is true of other propagating light fields in the theory. 

Therefore, all of these observables are well defined classical functions on the phase space. This is an important point.  We now extend the discussion above to gravity to show that, explicitly,  within the linearized theory, we may indeed expect such questions to be answered by state-independent operators. 

\subsection{Coherent states in linearized gravity}
We now turn to an analysis of gravity. Here we are interested in establishing the following. \\
\begin{quote}
{\bf Claim:} If we consider two nearby points in the gravitational phase space with metrics $g^{\text{b}}_{\mu \nu}(\vect{x})$ and $g^{\text{e}}_{\mu \nu}(\vect{x})$ then one can define a covariant inner product on the corresponding coherent states in the Hilbert space which behaves like $e^{-\nc \distphase(g^{\text{b}}, g^{\text{e}})}$ where we can compute the function $\distphase$ in the linearized approximation. 
\end{quote}

First we remind the reader how the discussion of \eqref{secsemireview} generalizes to linearized gravity. We are only able to work in the linearized setting, and although it would be interesting to explore this construction further in a  fully non-linear setting, we do not know how to do this.

We consider fluctuations of the metric, about a background metric, defined by
\be
g_{\mu \nu} = g_{\mu \nu}^{\text{b}} + \sqrt{8 \pi G_N} h_{\mu \nu},
\ee
and the normalization is chosen so that the kinetic term of $h_{\mu \nu}$ is canonically normalized. Here $g_{\mu \nu}^{\text{b}}$ may be 
any background metric, that is a solution of the equations of motion and is asymptotically AdS. We do not take it to be necessarily the AdS-Schwarzschild solution. 

Now, on general grounds, we expect that solutions to the classical equations of motion will be given  by
\be
h_{\mu \nu}(\vect{x}) = \sum_{i,\omega} a^{i}_{\omega} g^{(i)}_{\mu \nu}(\omega,  \vect{x}) + {\rm h.c},
\ee
where $i$ runs over the different ${(d+1)(d-2) \over 2}$ possible polarizations of the graviton, where $d$ is the boundary dimension and $a^i_{\omega}$ are just linear coefficients at the moment. The 
different eigenfunctions are denoted by $\omega$. In empty AdS or AdS Schwarzschild, for example, this would constitute a set of integers to pick
out the spherical harmonic on the $S^{d-1}$ and a ``radial momentum''. We will not require the detailed form of these eigenfunctions, or even
of their eigenvalues.  We are not assuming that there is a timelike isometry in the space, and so, in principle, $\omega$ may not correspond intuitively to a ``frequency.'' 

We also assume that we have picked a basis set of distinct solutions $g^{(i)}_{\mu \nu}$, which are not equivalent
under gauge transformations, and we normalize the functions $g^{(i)}_{\mu \nu} (\omega, \vect{x})$ so that the canonical Poisson brackets translate into the statement
\be
\label{normalizeigen}
\{a^{i}_{\omega}, a^{j,\dagger}_{\omega'} \}_{\text{P.B}} = -i \delta^{i j} \delta_{\omega, \omega'}.
\ee

We quantize the theory  and obtain a vacuum state $a^{i}_{\omega} |\Omega \rangle = 0$. Note  that now $a^{i}_{\omega}$ is an operator on the Hilbert space of the linearized theory.  We then define coherent states by labelling them with a set of functions $\chi^i(\omega)$. Starting with the vacuum, 
\be
|\chi \rangle \equiv {\cal N}_{\chi} e^{\sum_{i,\omega} a^{i,\dagger}_{\omega} \chi_{\omega}^i}  | \Omega \rangle.
\ee
where ${\cal N}_{\chi}$ is a normalization factor. We see that
\be
\langle \chi | \chi \rangle = |{\cal N}_{\chi}|^2 e^{\sum_{i,\omega} |\chi_{\omega}^i|^2}.
\ee
So for the state to be normalized, we should set
\be
\label{distance}
{\cal N}_{\chi} = e^{-{1 \over 2}\sum_{i,\omega}  |\chi_{\omega}^i|^2}.
\ee
Note that $|\chi_{\omega}^i|^2$ can also be interpreted as the ``occupation number'' in the mode $\omega$; so the exponent in the normalization factor is just the total occupation number in the state.

One measure of how large the deviation of the field is from the background metric is given by
\be
\label{diffvac}
\langle \Omega| \chi \rangle = {\cal N}_{\chi}.
\ee
Here the vacuum is just the original background metric. So we see that this coherent state is substantially different from the original background metric, as a quantum state, if the occupation number is large. 
In this state the metric  has an expectation value
\be
\label{gegb}
\begin{split}
g^{\text{e}}_{\mu \nu}=\langle \chi | g_{\mu \nu}(\vect{x}) | \chi \rangle &= \langle \chi | g^{\text{b}}_{\mu \nu}(\vect{x}) + \sqrt{8 \pi G_N} h_{\mu \nu}(\vect{x}) | \chi \rangle  \\ 
&= g^{\text{b}}_{\mu \nu}(\vect{x})  + \sqrt{8 \pi G_N} \sum_{i,\omega} \left( \chi^i_{\omega} g^{(i)}_{\mu \nu}(\omega, \vect{x}) + \text{h.c} \right).
\end{split}
\ee
So we see that the space $|\chi \rangle$ represents a nearby point in phase space, where the value of the metric has changed to $g^{\text{e}}_{\mu \nu}(\vect{x})$.  Therefore \eqref{distance} shows how the corresponding inner-product in Hilbert space varies. 

Now, in deriving \eqref{distance} we made explicit reference to a set of mode functions. But we would like it to depend only on the two metrics $g^{\text{e}}_{\mu \nu}(\vect{x})$ and $g^{\text{b}}_{\mu \nu}(\vect{x})$. To check that this is covariant, let us consider how this changes under a Bogoliubov transformation of the modes.  We make a canonical transformation of the  $a_{\omega}^i$ variables to 
\be
\label{bogoliubovtransform}
\begin{split}
b_{\omega}^i = \sum_{\omega'} \left(\beta_{\omega \omega'} a_{\omega'}^i + \gamma_{\omega, \omega'} a_{\omega'}^{\dagger, i} \right), \\
b_{\omega}^{\dagger,i} = \sum_{\omega'} \left(\beta^*_{\omega \omega'} a_{\omega'}^{\dagger, i} + \gamma_{\omega, \omega'}^* a_{\omega'}^i \right).
\end{split}
\ee
In this analysis, we will assume that the polarization index $i$ does not enter the Bogoliubov coefficients. This is just to lighten the notation and does not represent any loss of generality.

For the new modes to have the canonical commutators
\be
[b^i_{\omega}, b_{\omega'}^{\dagger,i}] = \delta_{\omega, \omega'} ,
\ee
we see that we must have
\be
\label{bogoliubovcomplete}
\sum_{\omega''} \left( \beta_{\omega, \omega''} \beta^*_{\omega', \omega''} - \gamma_{\omega, \omega''} \gamma^*_{\omega', \omega''} \right) = \delta_{\omega, \omega'}.
\ee
An observer using these creation and annihilation operators would also use
a new basis of modes to represent the metric fluctuations that we call $\tilde{g}^{(i)}(\omega, \vect{x})$. In particular,
we have
\be
\label{newmodes}
\begin{split}
&\sum_{\omega} \beta_{\omega \omega'} \tilde{g}^{(i)}(\omega, \vect{x}) + \gamma^*_{\omega, \omega'} (\tilde{g}^{(i)}(\omega, \vect{x}))^* = g^{(i)}(\omega', \vect{x}), \\
&\sum_{\omega} \beta_{\omega \omega'}^* (\tilde{g}^{(i)}(\omega, \vect{x}))^* + \gamma_{\omega, \omega'} (\tilde{g}^{(i)}(\omega, \vect{x})) = (g^{(i)}(\omega', \vect{x}))^*.
\end{split}
\ee

Such an observer would set up a different set of coherent states
\be
|\tilde{\chi} \rangle_{\text{Bog}} = e^{\tilde{\chi}^i_{\omega} b_{\omega}^{\dagger,i}} |\Omega \rangle_{\text{Bog}},
\ee
where the vacuum  is now defined to satisfy $b^i_{\omega} |\Omega \rangle_{\text{Bog}} = 0$.
To get the same expectation value for the metric field, this observer could use a coherent state excitation with parameters $\tilde{\chi}^i_{\omega}$ so that
\be
\sum_{\omega} \tilde{\chi}^i_{\omega} \tilde{g}^i(\omega, \vect{x}) + (\tilde{\chi}^i_{\omega})^* (\tilde{g}^i(\omega, \vect{x}))^* = \sum_{\omega'} \chi^i_{\omega'} g^{(i)}(\omega',\vect{x}) + (\chi^i_{\omega})^* g^{(i)}(\omega',\vect{x}).
\ee
Using \eqref{newmodes}, we see that we need
\be
\tilde{\chi}^i_{\omega} = \sum_{\omega'} \left(\beta_{\omega \omega'} \chi_{\omega'}^i + \gamma_{\omega, \omega'} (\chi_{\omega'}^{i})^* \right).
\ee
Therefore we see that
\be
\label{tildesum}
\begin{split}
\sum_{i,\omega} |\tilde{\chi}^i_{\omega}|^2  = \sum_{i,\omega, \omega',\omega''}\Big[&\beta_{\omega \omega'} \beta^*_{\omega \omega''} \chi_{\omega'}^i (\chi_{\omega''}^i)^* + \gamma_{\omega, \omega'} \gamma^*_{\omega, \omega''} \chi_{\omega''}^{i} (\chi_{\omega'}^{i})^* 
\\
&+ \beta_{\omega \omega'} \gamma_{\omega \omega''} \chi_{\omega}^i \chi_{\omega''}^i + \beta_{\omega \omega'}^* \gamma_{\omega \omega''}^* (\chi_{\omega'}^i)^* (\chi_{\omega''}^i)^* 
\Big].
\end{split}
\ee
For a general Bogoliubov transformation therefore 
\be
\label{tildenormdiff}
\sum_{i,\omega} |\tilde{\chi}^i_{\omega}|^2 =  \sum_{i, \omega} |\chi^i_{\omega}|^2 + R,
\ee
where the remainder $R$ does not vanish.

However, in AdS/CFT we have an additional advantage: the presence of the boundary Hamiltonian. So we can define positive and negative energy with respect to the boundary Hamiltonian and demand that in terms of boundary energy eigenstates, both the sets of creation operators have strictly positive energy and the annihilation operators have negative energy.\footnote{Here, we are not concerned with the small tails that we discussed in the text, which may appear in these relations because we restrict observations to a finite time on the boundary.}
\be
\label{adscftconstraint}
\begin{split}
P_{E^+} a_{\omega}^i |E \rangle = 0, \quad P_{E^{+}} b_{\omega}^i | E \rangle = 0, \\
P_{E^{-}} a_{\omega}^{i,\dagger} | E \rangle = 0, \quad P_{E^{-}} b_{\omega}^{i,\dagger} | E \rangle = 0,
\end{split}
\ee
where $P_{E^+}$ ($P_{E^-}$) indicates the projector on the subspace formed by eigenstates with energy larger (smaller) than $E$.
If we restrict to such operators then we see that $\gamma_{\omega \omega'}$ in \eqref{bogoliubovtransform} must vanish. From \eqref{bogoliubovcomplete}, we then find that $\beta_{\omega \omega'}$ must be {\em unitary}. For this set of transformations, which obeys the natural AdS/CFT constraint \eqref{adscftconstraint}, we
see from \eqref{tildesum} that $R = 0$ in \eqref{tildenormdiff}.

To summarize, the conclusion is that using the AdS/CFT Hamiltonian to define positive energy, the notion of the distance of a coherent excitation from the background is robust in linearized gravity.

Now, let us examine this ``distance'' a little more closely.
Let us write the initial metric in a ``nice'' coordinate system so that all its components are of order the AdS radius squared $\ell^2$. In this case, we see that to make a substantial perturbation, we must take $h_{\mu \nu} \sim {\alpha \ell^2 \over \sqrt{8 \pi G_N}} = \alpha \nc$, where $\alpha$ is an $\Or[1]$ parameter that we have introduced.   At this point, the linearized theory is still valid if we keep $\alpha \ll 1$. If we apply  \eqref{diffvac} to such a perturbation, we see that
the coherent state construction predicts the following. The semi-classical states in the quantum theory, corresponding to two distinct 
solutions $g_{\mu \nu}^{e}(\vect{x})$ and $g_{\mu \nu}^{\text{b}}(\vect{x})$ are almost orthogonal and have an inner product
\be
\label{distancemetrics}
\langle g_{\mu \nu}^{\text{e}}(\vect{x}) | g_{\mu \nu}^{\text{b}}(\vect{x}) \rangle = e^{-\nc \distphase(g^{\text{e}}, g^{\text{b}})},
\ee
where $\distphase$ is a smooth $\Or[1]$ functional on the space of metrics. To compute this function, we write $g^{\text{e}}_{\mu \nu}(\vect{x})$ as an excitation over $g^{\text{b}}_{\mu \nu}(\vect{x})$ using \eqref{gegb} and compute the inner product given in (\ref{distance},\ref{diffvac}). The choice of mode functions that we use to express the excited state in terms of the background is unimportant by the argument above.

\subsubsection{Difficulties with state-independent operators}
Now the formula for the inner product \eqref{distancemetrics} above might seem encouraging. It may suggest the following naive program. In the full theory of quantum gravity, we identify points on the phase space with coherent states $|g \rangle$, write down a completeness relation analogous to \eqref{coherentcomplete} and then write a full state-independent metric operator as in \eqref{semiclasstateindg}: $\cop{g}_{\mu \nu}(\vect{x}) = \sum_g g_{\mu \nu}(\vect{x}) | g \rangle \langle g|$. This is the  basis for the expectation that we can find state-independent operators to represent the metric and other bulk fields.

However, recall that \eqref{coherentcomplete} was consistent only because the inner product \eqref{coherentinner} died off to arbitrarily small values to compensate for the infinite volume of phase space. It appears that this does not happen for the case of gravity: rather, intuition from the CFT suggests that in some cases the inner-product between different coherent states may saturate at a small but finite value even when the corresponding volume in classical phase space is very large.

We have seen an example of this in the case of the thermofield double. There the states $|\tfdT \rangle$ all represented metrically distinct geometries. If we identify these states with points on the phase space, then the parameter $\capt$ parameterizes an infinite direction in the classical phase space. However, even if we take $\capt$ to be large, the inner product saturates at  $\langle \tfd | \tfdT \rangle = \Or[e^{-{S \over 2}}]$  where $S$ is the entropy.

This suggests that the classical limit in AdS/CFT emerges somewhat differently than the intuition from canonical gravity would suggest.  Specifically, the following phenomenon occurs.  We can identify states in the CFT dual to metrics $|\psidual{g} \rangle \leftrightarrow |g \rangle$. However, when the distance between these states becomes ``large'', the inner-product in the CFT differs from the inner product predicted by semi-classical gravity. We have only been able to compute this semi-classical inner product reliably for small separations on the phase space. If we extrapolate this to the entire phase space then we can find cases where the semi-classical inner product is exponentially different from the CFT inner product.
\be
\label{badratio}
 {e^{-\nc \distphase(g^{\text{e}}, g^{\text{b}})}  \over |\langle \psidual{g^{\text{e}}} | \psidual{g^{\text{b}}} \rangle|}  = \Or[e^{-\nc}].
\ee

Returning to the example of the thermofield double, which is the source of our intuition, we note that the formula \eqref{statedependental} is precisely analogous to \eqref{projectorrep}. In both cases we know the action of an operator on a set of states that almost orthogonal to one another. However, while in \eqref{projectorrep} we are able to extend the integral to all of phase space and thereby obtain a state-independent operator, we cannot extend the limits on $\capt$ in \eqref{statedependental} to $\pm \infty$ because of the saturation of the inner product. 

Another manifestation of this obstacle is as follows. In the thermofield double, given a sequence $e^{S}$ states shifted by $\{\capt_1 \ldots \capt_{e^{S}} \}$, so that all of them are pairwise distinct,  we can still find coefficients $\coeff[i]$ so that
\be
\label{tfdaslinear}
\left| |\tfd \rangle - \sum_{i=1}^{e^{S}} \coeff[i] e^{i \hleft \capt_i} |\tfd \rangle \right|^2 = \Or[e^{-\nc}].
\ee
Note that \eqref{tfdaslinear} is {\em not} due to Poincare recurrence, which occurs after a much longer time scale $e^{e^S}$. The linear dependence
indicated in \eqref{tfdaslinear}  means that one geometry can be written as a linear combination of $e^{S}$ completely different geometries. The semi-classical theory
does not see any signs of \eqref{tfdaslinear}. This prevents a naive use of projectors on coherent states to build up a state-independent operator.

\begin{figure}[!h]
\begin{center}
\resizebox{0.4\textwidth}{!}{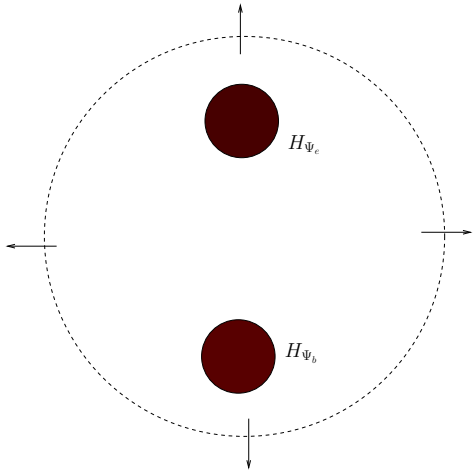}
\caption{When we quantize the theory we can put states in the Hilbert space in correspondence with the classical phase space. However, we may have to use different operators in different regions of phase space to represent a single classical function. }\label{figphaseschematic}
\end{center}
\end{figure}

\paragraph{Summary \\}
The picture that we get in this manner is shown in Fig \ref{figphaseschematic}.
  A slogan that would summarize this Appendix is that ``coherent states are always overcomplete, but the states in the CFT that correspond to coherent states of the metric are even more overcomplete than one would expect from a semi-classical analysis.'' This is what prevents us from lifting some well defined classical observables to state-independent operators. This issue is important and interesting and deserves further investigation.


\section{Mirror modes from bulk evolution}
One possible proposal to define the mirror operators may proceed as follows. Consider black holes formed by collapse in AdS. In each such classical solution,
we can trace the right moving modes behind the horizon to their origin to their support on the boundary of AdS in the past. This is what was done by Hawking in flat space \cite{Hawking:1974sw} using a geometric optics approximation.

\begin{figure}[!h]
\begin{center}
\includegraphics[height=0.3\textheight]{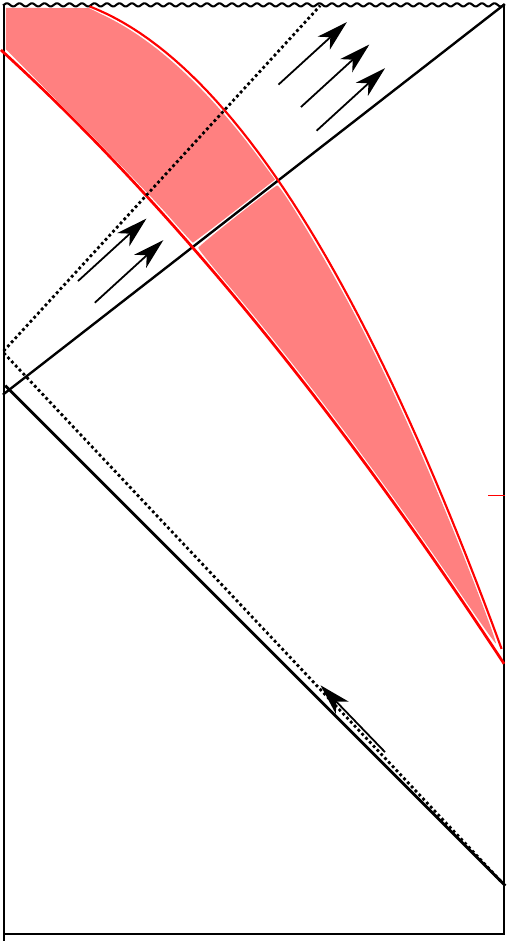}
\caption{Tracing the mirrors back to their origin on the boundary is difficult because of the trans-Planckian problem. However, even neglecting this issue does not help in constructing state-independent operators because of the ``fat tail'' in the inner product of different solutions. }\label{raytracing}
\end{center}
\end{figure}

Hawking's computation suffers from a trans-Planckian problem because the geometric optics calculation tells us that, at late times,  even low frequency right moving modes behind the horizon come from an extremely small time-band on the boundary. (See Figure \ref{raytracing}.) Therefore, in the past these low frequency modes must have had ultra-Planckian frequencies. 

Even if we ignore this issue and proceed with the naive calculation, we find that we can only attain a small number of microstates by considering black holes
formed from collapse. Page and Phillips estimated the number of possible configurations of massless radiation inside anti-de Sitter space \cite{page1985self}. Their calculation can be summarized as follows. Consider a gas of radiation in AdS$_{d+1}$ and, as usual, we set its radius to $1$. Then, Page and Phillips considered a self-gravitating gas of radiation assuming that it was locally in thermal equilibrium at all points. Their conclusion was that one recovers the standard
thermodynamic relation between the entropy and the energy at high energies for a gas in $d+1$ dimensions
\be
\label{entenrad}
S_{\text{rad}} =  \kappa_{\text{rad}} E^{{d \over d  + 1}},
\ee
where $\kappa_{\text{rad}}$ is an $\Or[1]$ constant which depends on the number of light degrees of freedom in the theory. 
On the other hand for high energies $E \gg \nc$, we know that the entropy of the black hole is given by
\be
\label{entenbh}
S_{\text{bh}} = \nc \kappa_{\text{bh}} \left({E \over \nc} \right)^{d-1 \over d},
\ee
which is the result for a gas with $\nc$ degrees of freedom in $d$-dimensions. We remind the reader that $\nc$ is the central charge, and so $\nc = N^2$ in the SU(N) supersymmetric Yang-Mills theory. 

Comparing \eqref{entenbh} with \eqref{entenrad} for energies of order $E \propto \nc$, we find that
\be
{S_{\text{bh}} \over S_{\text{rad}}} = {\kappa_{\text{bh}} \over \kappa_{\text{rad}} } \nc^{1 \over d}  E^{-1 \over d (d +1)} \propto \nc^{1 \over d+1}.
\ee
Therefore the entropy of the radiation is always subleading in this range. 

We caution the reader that \eqref{entenrad} is a little artificial in the regime in which we have applied it because the temperature that follows from \eqref{entenrad} is 
\be
T_{\text{rad}} = {1 \over \left({\partial S_{\text{rad}} \over \partial E} \right)} = \kappa_{\text{rad}}^{-1} E^{1 \over d+1}.
\ee
If we consider the case of the duality between AdS$_5$ and supersymmetric
Yang-Mills theory, with a 't Hooft coupling $\lambda$, then we do not expect the result \eqref{entenrad} to be valid beyond the string scale $\lambda^{1 \over 4}$,  at which point we expect to find a Hagedorn transition in the bulk.  So, in reality we do not even expect to be able to attain as many microstates as we considered above for the radiating star.

This is a rather robust result: following the collapse of black holes from reasonable geometric configurations allows us to explore only a small fraction of the Hilbert space at high energies. Now if we do decide to restrict to such a sector of the Hilbert space, the firewall paradoxes vanish since they can only make reference to generic states. Correspondingly, there is no difficulty in obtaining state-independent mirror operators that have the correct behaviour on this sector. 

We now note a second important point. In some cases, it may be possible to geometrize the microstates of the black hole as we did in section \ref{seceternal}. There, we were able to explore a significant fraction of the microstates of the eternal black hole classically by considering a one-parameter family of eternal black hole solutions. All of these were glued to the boundary with different time shifts, and we had to allow this time-shift to be exponentially large to ensure that the corresponding states in the CFT Hilbert space spanned a subspace of exponentially large dimension. 

However, in this situation we ran into the obstruction explored in section \eqref{secexpliciteternal} and also in Appendix \eqref{appcoherentgravity}. This obstacle is as follows. Any method of obtaining the mirror modes by analyzing classical solutions can, at most, specify these modes as functions on the classical phase space. For example in \eqref{secexpliciteternal}, in each solution left-shifted by the time $T$, the mirrors were the modes of $\Oleft[t+T,\Omega]$. However, in this situation we encountered the ``fat tail'' of \eqref{fattail}. This ``fat tail'' prevents us from lifting a classical function on this large phase space to a corresponding linear operator in the Hilbert space. 

Therefore, the study of classical solutions cannot help in obtaining state-independent mirror operators.


\bibliographystyle{JHEPmod}
\bibliography{references}

\end{document}